\documentclass[a4paper,fleqn,usenatbib,onecolumn]{mnras}

\usepackage[T1]{fontenc}
\usepackage{ae,aecompl}
\usepackage{longtable,dcolumn}

\usepackage{graphicx,booktabs} 
\usepackage{amsmath}	
\usepackage{amssymb}	

\title[Short title, max. 45 characters]{Empirical velocity profiles for galactic rotation curves}

\author[E. L\'{o}pez Fune et al.]{
E. L\'{o}pez Fune,$^{1}$ \thanks{E-mail: elpezfun@lpnhe.in2p3.fr}\\
\\
$^{1}$ LPNHE, 4 Place Jussieu, Tour 22, 1er \'{e}tage, 75005 Paris, France.\\
}

\date{Accepted XXX. Received YYY; in original form ZZZ}

\pubyear{2015}

\begin{document}
\label{firstpage}
\pagerange{\pageref{firstpage}--\pageref{lastpage}}
\maketitle

\begin{abstract}
A unified parametrization of the circular velocity, which accurately fits 850 galaxy rotation curves without needing in advance the knowledge of the luminous matter components, nor a fixed dark matter halo model, is proposed. A notable feature is that the associated gravitational potential increases with the distance from the galaxy center, giving rise to a length scale indicating a finite size of a galaxy, and after,  {the Keplerian fall-off of the parametrized circular velocity is recovered according to Newtonian gravity}, making possible the estimation of the total mass enclosed by the galaxy.
\end{abstract}

\begin{keywords}
dark matter -- rotation curves -- analytical
\end{keywords}



\section{Introduction}\label{Sec.0.0.0}

In the innermost regions of the galaxy rotation curves (hereafter RCs), the luminous matter contribution dominates the
dynamics as light traces the mass inferred from disk rotation \citep{Athanassoula:1987hp, Persic:1990uj, Palunas:2000qj} out to a radius ranging between 1 and 3 disk exponential length scales (depending upon the galaxy luminosity \citep{Salucci:1999qu}). Far distant from the galaxy centres, the circular velocities of stars and gas are found to be constant or even increasing with radius despite the sharp radial decrease of the stellar and gaseous surface brightnesses as the main effect of the dark matter (hereafter DM) halo hosting the galaxy. Both, the luminous and DM contributions to the RCs, can explain the observed flat circular velocities.

The RCs of spiral galaxies are one the main tracers of their mass distributions, as they represent observables of galaxy
formation and provide relevant clues to unveil the nature of the DM \citep{Bosma:1978phd, Bosma:1979aa, Rubin:1980zd}. A
detailed knowledge of the galaxy morphology and a high spatial resolution RC is clearly essential for numerical simulations on
galactic structure formation \citep{Cen:1992us, Navarro:1994zk, Evrard:1994ct}, as well as for testing
new DM models.

The dynamics of galaxy RCs considering a surrounding DM halo can be studied in the framework of the standard mass
modelling or with the local density method \citep{Salucci:2010qr}, which are widely spread in the literature. The standard mass
modelling consists in fitting directly the experimental RCs using a fixed stellar and gaseous disk model, such as a Freeman
disk, and global halo mass model such as the Navarro-Frenkel-White density profile (hereafter NFW) \citep{Navarro:1995iw, Navarro:1996gj} or the Burkert halo density profile (hereafter BRK) \citep{Burkert:1995yz} respectively, knowing in advance the stellar and
gaseous mass distributions from the galaxy's photometry. This method aims to globally fit the RCs with the total velocity
given in quadrature by $V^{2}(r)=V_{h}^{2}(r)+V_{D}^{2}(r)$, where $V(r)$ is the total circular velocity at a distance $r$ from the galaxy centre, $V_{h}(r)$ is the DM circular velocity contribution extracted from a specific DM halo density profile and $V_{D}^{2}(r)=V_{s}^{2}(r)+V_{g}^{2}(r),$ the disk velocity, with $V_{s}(r), V_{g}(r)$ the stellar disk and gaseous velocity components. The radial dependence of $V_{s}(r)$ and $V_{g}(r)$ is derived from Newtonian gravity equations, using the corresponding surface mass densities, which are obtained from the photometric observations.

In the local density method, the RC and the DM mass model are analysed only in regions where the luminous matter contribution is not relevant compared to the DM effects \citep{Salucci:2010qr}. This method, discussed in details in \citep{Salucci:2010qr}, has been applied first to estimate the local DM density in the Milky Way at the Sun's location; later, in \citep{Karukes:2015fma} to the spiral galaxy NGC 3198, and more recently in \citep{Fune:2016uvn} to determine the DM halo properties of the spiral galaxy M33. It allows to derive very precisely the DM density in the outskirts of a galaxy provided that the stellar and gas surface densities are well known and their circular velocities and their first derivatives are accurately determined, so that $\delta V/V < 0.05,$ $\delta d \log V/d \log r < 0.1.$ The idea is to resort the equation of centrifugal equilibrium, which holds the spiral galaxies:
\begin{align}\label{Eq.12}
\dfrac{V^{2}(r)}{r}=a_{h}(r)+a_{s}(r)+a_{g}(r),
\end{align}
\noindent where $a_{h}(r), a_{s}(r),$ and $a_{g}(r)$ are the radial accelerations, generated by the DM halo, stellar and gaseous disks respectively.

Under the approximation of a spherical DM halo, the densities are
\begin{align}
\rho_{h}(r)&=\rho(r)-\rho_{s}(r)-\rho_{g}(r)=\dfrac{X_{q}}{4\pi G r^{2}}\dfrac{d}{dr}\left(rV^{2}-rV^{2}_{s}-rV^{2}_{g}\right),\label{Eq.13}\\
\rho(r)    &=\dfrac{1}{4\pi G r^{2}}\dfrac{d}{dr}\left(rV^{2}\right),\label{Eq.14}\\
\rho_{s}(r)&=\dfrac{1}{4\pi G r^{2}}\dfrac{d}{dr}\left(rV^{2}_{s}\right),\label{Eq.15}\\
\rho_{g}(r)&=\dfrac{1}{4\pi G r^{2}}\dfrac{d}{dr}\left(rV^{2}_{g}\right),\label{Eq.16}
\end{align}
\noindent where $X_{q}$ is a factor correcting the spherical Gauss law used above in case of an oblate DM halo and it takes values between
1.05 and 1.00 (see details in \citet{Salucci:2010qr}), $V(r)$ is the velocity given by the RC, $V_{s}(r)$ and $V_{g}(r)$ are the stellar and
gas velocity contributions. The strength of this method lies in the fact that one has transformed the surface mass density of the stellar and gaseous disks in effective bulk densities with the aid of the spherical Gauss's law. Since the velocities induced by the stellar and gaseous disks decrease as $r^{-1/2}$ after certain length scale $r_{0}$, one expects a sharp decrease for their effective densities $\rho_{s}(r)$ and $\rho_{g}(r),$ and one is left only with the DM contribution to the observed RC. From this fact, one can infer the DM halo properties directly from the experimental data and not relaying on the luminous matter contributions.

A very important issue left to treat in the local density method is to derive an analytical expression for the total circular velocity $V(r)$ from the available discreet set of points from the experimental RC. An analytic expression for $V(r)$, will make easier the numerical computation of the total density as given by Eq.\eqref{Eq.14}. In order to alleviate this issue, in \citep{Fune:2016uvn} was proposed an empirical velocity profile to fit the total RC of the spiral galaxy M33, for the later use of the local density method. This empirical velocity profile has three free parameters and it is the simplest velocity profile that reproduces an asymptotically flat RC at large galactocentric distances:

\begin{align}
V(r)=V_{0}\dfrac{r/r_{c}+d}{r/r_{c}+1}\label{Eq.3.2.1},
\end{align}
\noindent where $V_{0}$ is the constant circular velocity of the outskirts of the galaxy, $r_{c}$ is a radial length-scale and $d$ is a dimensionless parameter different from 1 in order to prevent $V(r)$ to be constant. Moreover, in the innermost parts of the galaxy, the circular velocity increases, which means that the parameter $d$ is constrained to the range $0\leq d<1.$ To compute the corresponding
density, it was assumed by extrapolation, that the extension of the experimental RC between any two data points was given by a point $(r, V(r)).$

\bigskip

The empirical velocity profile given by Eq.\eqref{Eq.3.2.1} has a limited use since it doesn't show the usual bumps of the bulge-to-disk
transition in the innermost regions of the galaxy RCs. Despite this issue, its application in the local density method is valid since one only
needs to know the velocity in the distant regions of the galaxy centres, free from the influences of the luminous matter. The motivation in this article is to introduce a more generalized model-independent empirical velocity profile that overcomes this issue and could help to express the total local density from the experimental RC, so that one can study in a deeper way the nature of the DM that surrounds galaxies.

This article is organized in three sections. In Section \eqref{Sec.1.0} are discussed the main properties of the introduced circular
velocity profile, the gravitational potential that induces such a circular velocity and the total local density. In Section \eqref{Sec.2.0} are
presented the results from the fittings of 850 galaxy RCs with the corresponding discussions on the values of the best fitting
parameters and their possible correlations. In Section \eqref{Sec.3.0} are presented the conclusions and acknowledgements of this article.

\section{Empirical velocity profile} \label{Sec.1.0}

There are many situations in which one needs a simple fitting formula for a RC between the initial and the final experimental points such that by extrapolation, it converges to the Keplerian circular velocity at very large distances. In order to do so, a new empirical formula for the circular velocity is proposed:
\begin{align}
V_{n}(r)&=\dfrac{V_{0}d^{n}(r/r_{c})^{\frac{3}{2}}(1+r/r_{c})^{-(n+\frac{1}{2})}}{\sqrt{1+(r/r_{c})^{2}}}\sum_{i=0}^{n}\left(\dfrac{r/r_{c}}{d}\right)^{i}\label{Eq.2.2},
\end{align}
\noindent where as in Eq.\eqref{Eq.3.2.1}, $V_{0}$ is the constant circular velocity of the outskirts of the galaxy, $r_{c}$ is a radial length-scale, $d$ is a dimensionless parameter and an additional free parameter $n\geq1$ has been introduced, which is a positive integer number. This empirical velocity profile enlarges the number of RCs that one can analyse with respect to the main approach, that is: the mass modelling method, and although the functional fit might not have a direct physical meaning, mathematically it carries the properties of the RCs.

\bigskip

Similarly, as in Eq.\eqref{Eq.3.2.1}, at large distances $r\gg r_{c}$ the velocity profile approaches asymptotically the constant value $V_{0},$ independent of $r_{c}, d$ and $n,$ and reproduces a flat RC. At small distances $r\ll r_{c}$ the behaviour is $V_{n}(r\ll r_{c})\sim V_{0}d^{n}(r/r_{c})^{\frac{3}{2}},$ which is always increasing and the RC starts at $r=0$ regardless the values of $d$ and $n.$  {It worth also to notice that as in Eq.\eqref{Eq.3.2.1} was needed $0\leq d<1$ to reproduce an increasing RC near $r=0$, for Eq.\eqref{Eq.2.2}, the parameter $d$ can take any real positive value since this empirical velocity profile always increases near the galaxy centres.}


\subsection{Gravitational potential}\label{Sec.1.1}

The introduced circular velocity profile aims to fit globally any galactic RC, which means that it already contains the information from the total luminous and DM contributions. The total gravitational potential that induces such a circular velocity on a massive point-like particle can be obtained by solving the differential equation:
\begin{align}
\left.\dfrac{\partial\Phi_{n}(\vec{r})}{\partial r}\right|_{z=0}=-\dfrac{V^{2}_{n}(r)}{r}.\label{Eq.10}
\end{align}
\noindent Assuming spherical symmetry, the Newtonian equations of motion for massive particles confine the stable orbits to the $\theta=\pi/2$ plane ( {$z=0$ plane}), then, one can get rid of the partial derivative and write Eq.\eqref{Eq.10} as
\begin{align}
\dfrac{d\Phi_{n}(r)}{d\,r}&=-\dfrac{V^{2}_{0}d^{2n}}{r_{c}}\dfrac{(r/r_{c})^{2}(1+r/r_{c})^{-(2n+1)}}{1+(r/r_{c})^{2}}\sum_{i=0}^{2n}C_{(i,n)}\left(\dfrac{r/r_{c}}{d}\right)^{i},\\\label{Eq.11}
C_{(i,n)}&=
\left\{
  \begin{array}{ll}
    i+1, & \hbox{if $0\leq i\leq n$;} \\
    2n-i+1, & \hbox{if $n\leq i\leq 2n$.}
  \end{array}
\right.
\end{align}

Notice that to the gravitational potential contribute $2n+1$ components, but there is one which is not conventional to the known Newtonian physics; to see this more in details, one can expand $\Phi_{n}(r)$ as:
\begin{align}
\Phi_{n}(r)=\Phi_{n}(+\infty)&+\int_{r/r_{c}}^{+\infty}\dfrac{G M_{0}d^{2n}}{r_{c}}\dfrac{s^{2}(1+s)^{-(2n+1)}}{1+s^{2}}\sum_{i=0}^{2n-1}C_{(i,n)}\left(\dfrac{s}{d}\right)^{i}ds+\int_{r/r_{c}}^{+\infty}\dfrac{G M_{0}}{r_{c}}\dfrac{s}{1+s^{2}}\left(\dfrac{s}{1+s}\right)^{2n+1}ds,
\end{align}
\noindent being $M_{0}=V_{0}^{2}r_{c}/G$ a constant with mass unit and $G$ is the universal gravitational constant. Assuming the factor $G M_{0}/r_{c}$ to be a constant for all $r$, the gravitational potential can thus be written as
\begin{align}
\Phi_{n}(r)&=\Phi_{n}(+\infty)+\Phi_{n\,c}(r)+\int_{r/r_{c}}^{+\infty}\dfrac{G M_{0}}{r_{c}}\dfrac{s}{1+s^{2}}ds,\label{EMPVELGRAVPOT}
\end{align}
\noindent with
\begin{align}
\Phi_{n\,c}(r)&=\int_{r/r_{c}}^{+\infty}\dfrac{G M_{0}d^{2n}}{r_{c}}\dfrac{s^{2}(1+s)^{-(2n+1)}}{1+s^{2}}\sum_{i=0}^{2n-1}C_{(i,n)}\left(\dfrac{s}{d}\right)^{i}ds+\int_{r/r_{c}}^{+\infty}\dfrac{G M_{0}}{r_{c}}\sum_{j=1}^{2n+1}\binom{2n+1}{j}(-1)^{j}\dfrac{s}{(1+s)^{j}(1+s^{2})}ds
\end{align}
\noindent the convergent part of $\Phi_{n}(r)$. Under the assumption that $G M_{0}/r_{c}$ is constant throughout the galaxy, the logarithmically divergent part of the total gravitational potential, responsible for the flatness of the RC, gives  {an} ill-defined expression, that will always bound the orbit of any massive particle and no possibility of scape will be possible, which in turn means that the Universe should be populated by a single galaxy and everything should be bounded to it. This is clearly not true as there are no observations that could support this assertion, and therefore, a finite scape velocity should exist even in the presence of a large DM component.

\bigskip
This issue can be easily solved by relaxing the condition of constancy of the parameters $r_{c}$ and $V_{0}$ at large galactocentric distances. Indeed, to guarantee the convergence of every integral in the gravitational potential, it is enough to suppose that $r_{c}$ is constant, or slowly varying with the galactic radius till some characteristic length scale  $r_{\text{edge}},$ and after, increases linearly with $r$, while keeping constant $d, n$ and the factor $GM_{0}.$ Notice that higher powers of $r$ will improve faster the convergence, but the minimum power required to make the total gravitational potential well defined, is at the linear order. If one attains to this prescription, since $V_{0}^{2}=GM_{0}/r_{c},$, then, for $r>r_{\text{edge}}$ one expects the Keplerian fall of the circular velocity
as $V_{n}(r>r_{\text{edge}})\varpropto\sqrt{GM_{0}/r},$, where the proportionality constant depends on $n,\,d,$ $r_{\text{edge}}$ and $\langle r_{c}\rangle$ (the constant part of $r_{c}$ for $r\ll r_{\text{edge}}$). The power of this empirical method relies in the fact that Eq.\eqref{Eq.2.2}, altogether with this ``renormalization'' prescription, one is able to reproduce analytical flat RCs, which after a length scale $r_{\text{edge}}$, the Keplerian regime is recovered for such a circular velocity and also allows to make predictions on the masses of galaxies.


\bigskip

\begin{figure}
  \centering
  \includegraphics[width=0.45\textwidth]{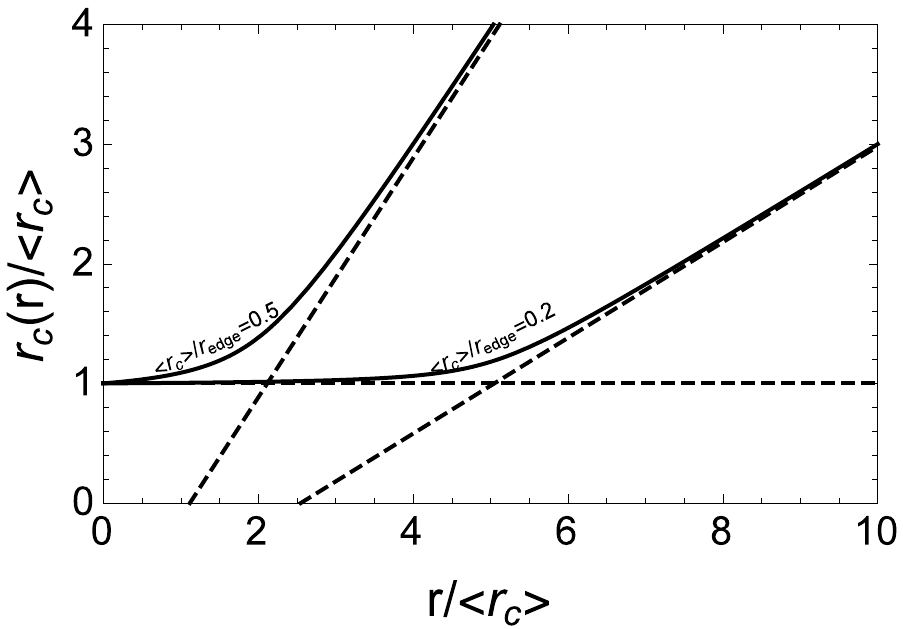}
    \includegraphics[width=0.46\textwidth]{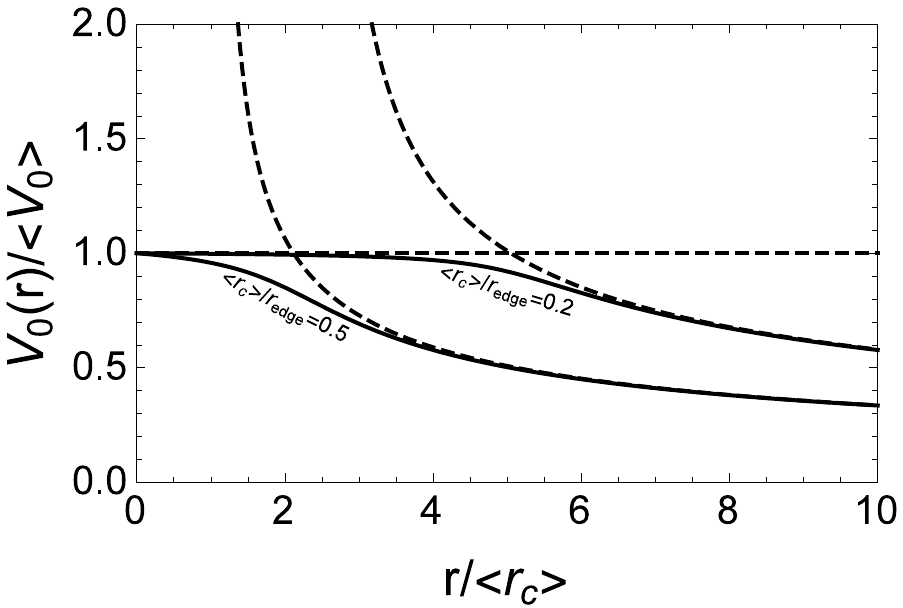}\\
  \caption{The left panel shows the dependence with $r/\langle r_{c}\rangle$ of $r_{c}(r)/\langle r_{c}\rangle$ as given by Eqs.\eqref{keplerianrenormalization}, for different values of the ratio $\langle r_{c}\rangle/r_{\text{edge}}$: 0.2 and 0.5 respectively (see labels); the dashed lines corresponds to their linear asymptotes. The right panel shows as well the dependence with $r/\langle r_{c}\rangle$ of $V_{0}(r)/\langle V_{0}\rangle$ as given by Eqs.\eqref{keplerianrenormalization}, taking again the ratio $\langle r_{c}\rangle/r_{\text{edge}}$: 0.2 and 0.5 (see labels); as before, the dashed lines correspond to their $\sim1/\sqrt{r}$ asymptotes.}\label{newtonrenormalizedrcV0}
\end{figure}

\begin{figure}
  \centering
  \includegraphics[width=0.45\textwidth]{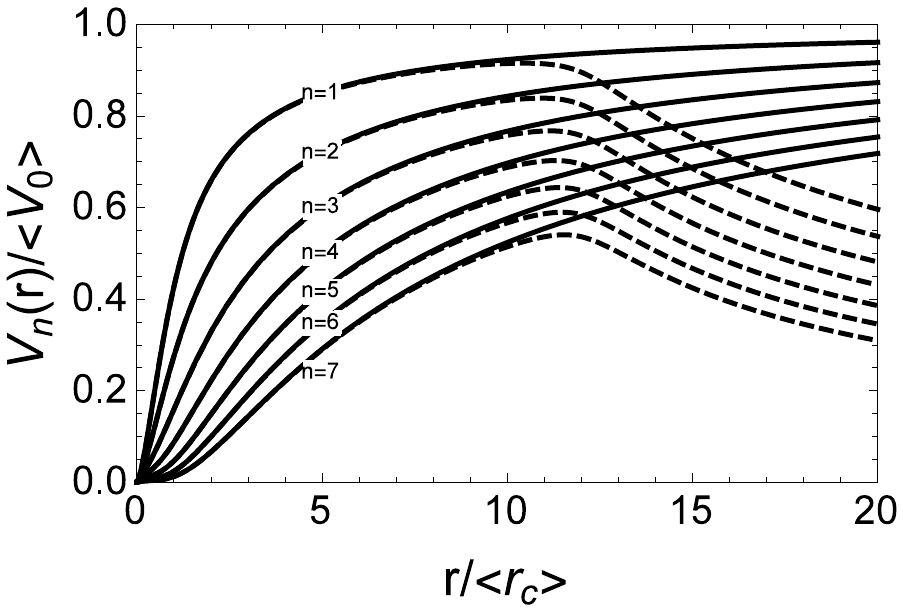} \includegraphics[width=0.45\textwidth]{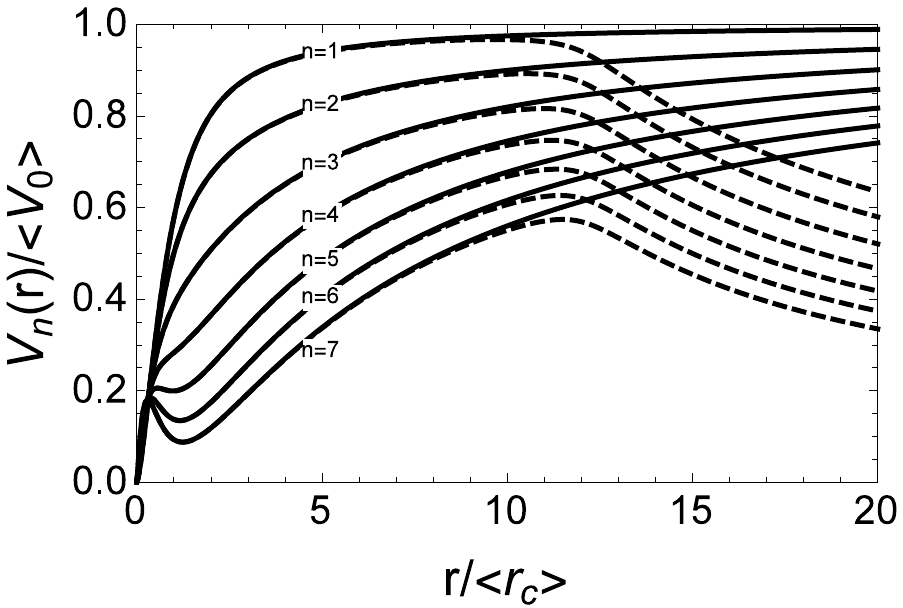}\\
  \caption{In both panels are shown the velocity formula Eq.\eqref{Eq.2.2} for $d=0.7$ (left) and $d=1.3$ (right) for several integer values of $n:$ $1\leq n\leq7$ with their corresponding labels for the prescription Eqs.\eqref{keplerianrenormalization}. The continuous lines are given by Eq.\eqref{Eq.2.2}, while the dashed lines correspond to the newtonian renormalized ones with $r_{\text{edge}}/\langle r_{c}\rangle=12$.}\label{velocityprofiles}
\end{figure}

As a toy example of such a variation for $r_{c}$ and $V_{0},$ which respects the continuity of the velocity and its derivative, one
could propose, among many others:
\begin{align}
r_{c}(r)&=\langle r_{c}\rangle f(r),\;\; V_{0}(r)=\langle V_{0}\rangle/\sqrt{f(r)},\;\;\;f(r)=\left(1+r/r_{\text{edge}}+\sqrt{(r/r_{\text{edge}}-1)^{2}+\langle r_{c}\rangle^{2}/r_{\text{edge}}^{2}}-\sqrt{1+\langle r_{c}\rangle^{2}/r_{\text{edge}}^{2}}\right),\label{keplerianrenormalization}
\end{align}
\noindent where $GM_{0}=\langle V_{0}\rangle^{2}\langle r_{c}\rangle$, $\langle V_{0}\rangle$ and $\langle r_{c}\rangle$ are constants. Recall that $\langle V_{0}\rangle$ and $\langle r_{c}\rangle$ are the constant parts of $V_{0}(r)$ and $r_{c}(r)$ for $r\ll r_{\text{edge}}$ respectively. In Figs.\eqref{newtonrenormalizedrcV0} are shown the radial dependence of $r_{c}(r)$ and $V_{0}(r)$ for the above prescription. Notice that
for small values of $r/\langle r_{c}\rangle,$ $r_{c}(r)$ and $V_{0}(r)$ are almost constants and then, they start to vary such to make the gravitational potential finite at large distances as needed. Moreover, in Figs.\eqref{velocityprofiles} are shown the circular velocity profiles as given by Eq.\eqref{Eq.2.2} (continuous lines) and the same Eq.\eqref{Eq.2.2} but with the renormalized coefficients as given by the prescription Eqs.\eqref{keplerianrenormalization} (dashed lines) for $d=0.7$ (left panel) and $d=1.3$ (right panel) respectively. The Newtonian gravitational potential is recovered for large values of $r/\langle r_{c}\rangle$ and how quick this regime is reached, it depends on the ratio $\langle r_{c}\rangle/r_{\text{edge}}.$

\begin{figure}
  \centering
  \includegraphics[width=0.45\textwidth]{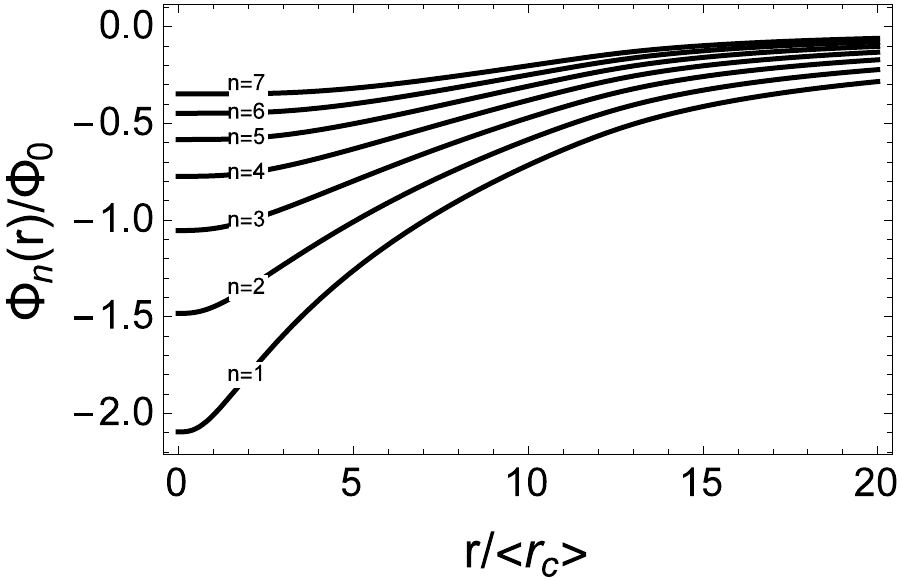} \includegraphics[width=0.45\textwidth]{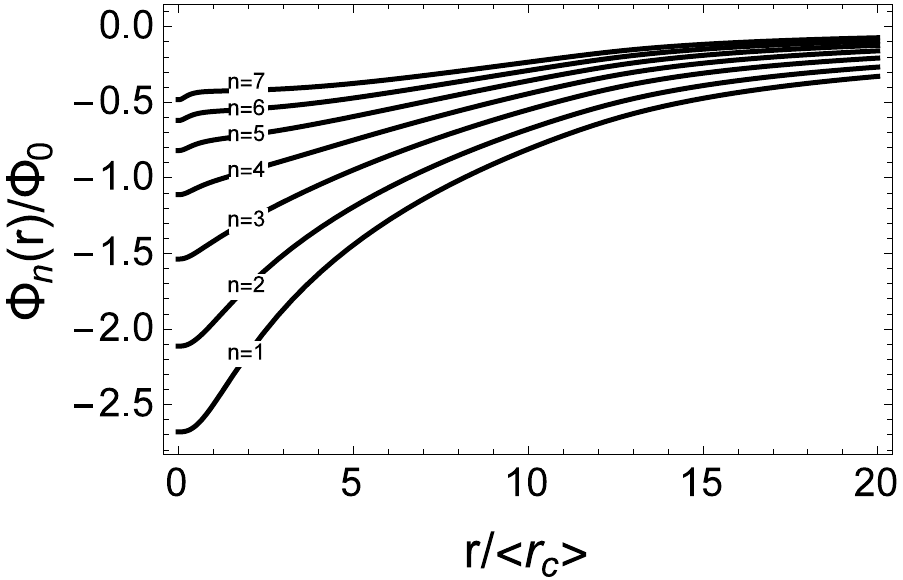}\\
  \caption{In both panels are shown the gravitational potential in units of $\Phi_{0}=\langle V_{0}\rangle^{2}$ for $d=0.7$ (left panel) and $d=1.3$ (right panel) for several integer values of $n:$ $1\leq n\leq7$ with their corresponding labels for the prescription Eqs.\eqref{keplerianrenormalization} and $r_{\text{edge}}/r_{c}=12$.}\label{EMPVELGRAVPOTGRAPHICS}
\end{figure}

In Figs.\eqref{EMPVELGRAVPOTGRAPHICS} on the other hand, are shown the graphics of the gravitational potential for $d=0.7$ (left panel) and $d=1.3$ (right panel), for $n$ from 1 to 7, by setting $\Phi_{n}(+\infty)= 0.$

As the Keplerian regime is reached for $r\gg r_{\text{edge}},$ this means that beyond $r_{\text{edge}},$ the circular velocity of any massive
point-like particle is the same as if it is moving in vacuum, influenced by the gravitational field created by a point-like particle of mass
$M_{T},$ which can be computed from Eq.\eqref{Eq.2.2} substituting the renormalized coefficients $r_{c}\mapsto r_{c}(r)$ and $V_{0}\mapsto V_{0}(r)$ and taking the limit $r\to+\infty$ of the expression $rV^{2}(r)/G$ (keeping $GM_{0}, d$ and $n$ constants) with a suitable election of the renormalization conditions for $r_{c}(r)$ and $V_{0}(r).$ For example, for the renormalization prescription given by Eqs.\eqref{keplerianrenormalization}, $M_{T}$ is given by:

\begin{align}\label{Eq.20}
M_{T}=M_{0}\dfrac{d^{2n}x^{4}\left(1+x\right)^{-(2n+1)}}{1+x^2}\sum_{i=0}^{2n}C_{(i,n)}\left(\dfrac{x}{d}\right)^{i},\;M_{0}=\dfrac{\langle V_{0}\rangle^{2}\langle r_{c}\rangle}{G},\;x=\frac{r_{\text{edge}}}{2\langle r_{c}\rangle}.
\end{align}

Worth to mention that the deduced mass will depend on how one chooses the slope of the linear dependence with $r$ of the function $f(r)$ given by Eqs.\eqref{keplerianrenormalization}, therefore $M_{T}$ serves as only an estimation of the total mass enclosed by the galaxy.

\bigskip
\bigskip

The interpretation of the length-scale $r_{\text{edge}},$ as mentioned above, is that any point-like massive particle moving in orbits beyond this distance, will feel like it is moving in vacuum by the effects of the gravitational field produced by a point-like source with mass $M_{T}.$ This means that beyond the scale $r_{\text{edge}},$ no luminous nor DM influence the state of motion and one can identify $r_{\text{edge}}$ with the DM halo virial radius $R_{\text{vir}}$, as it represents a measure of the ``edge'' of a galaxy. Moreover, since the total mass $M_{T}$ is the sum of the DM halo virial mass $M_{\text{vir}}$ and luminous matter mass $M_{L}$ components, and if one uses the relation between the virial mass and virial radius,
\begin{align}
&r_{\text{edge}}=\left(\dfrac{3}{97.2}\dfrac{\text{M}_{\text{vir}}}{4\pi\rho_{\text{crit}}}\right)^{1/3}\;\;\text{kpc}.
\end{align}
\noindent then:
\begin{align}
\dfrac{M_{T}}{\frac{4\pi}{3}\rho_{\text{crit}} r_{\text{edge}}^{3}}=\dfrac{\text{M}_{\text{vir}}+\text{M}_{L}}{\frac{4\pi}{3}\rho_{\text{crit}} r_{\text{edge}}^{3}}=\dfrac{\text{M}_{\text{vir}}}{\frac{4\pi}{3}\rho_{\text{crit}}r_{\text{edge}}^{3}}\left(1+\dfrac{\text{M}_{\text{L}}}{\text{M}_{\text{vir}}}\right).
\end{align}
\noindent Since the ratio $\text{M}_{\text{L}}/\text{M}_{\text{vir}}$ is in principle constrained to $0<\text{M}_{\text{L}}/\text{M}_{\text{vir}}<1,$ then the scale length $r_{\text{edge}}$ is constrained to satisfy the inequalities
\begin{align}
97.2<\dfrac{M_{T}}{\frac{4\pi}{3}\rho_{\text{crit}} r_{\text{edge}}^{3}}<2\times97.2,\label{redgeconstrain}
\end{align}
\noindent where recall that $M_{T}$ depends on $r_{\text{edge}}$ as well.

\subsection{The total local density}\label{Sec.1.2}

The total local density can be computed from Eq.\eqref{Eq.14} by substituting the analytical expression for $V(r)$ given by Eq.\eqref{Eq.2.2}. The essence of this empirical velocity profile method is to take the observational information given by the galaxy RC, convert it to an analytical expression using Eq.\eqref{Eq.2.2} (with the corresponding error propagation) and compute the local density assuming an extrapolation between any two observational points. There are two physical regimes where one doesn't have enough data to keep testing the empirical velocity formula. The first regime is the outskirts of the galaxy RCs, which in the previous subsection, by extrapolation was assumed that after the length scale $r_{\text{edge}}$, the Keplerian regime is reached. The second regime where many galaxy RCs have a poor spatial resolution is in the innermost regions, where precisely the cusp-core issue holds \citep{deBlok:2002vgq, deBlok:2008wp, deBlok:2009sp, Salucci:2000ps, Moore:1994yx, Gentile:2004tb, Gentile:2005de} (and references therein). The discussion on this cusp-core controversy is out of the scope of the present article, however, in the inner parts of the galaxy RCs, the logarithmic slope of the associated total local
density can be modified via renormalization of the other free parameter $d$ at the length scale $\langle r_{c}\rangle.$ Indeed, as already discussed, for $r\ll \langle r_{c}\rangle$ , the behaviour of the circular velocity profile Eq.\eqref{Eq.2.2} is $V_{n}(r)\sim r^{\frac{3}{2}},$ giving a
logarithmic slope of the total local density as $d\log\rho(r)/d\log r=1;$ recall that in the regime $r\sim 0$ kpc, the logarithmic slopes of
the NFW and BRK DM density profiles are $d\log\rho_{NFW}(r)/d\log r=-1$ and $d\log\rho_{BRK}(r)/d\log r=0$ respectively. In this galactic central regime, the stellar and gaseous contributions are relevant, so the logarithmic slope of the total local density may change according to the logarithmic slope of the luminous matter distribution as well, but regarding the parameter $d$ in Eq.\eqref{Eq.2.2}, a renormalization procedure could absorb the powers of $r$ such to make coincide both logarithmic slopes as long as beyond the length scale $\langle r_{c}\rangle$ the parameter $d$ becomes constant again. As a toy example, a kind of renormalization of $d$ like
\begin{align}
&d(r)=\langle d\rangle (1+(r/\langle r_{c}\rangle)^{-\frac{3-2\alpha}{2n}}),\label{drenorm}
\end{align}
\noindent being $\langle d\rangle$ the constant part of $d(r)$ at long distances and $\alpha$ the renormalizing power-law for $d(r)$, that will give a cored total local density for $\alpha=1,$ increasing density for $1<\alpha\leq 3/2$ and cuspy density profiles for $\alpha<1$ ($\alpha=1/2$ for the NFW density profile). Notice that for the velocity profile Eq.\eqref{Eq.2.2}, the coefficient $\alpha$ can't be larger than $3/2$ since $d$ will no longer be a constant for such a prescription at large distances compared to $\langle r_{c}\rangle$. The renormalization of the parameter $d$ could be a long subject of discussion, which is out of the scope of the present article.  {Future measurements on galaxy RCs with a better spatial resolution near their centres could help to clarify the sign of the logarithmic slope of the local density near $r=0$ kpc.} In what follows in this article, no renormalization procedure is taken into account for the parameter $d.$

\bigskip

\begin{figure}
  \centering
  \includegraphics[width=0.45\textwidth]{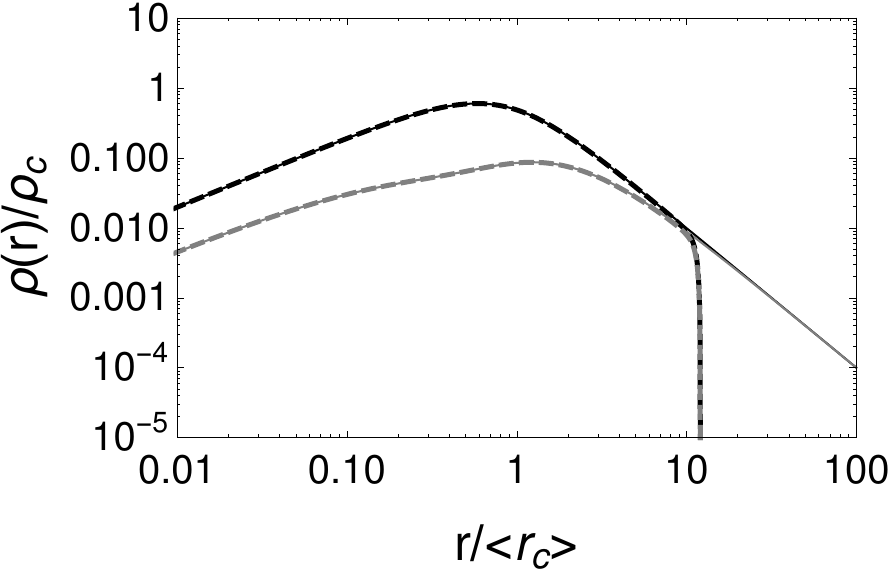} \includegraphics[width=0.45\textwidth]{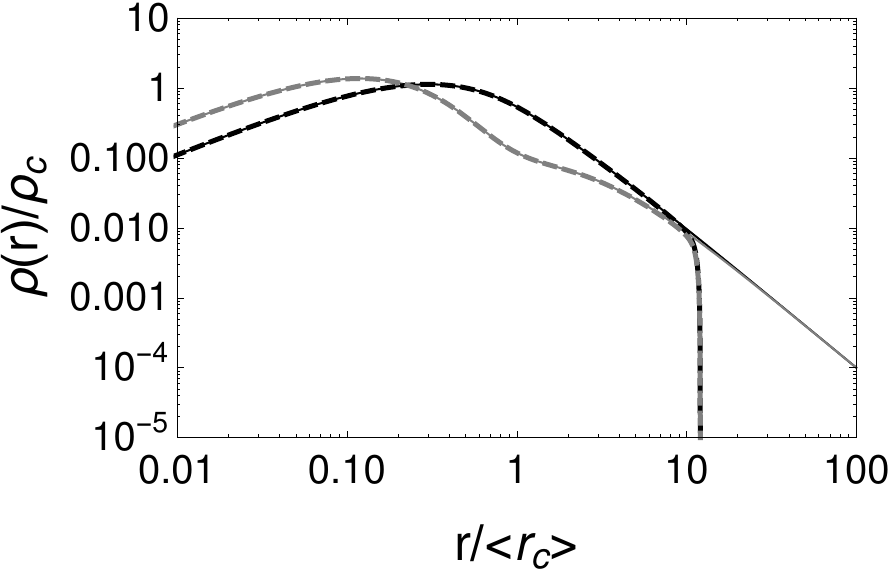}\\
  \caption{In both panels are shown the total local density in units of $\rho_{c}=M_{0}/4\pi/3\langle r_{c}\rangle^{3}$ for $d=0.7,$ $n=1, 3$ (left panel) and $d=1.3,$ $n=2, 4$ (right panel). The black and gray lines in the left panel correspond to $n=1$ and $n=3,$ and in the right panel to $n=2$ and $n=4$ respectively. The continuous lines correspond to the local density computed directly from Eq.\eqref{Eq.14} by substituting the analytical expression for $V_{n}(r)$ given by Eq.\eqref{Eq.2.2}, and dashed lines correspond to the local density but with the renormalization prescription Eqs.\eqref{keplerianrenormalization}, where in both cases the ratio $r_{\text{edge}}/\langle r_{c}\rangle=12$ was fixed.}\label{LOCALDEN}
\end{figure}

In Figs.\eqref{LOCALDEN} are shown the ratio $\rho(r)/\rho_{c}$ of the total local density $\rho(r)$ and the constant $\rho_{c}=M_{0}/4\pi/3\langle r_{c}\rangle^{3}$ as a function of $r/\langle r_{c}\rangle,$ where $M_{0}$ is given in Eq.\eqref{Eq.20}, for $d=0.7,$ $n=1, 3$ (left panel) and $d=1.3,$ $n=2, 4$ (right panel). The continuous lines, correspond to the total local density computed directly from Eq.\eqref{Eq.14} by substituting the analytical expression for $V(r)$ given by Eq.\eqref{Eq.2.2} and they show a positive logarithmic slope for distances below the length scale $\langle r_{c}\rangle$ as discussed previously. They arrive to a local maximum near $r=\langle r_{c}\rangle$ and after, they decline with logarithmic slope $d\log\rho(r)/d\log r = −2.$ The dashed lines, which correspond to the total local density computed with $r_{c}(r)$ and $V_{0}(r)$ scale-dependent as in Eqs.\eqref{keplerianrenormalization}, have the same behaviour as the continuous lines in the innermost regions, but around the length scale $r\simeq r_{\text{edge}}$, they present a sharp fall because the Keplerian regime of the circular velocity is reached.

\section{Results from the fittings with 850 galaxies.}\label{Sec.2.0}

In the Figs.\eqref{Fig.01} are shown\footnote{The complete set of fitted RCs can be found in this link: https://www.dropbox.com/s/vlfzbtpszw4uk12/graphics.tar.gz?dl=0} the experimental RCs and the corresponding fits using Eq.\eqref{Eq.2.2} (continuous lines) and with the renormalization prescription described in the last section to get the Keplerian fall of the circular velocity (dashed black lines). For such a study, a sample of 850 different RCs was taken from Refs.\citep{Oh:2015xoa, deBlok:2001qy, Fune:2016uvn, Corbelli:2014lga, Karukes:2015fma, Lelli:2014sxa, Weldrake:2002ri, Moiseev:2014wha, Swaters:2009by, Persic:1995tc, KuziodeNaray:2006wh, KuziodeNaray:2007qi}. The RCs sample cover a wide range of different morphological types of galaxies, which include from dwarf spiral galaxies such as DDO154, DDO52 and NGC2366 \citep{Oh:2015xoa}, low surface brightness (LSB) spiral galaxies such as F5631, F5683, F5718 \citep{deBlok:2001qy}, till barred and lenticular spirals \citep{Persic:1995tc}. In the RC plots Figs.\eqref{Fig.01}, there are RCs with error bars and with open symbols. All galaxies with error bars meet the excellence criterion since their approaching and receding arms are very symmetric and their RCs are extended out to (at least) the optical radius. The RCs with open symbols, as explained in \citep{Persic:1995tc}, they fail in at least one of the criteria stated above and therefore may be not suitable for accurate and direct mass modelling. However, they constitute a large database for those methods able to recover the DM properties by needing less stringent requirements on the RC quality. Since there are 4 free parameters to fit with Eq.\eqref{Eq.2.2}, from the list of 900 different galaxies analysed in \citep{Persic:1995tc}, another selection criterion was used such that the RCs fitted contain 8 or more experimental points.

In this article was performed first a $\chi^{2}$ minimization procedure to determine the parameters $(V_{0}, r_{c}, d, n)$ regarding the formula given by Eq.\eqref{Eq.2.2} which throughout all the computation procedure they were considered as free or uncorrelated. Given the goodness of the first fits with Eq.\eqref{Eq.2.2}, a second fit was performed to the above mentioned experimental RCs by using again Eq.\eqref{Eq.2.2}, this time, $r_{c}$ and $V_{0}$ are scale-dependent as given in Eqs.\eqref{keplerianrenormalization} but fixing the new parameters as equals to those of the first fits $(V_{0}, r_{c}, d, n)=(\langle V_{0}\rangle, \langle r_{c}\rangle, d, n),$ and constrained $r_{\text{edge}}$ in the range given by Eq.\eqref{redgeconstrain}. For the error estimation in each parameter, as described in \citep{Bevington:2003dr} for non-linear fitting methods, was used the formula
$\sigma_{i}^{2}=2(\partial^{2} \chi^{2}/\partial^{2}a_{i})^{-1},$ being $a_{i}$ any of the parameter and $\sigma_{i}$ its corresponding error. In Table.\eqref{Tab.2} are tabulated the best fitting values (and their errors) of $(\langle V_{0}\rangle, \langle r_{c}\rangle, d, n, r_{\text{edge}})$ for each galaxy analysed, including the total predicted mass $M_{T}$ and the values of the reduced $\chi^{2}$, i.e. $\chi^{2}/\# d.o.f$, where $\# d.o.f$ equals the number of experimental points minus the number of free parameters.

\bigskip

 {From the sample of galaxies with asymmetric RCs, as can be seen from Figs.\eqref{Fig.01}, a large scattering from Eq.\eqref{Eq.2.2} is observed, being more frequent the data points to scatter to velocities higher than the model than to the ones scattered below when one would expect approximately the same compromise. This asymmetry is due to the fact that since no error bars are present, when minimizing the $\chi^{2}$ distribution, all the experimental points have the same statistical significance, so when determining the best fitting value of the parameter $V_{0}$, which have to be bounded from above to a few hundreds km s$^{-1}$, the innermost points of the RC weight more than the ones in the outskirts of the galaxy (where $V_{0}$ is relevant), tending to give a lower bound for $V_{0}.$ However, as already discussed above, these galaxy RCs with large asymmetries may be not suitable for accurate modelling, but they constitute a large database for statistical studies of the DM.}

\bigskip

\begin{figure}
\flushleft
\includegraphics[width=0.45\textwidth]{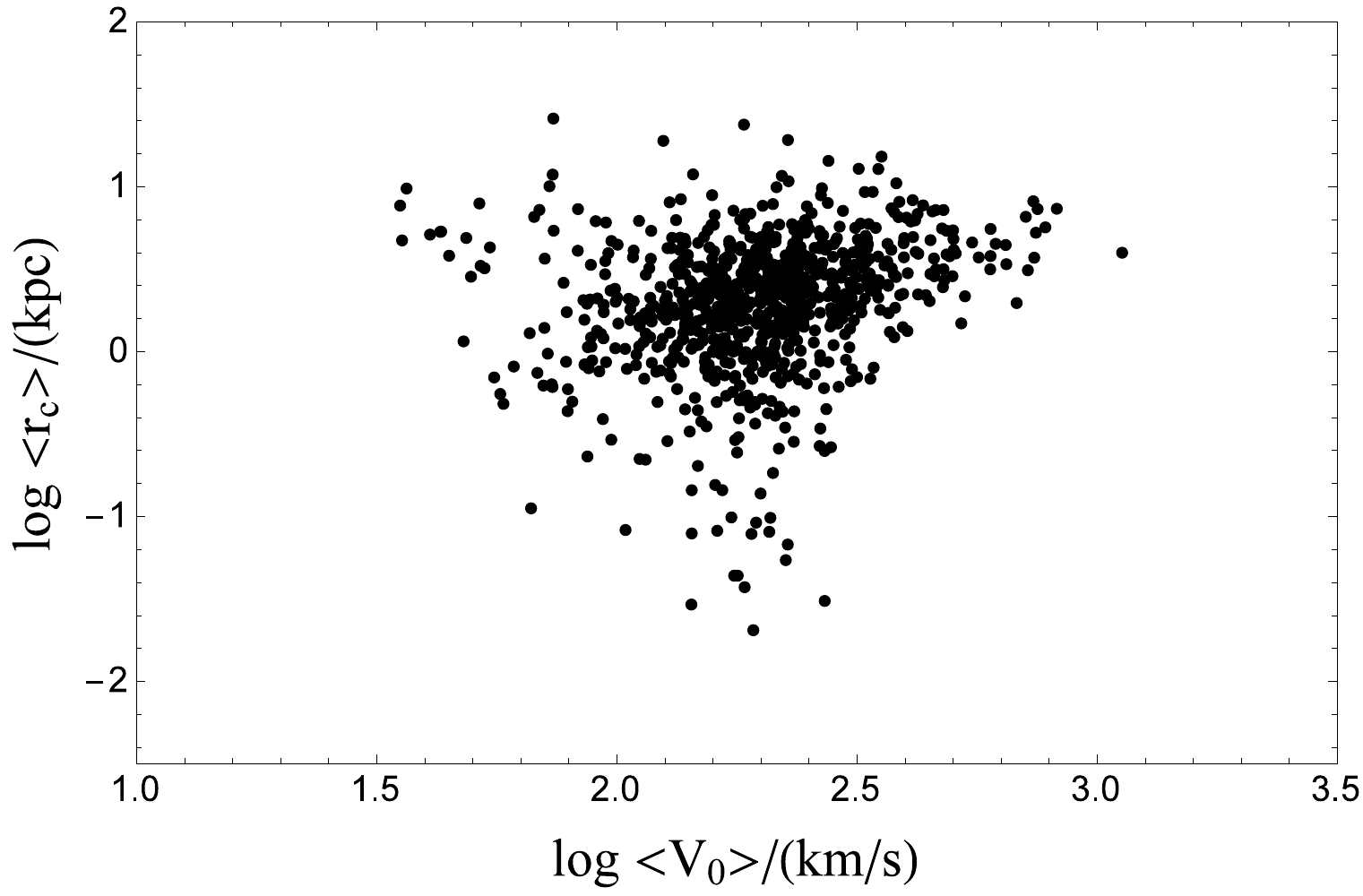}
\includegraphics[width=0.45\textwidth]{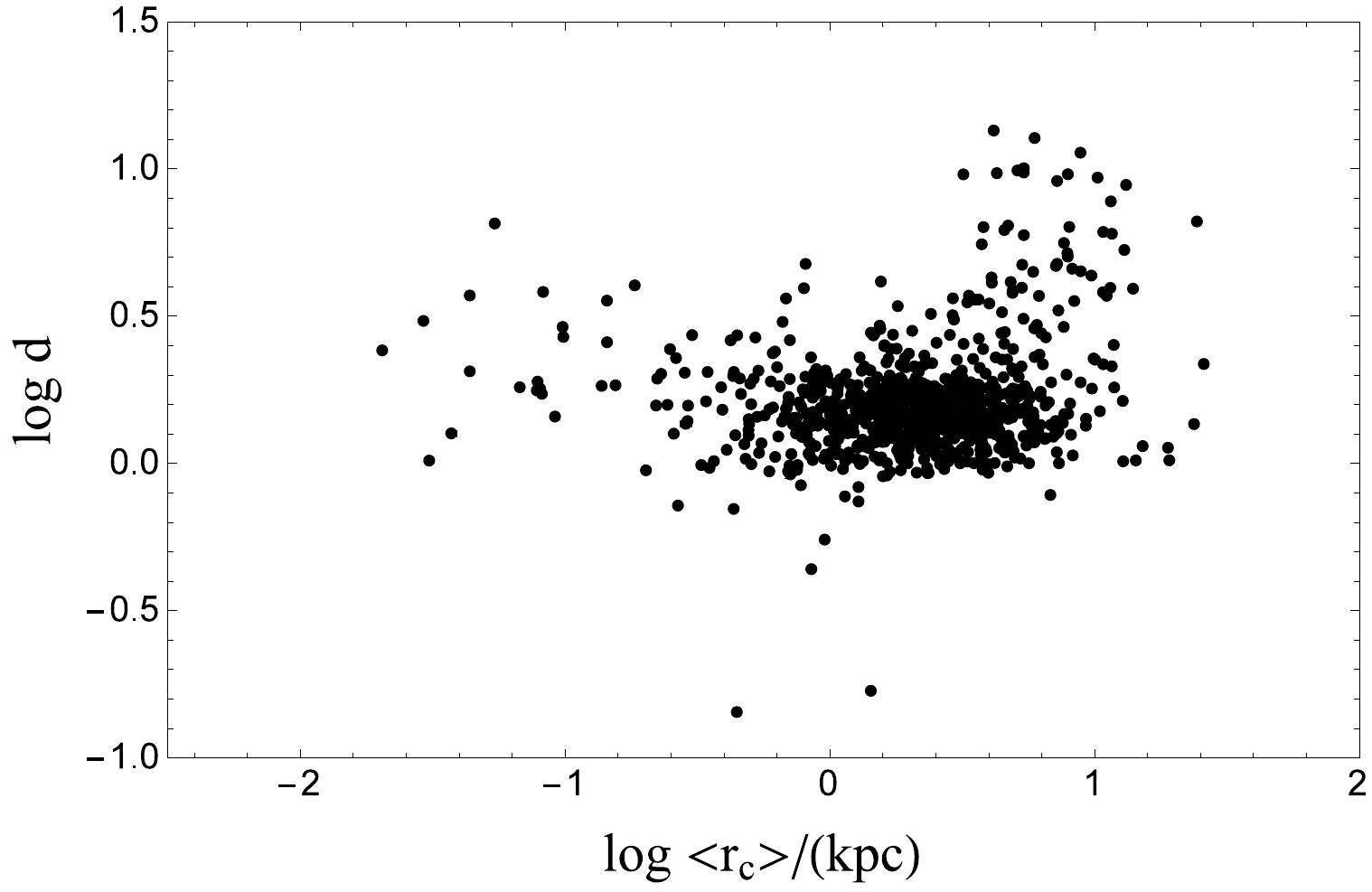}\\
\includegraphics[width=0.45\textwidth]{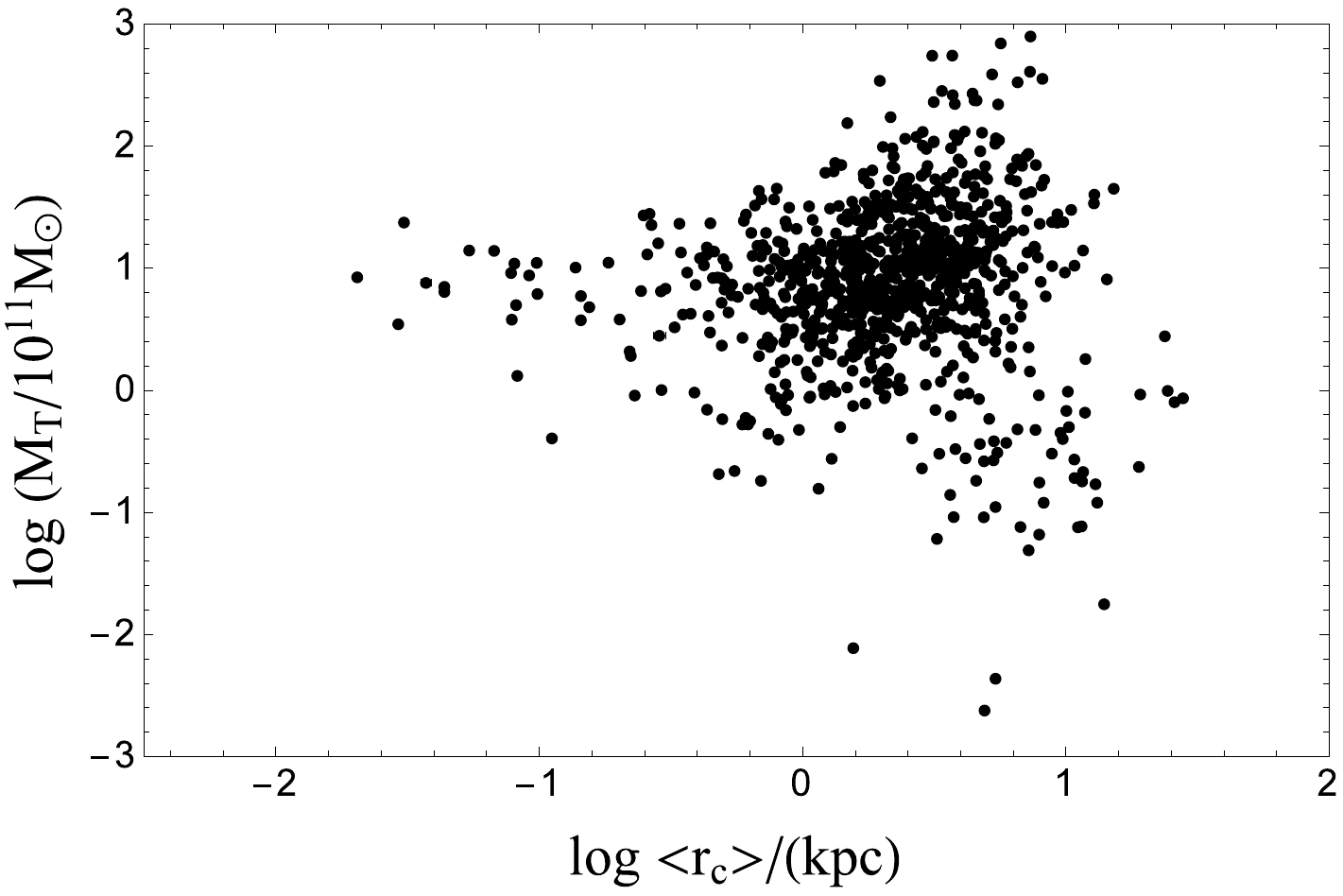}
\includegraphics[width=0.45\textwidth]{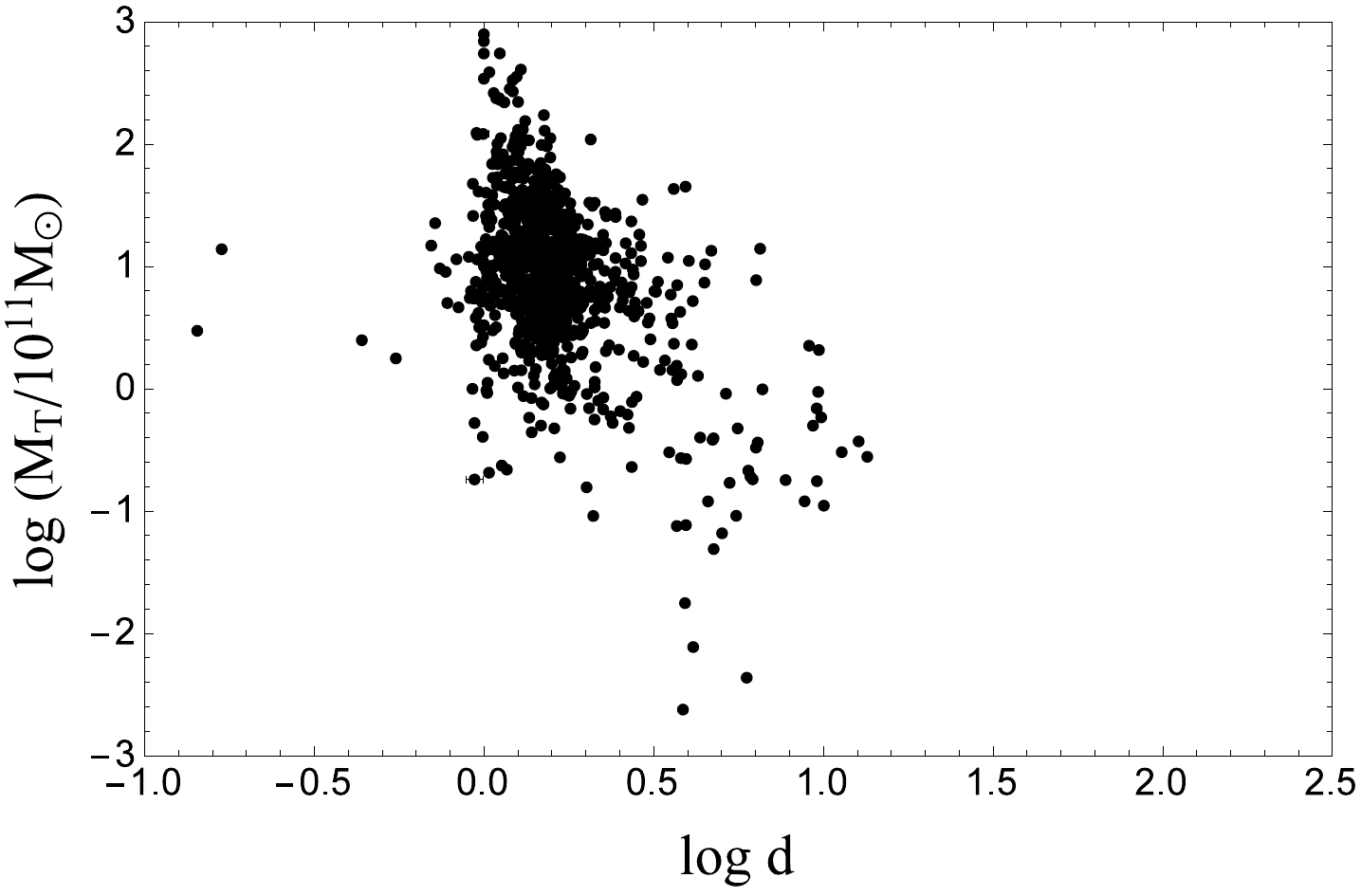}\\
\caption{Best fitting parameters plots for all value of $n.$ Large dispersions are observed in the $\langle V_{0}\rangle-\langle r_{c}\rangle$ (top-left panel), $\langle r_{c}\rangle-d$ (top-right panel), $\langle r_{c}\rangle-M_{T}$ (bottom-left panel) and $d-M_{T}$ (bottom-right panel) planes.}\label{Correlations}
\end{figure}

The relatively large amount of parameters is justified by the fact that the circular velocity profile used during the fitting process contains already all the information from the DM halo, bulge, stellar and gaseous disks and possible bars or other structures that will contribute to the experimental RCs by their composite probabilities. Moreover, in Figs.\eqref{Correlations} are shown the best fitting values as points (with error bars) in a two dimensional graphics of $\langle V_{0}\rangle-\langle r_{c}\rangle,$ $\langle r_{c}\rangle-d,$ $\langle r_{c}\rangle-M_{T}$ and $d-M_{T}$ for all values of the integer parameter $n$. As one may see in Figs.\eqref{Correlations}, there exist large dispersions between the best fitting values of the pairs of parameters $(\langle V_{0}\rangle,\langle r_{c}\rangle)$ (top-left), $(\langle r_{c}\rangle,d)$ (top-right), and $(\langle r_{c}\rangle,M_{T})$ (bottom-left). For the pair $(d,M_{T})$ (bottom-right) instead, one may see a weak sign of a negative correlation, however, a large dispersion persists.

\bigskip

\begin{figure}
\centering
\includegraphics[width=0.45\textwidth]{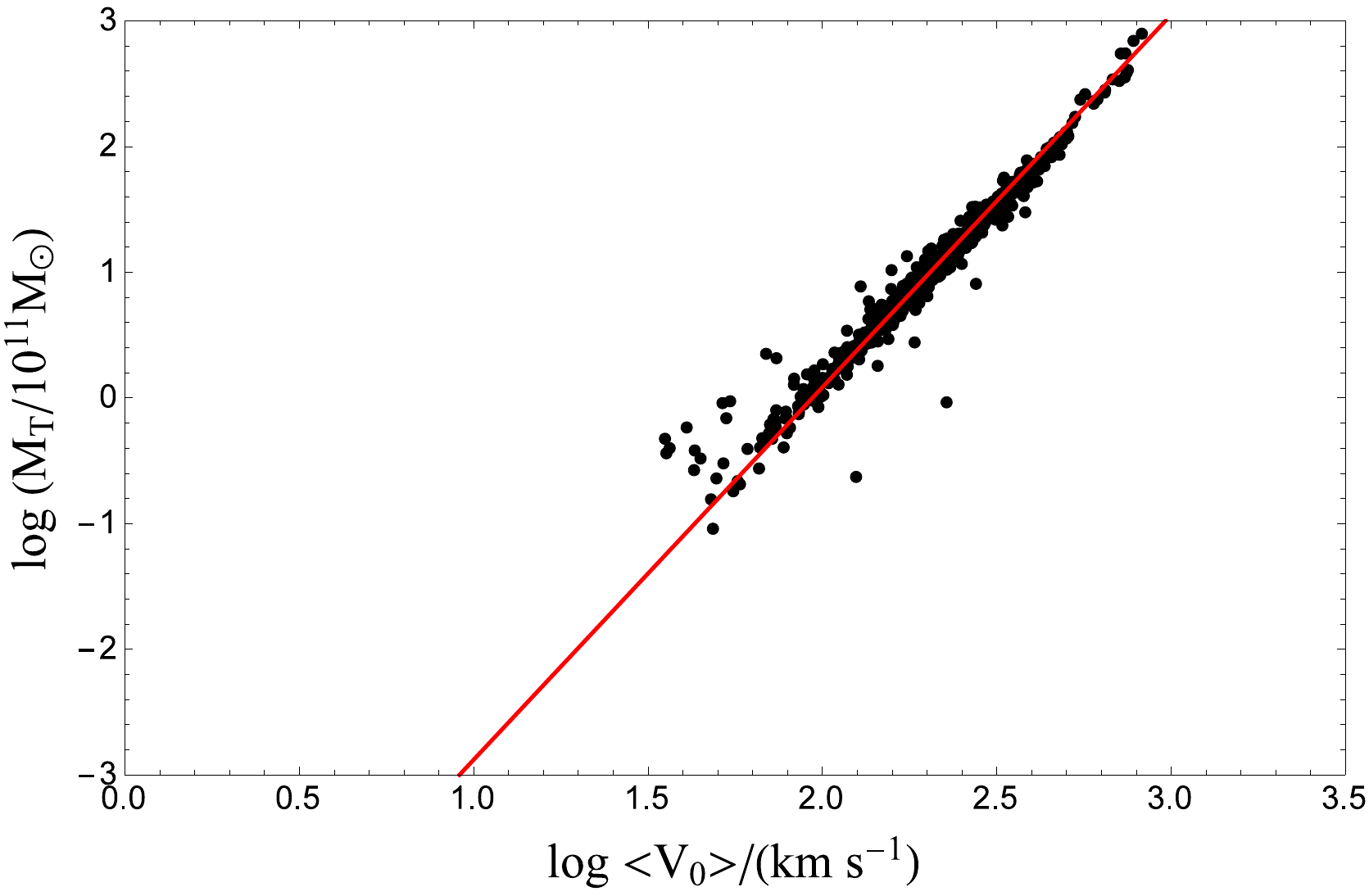}
\includegraphics[width=0.45\textwidth]{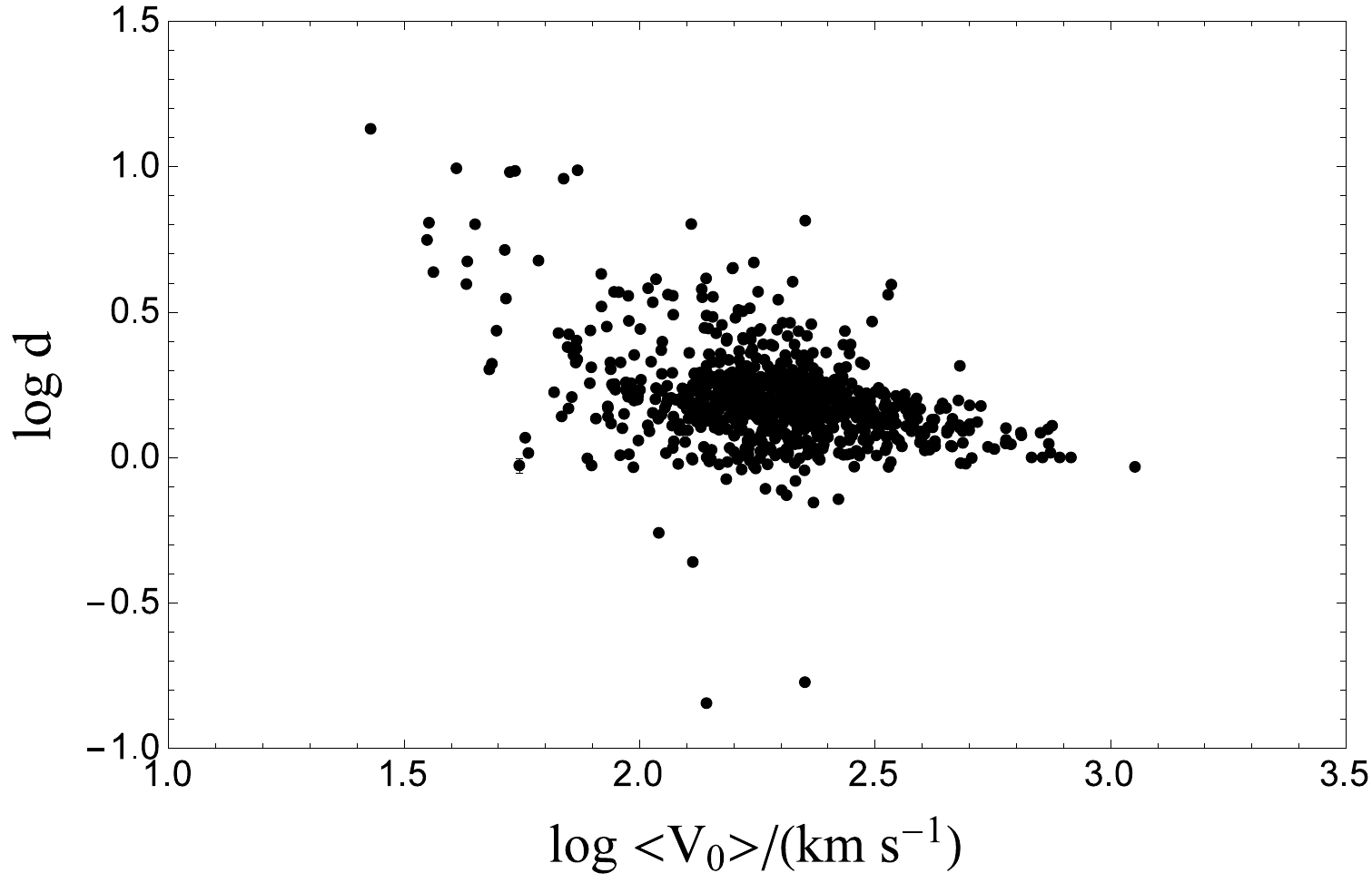}\\
\caption{Best fitting parameters plots for all value of $n$ of the pairs $(\langle V_{0}\rangle, M_{T})$ (left panel) and $(\langle V_{0}\rangle, d)$ (right panel). Some dispersion is observed in the $\langle V_{0}\rangle-d$ plane, however, there is clearly a power-law correlation between the parameters $\langle V_{0}\rangle$ and $M_{T}$, where the best fitting straight line (in red colour) corresponds to Eq.\eqref{TULLYFISHERLINEAR}.} \label{TULLYFISHER}
\end{figure}

In Figs.\eqref{TULLYFISHER} are the plots of the best fitting values of $\langle V_{0}\rangle$ and the total predicted mass $M_{T}$ (left panel), as well as the pair $(\langle V_{0}\rangle, d)$ (right panel). For the pair $(\langle V_{0}\rangle, M_{T}),$ one finds a positive correlation relation compatible with a power-law as
a Tully-Fisher-like formula Eq.\eqref{TULLYFISHERLINEAR}:
\begin{align}
\log\left(\dfrac{M_{T}}{10^{11}M_{\odot}}\right)=-(5.84\pm0.01)+(2.96\pm0.01)\log\left(\dfrac{\langle V_{0}\rangle}{km/s}\right).\label{TULLYFISHERLINEAR}
\end{align}
\noindent In the right panel of Figs.\eqref{TULLYFISHER} one can see also a week sign of a negative correlation between the parameters $d$ and $\langle V_{0}\rangle$ but a large dispersion still exists. This means that despite there are five free parameters to make the general fit, only four of them are independent. More extended and higher quality RCs are needed in order to arrive to conclusions on the correlation
of the pairs $(\langle V_{0}\rangle, d)$ and $(d, M_{T}).$

Moreover, in \citep{Corbelli:2014lga, Fune:2016uvn} was computed the virial mass of the galaxy M33 using the standard mass modelling and the local density method. Along with the stellar mass reported by \citet{Corbelli:2014lga}, the total mass in the mass modelling method corresponding to the NFW DM halo density profile as the best fit is $M_{T}=(4.36\pm 1.00)\times 10^{11}$ M$_{\sun}$, while for the BRK DM halo density profile in the local density method \citet{Fune:2016uvn} reported $M_{T}=(3.06\pm 0.80)\times 10^{11}$ M$_{\sun}$.  From both computed masses, with different methods, only the BRK profile in the local density method is compatible with the total mass computed via Eq.\eqref{Eq.20}: $M_{T}=(2.785\pm 0.003)\times 10^{11}$ M$_{\sun}$.

\section{Conclusions}\label{Sec.3.0}

The main limitation of the local density method is that one needs to provide an analytical and model-independent formula to represent the experimental data of the RC to compute the total local density. For the galaxy M33, it was successfully introduced Eq.\eqref{Eq.3.2.1} which solved this problem. However, there are many situations in which this formula doesn't provide a good representation of the experimental data, for example, when dealing with galaxies with only increasing RCs, slightly decreasing ones, prominent bulges or other galactic substructures. In this paper was given a solution to this problem by postulating an empirical circular velocity profile Eq.\eqref{Eq.2.2} as a function of the galactic radius, which at large distances gives a flat RC and fulfils the necessities above mentioned. A $\chi^{2}$ fitting was performed with 850 different galaxy RCs with an unprecedented degree of accuracy. Given the goodness of the fits in the experimental data range, Eq.\eqref{Eq.2.2} was promoted as an extrapolation function to the regions where there are no more experimental points. Since an asymptotically flat RC will give rise to an ill-defined gravitational potential, a renormalization procedure of the length scale $r_{c}$ and velocity $V_{0}$ was implemented in such a way that for distances greater than a length scale $r_{\text{edge}}>\langle r_{c}\rangle$, the circular velocity profile converges to the Keplerian one, providing a well defined total gravitational potential, and hence, to estimate the mass enclosed by the galaxy (including de DM halo), which are in agreement with other galaxy mass method estimations.



\begin{figure}
\flushleft
{\tiny\renewcommand{\arraystretch}{.95}
\resizebox{!}{.64\textwidth}{%
\begin{tabular}{lllll}
\includegraphics[width=0.50\textwidth]{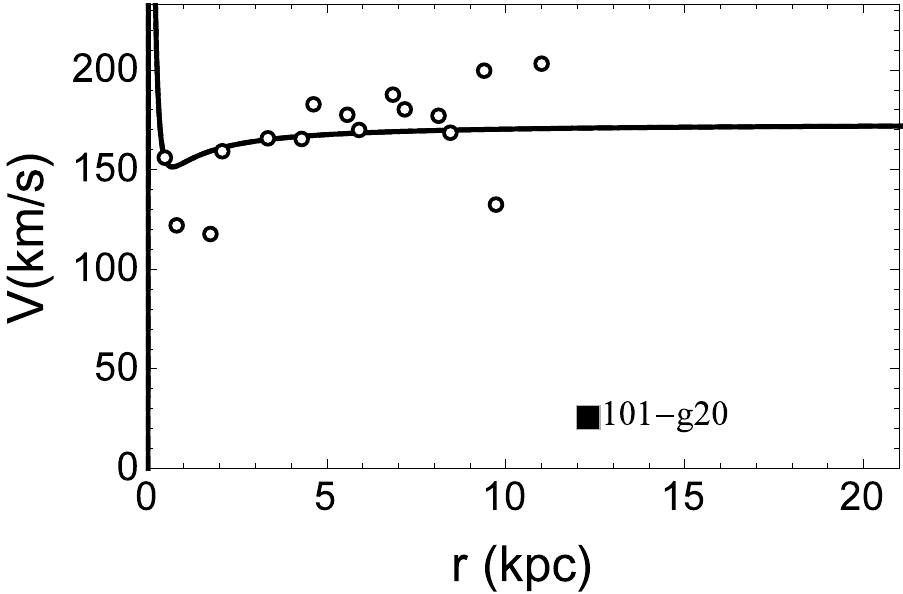}&
\includegraphics[width=0.50\textwidth]{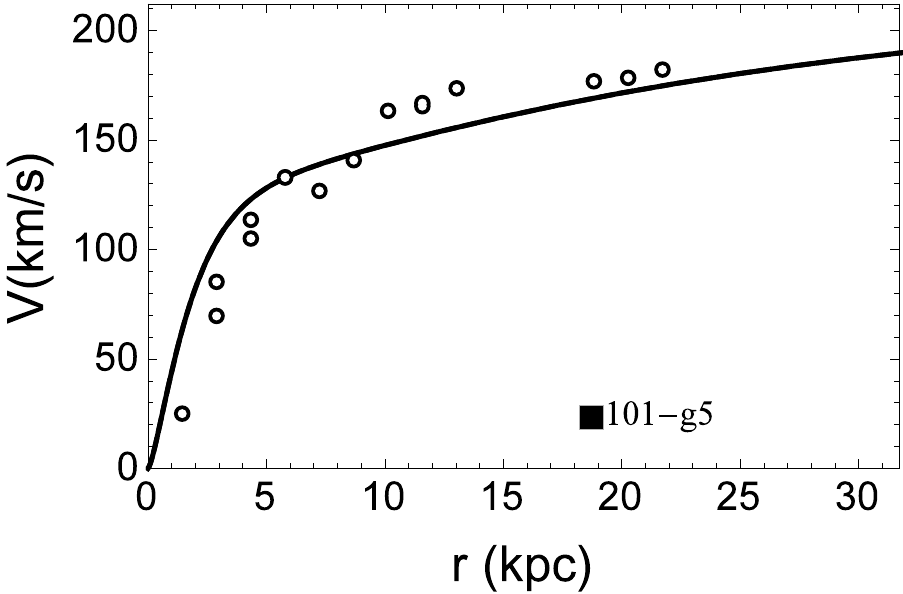}&
\includegraphics[width=0.50\textwidth]{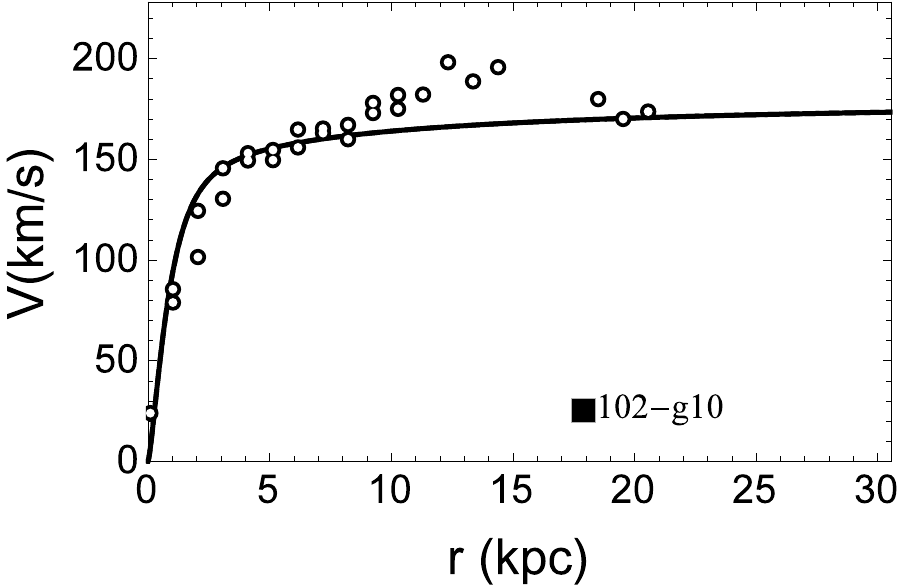}&
\includegraphics[width=0.50\textwidth]{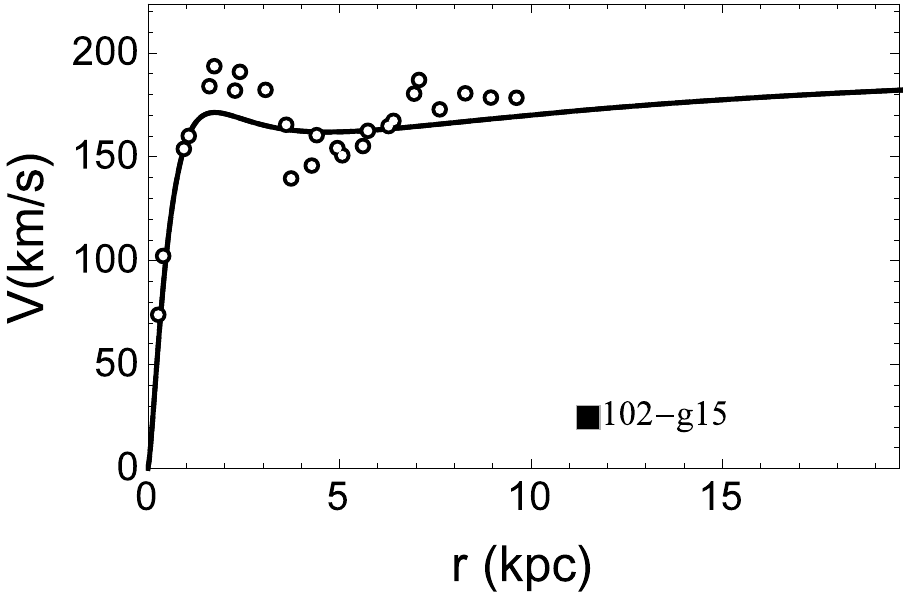}&
\includegraphics[width=0.50\textwidth]{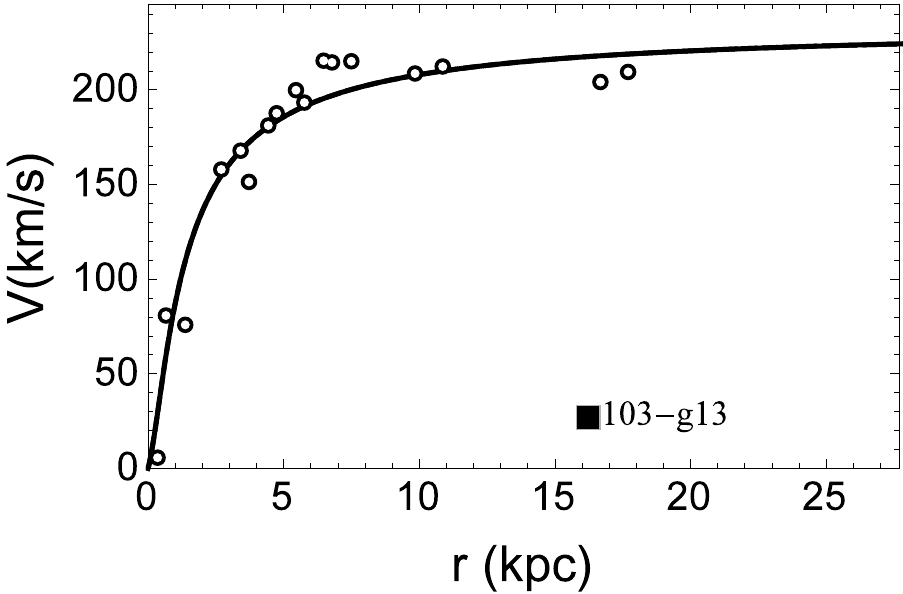}\\
\includegraphics[width=0.50\textwidth]{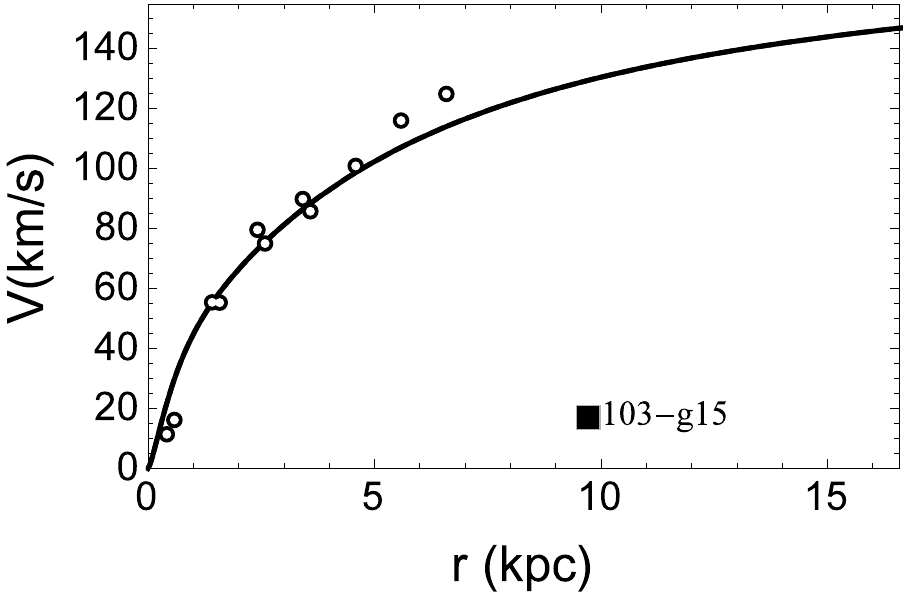}&
\includegraphics[width=0.50\textwidth]{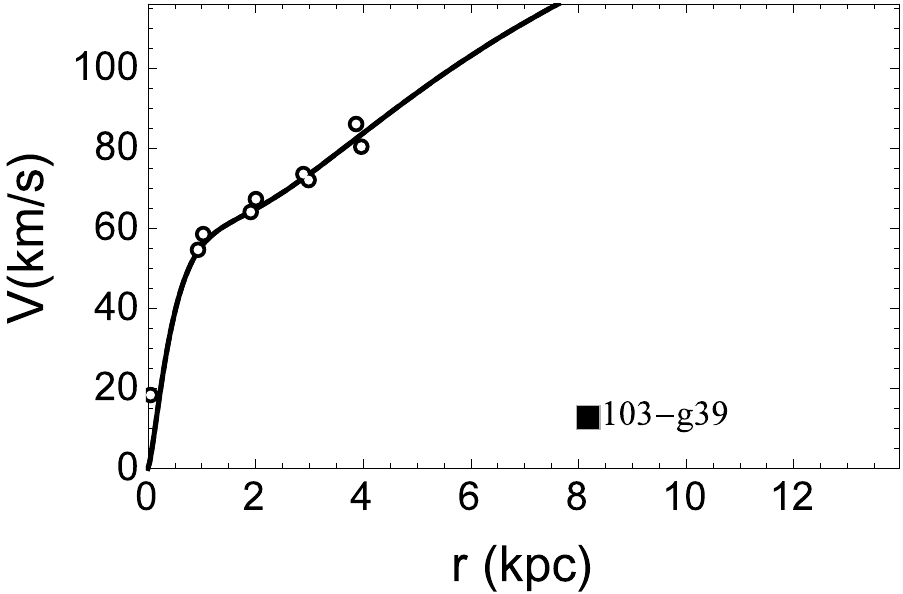}&
\includegraphics[width=0.50\textwidth]{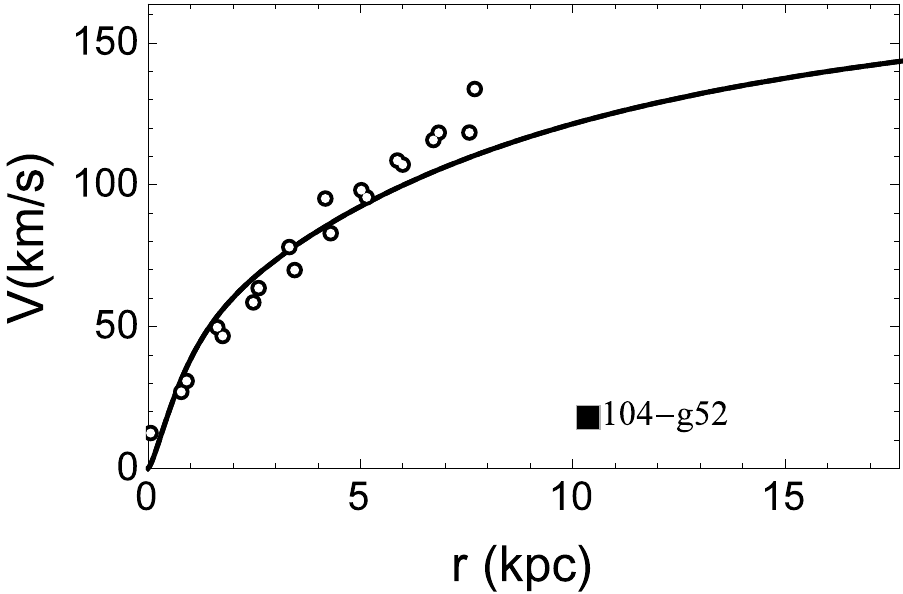}&
\includegraphics[width=0.50\textwidth]{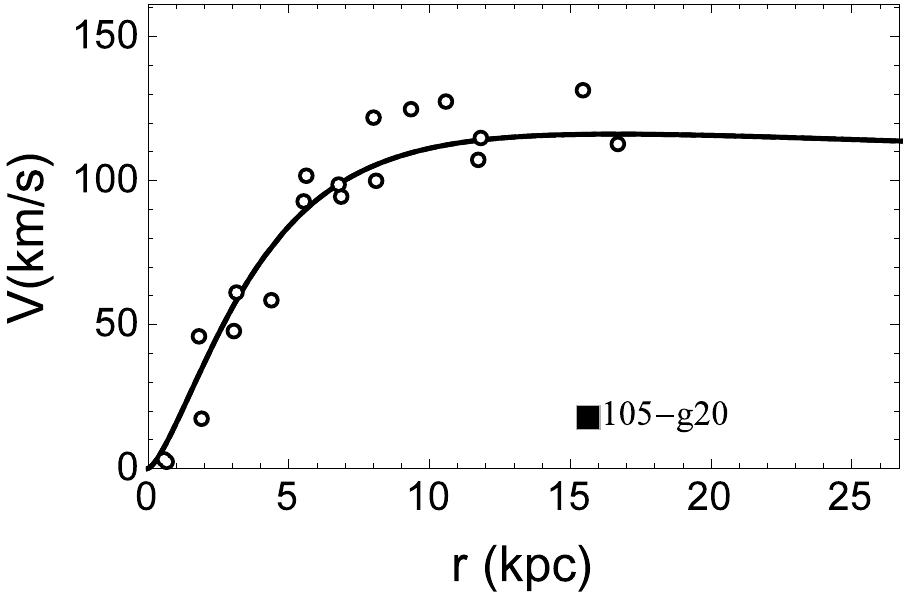}&
\includegraphics[width=0.50\textwidth]{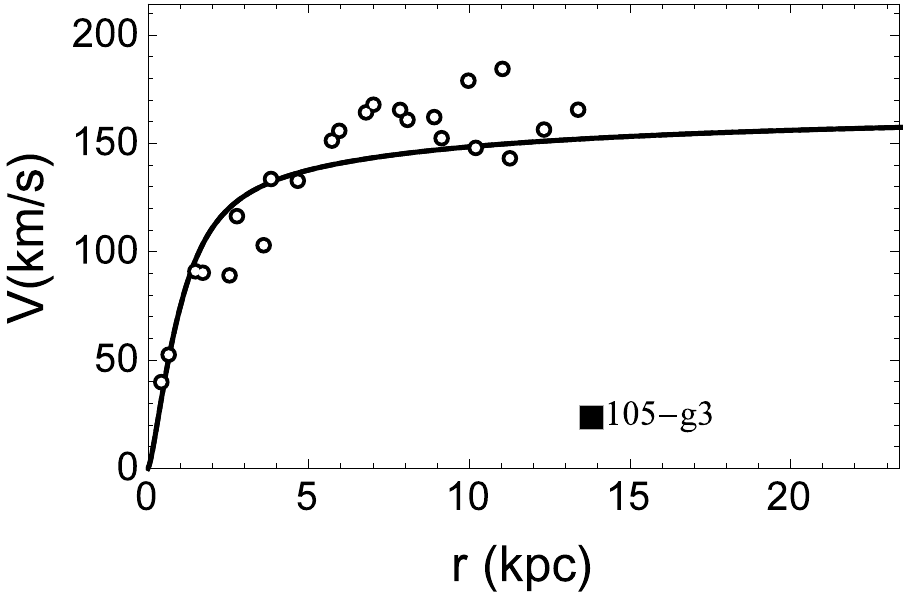}\\
\includegraphics[width=0.50\textwidth]{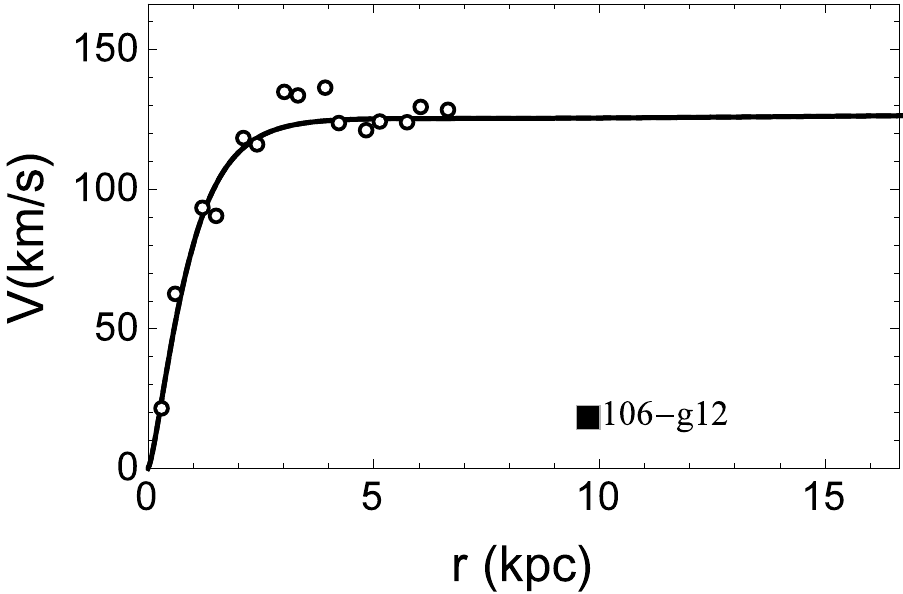}&
\includegraphics[width=0.50\textwidth]{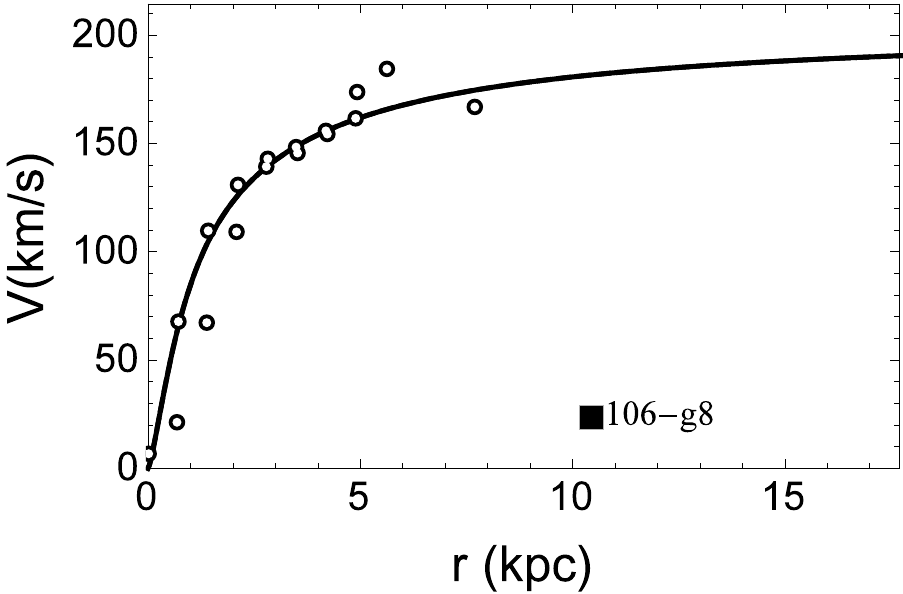}&
\includegraphics[width=0.50\textwidth]{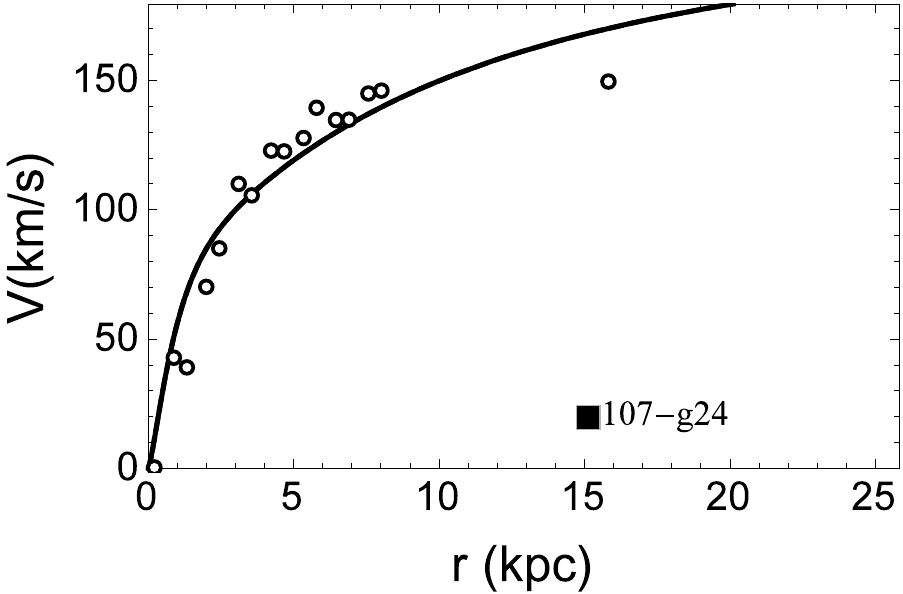}&
\includegraphics[width=0.50\textwidth]{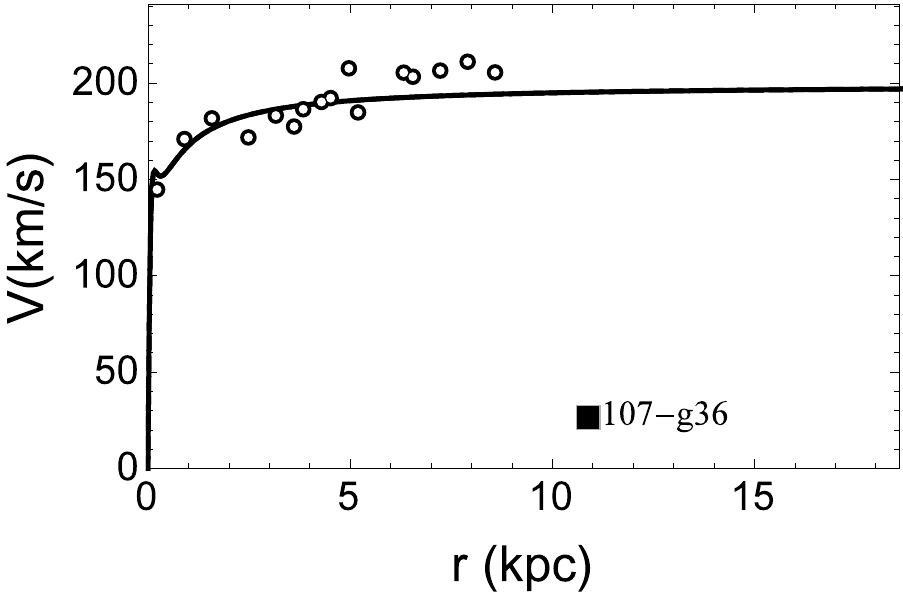}&
\includegraphics[width=0.50\textwidth]{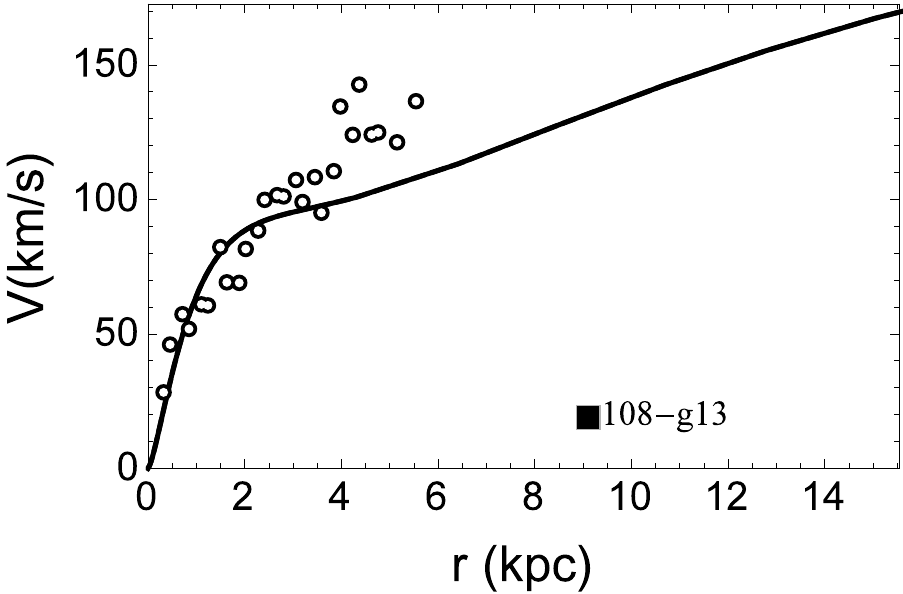}\\
\includegraphics[width=0.50\textwidth]{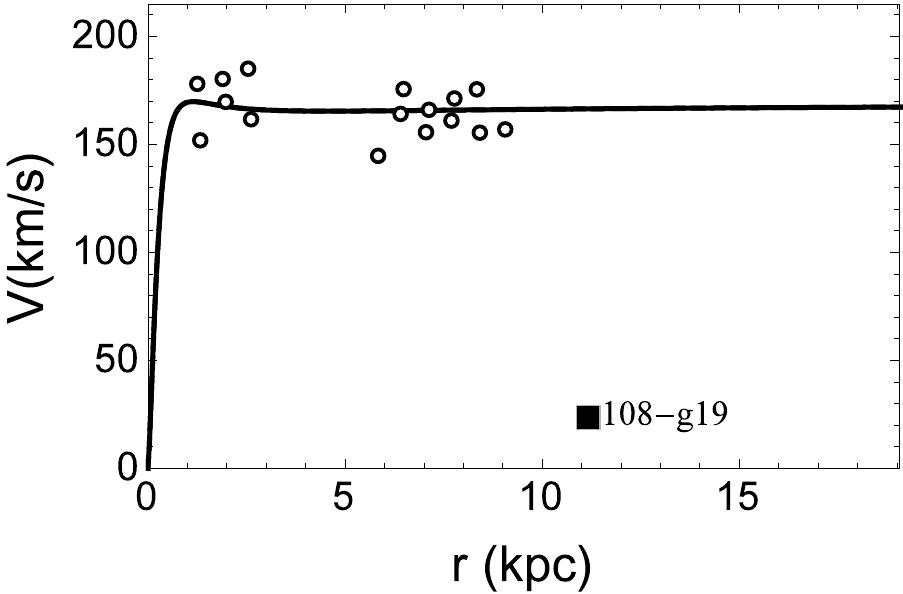}&
\includegraphics[width=0.50\textwidth]{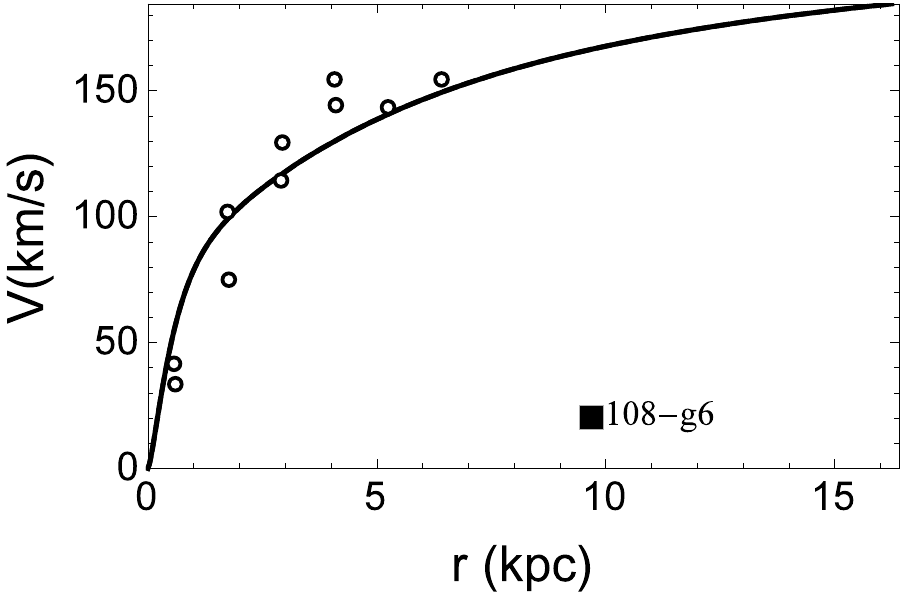}&
\includegraphics[width=0.50\textwidth]{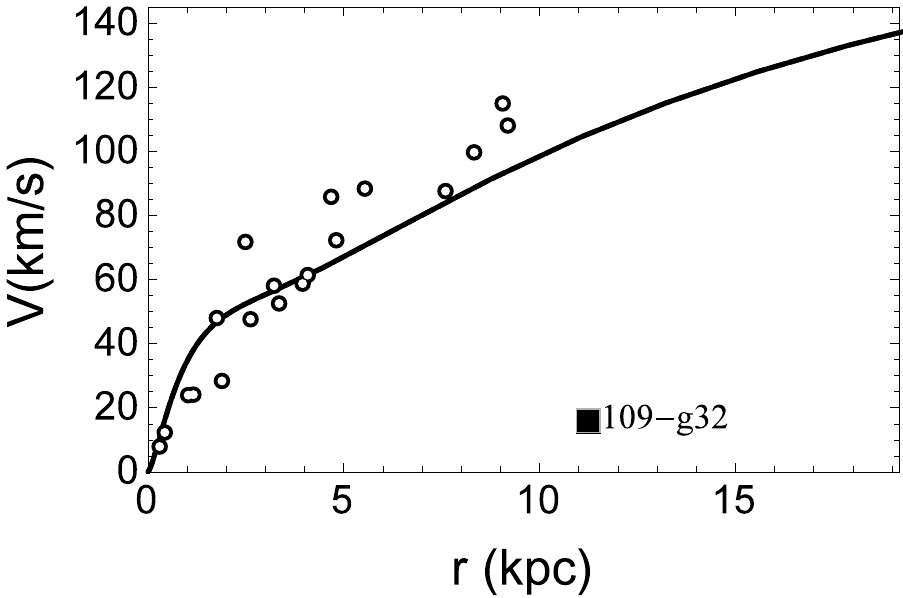}&
\includegraphics[width=0.50\textwidth]{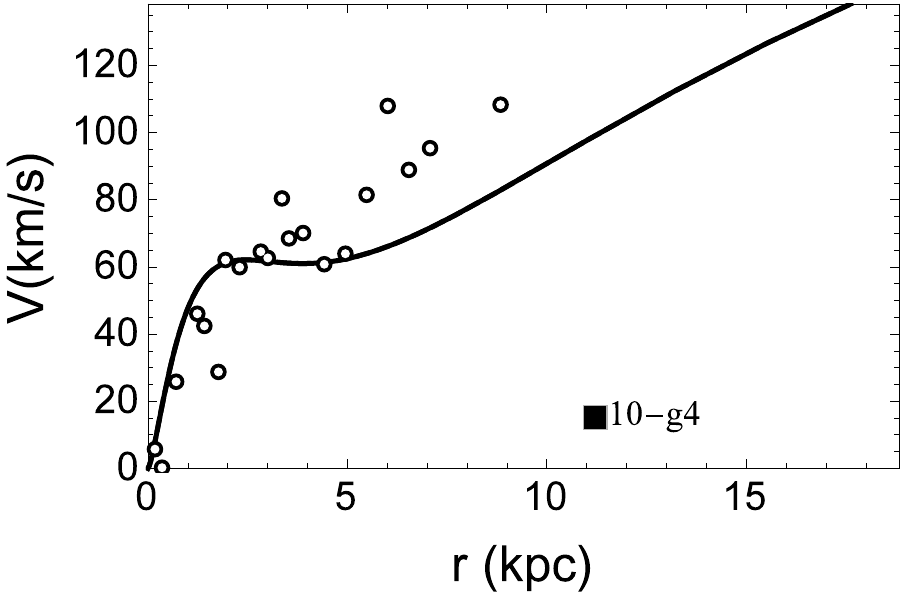}&
\includegraphics[width=0.50\textwidth]{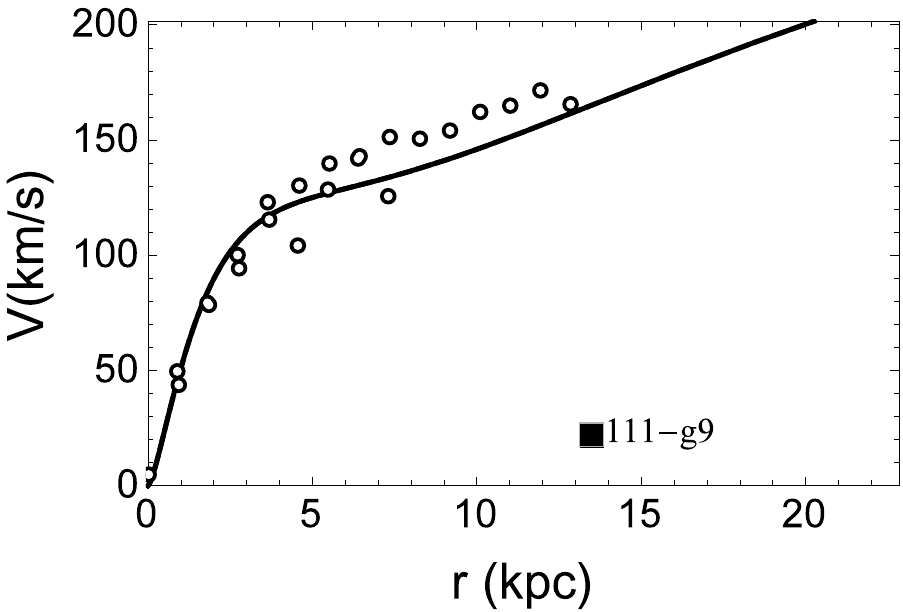}\\
\end{tabular}}}
\end{figure}

\clearpage
\begin{figure}
\flushleft
{\tiny\renewcommand{\arraystretch}{.95}
\resizebox{!}{.64\textwidth}{%
\begin{tabular}{lllll}
\includegraphics[width=0.50\textwidth]{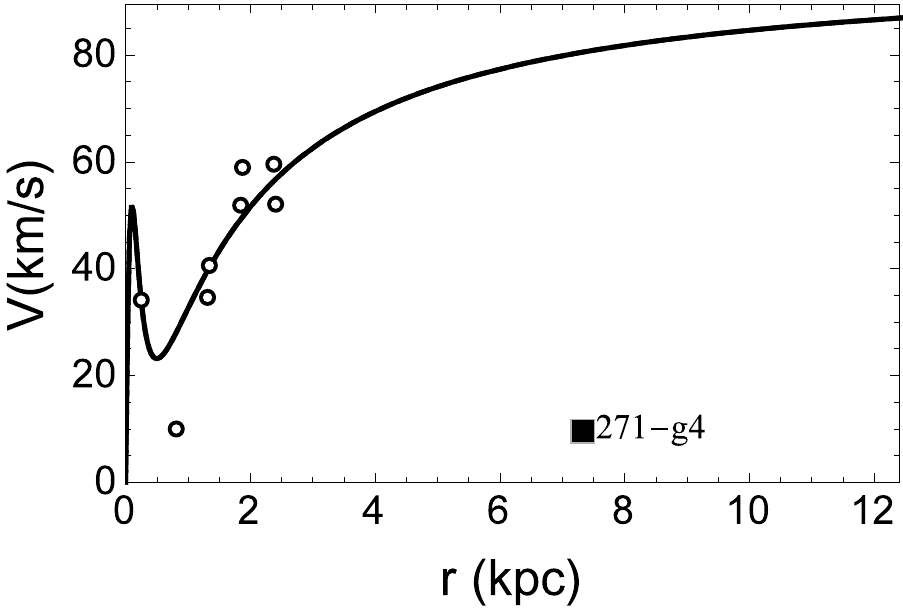}&
\includegraphics[width=0.50\textwidth]{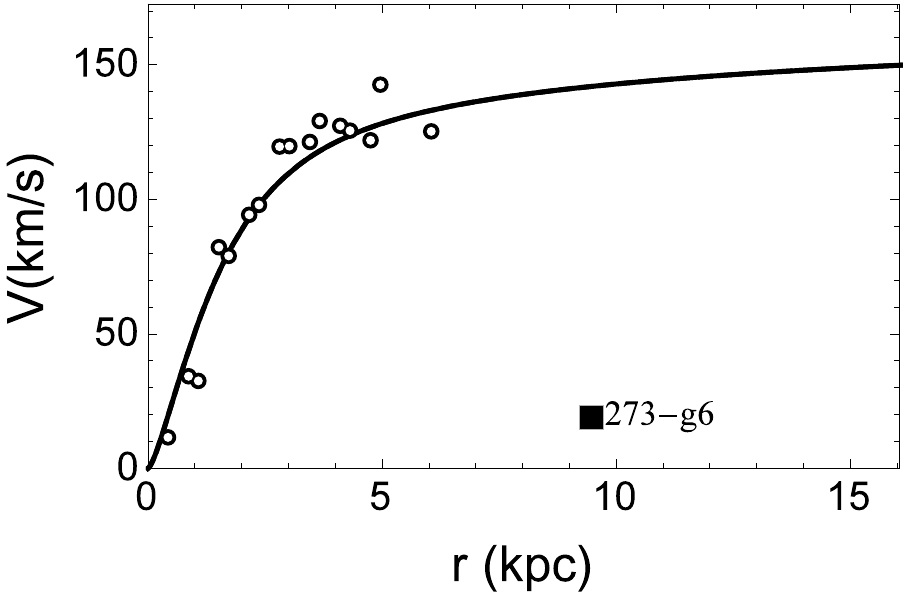}&
\includegraphics[width=0.50\textwidth]{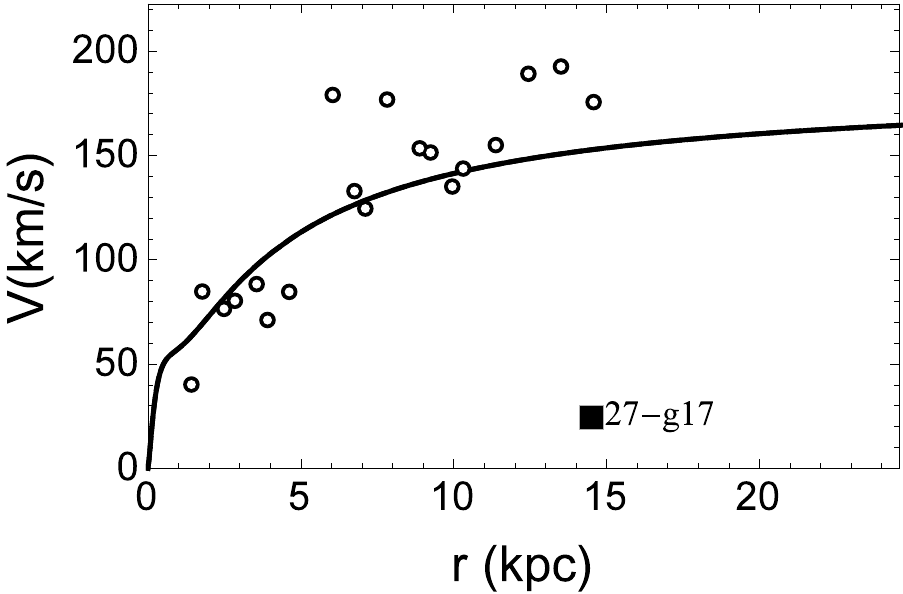}&
\includegraphics[width=0.50\textwidth]{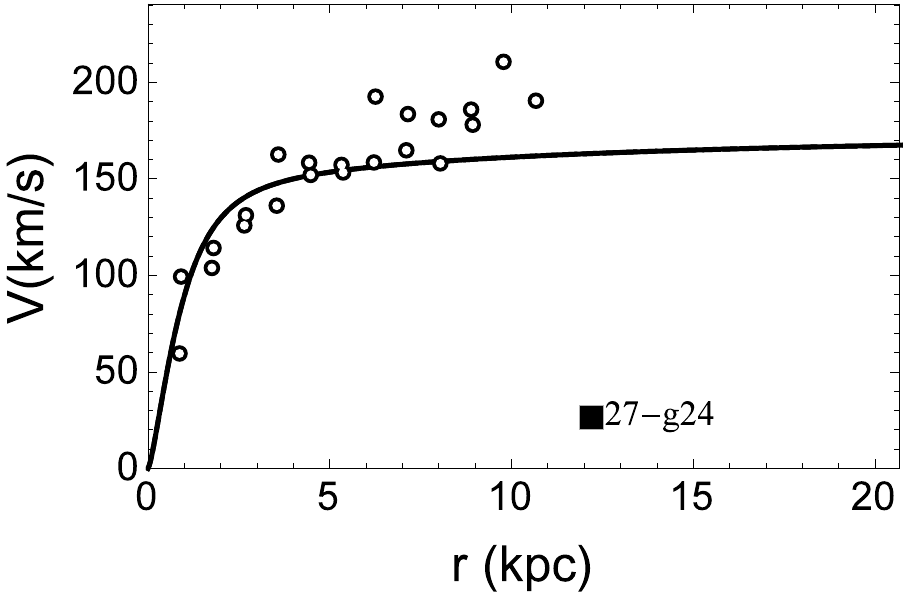}&
\includegraphics[width=0.50\textwidth]{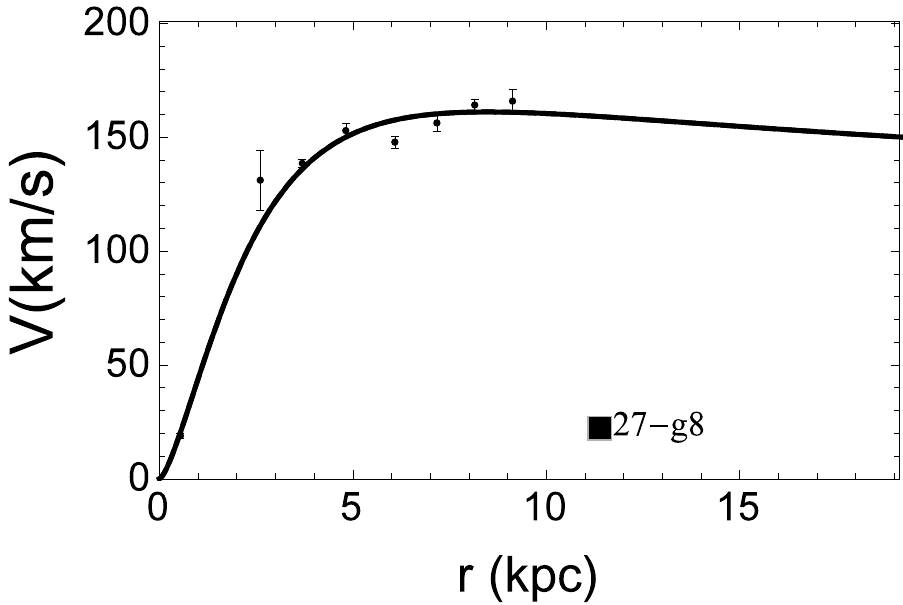}\\
\includegraphics[width=0.50\textwidth]{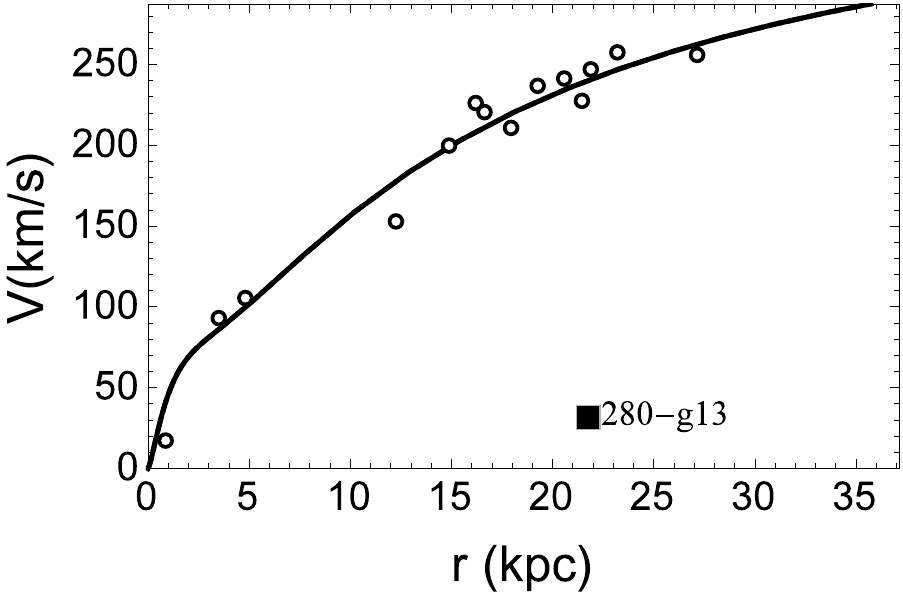}&
\includegraphics[width=0.50\textwidth]{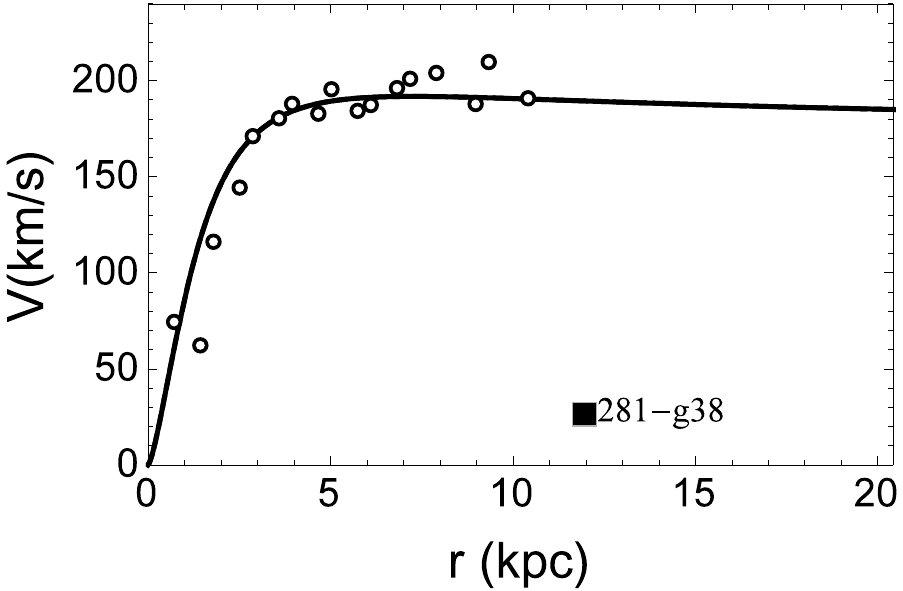}&
\includegraphics[width=0.50\textwidth]{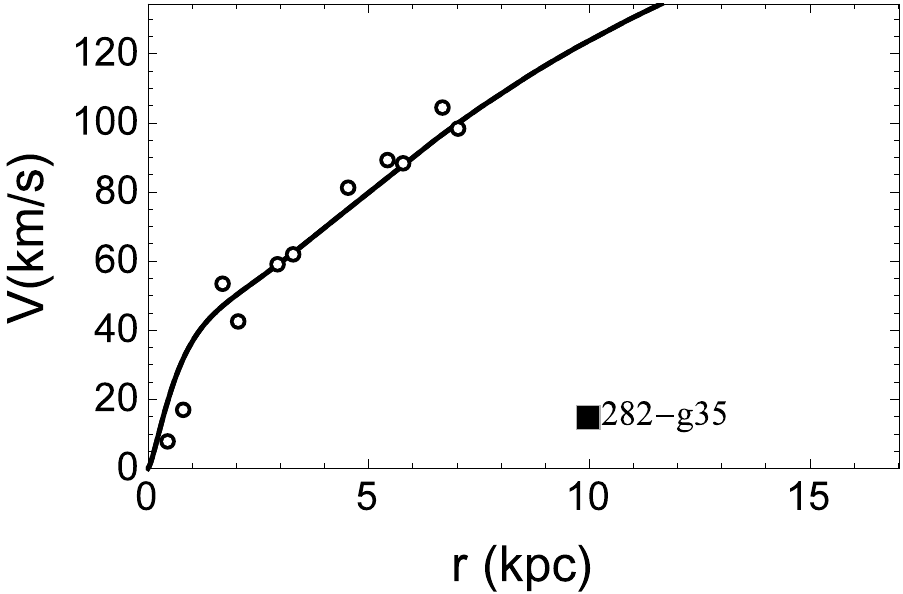}&
\includegraphics[width=0.50\textwidth]{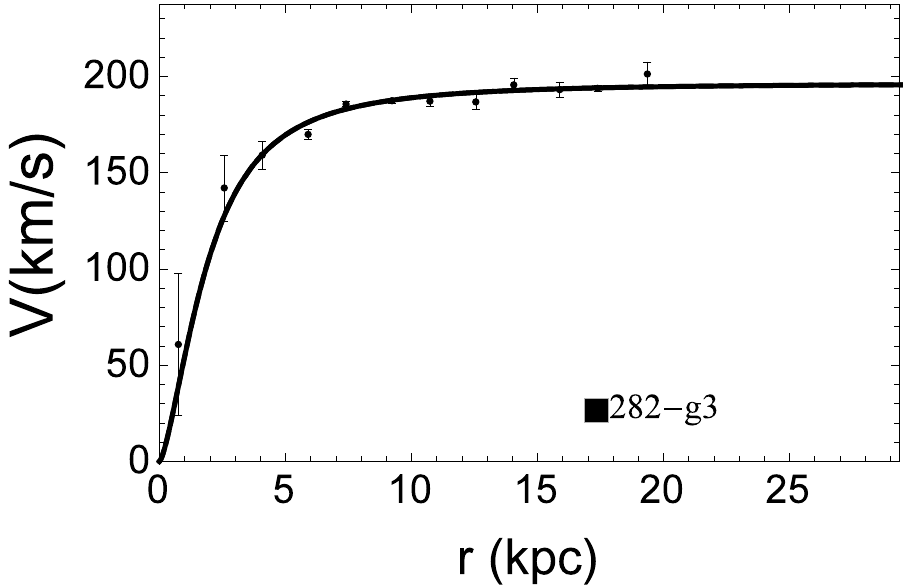}&
\includegraphics[width=0.50\textwidth]{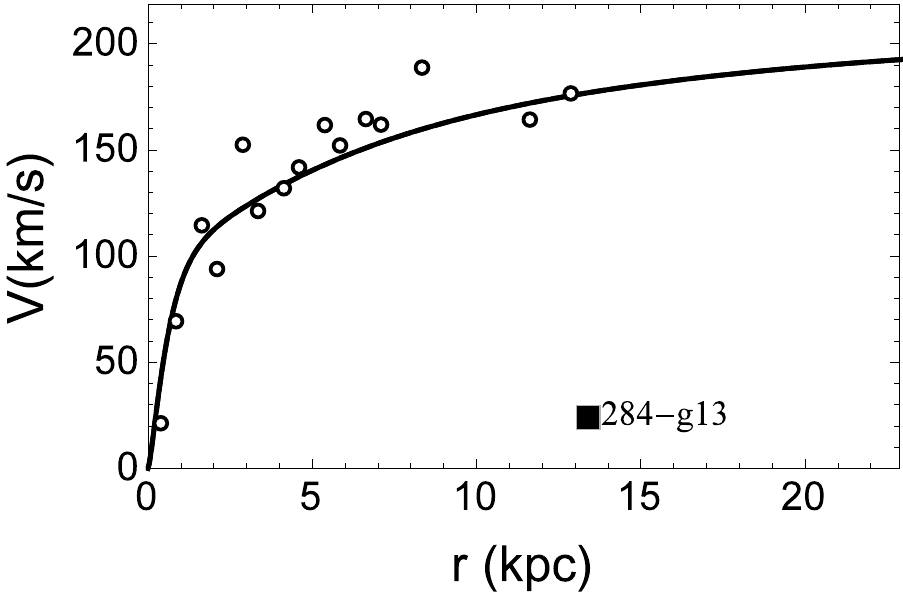}\\
\includegraphics[width=0.50\textwidth]{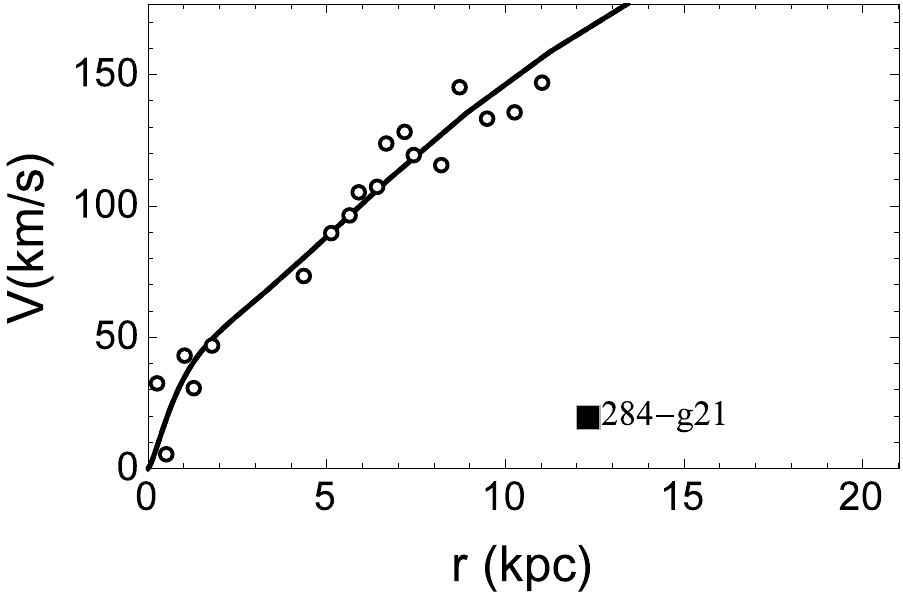}&
\includegraphics[width=0.50\textwidth]{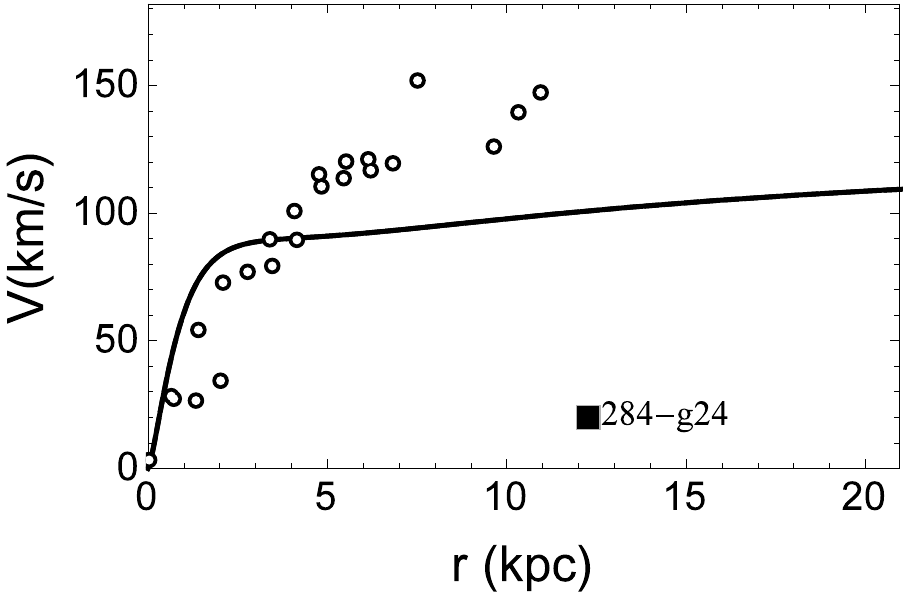}&
\includegraphics[width=0.50\textwidth]{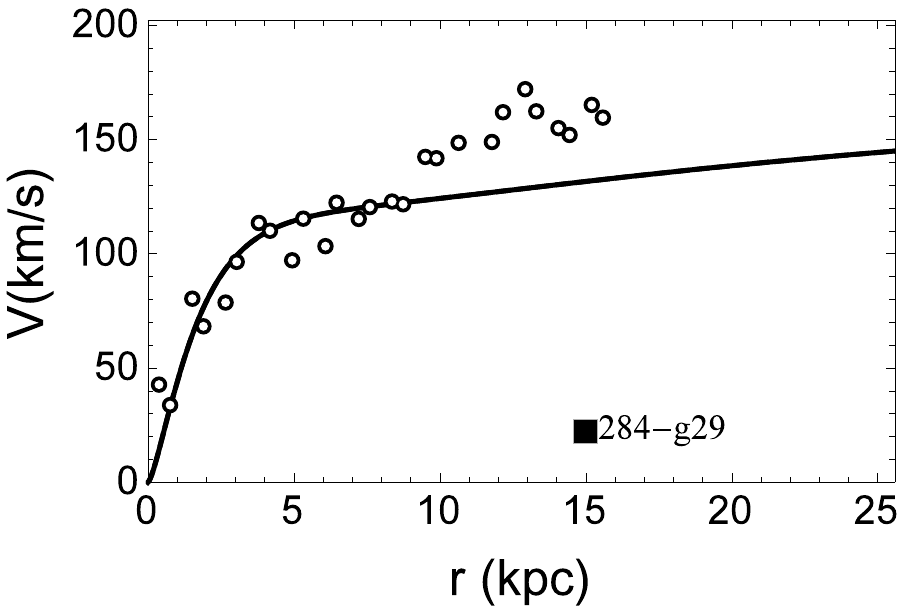}&
\includegraphics[width=0.50\textwidth]{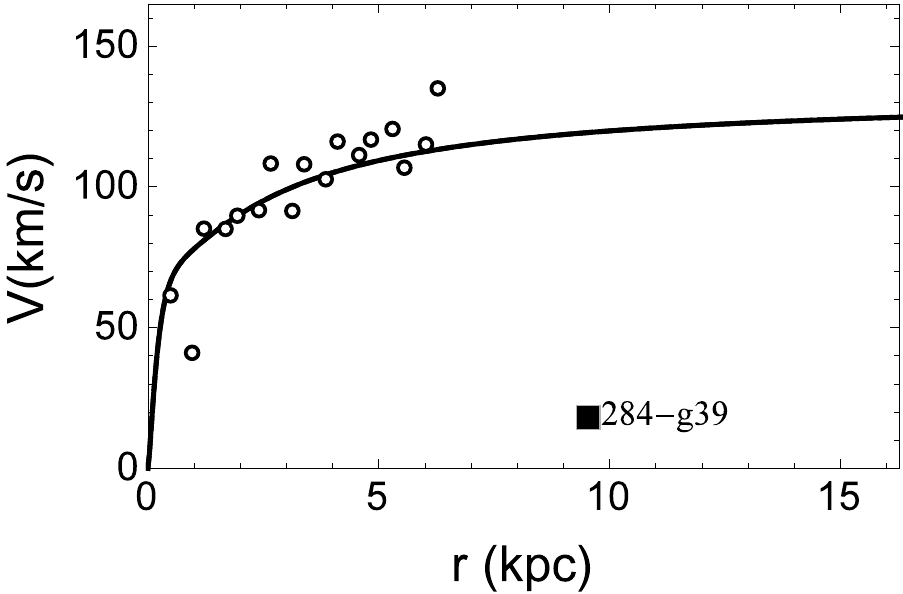}&
\includegraphics[width=0.50\textwidth]{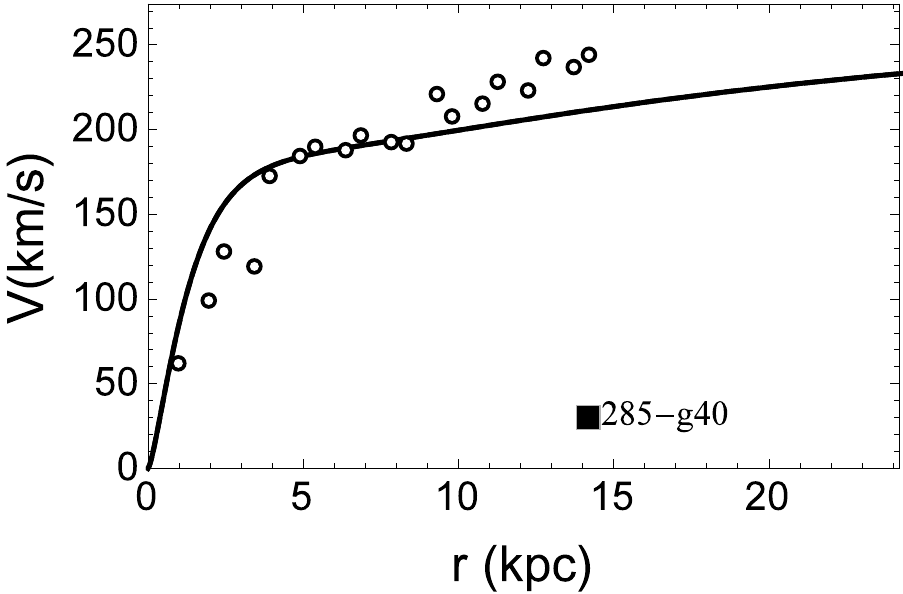}\\
\includegraphics[width=0.50\textwidth]{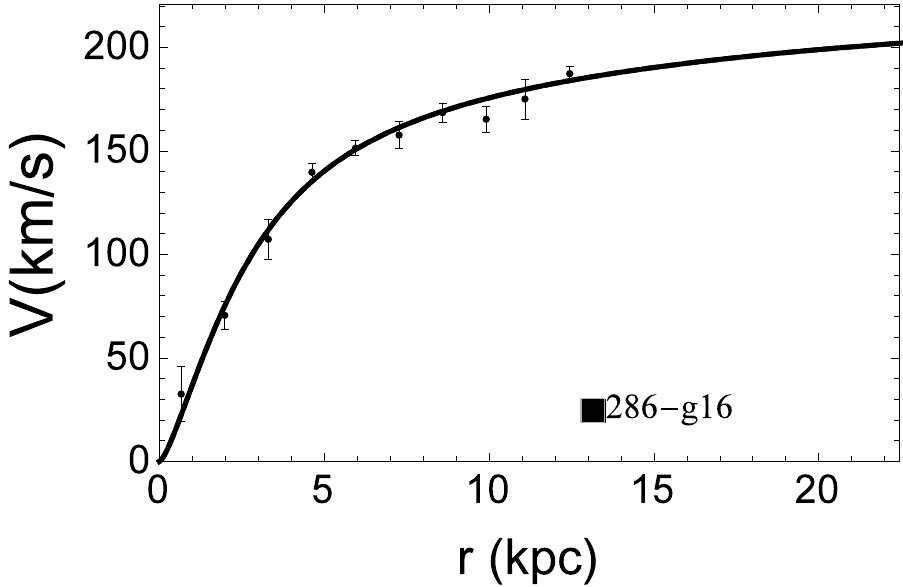}&
\includegraphics[width=0.50\textwidth]{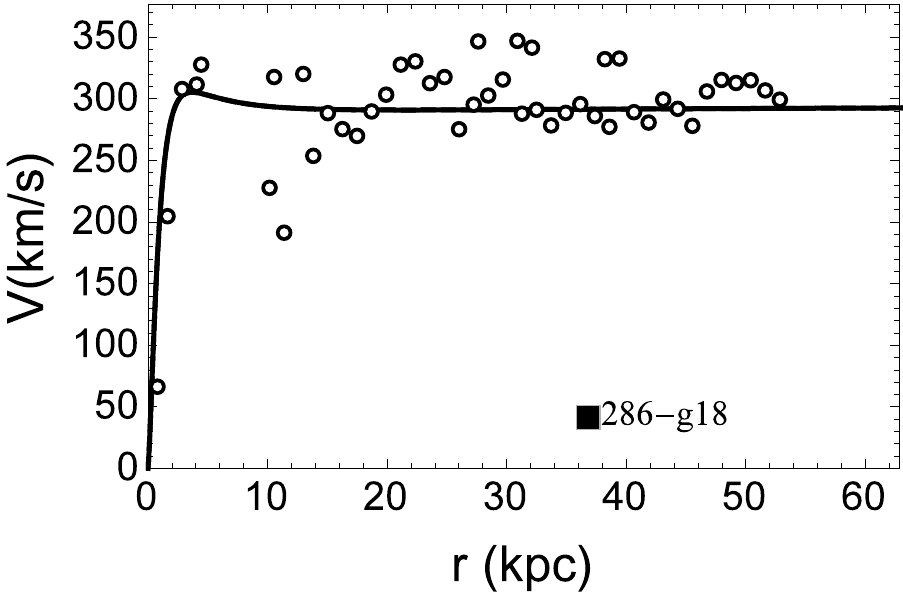}&
\includegraphics[width=0.50\textwidth]{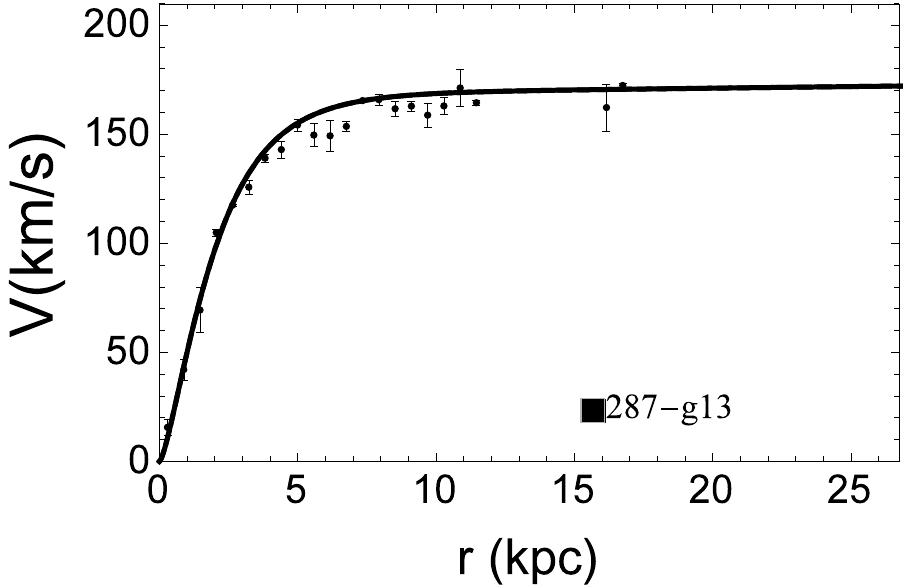}&
\includegraphics[width=0.50\textwidth]{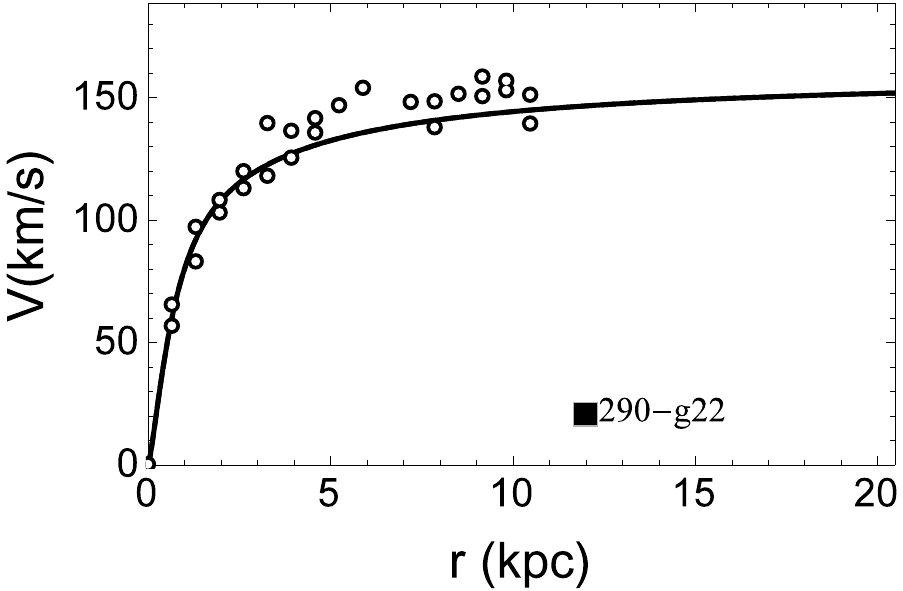}&
\includegraphics[width=0.50\textwidth]{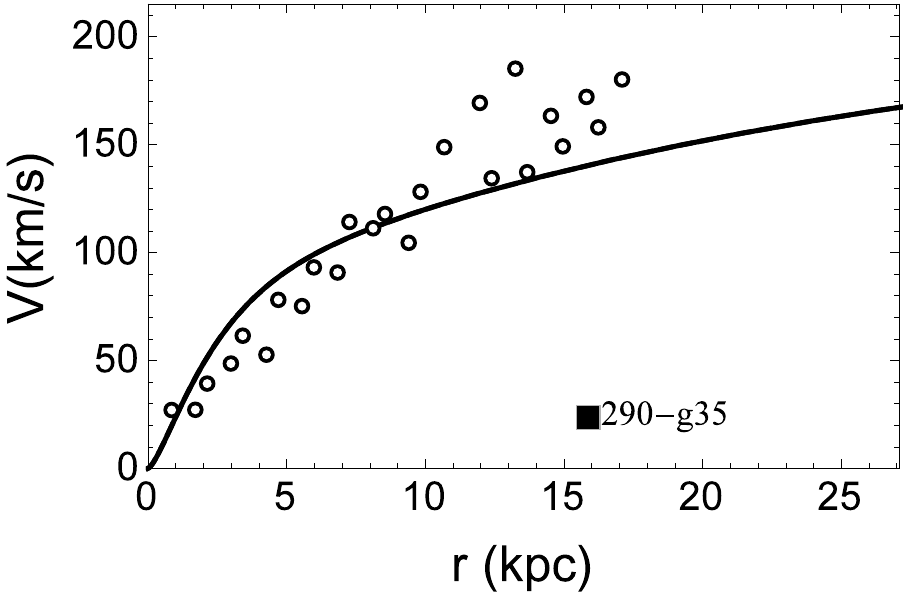}\\
\end{tabular}}}
\end{figure}

\clearpage
\begin{figure}
\flushleft
{\tiny\renewcommand{\arraystretch}{.95}
\resizebox{!}{.64\textwidth}{%
\begin{tabular}{lllll}
\includegraphics[width=0.50\textwidth]{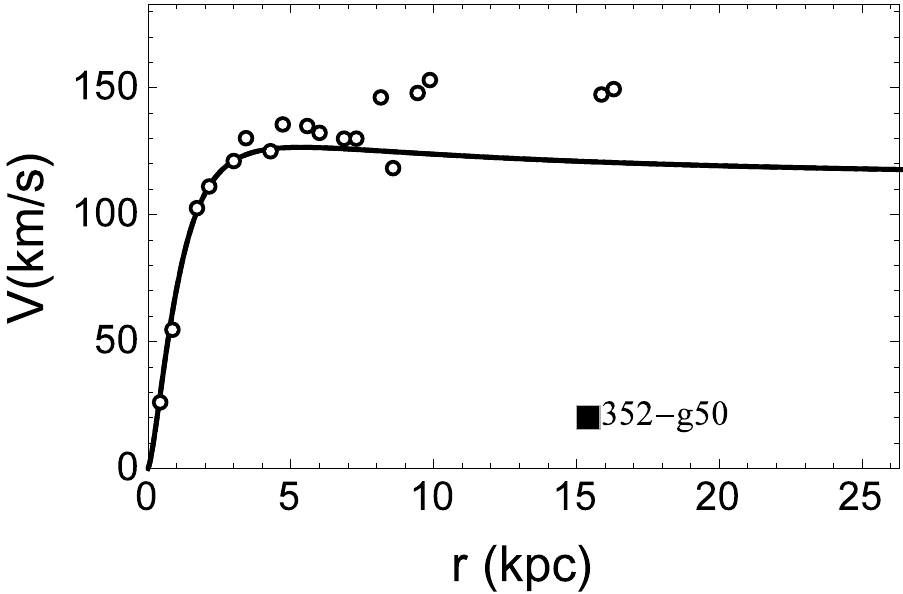}&
\includegraphics[width=0.50\textwidth]{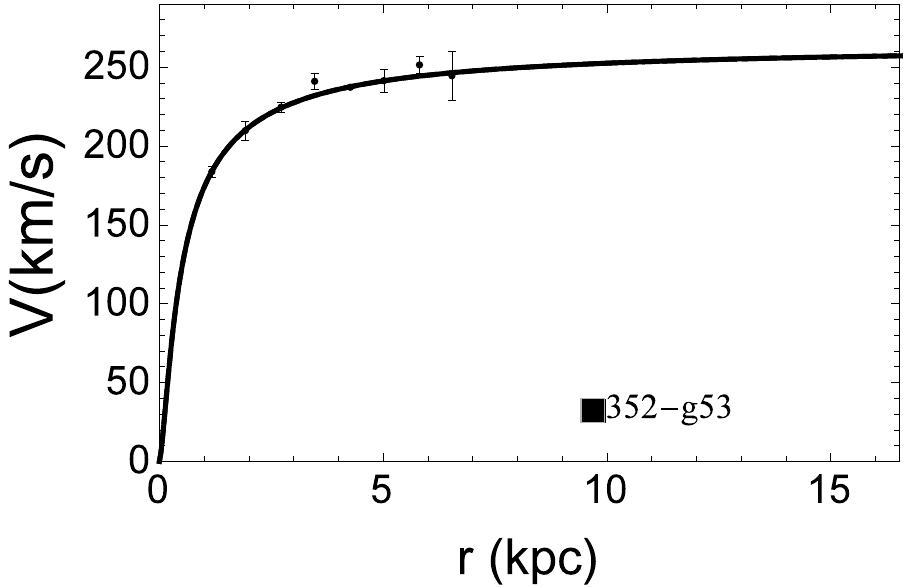}&
\includegraphics[width=0.50\textwidth]{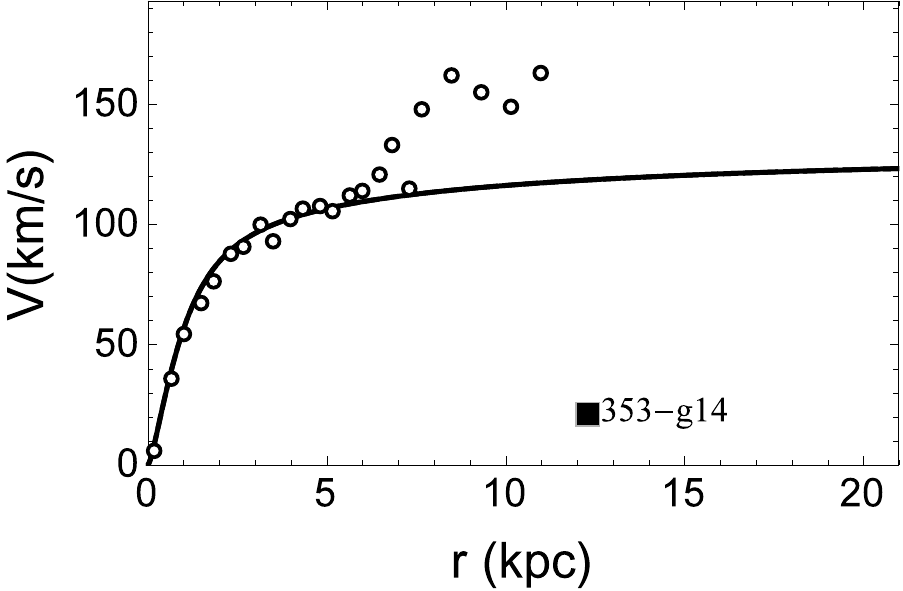}&
\includegraphics[width=0.50\textwidth]{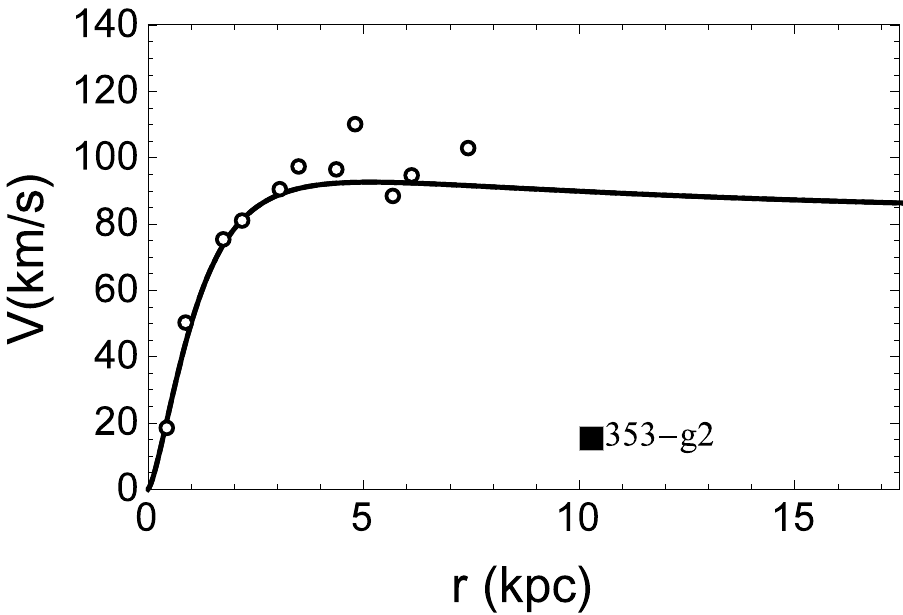}&
\includegraphics[width=0.50\textwidth]{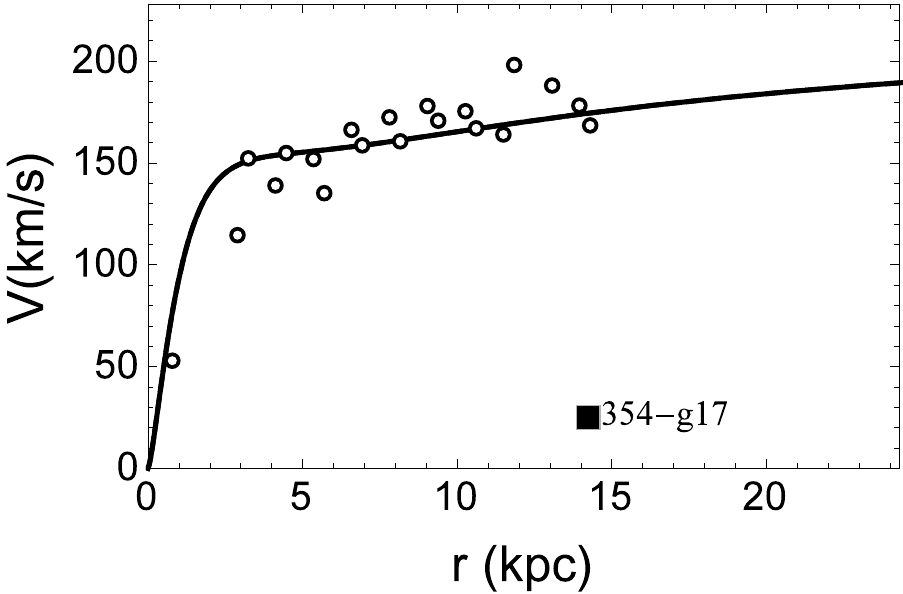}\\
\includegraphics[width=0.50\textwidth]{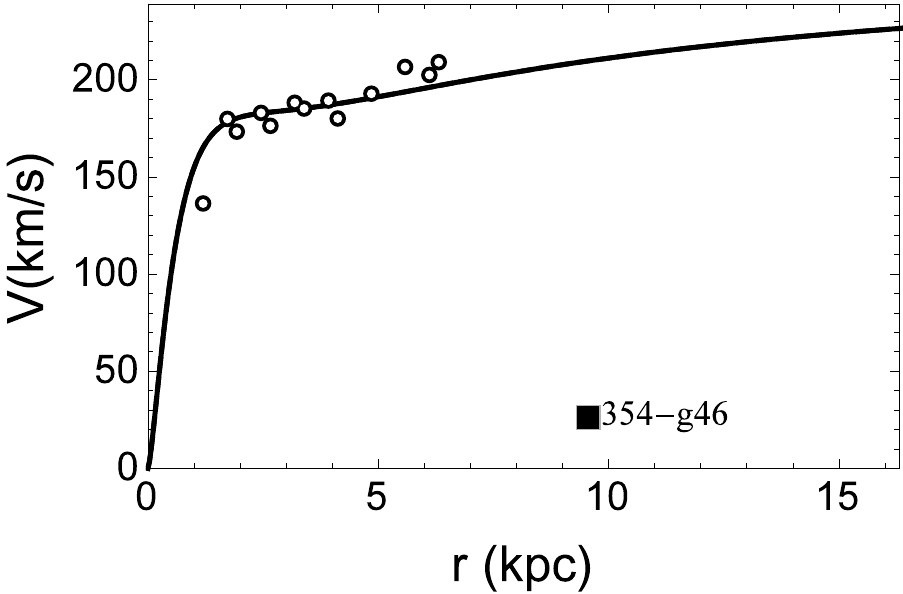}&
\includegraphics[width=0.50\textwidth]{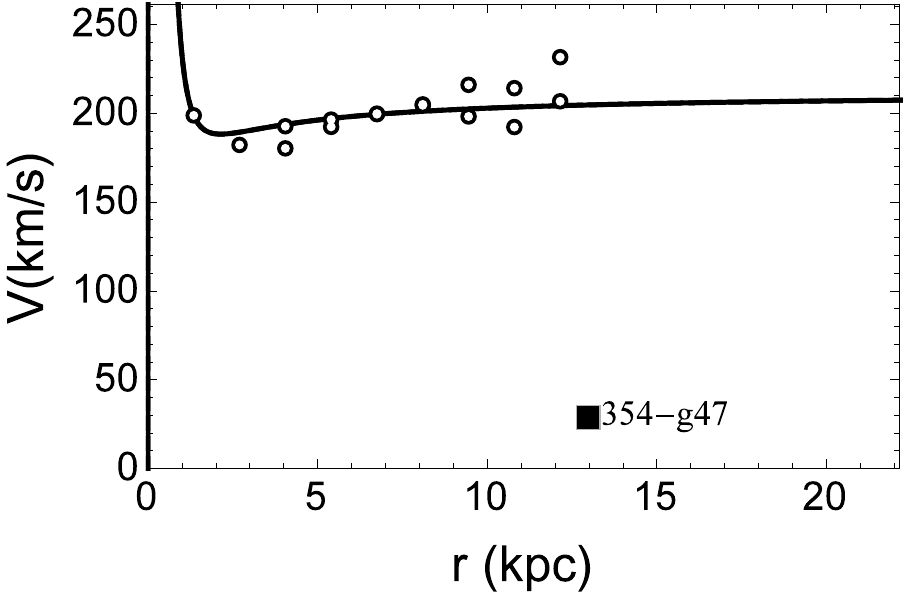}&
\includegraphics[width=0.50\textwidth]{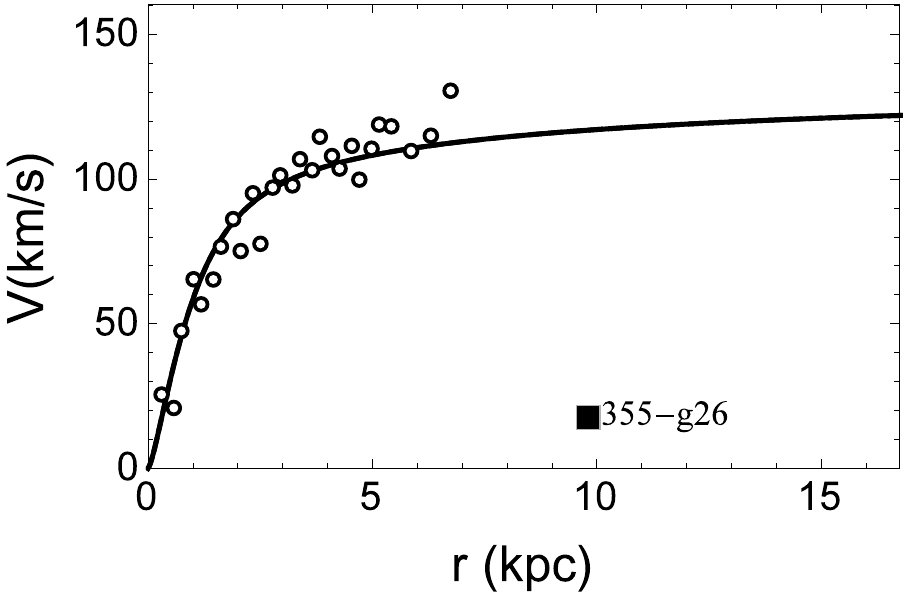}&
\includegraphics[width=0.50\textwidth]{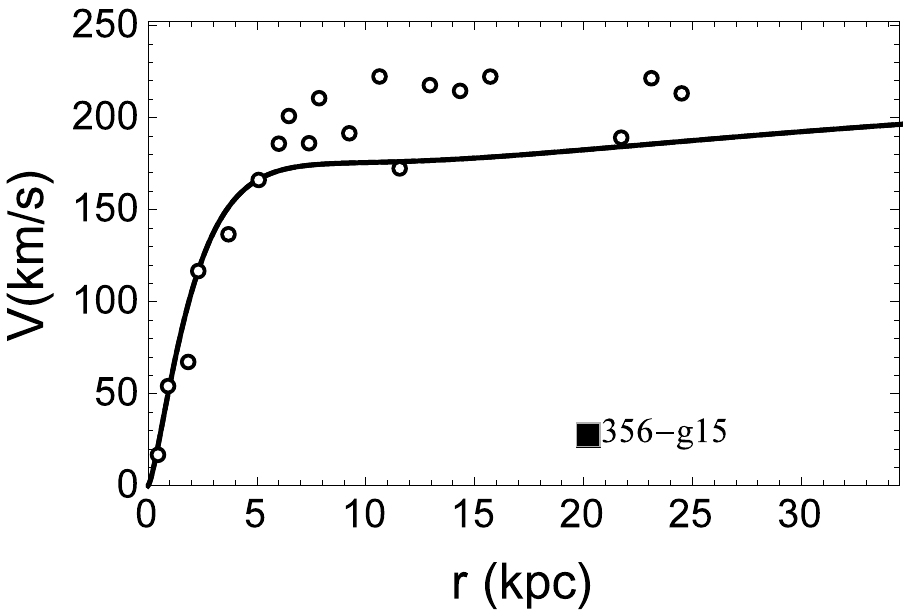}&
\includegraphics[width=0.50\textwidth]{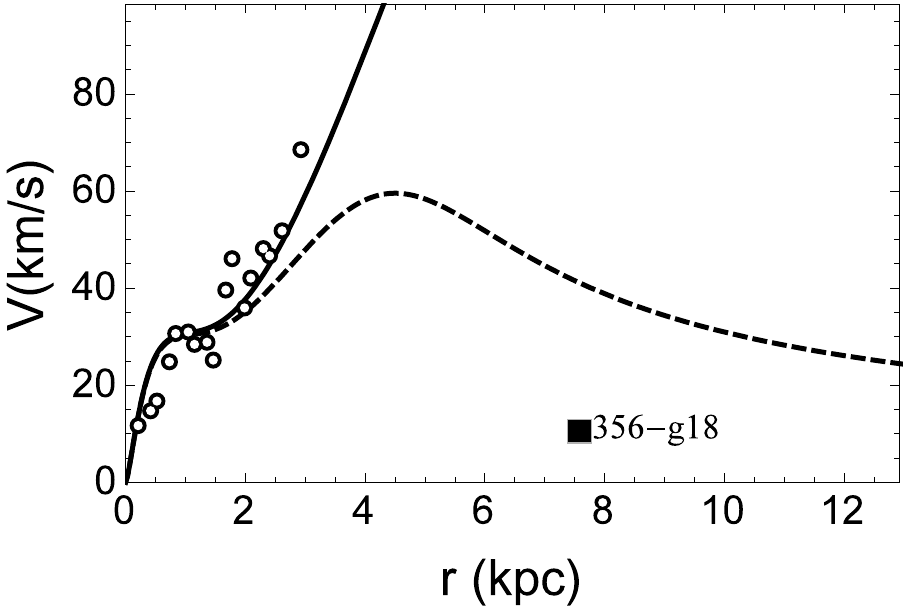}\\
\includegraphics[width=0.50\textwidth]{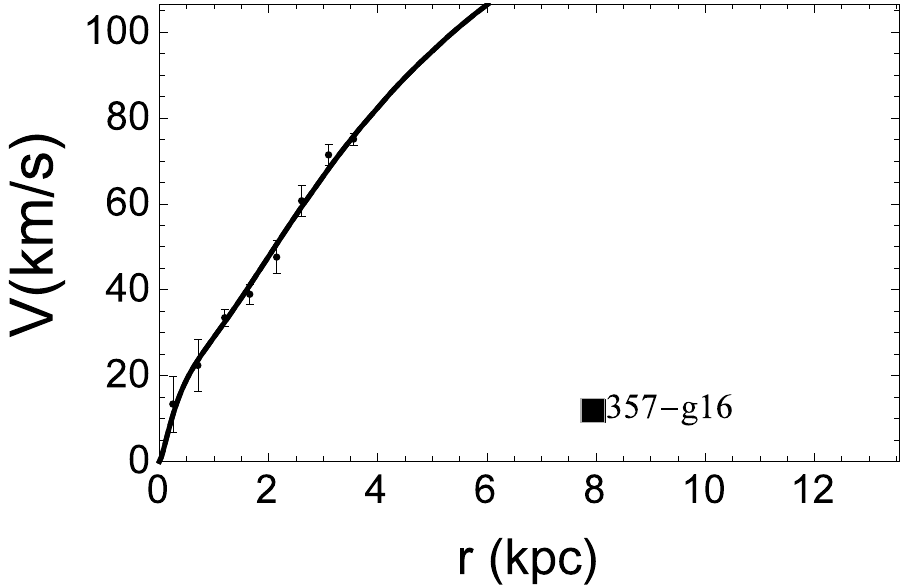}&
\includegraphics[width=0.50\textwidth]{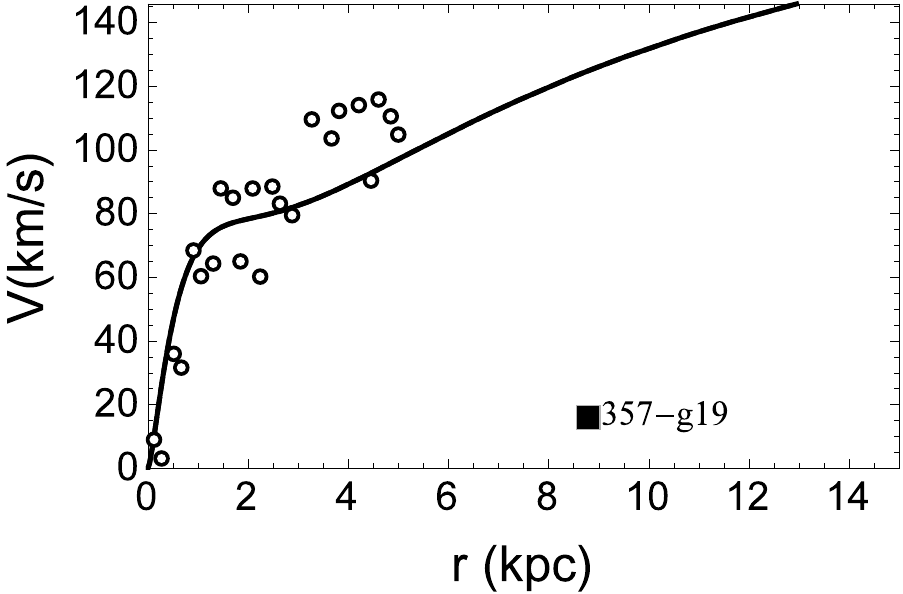}&
\includegraphics[width=0.50\textwidth]{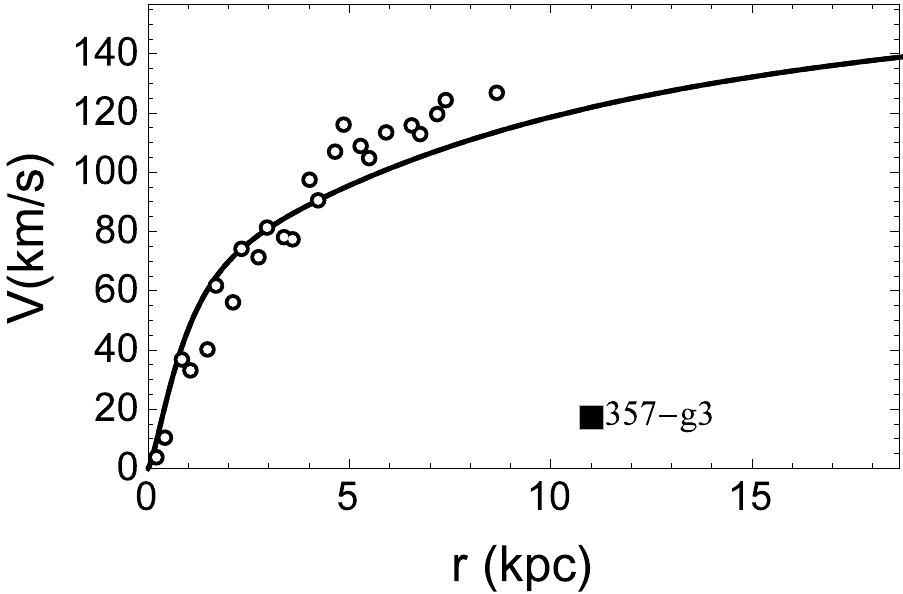}&
\includegraphics[width=0.50\textwidth]{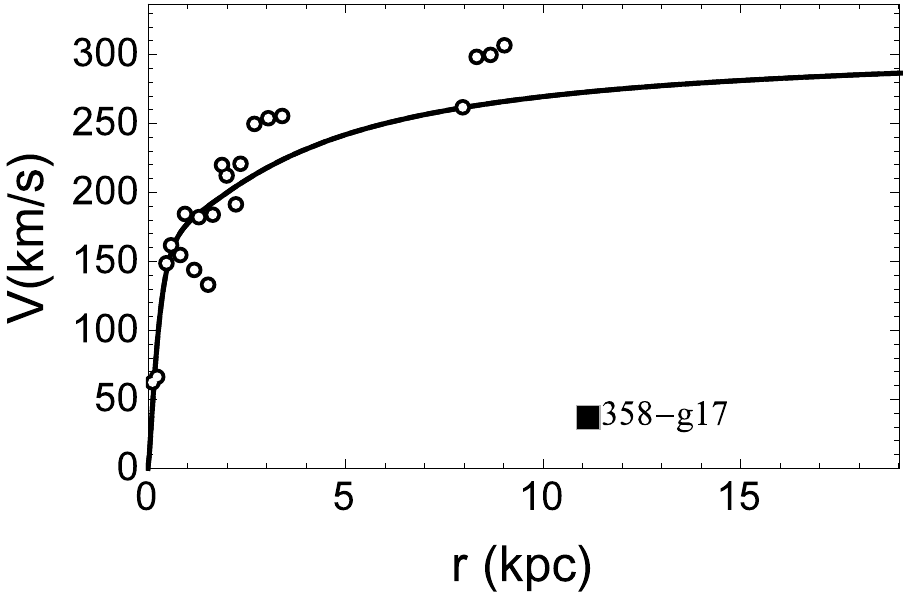}&
\includegraphics[width=0.50\textwidth]{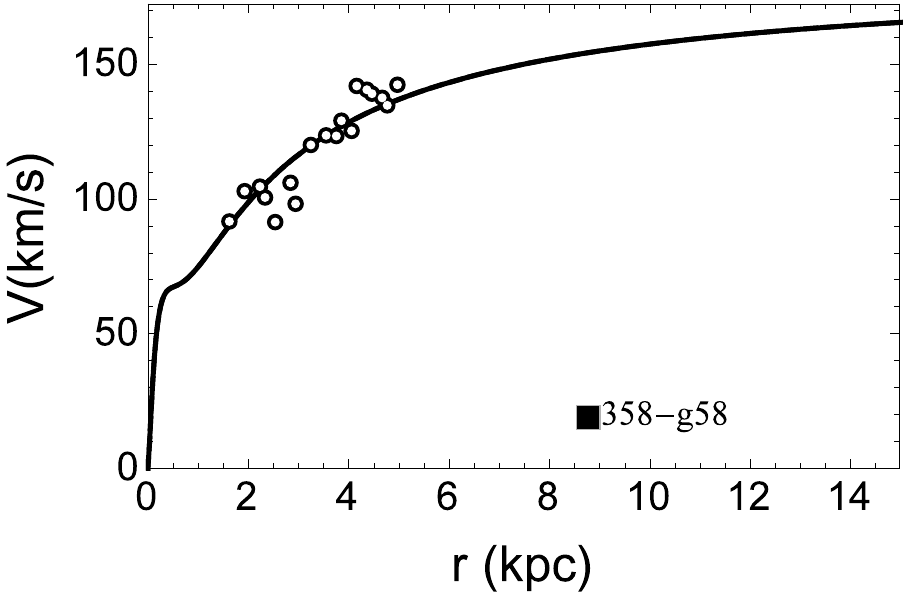}\\
\includegraphics[width=0.50\textwidth]{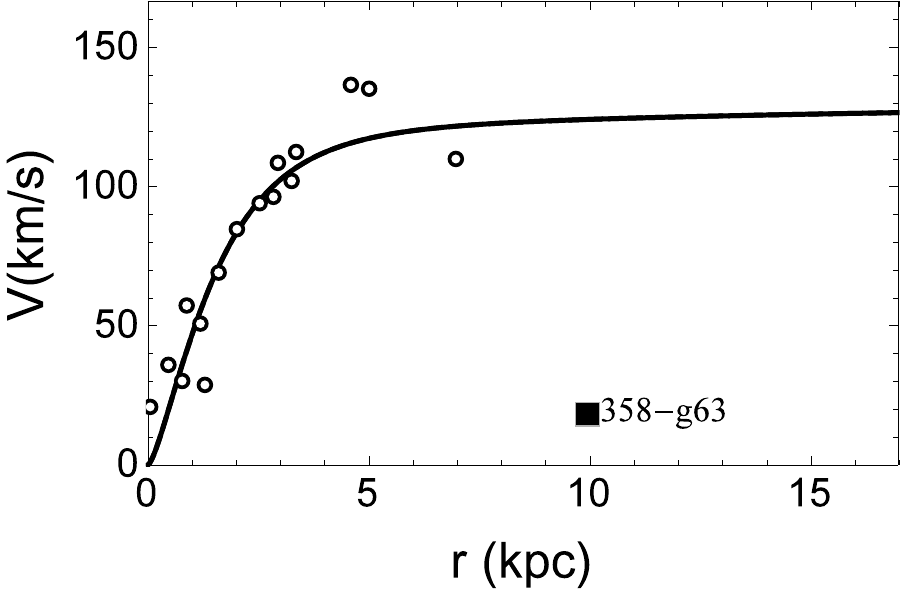}&
\includegraphics[width=0.50\textwidth]{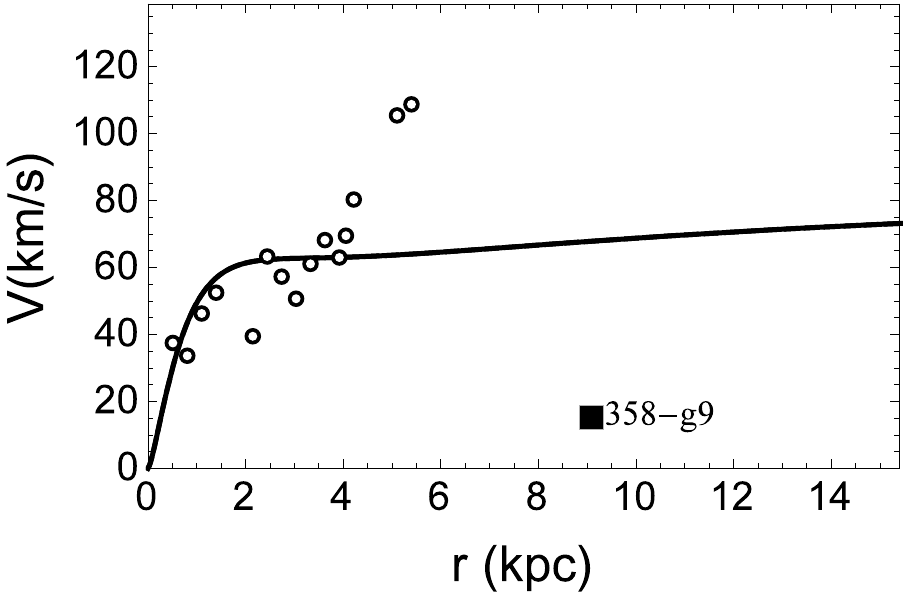}&
\includegraphics[width=0.50\textwidth]{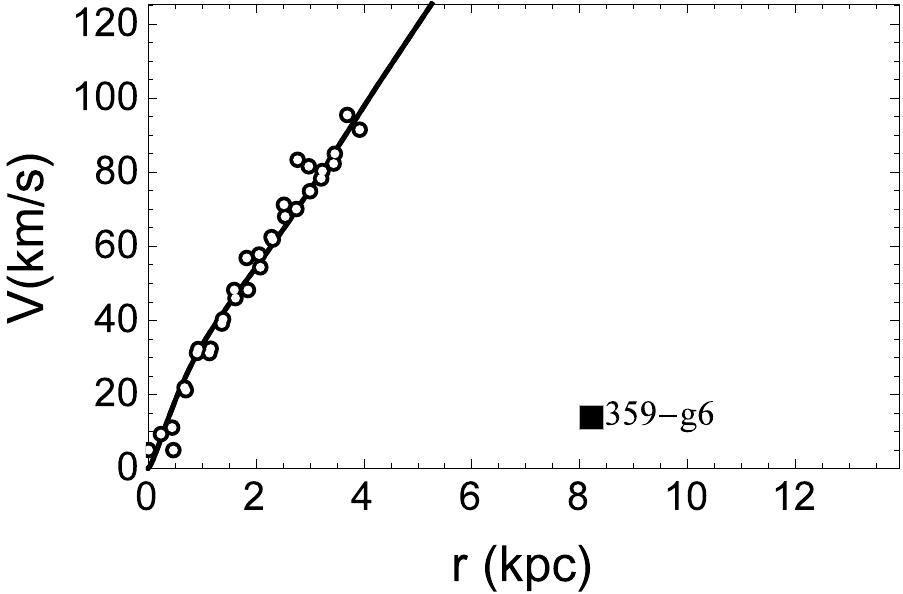}&
\includegraphics[width=0.50\textwidth]{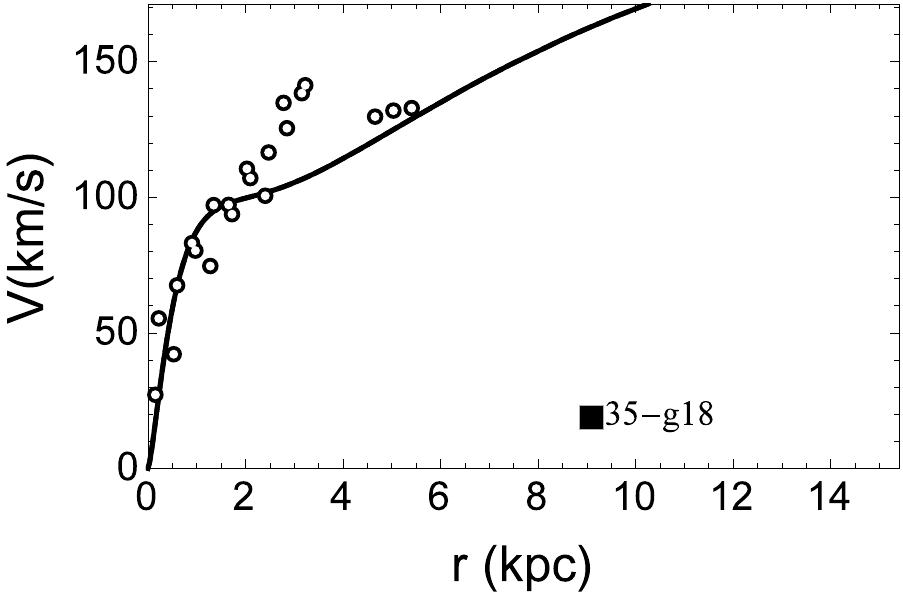}&
\includegraphics[width=0.50\textwidth]{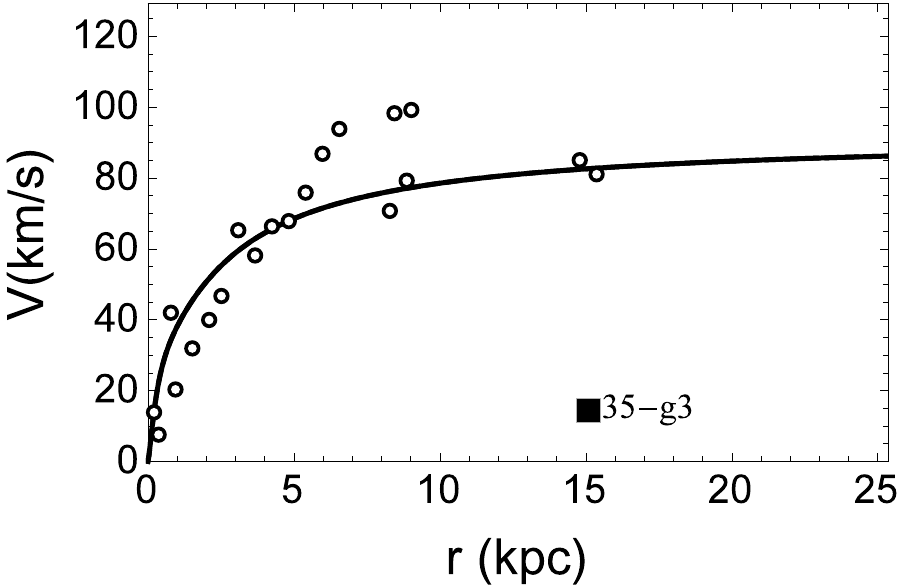}\\
\end{tabular}}}
\end{figure}

\clearpage
\begin{figure}
\flushleft
{\tiny\renewcommand{\arraystretch}{.95}
\resizebox{!}{.64\textwidth}{%
\begin{tabular}{lllll}
\includegraphics[width=0.50\textwidth]{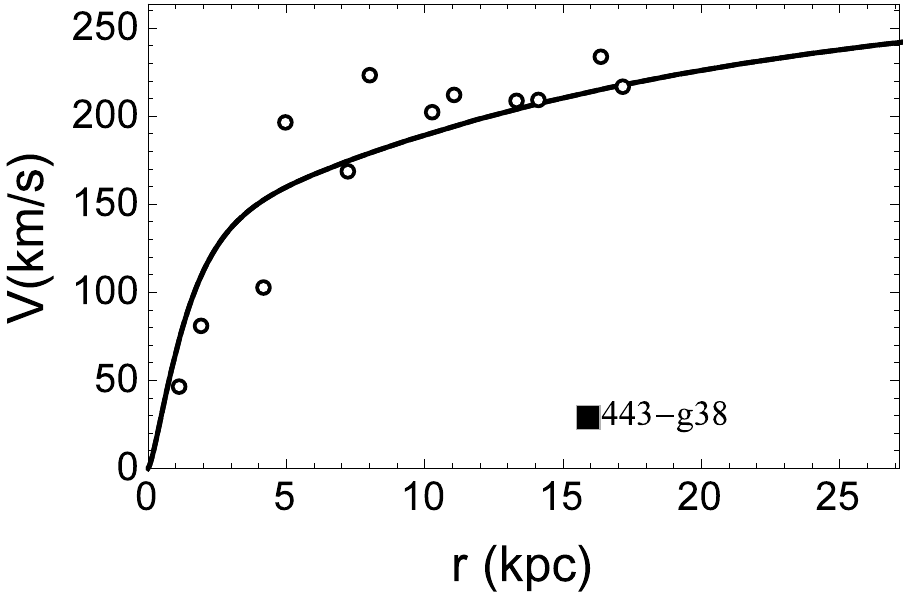}&
\includegraphics[width=0.50\textwidth]{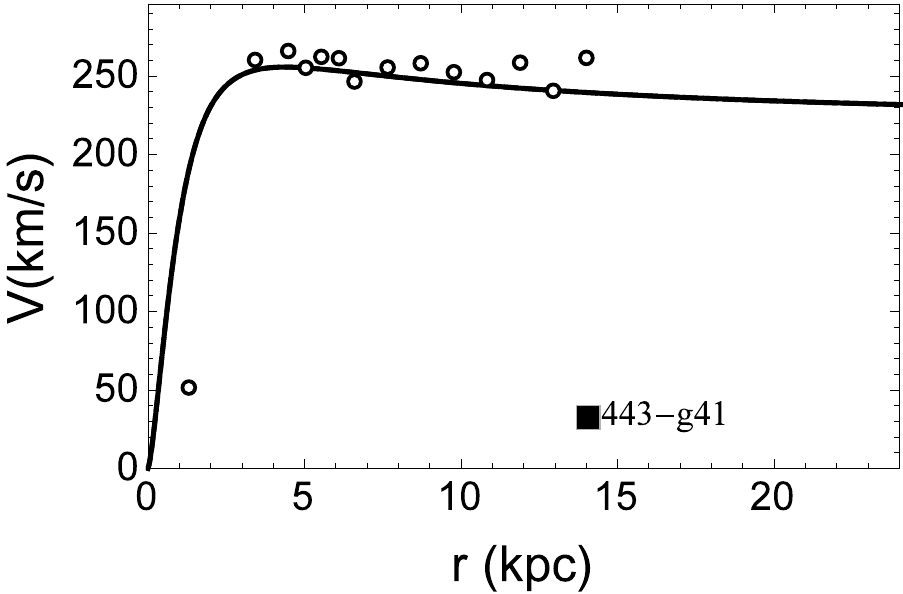}&
\includegraphics[width=0.50\textwidth]{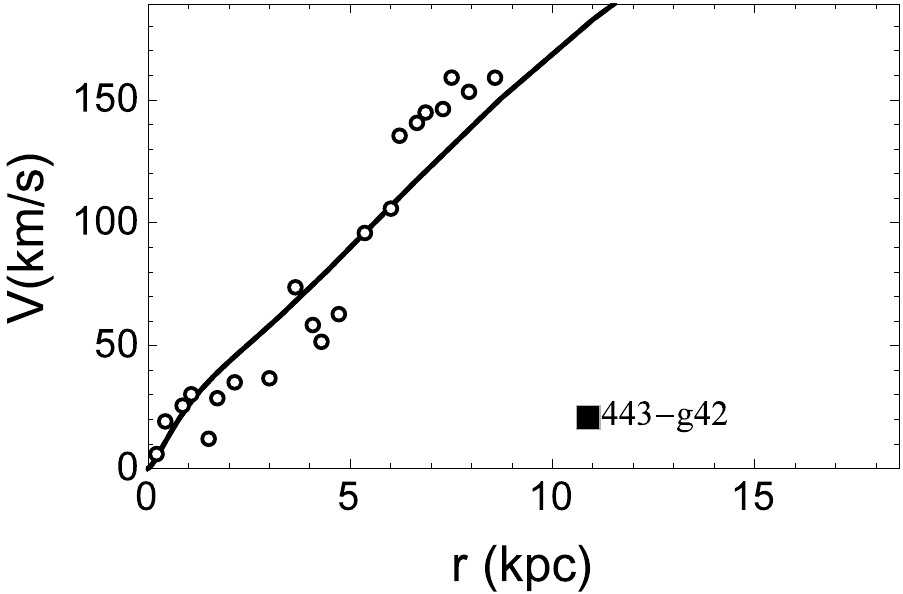}&
\includegraphics[width=0.50\textwidth]{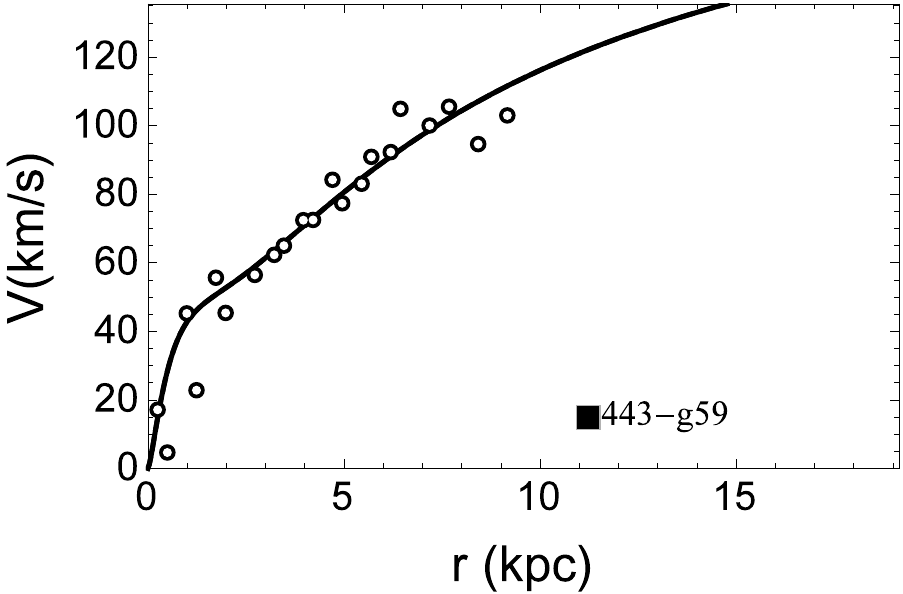}&
\includegraphics[width=0.50\textwidth]{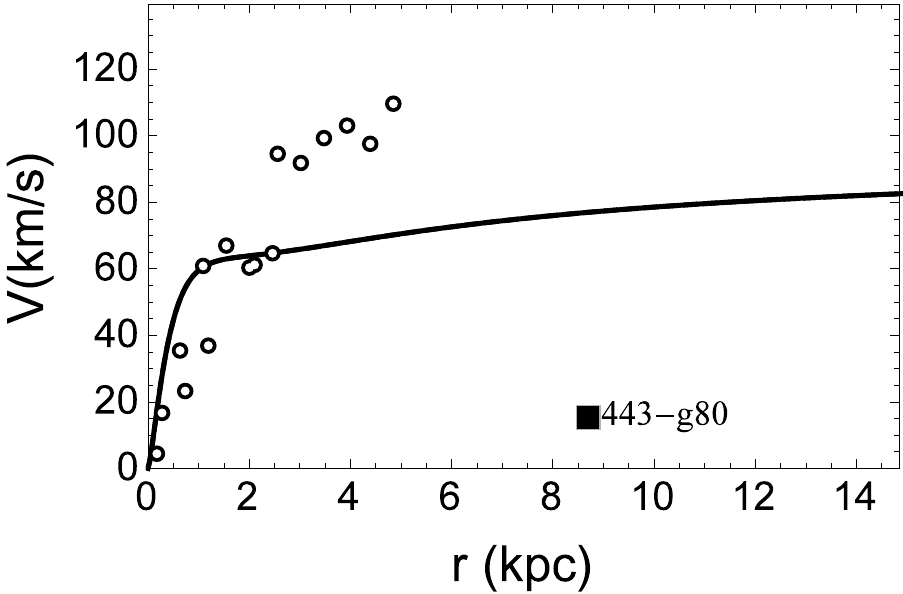}\\
\includegraphics[width=0.50\textwidth]{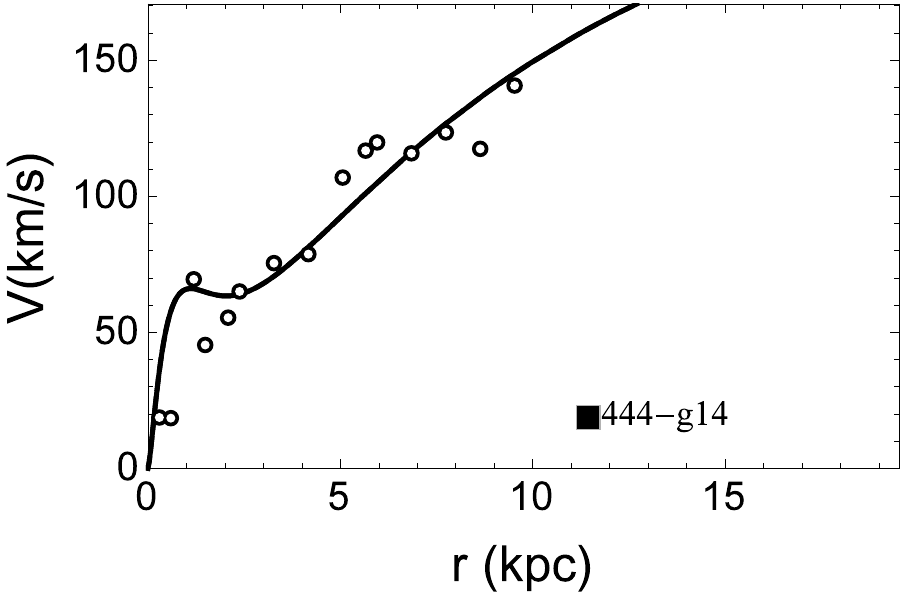}&
\includegraphics[width=0.50\textwidth]{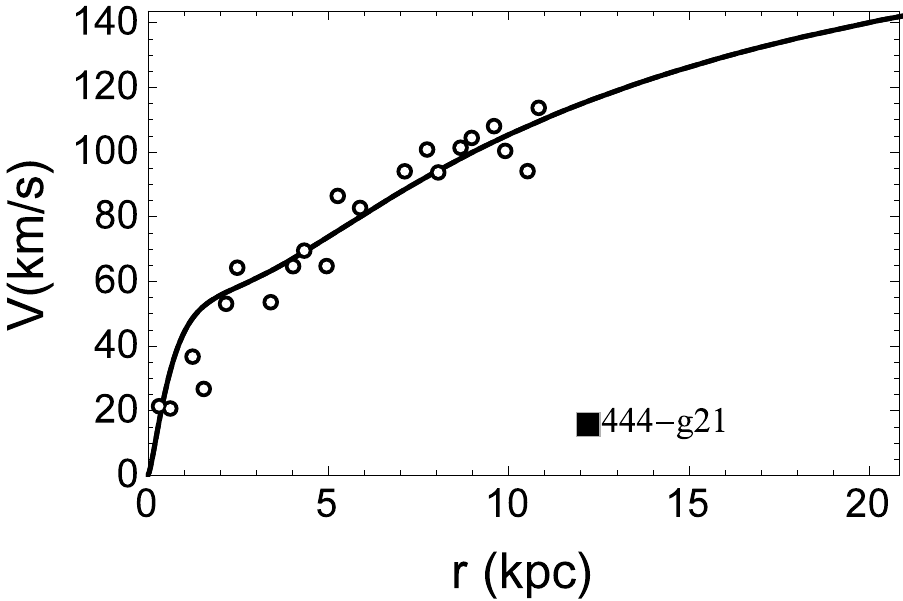}&
\includegraphics[width=0.50\textwidth]{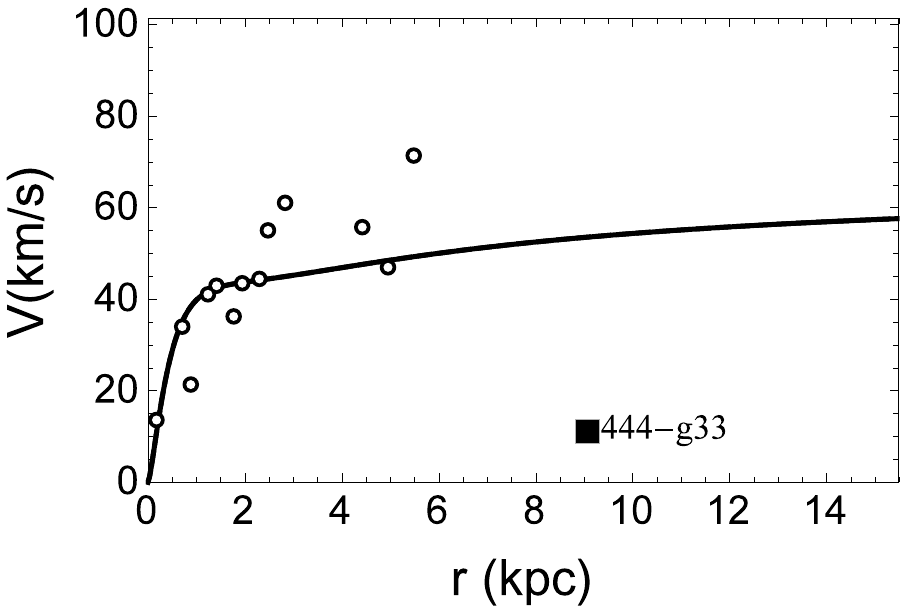}&
\includegraphics[width=0.50\textwidth]{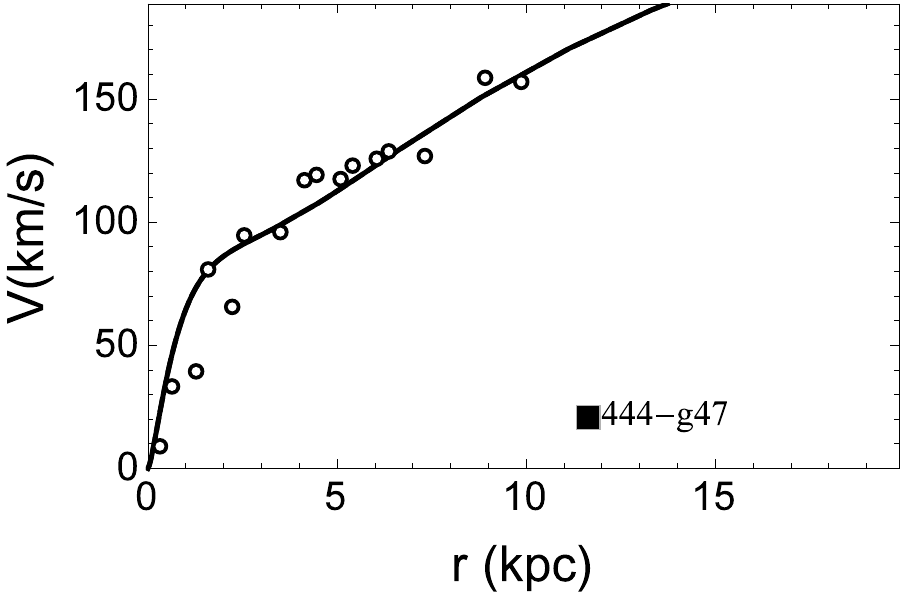}&
\includegraphics[width=0.50\textwidth]{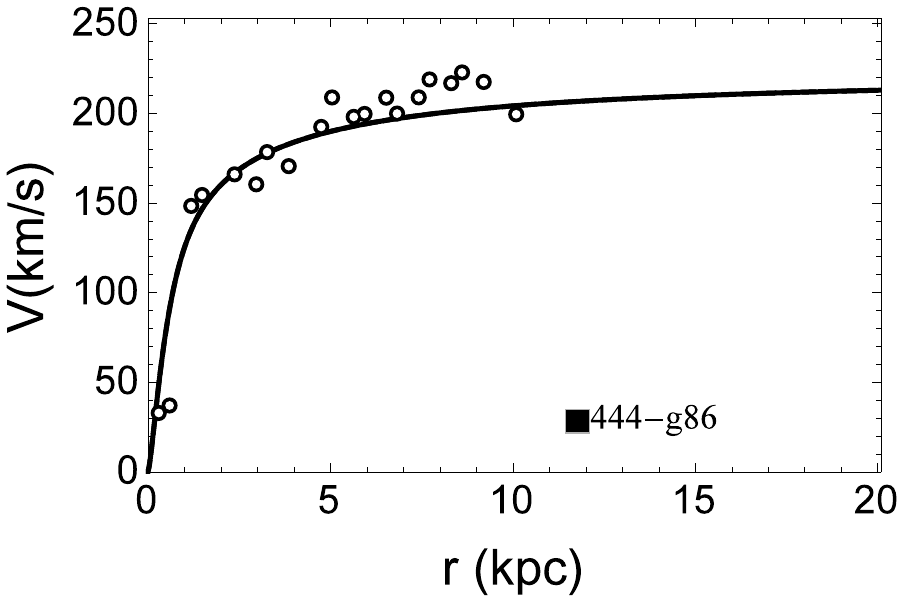}\\
\includegraphics[width=0.50\textwidth]{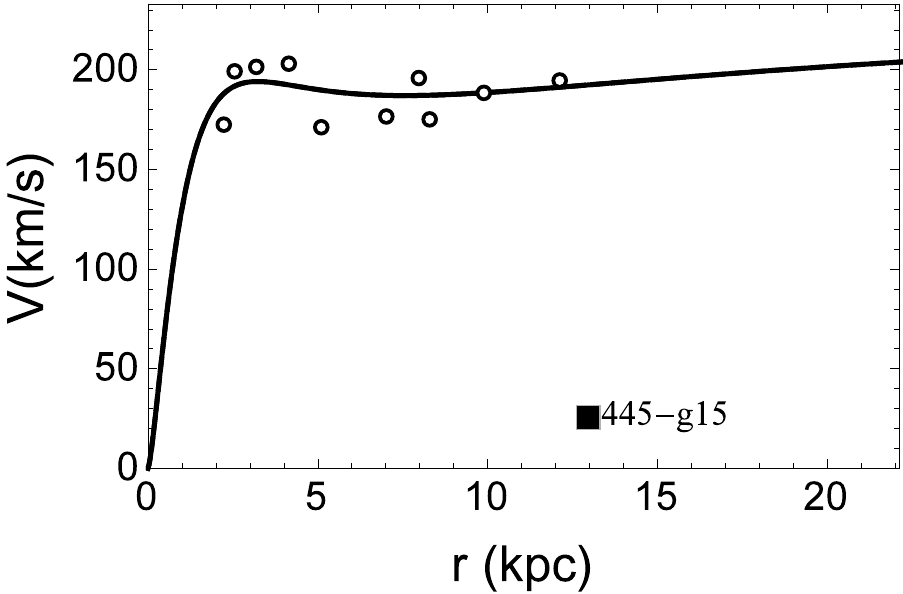}&
\includegraphics[width=0.50\textwidth]{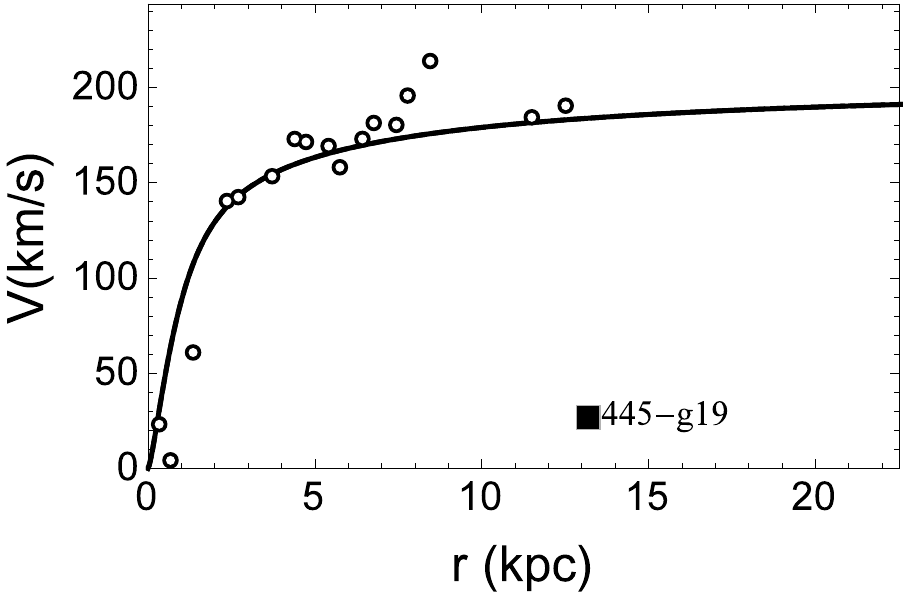}&
\includegraphics[width=0.50\textwidth]{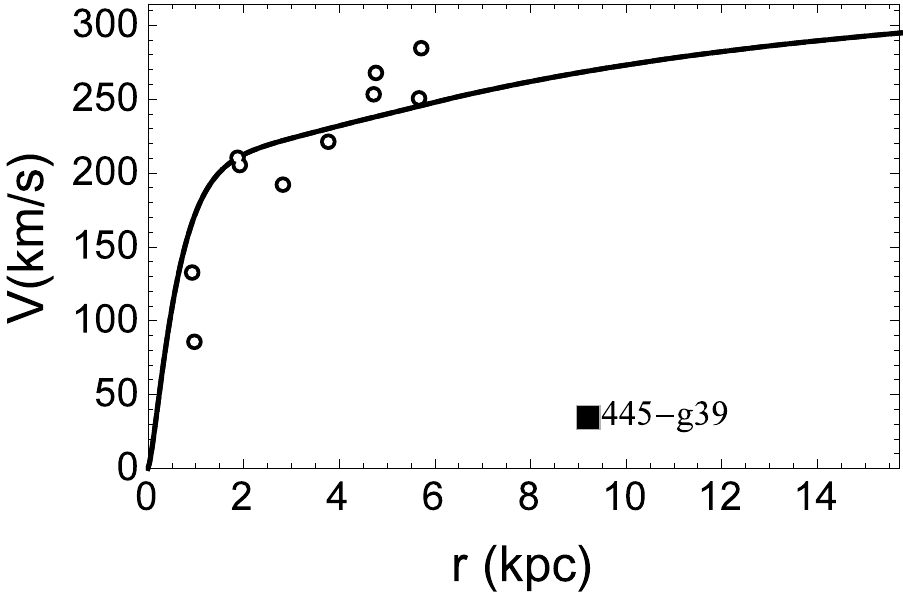}&
\includegraphics[width=0.50\textwidth]{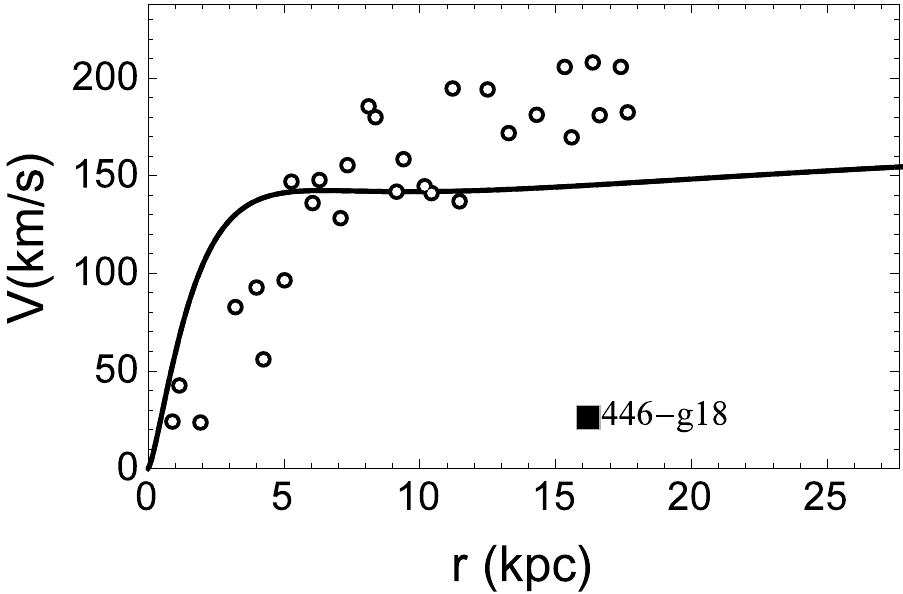}&
\includegraphics[width=0.50\textwidth]{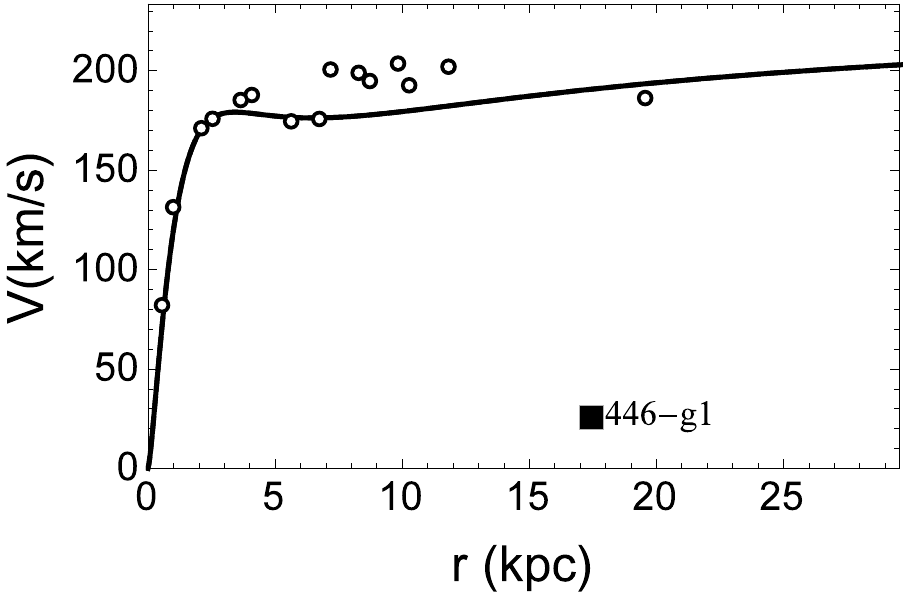}\\
\includegraphics[width=0.50\textwidth]{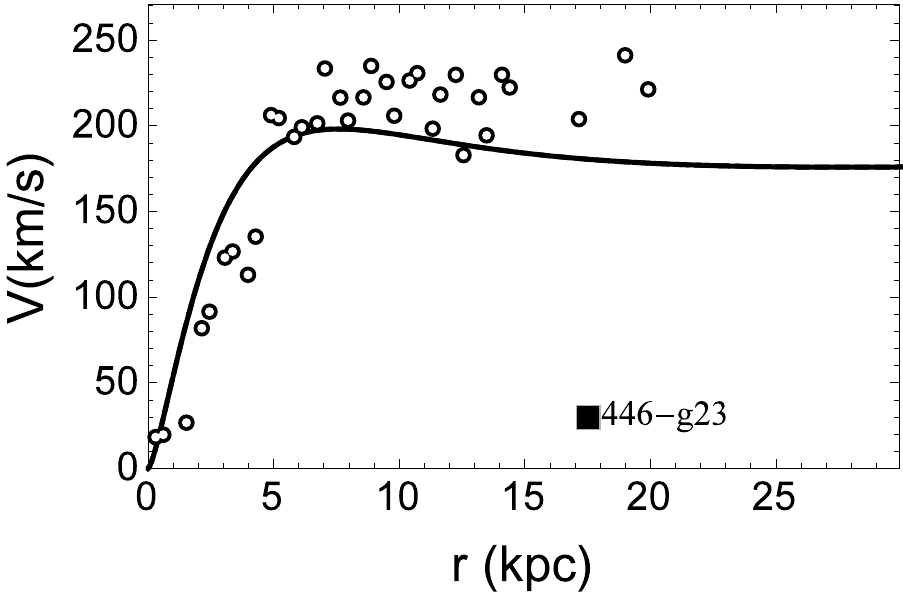}&
\includegraphics[width=0.50\textwidth]{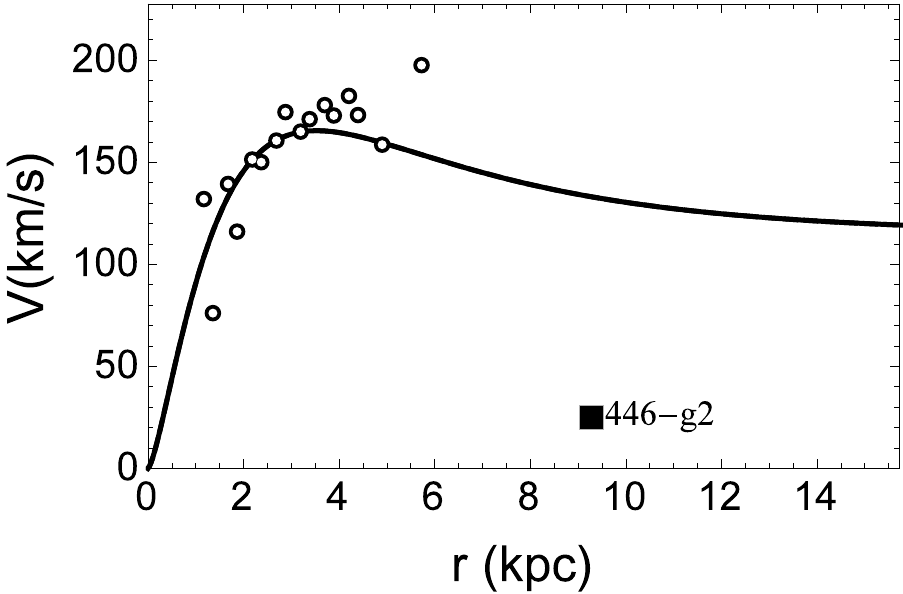}&
\includegraphics[width=0.50\textwidth]{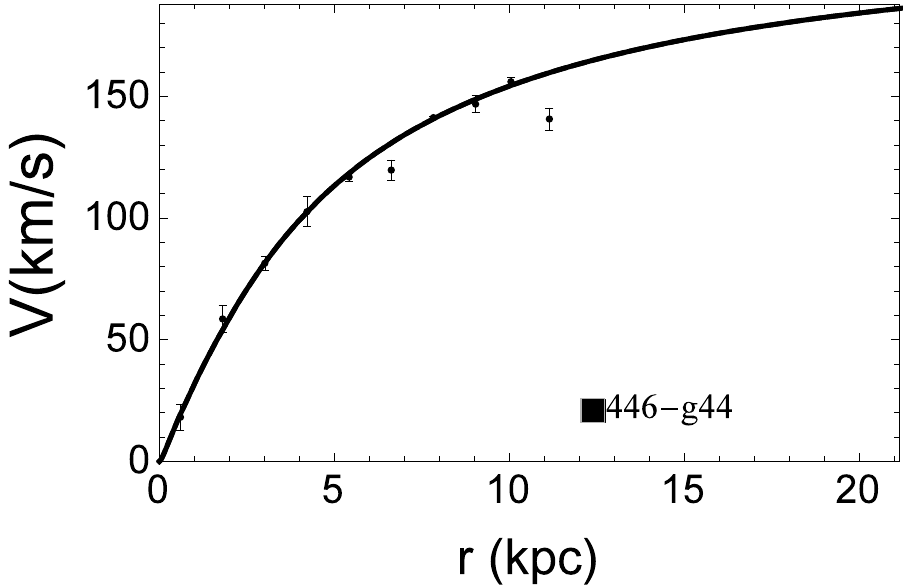}&
\includegraphics[width=0.50\textwidth]{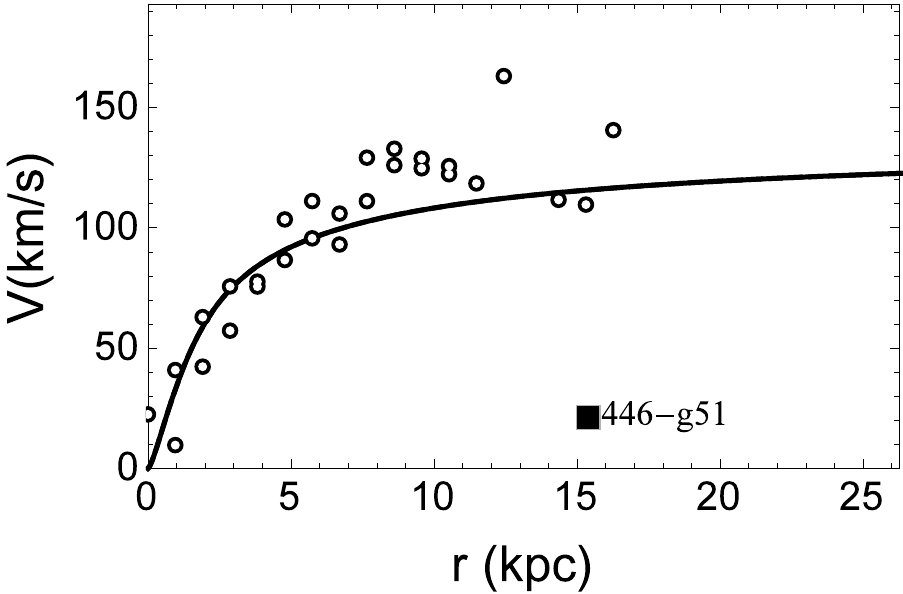}&
\includegraphics[width=0.50\textwidth]{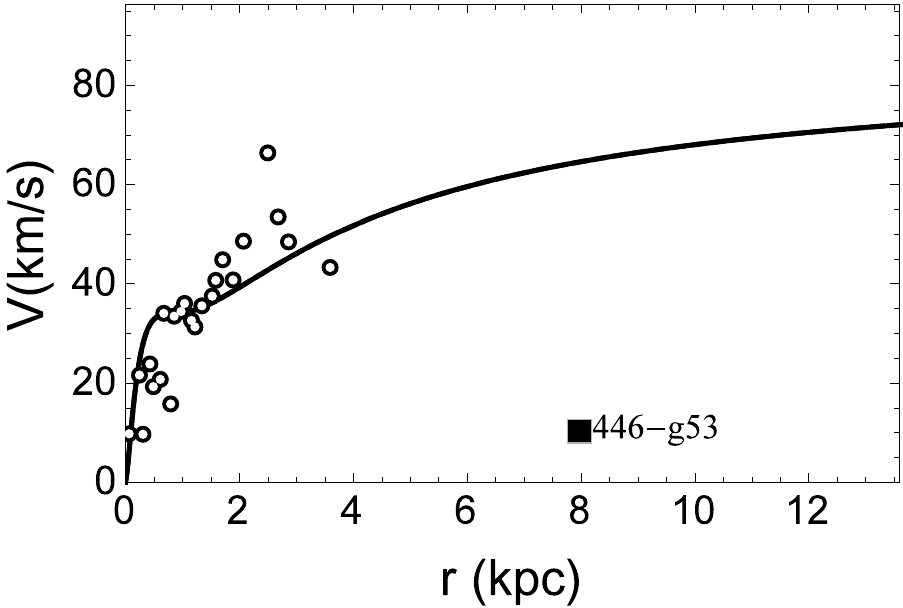}\\
\end{tabular}}}
\end{figure}

\clearpage

\begin{figure}
\flushleft
{\tiny\renewcommand{\arraystretch}{.95}
\resizebox{!}{.64\textwidth}{%
\begin{tabular}{lllll}
\includegraphics[width=0.50\textwidth]{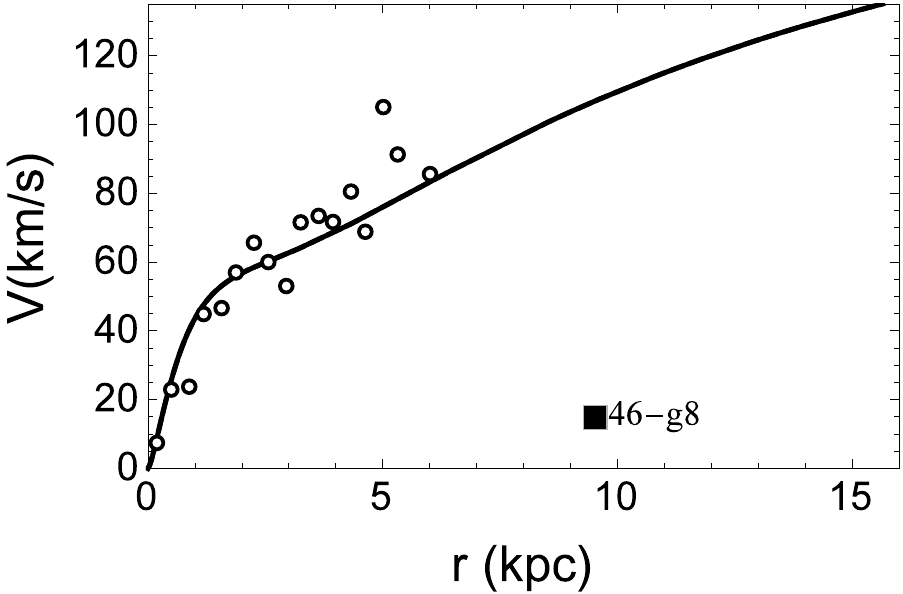}&
\includegraphics[width=0.50\textwidth]{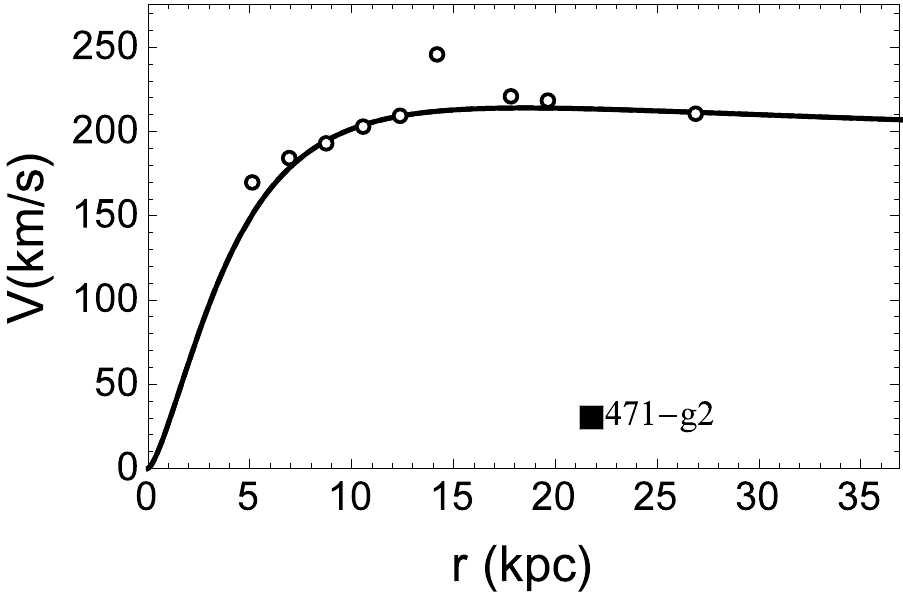}&
\includegraphics[width=0.50\textwidth]{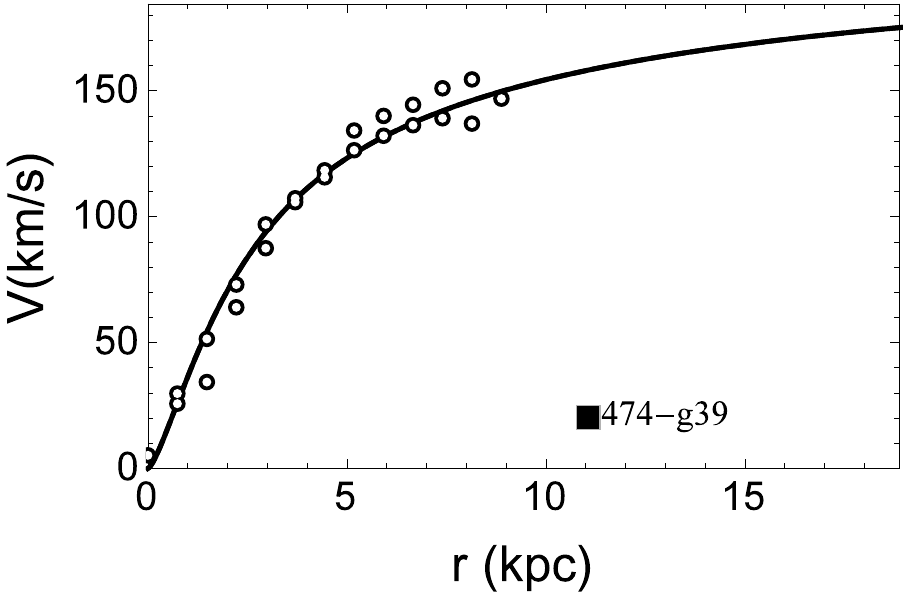}&
\includegraphics[width=0.50\textwidth]{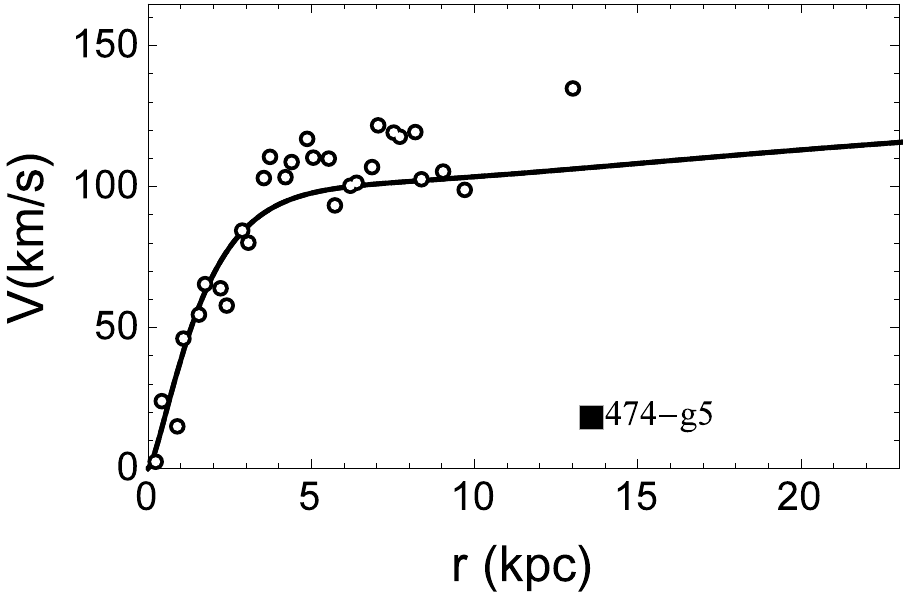}&
\includegraphics[width=0.50\textwidth]{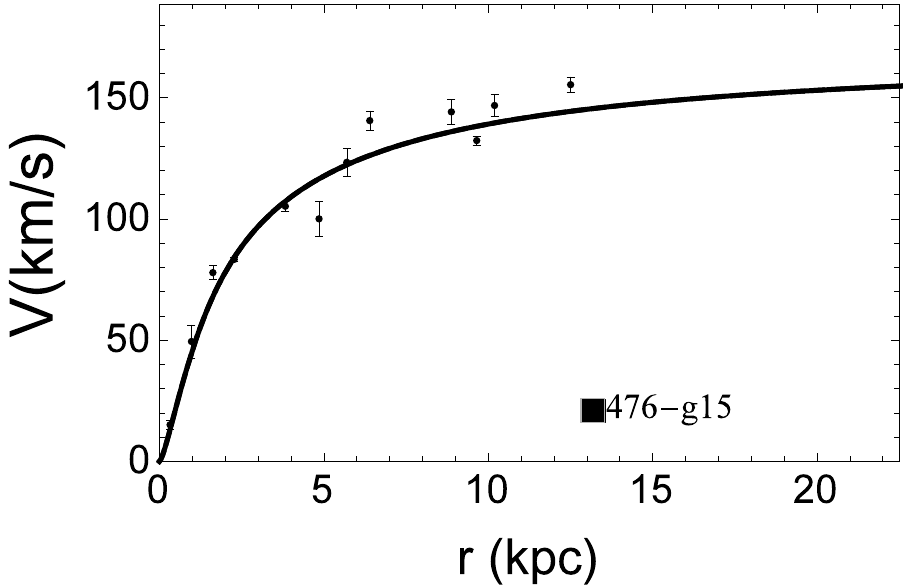}\\
\includegraphics[width=0.50\textwidth]{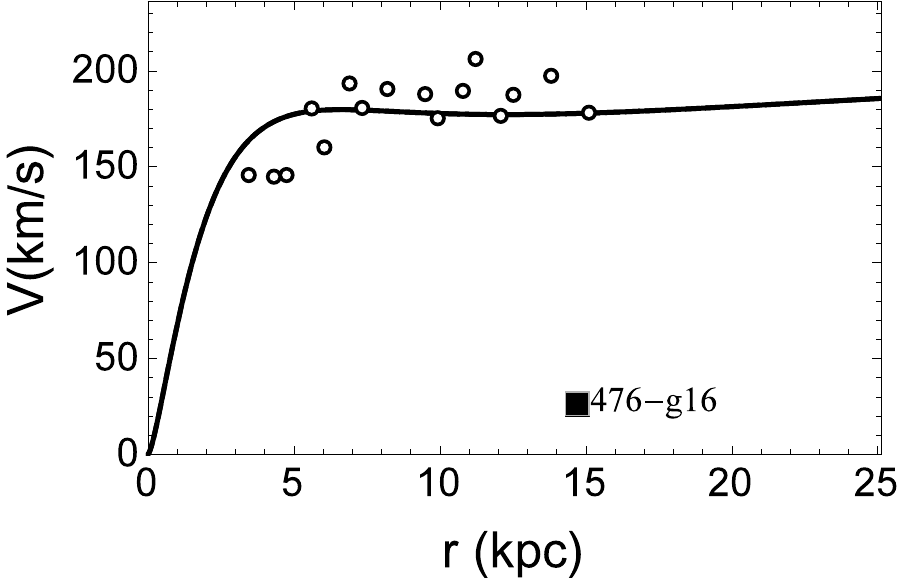}&
\includegraphics[width=0.50\textwidth]{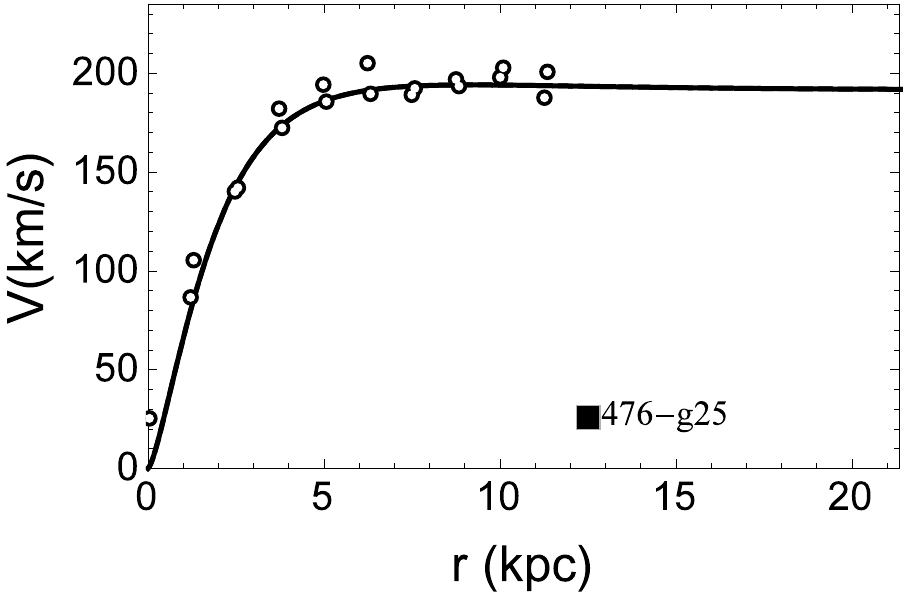}&
\includegraphics[width=0.50\textwidth]{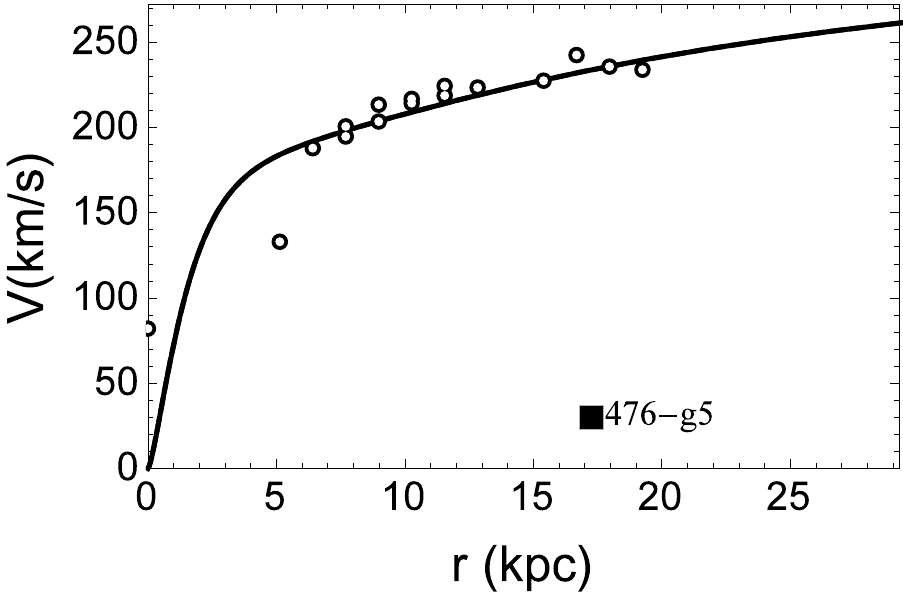}&
\includegraphics[width=0.50\textwidth]{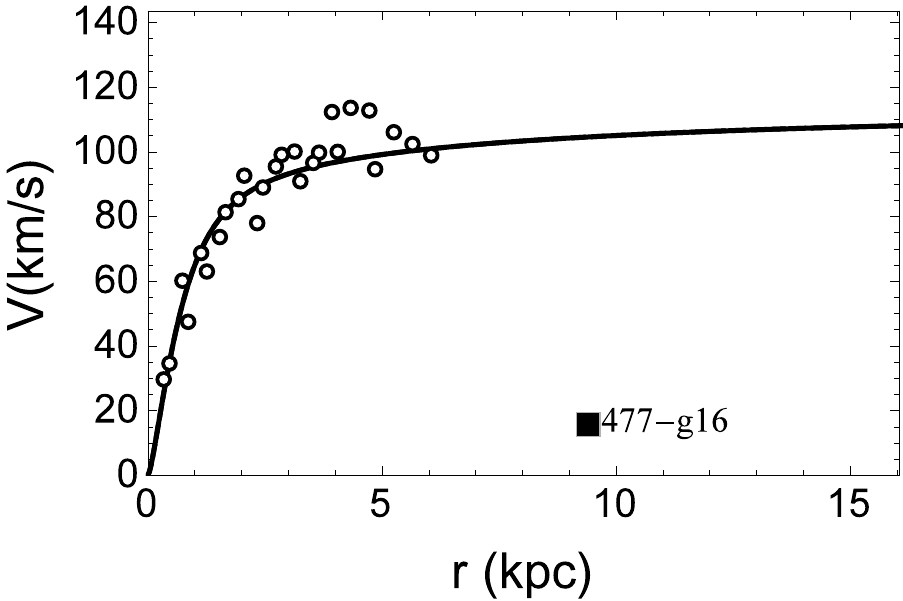}&
\includegraphics[width=0.50\textwidth]{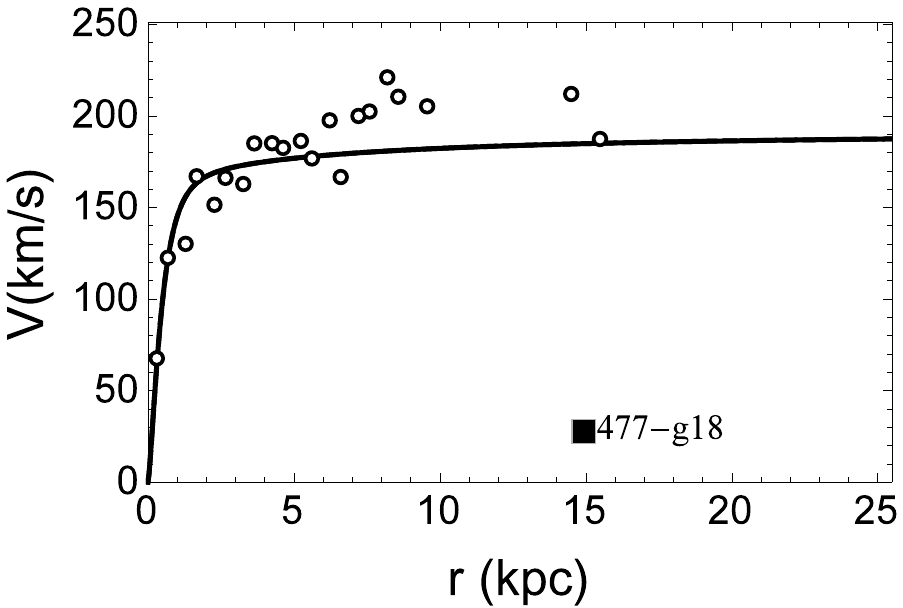}\\
\includegraphics[width=0.50\textwidth]{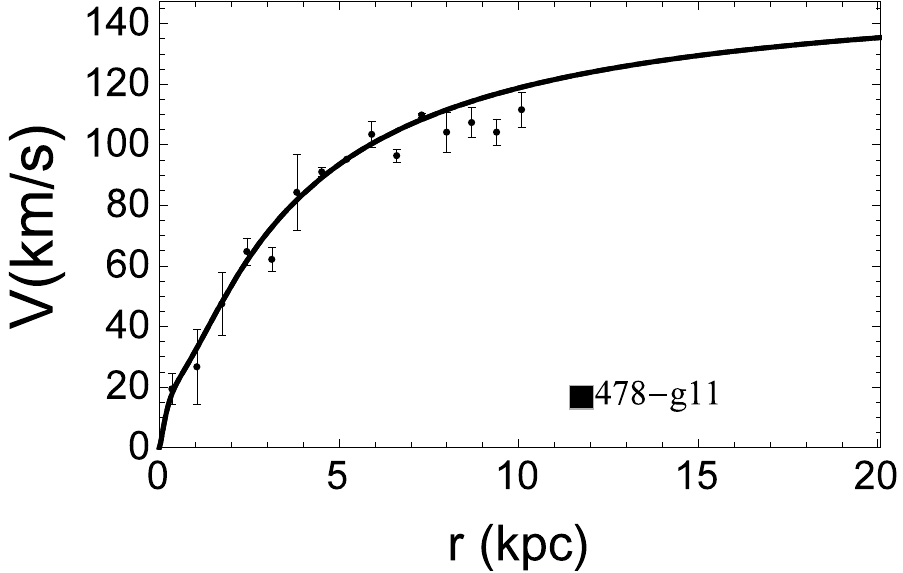}&
\includegraphics[width=0.50\textwidth]{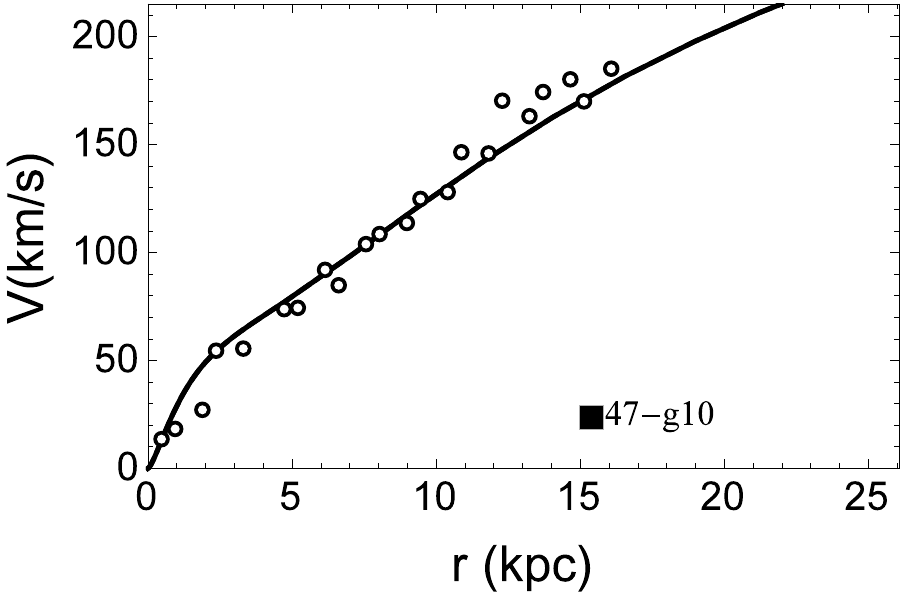}&
\includegraphics[width=0.50\textwidth]{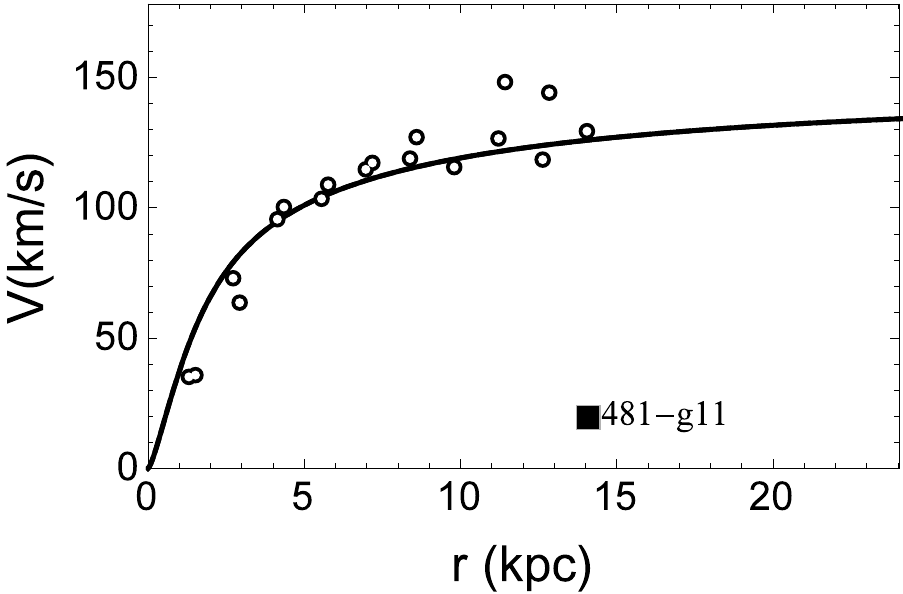}&
\includegraphics[width=0.50\textwidth]{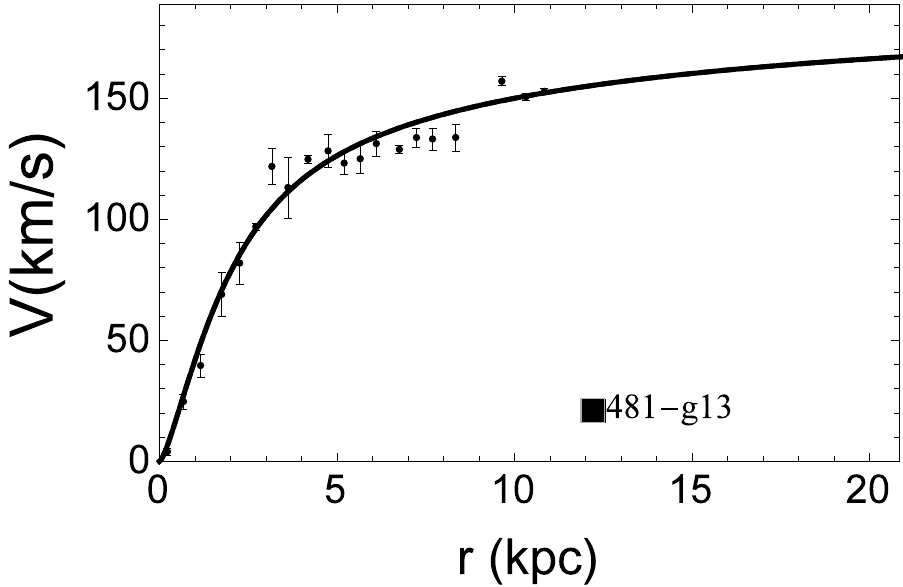}&
\includegraphics[width=0.50\textwidth]{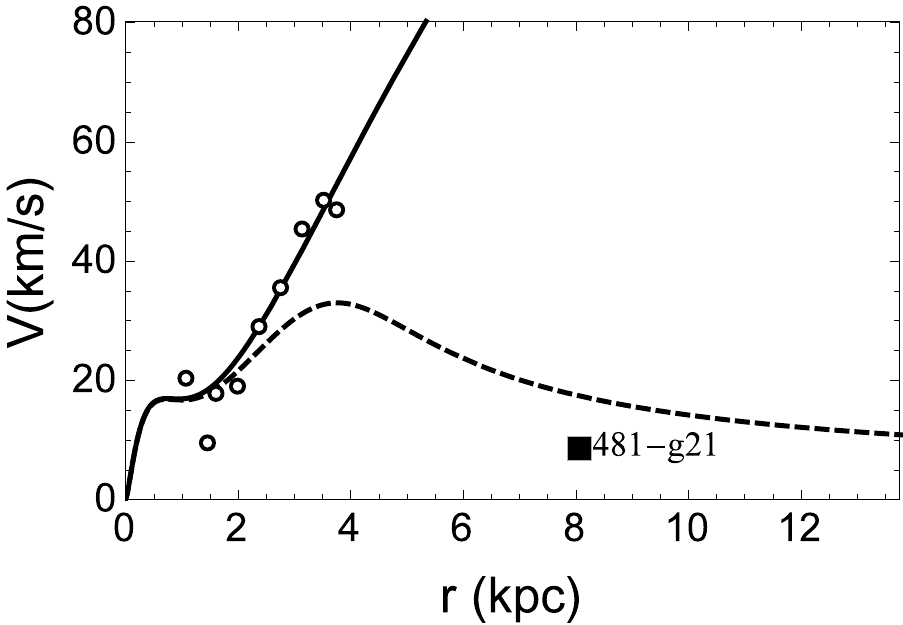}\\
\includegraphics[width=0.50\textwidth]{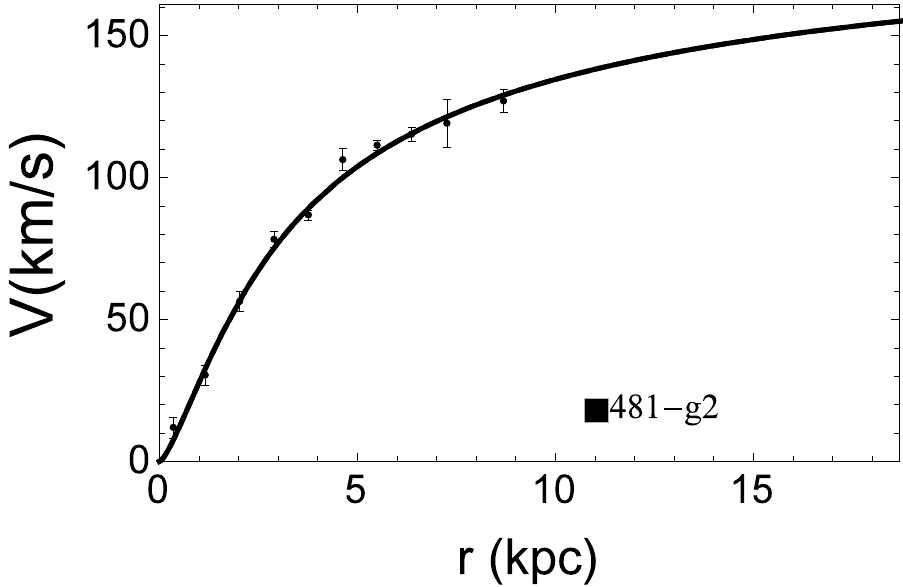}&
\includegraphics[width=0.50\textwidth]{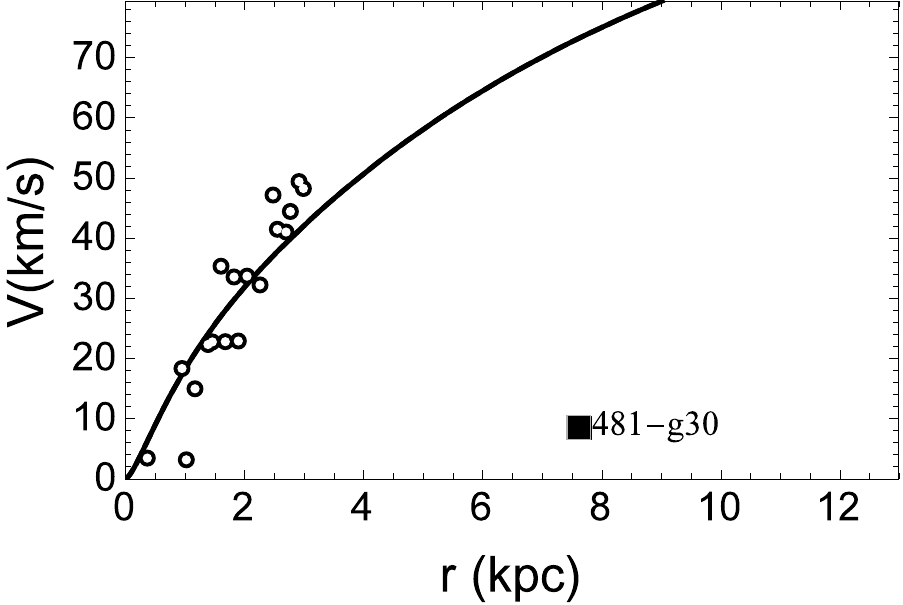}&
\includegraphics[width=0.50\textwidth]{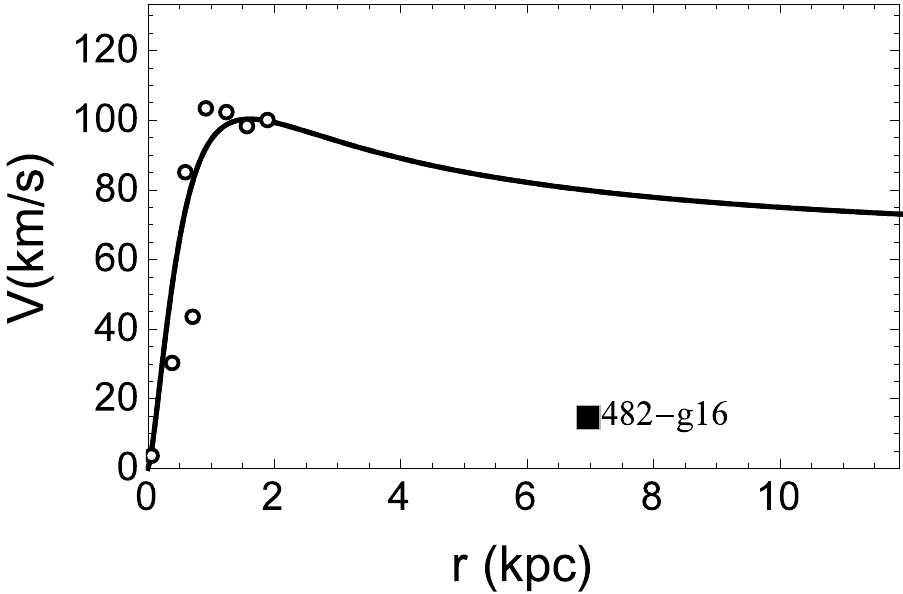}&
\includegraphics[width=0.50\textwidth]{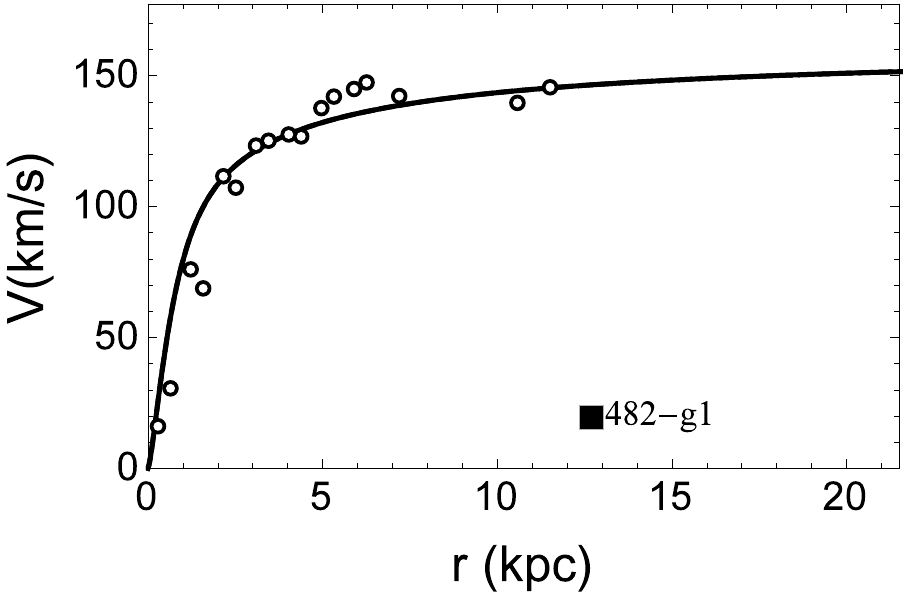}&
\includegraphics[width=0.50\textwidth]{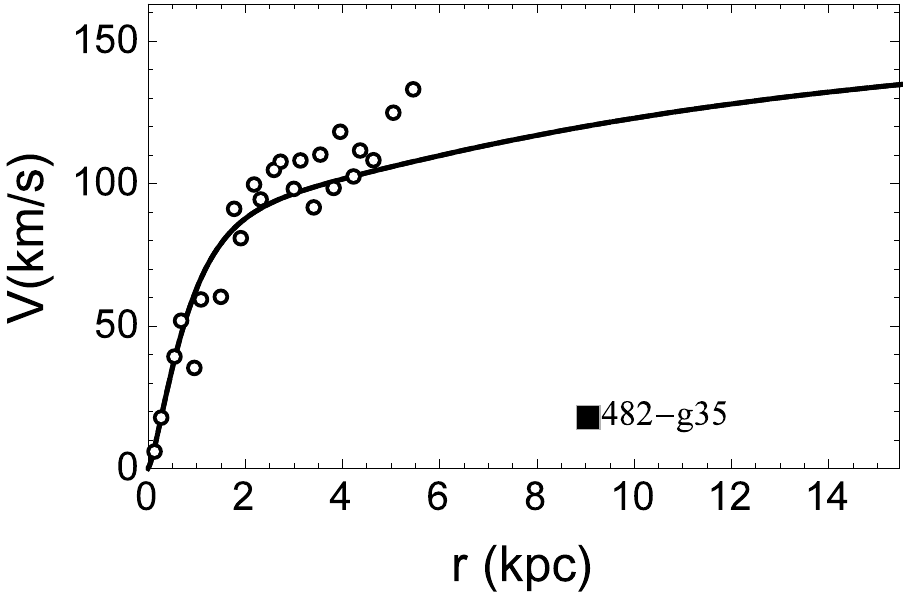}\\
\end{tabular}}}
\end{figure}

\clearpage

\begin{figure}
\flushleft
{\tiny\renewcommand{\arraystretch}{.95}
\resizebox{!}{.64\textwidth}{%
\begin{tabular}{lllll}
\includegraphics[width=0.50\textwidth]{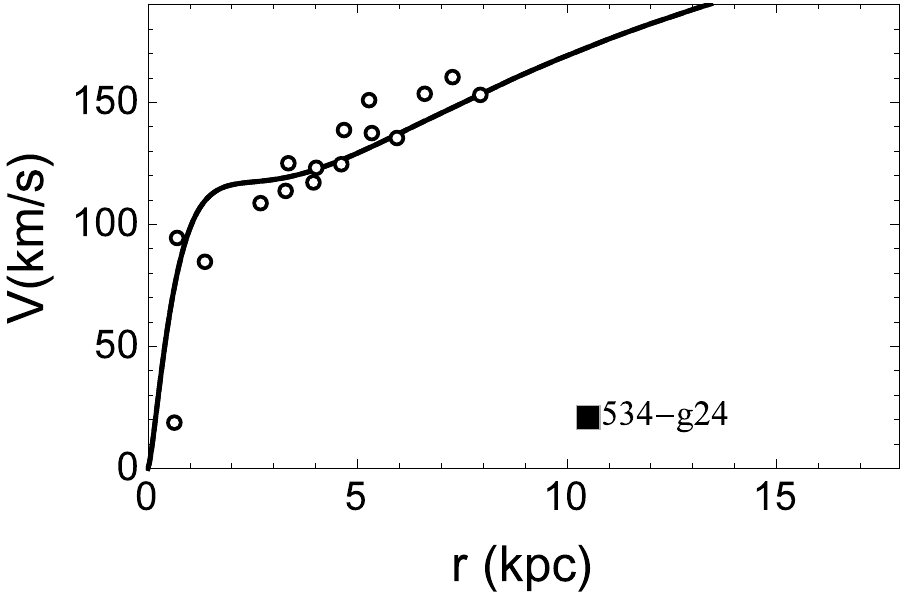}&
\includegraphics[width=0.50\textwidth]{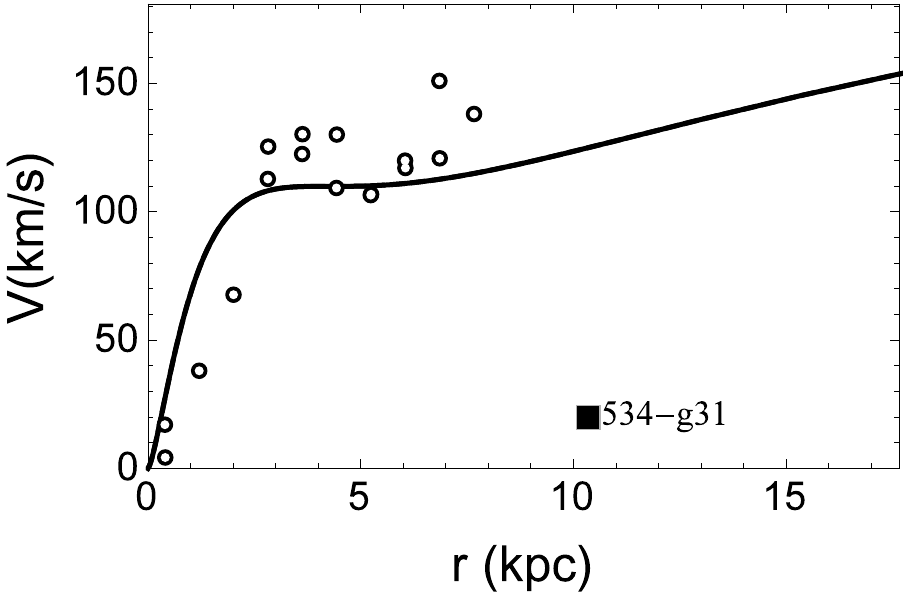}&
\includegraphics[width=0.50\textwidth]{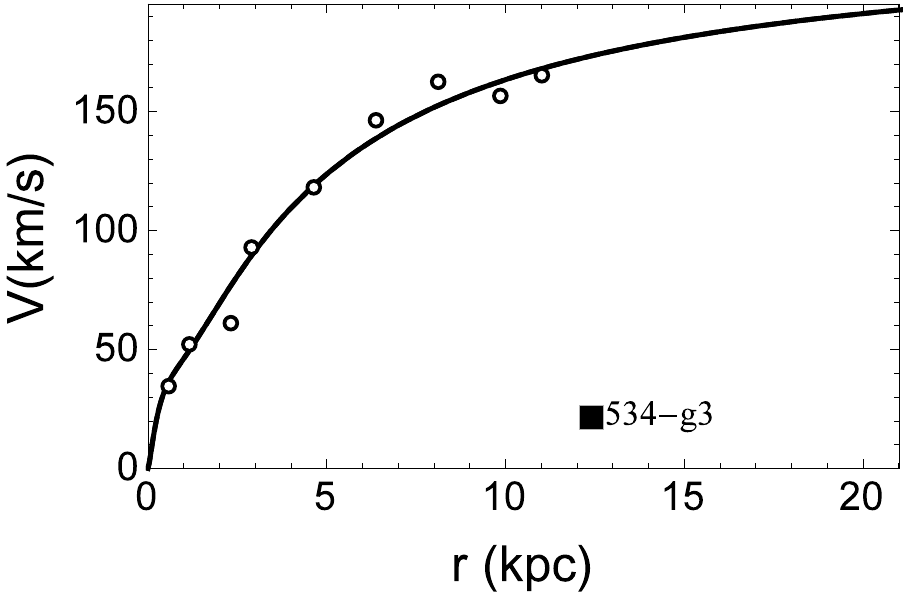}&
\includegraphics[width=0.50\textwidth]{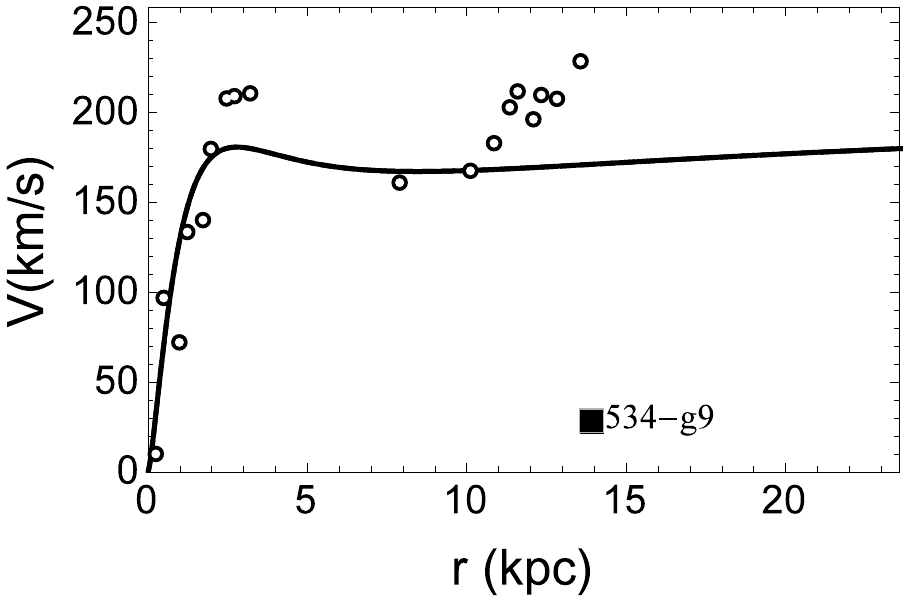}&
\includegraphics[width=0.50\textwidth]{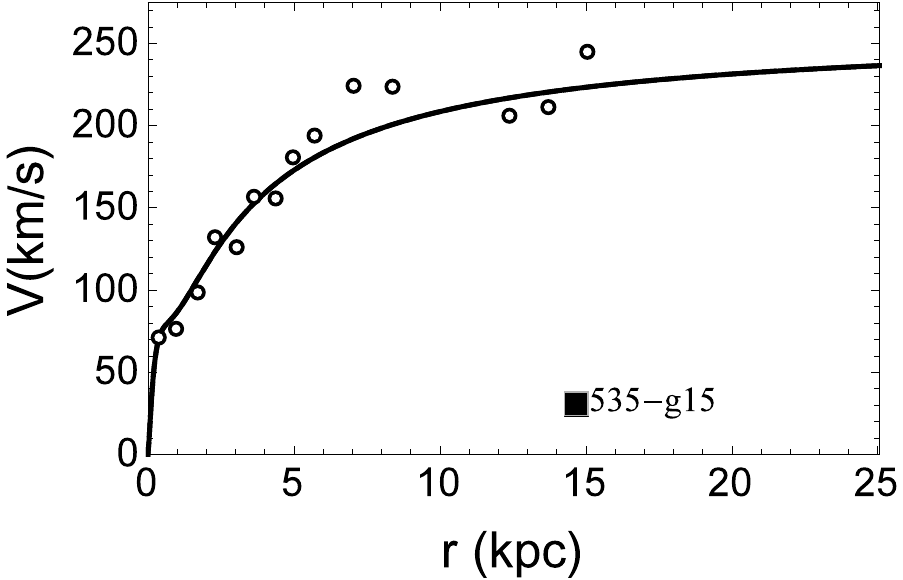}\\
\includegraphics[width=0.50\textwidth]{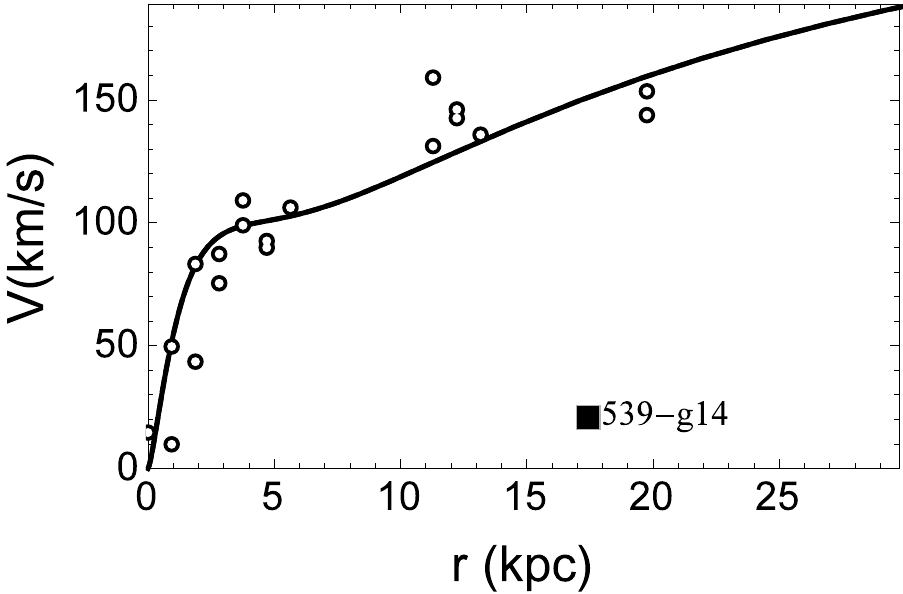}&
\includegraphics[width=0.50\textwidth]{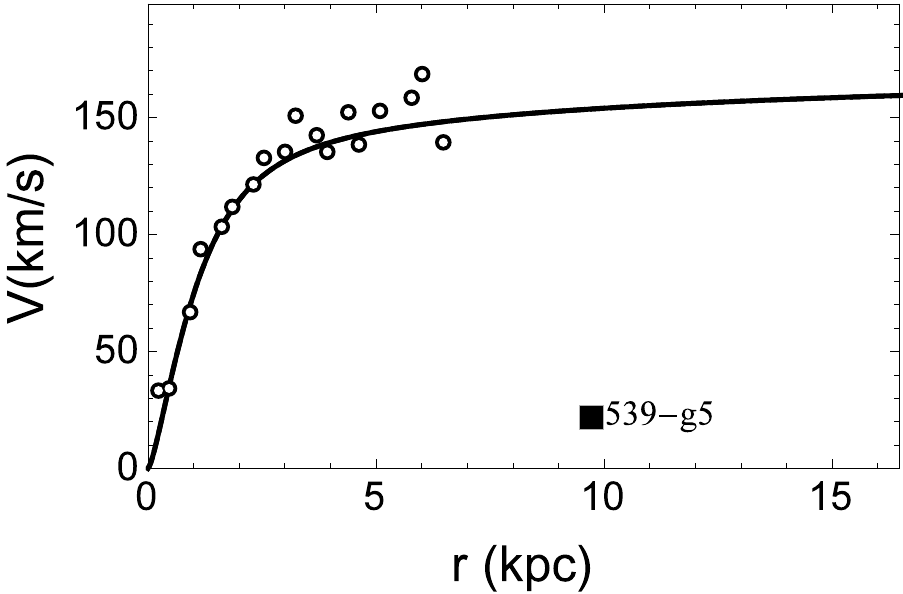}&
\includegraphics[width=0.50\textwidth]{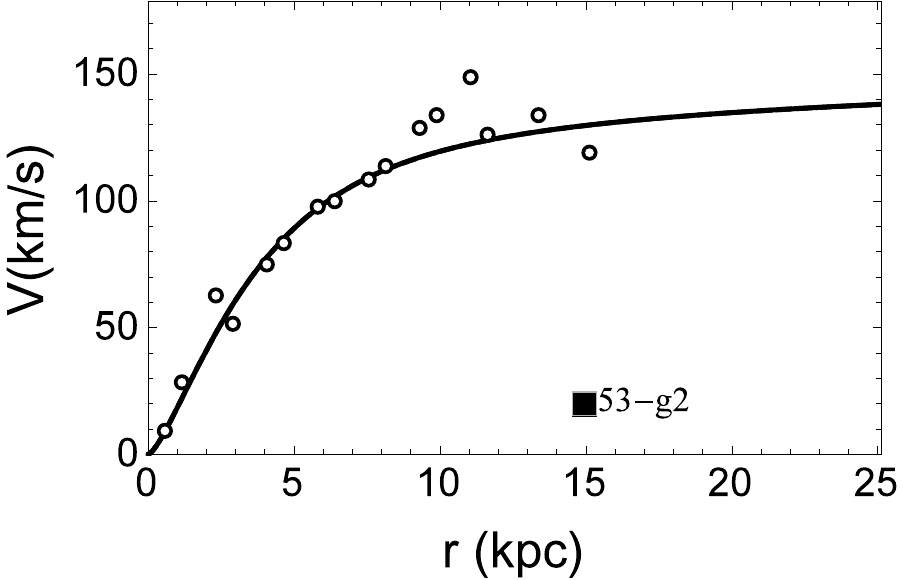}&
\includegraphics[width=0.50\textwidth]{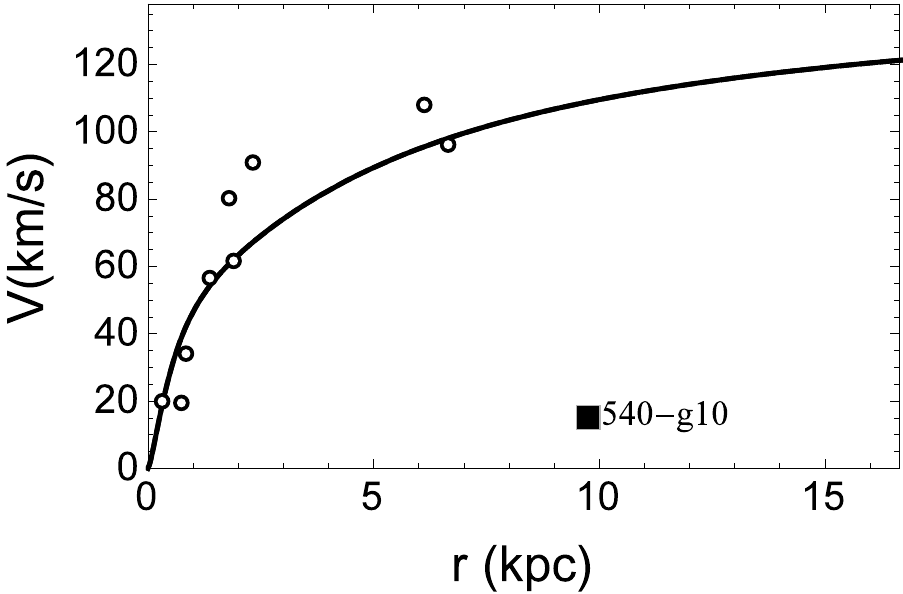}&
\includegraphics[width=0.50\textwidth]{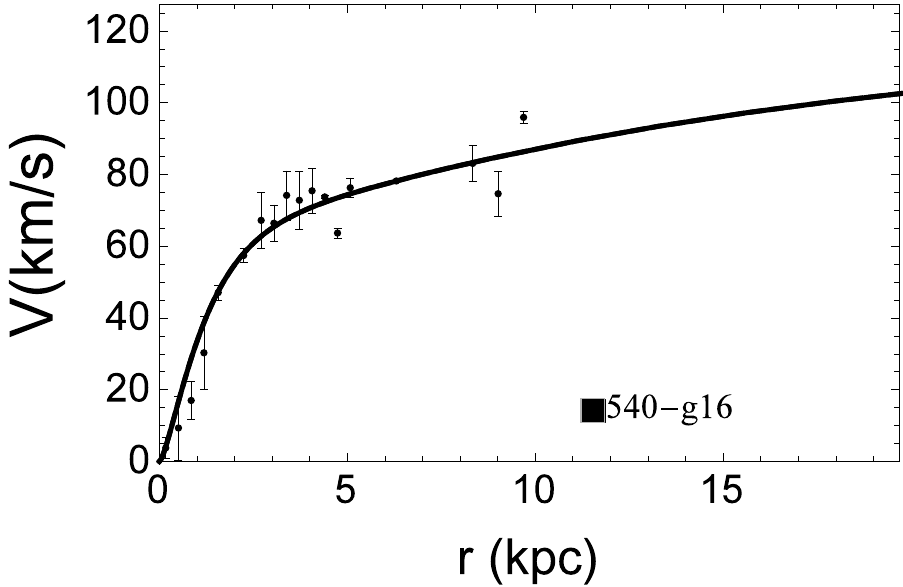}\\
\includegraphics[width=0.50\textwidth]{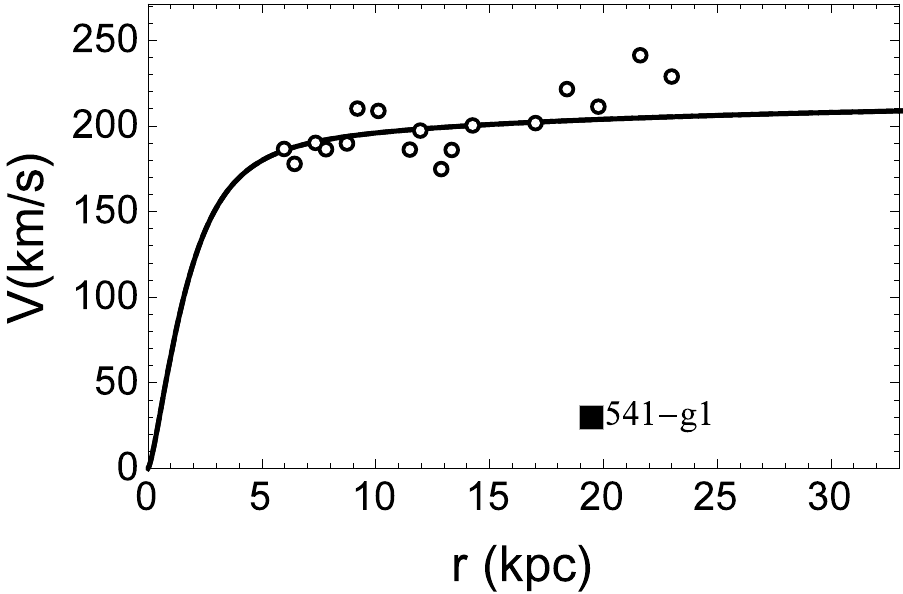}&
\includegraphics[width=0.50\textwidth]{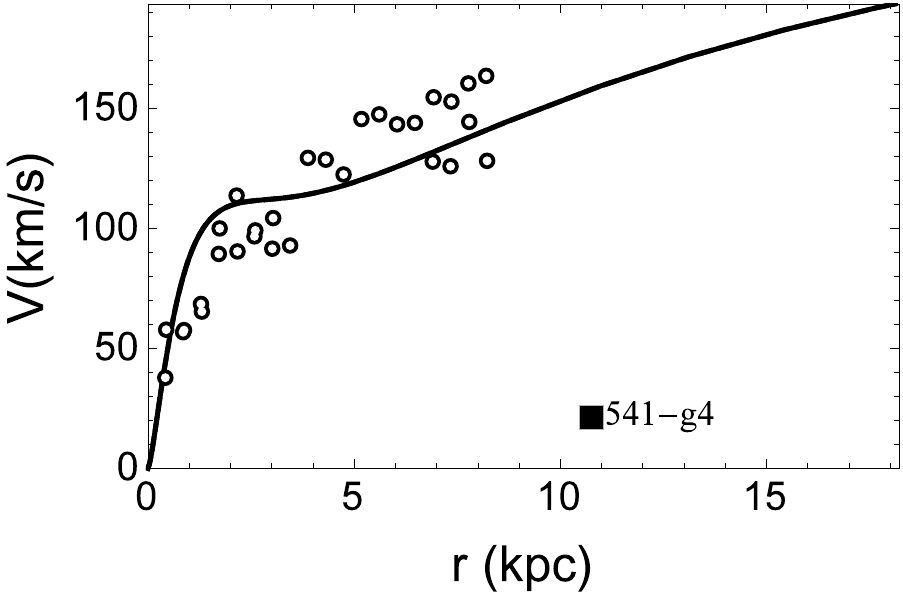}&
\includegraphics[width=0.50\textwidth]{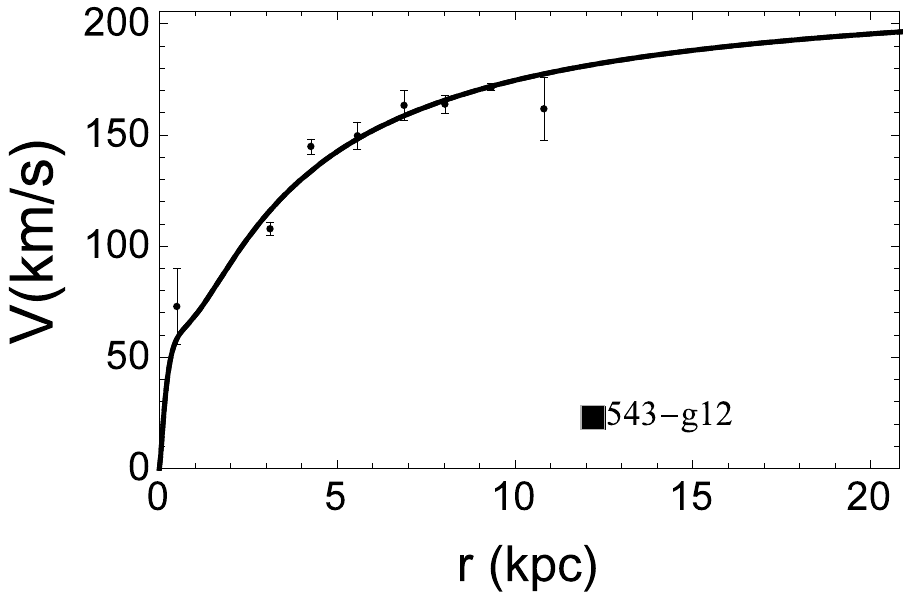}&
\includegraphics[width=0.50\textwidth]{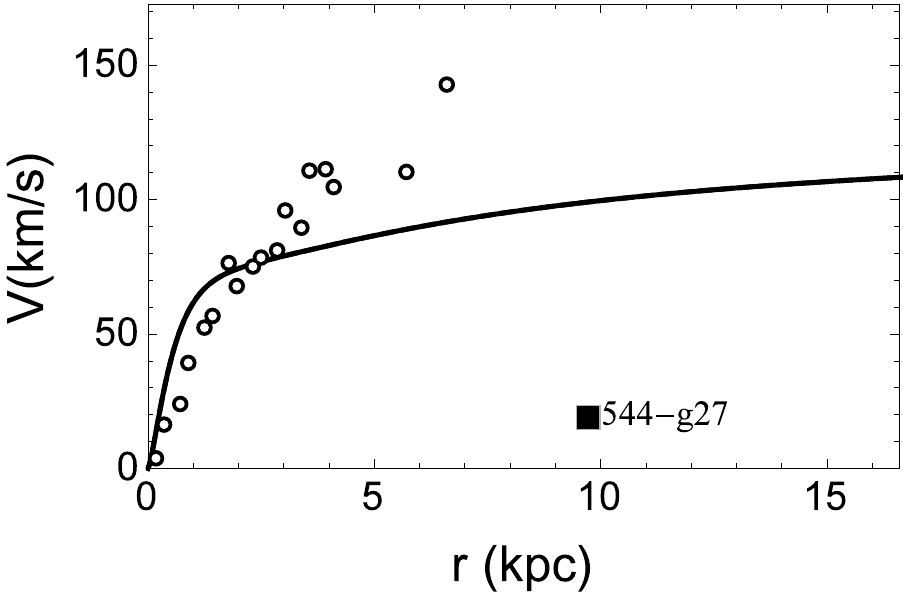}&
\includegraphics[width=0.50\textwidth]{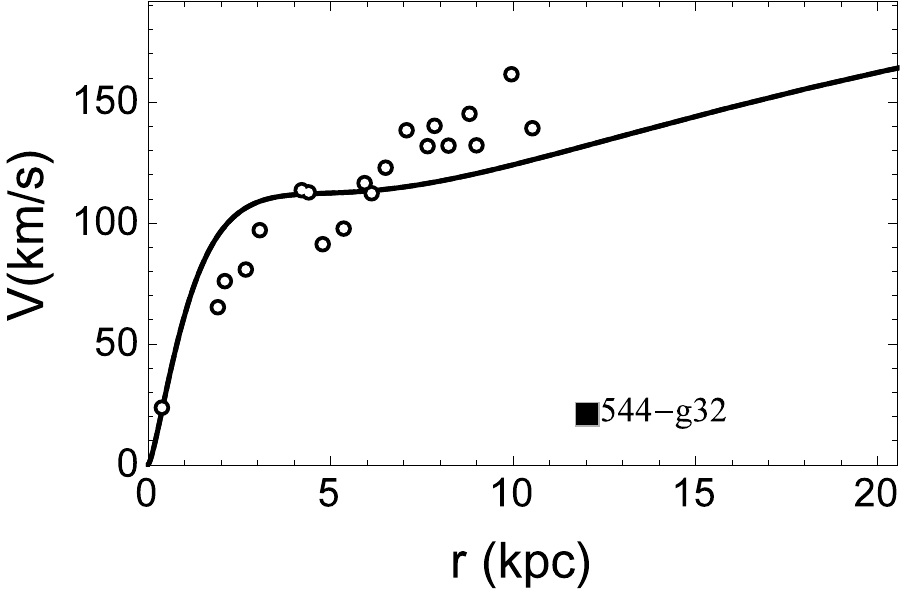}\\
\includegraphics[width=0.50\textwidth]{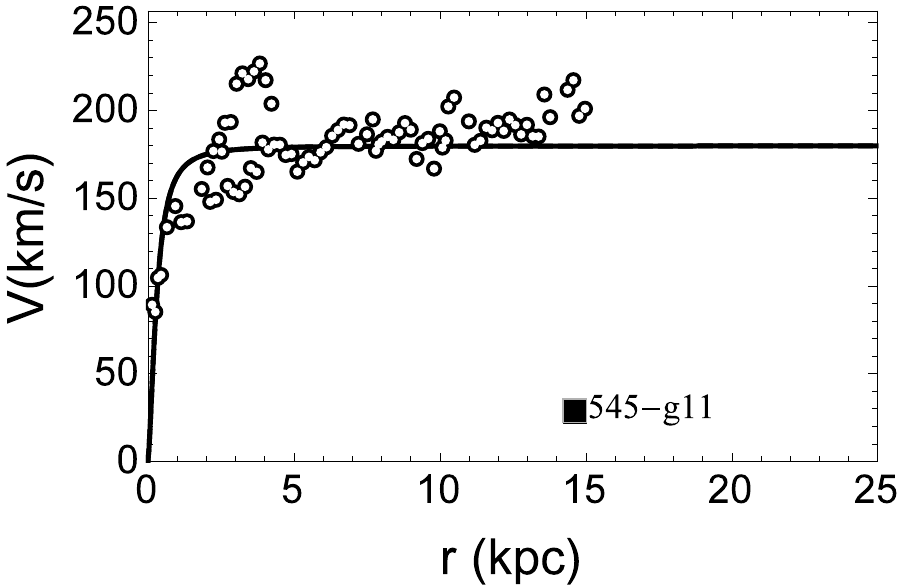}&
\includegraphics[width=0.50\textwidth]{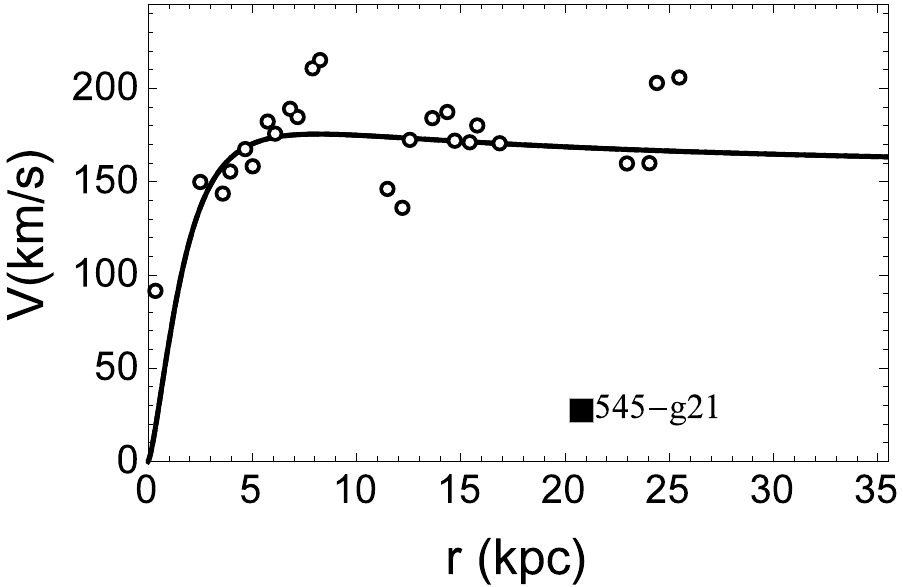}&
\includegraphics[width=0.50\textwidth]{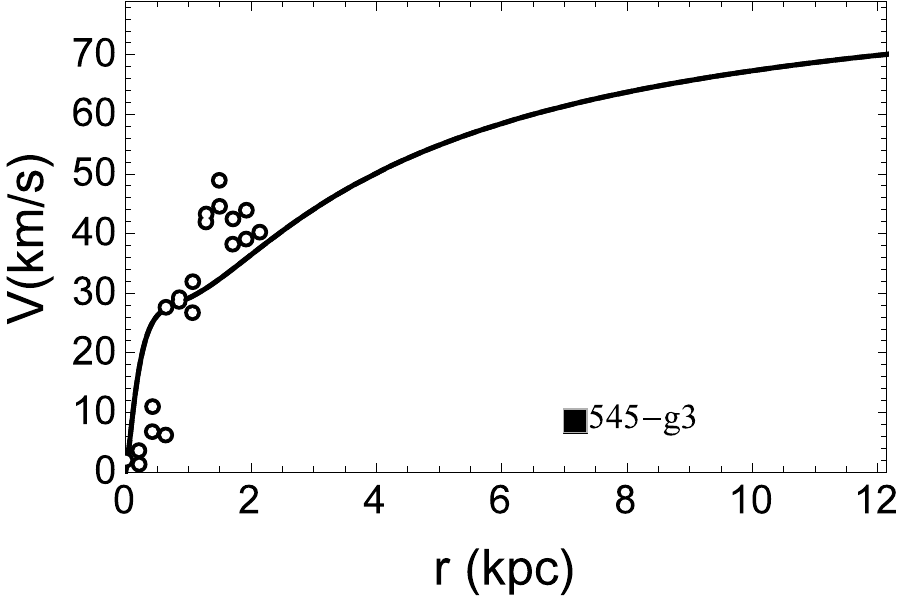}&
\includegraphics[width=0.50\textwidth]{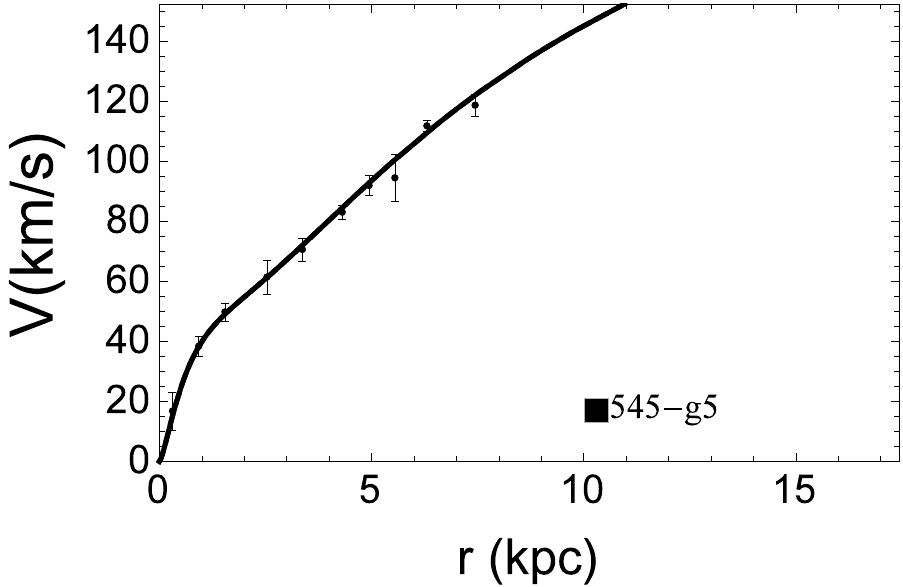}&
\includegraphics[width=0.50\textwidth]{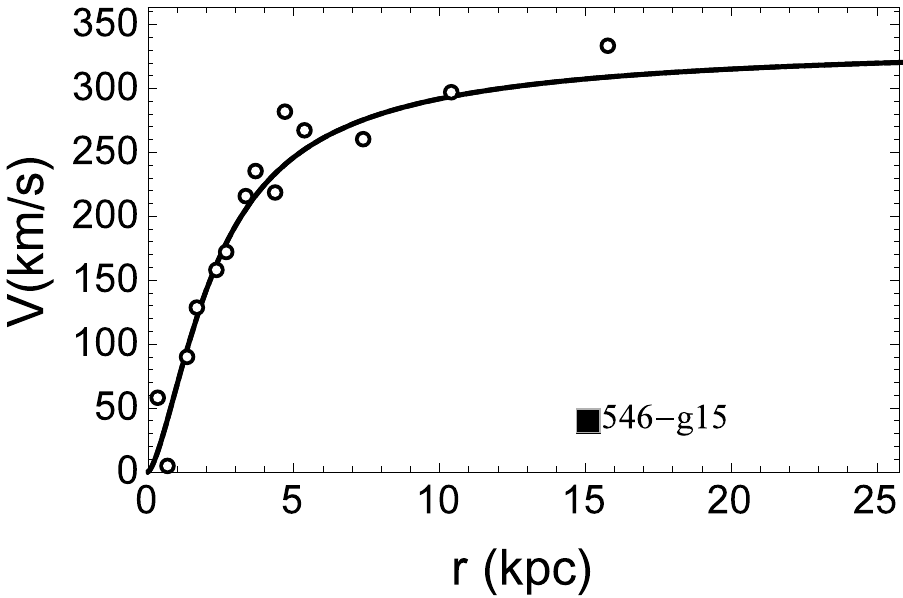}\\
\end{tabular}}}
\end{figure}

\clearpage

\begin{figure}
\flushleft
{\tiny\renewcommand{\arraystretch}{.95}
\resizebox{!}{.64\textwidth}{%
\begin{tabular}{lllll}
\includegraphics[width=0.50\textwidth]{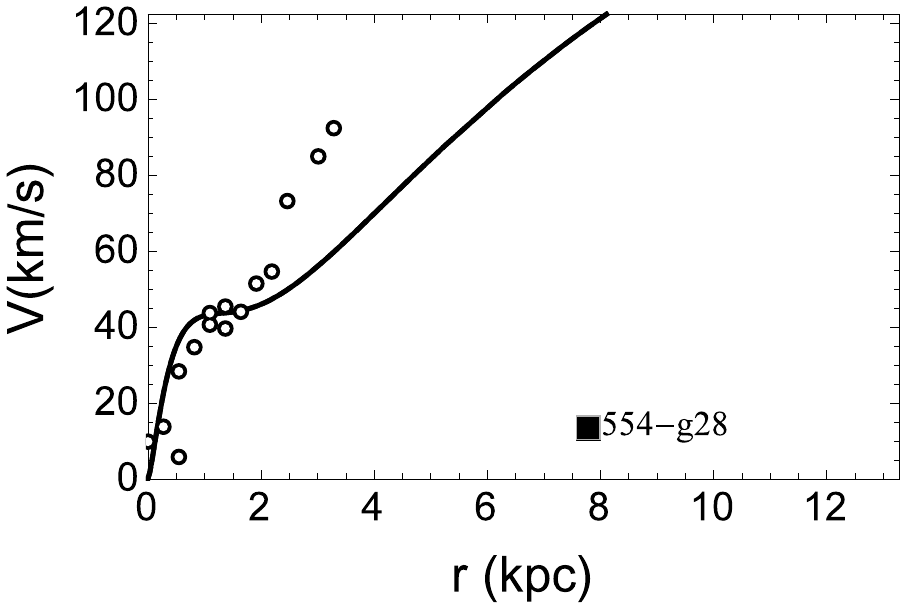}&
\includegraphics[width=0.50\textwidth]{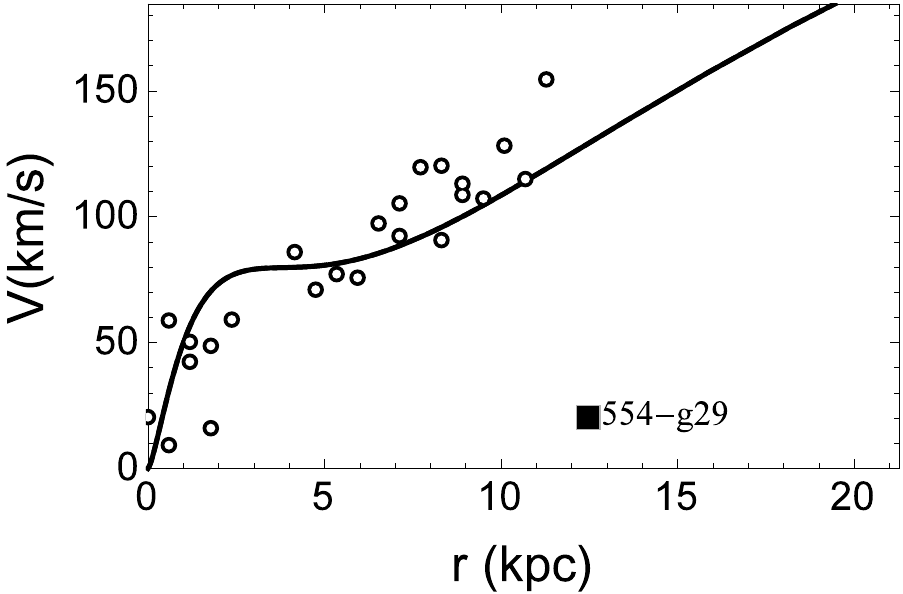}&
\includegraphics[width=0.50\textwidth]{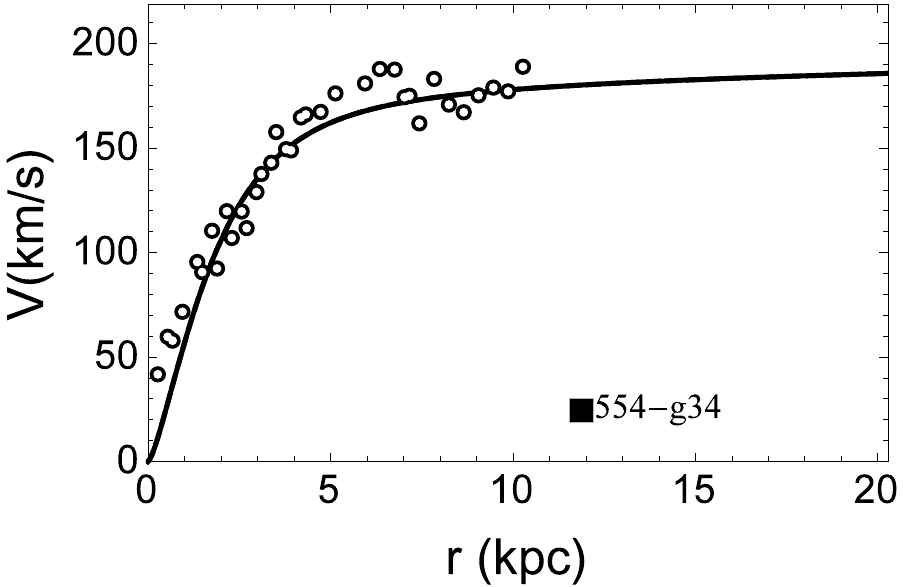}&
\includegraphics[width=0.50\textwidth]{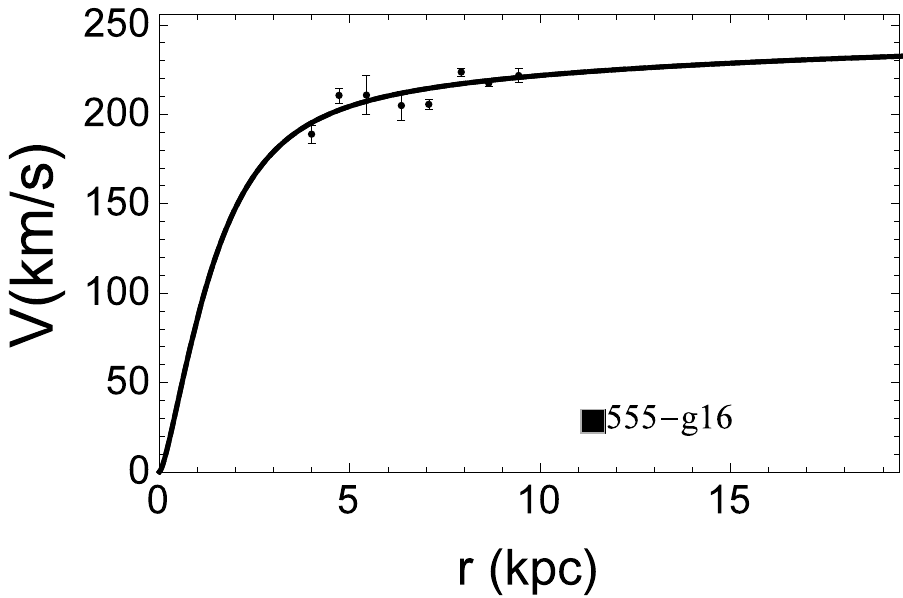}&
\includegraphics[width=0.50\textwidth]{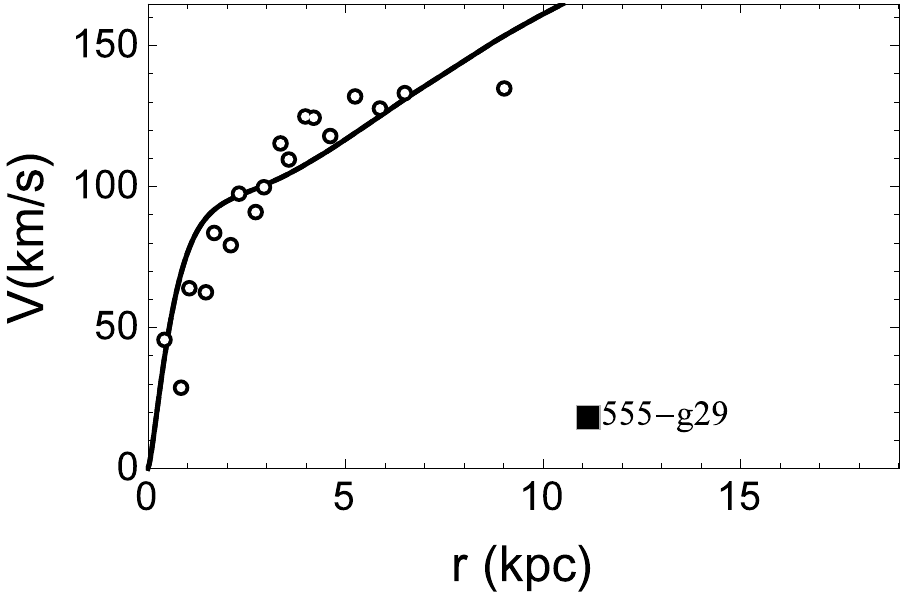}\\
\includegraphics[width=0.50\textwidth]{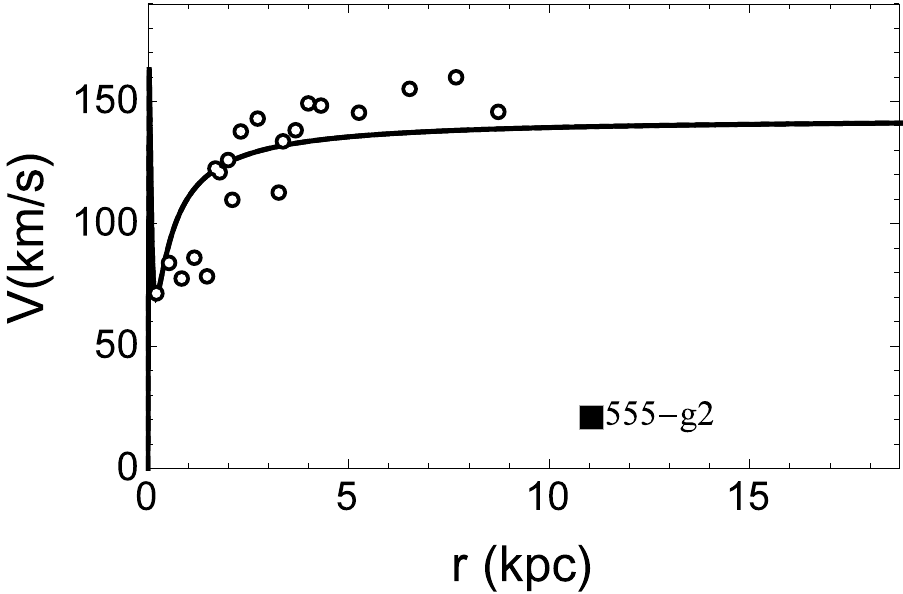}&
\includegraphics[width=0.50\textwidth]{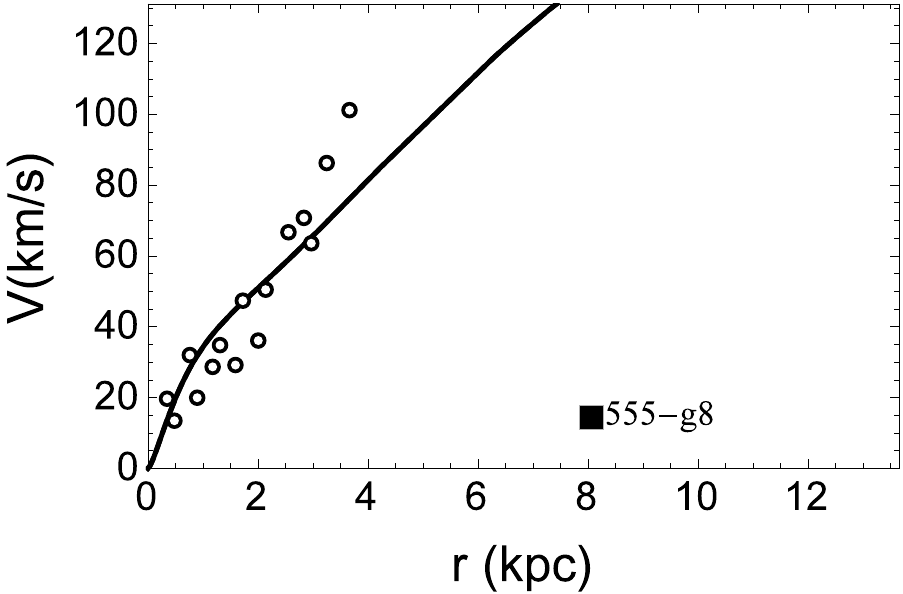}&
\includegraphics[width=0.50\textwidth]{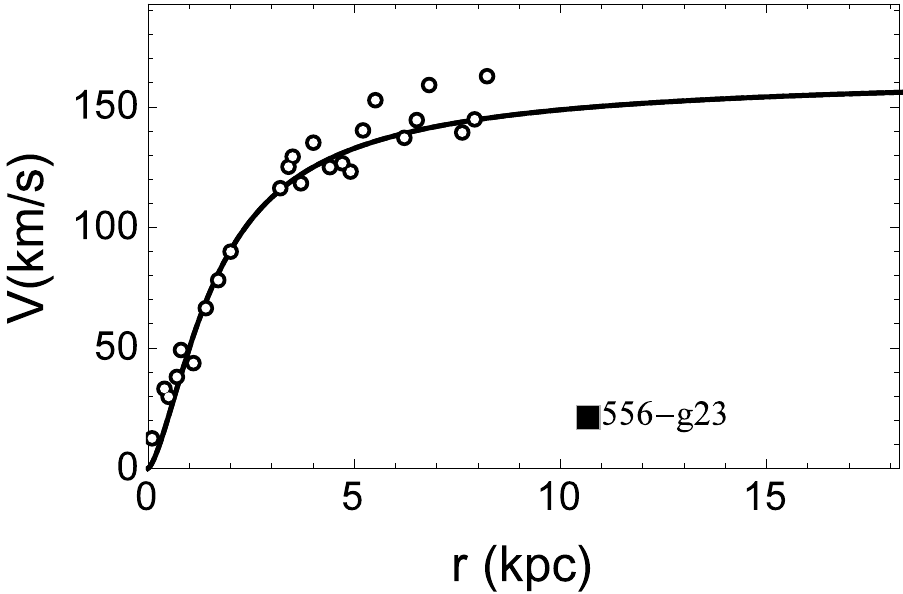}&
\includegraphics[width=0.50\textwidth]{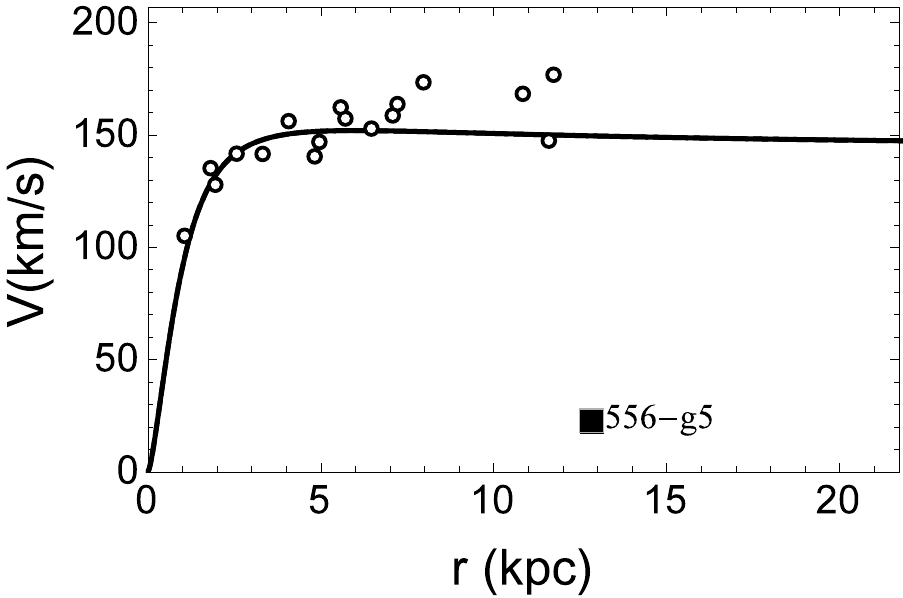}&
\includegraphics[width=0.50\textwidth]{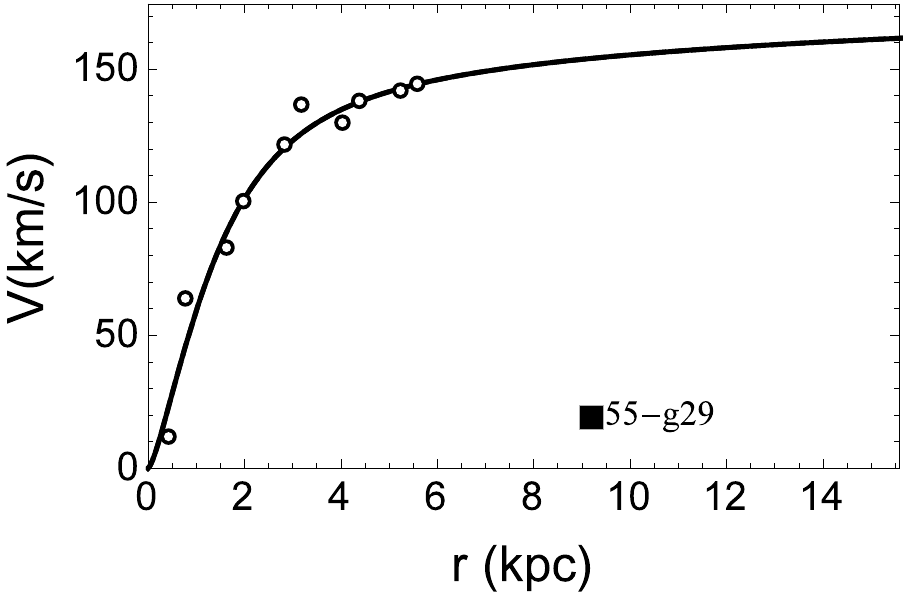}\\
\includegraphics[width=0.50\textwidth]{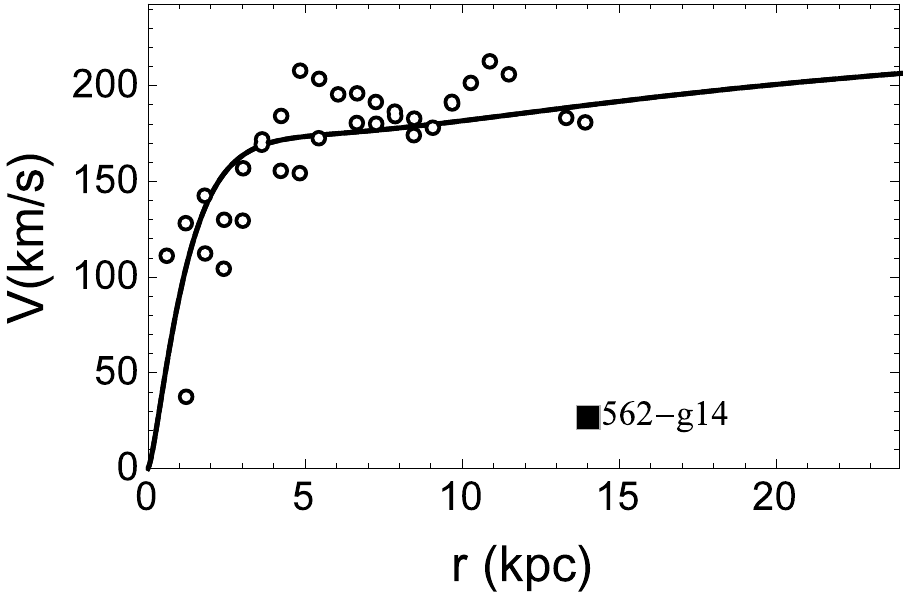}&
\includegraphics[width=0.50\textwidth]{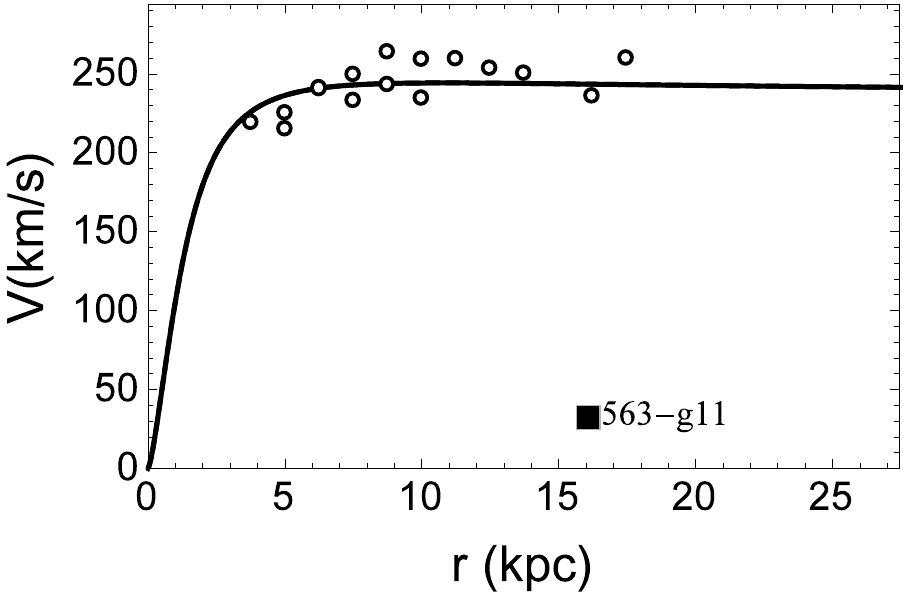}&
\includegraphics[width=0.50\textwidth]{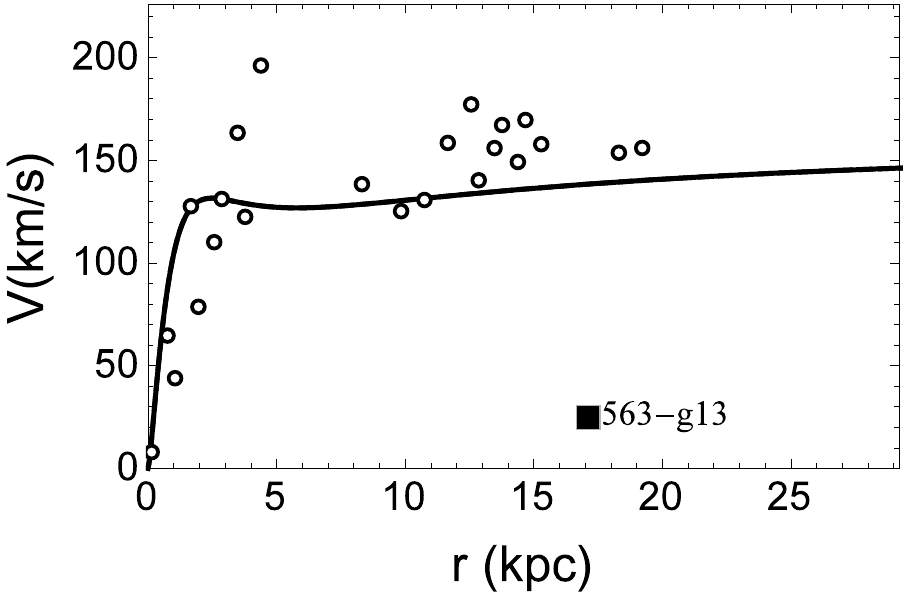}&
\includegraphics[width=0.50\textwidth]{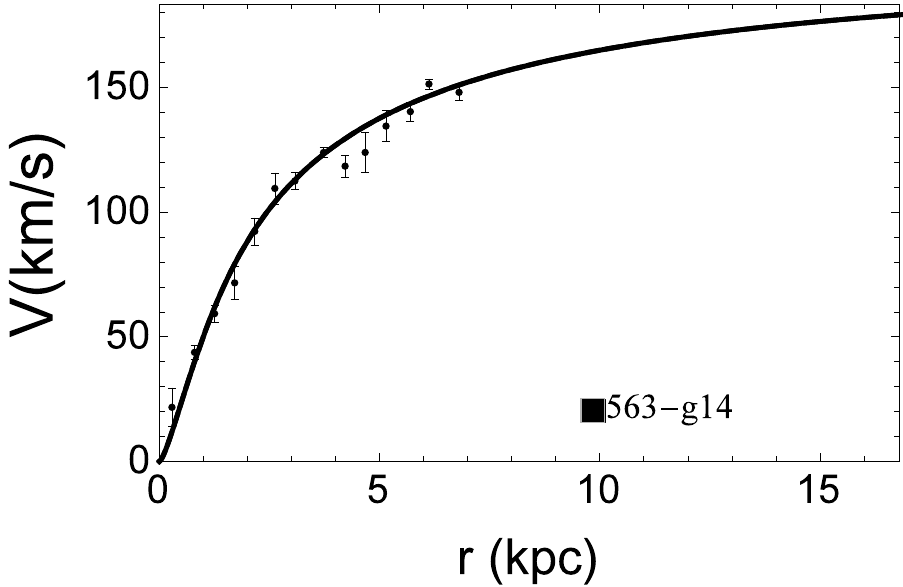}&
\includegraphics[width=0.50\textwidth]{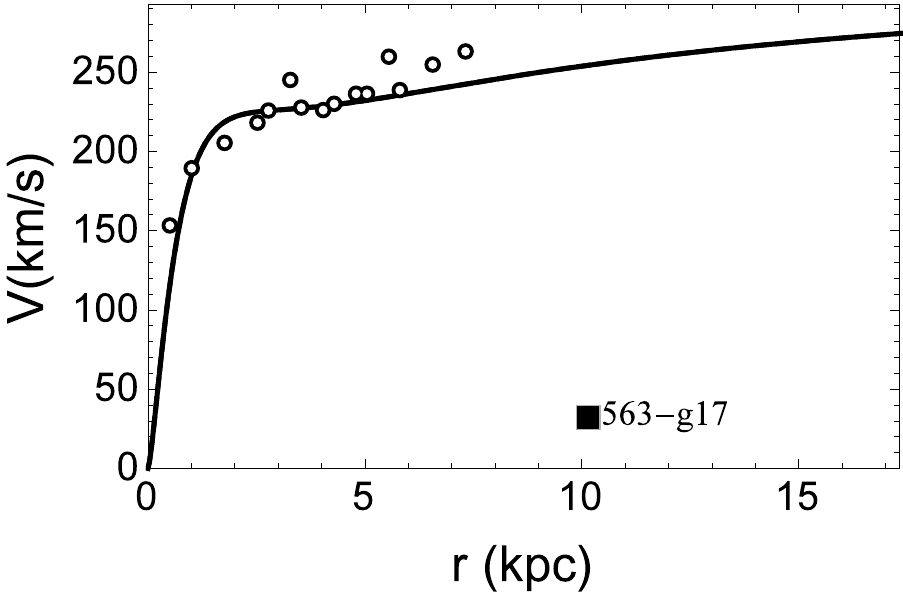}\\
\includegraphics[width=0.50\textwidth]{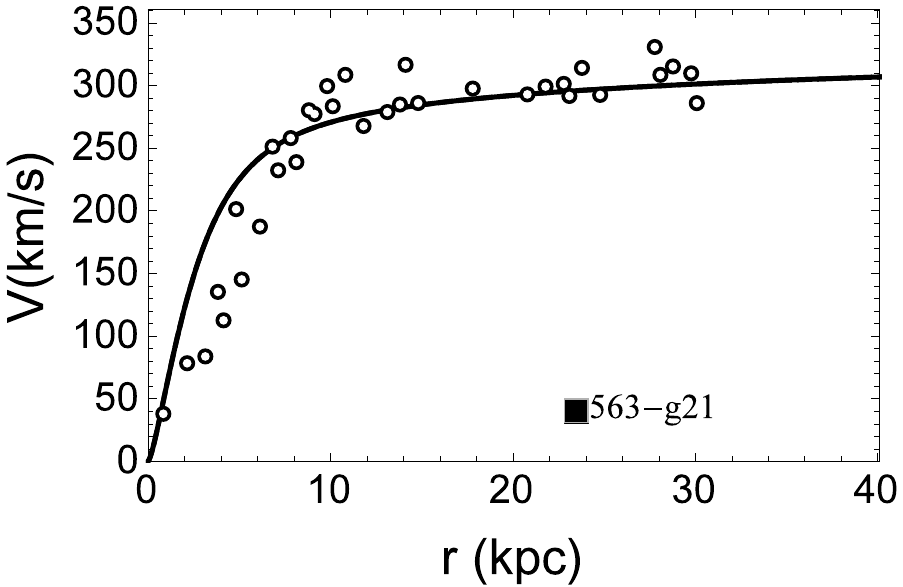}&
\includegraphics[width=0.50\textwidth]{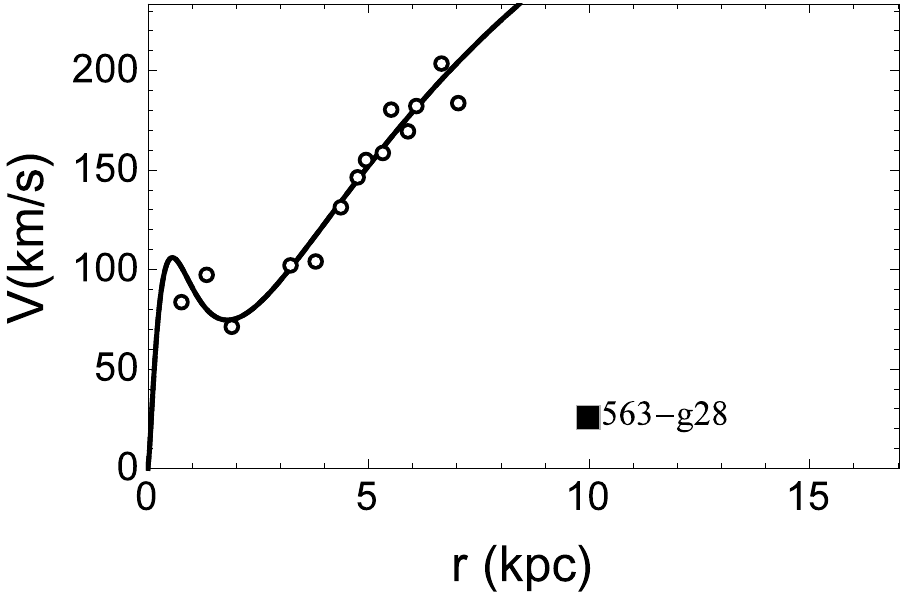}&
\includegraphics[width=0.50\textwidth]{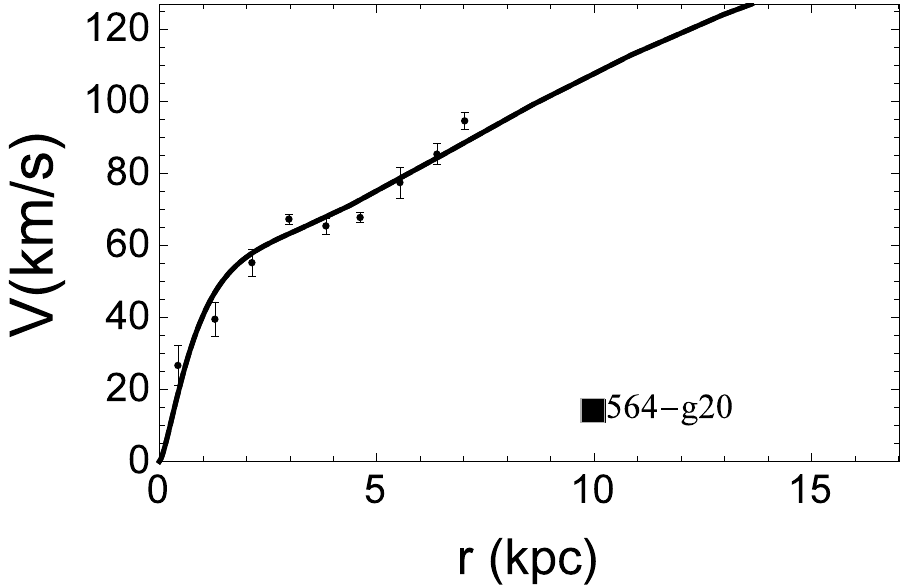}&
\includegraphics[width=0.50\textwidth]{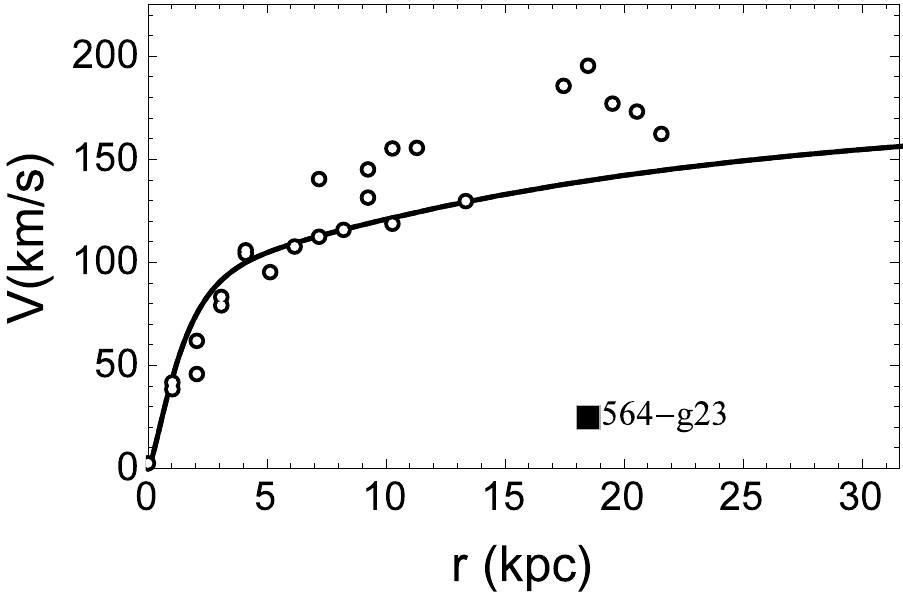}&
\includegraphics[width=0.50\textwidth]{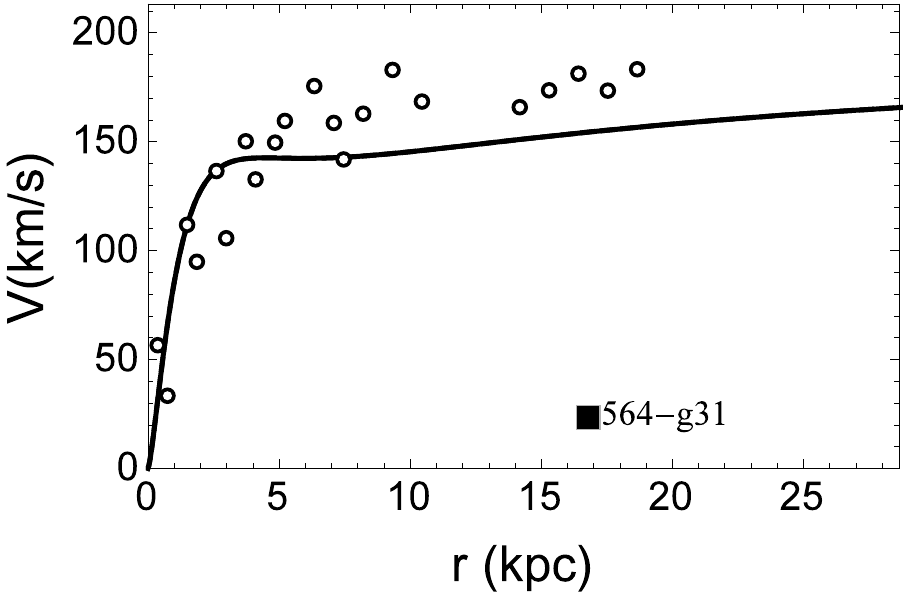}\\
\end{tabular}}}
\end{figure}

\clearpage

\begin{figure}
\flushleft
{\tiny\renewcommand{\arraystretch}{.95}
\resizebox{!}{.64\textwidth}{%
\begin{tabular}{lllll}
\includegraphics[width=0.50\textwidth]{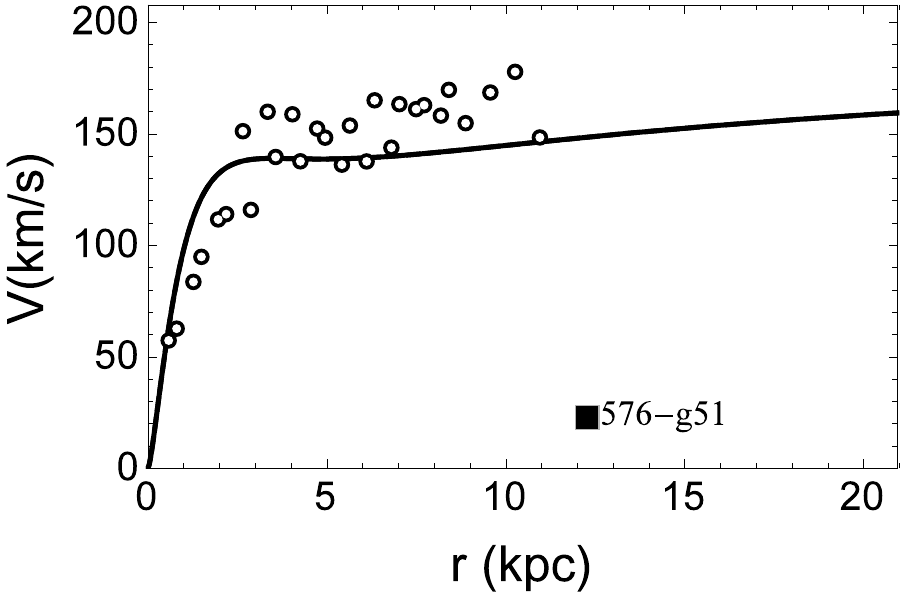}&
\includegraphics[width=0.50\textwidth]{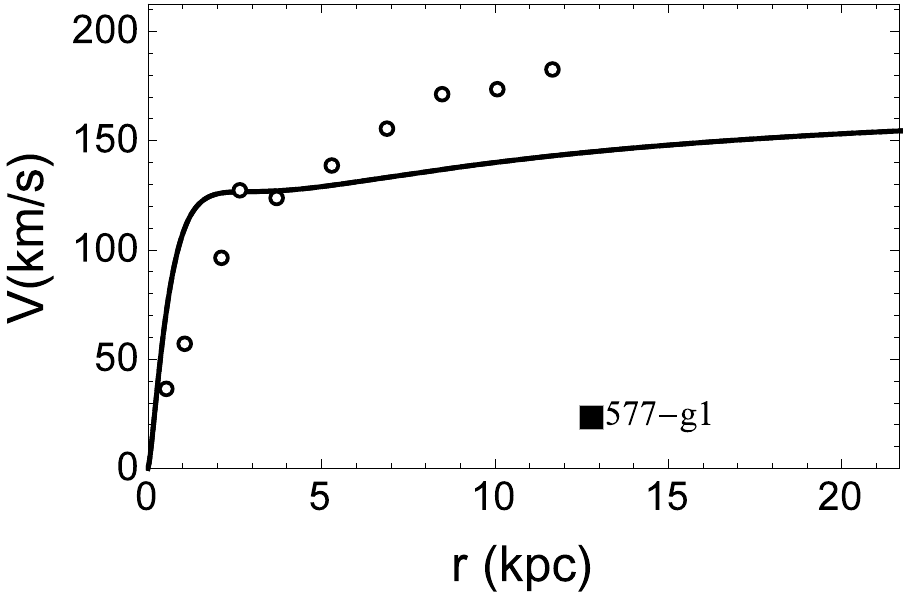}&
\includegraphics[width=0.50\textwidth]{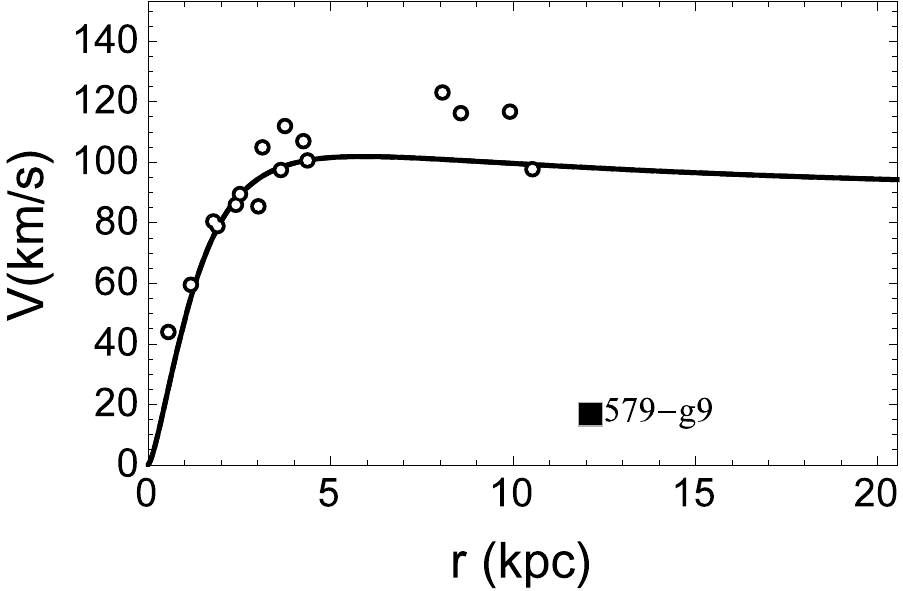}&
\includegraphics[width=0.50\textwidth]{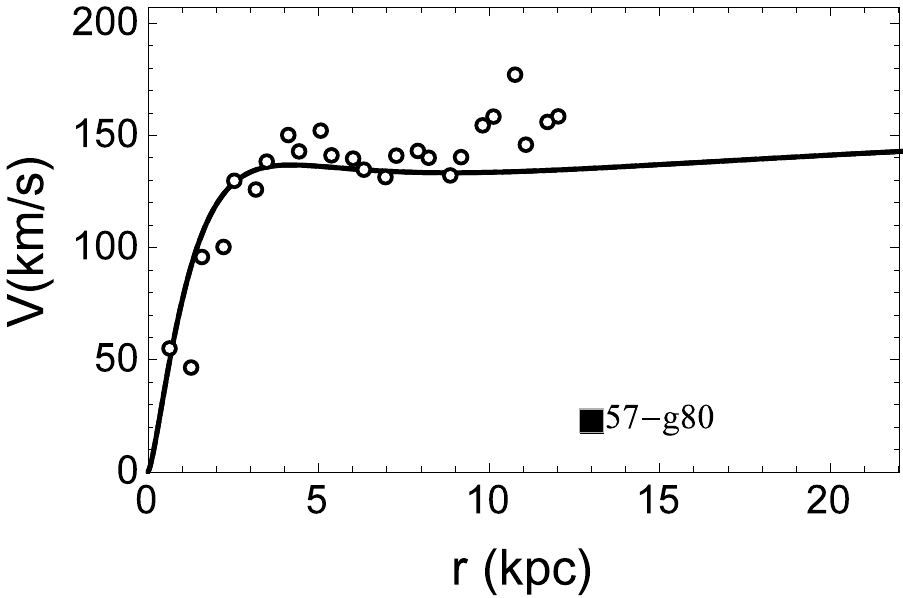}&
\includegraphics[width=0.50\textwidth]{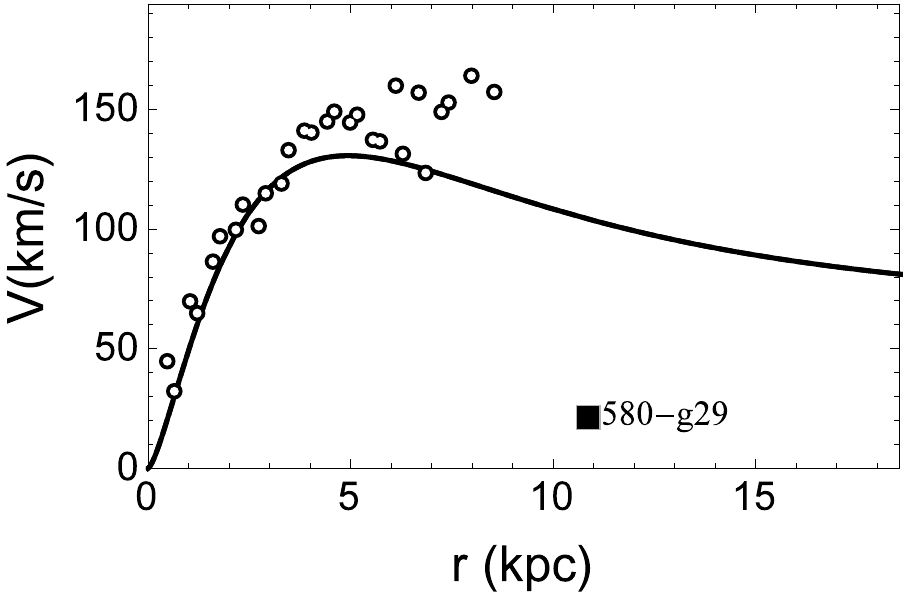}\\
\includegraphics[width=0.50\textwidth]{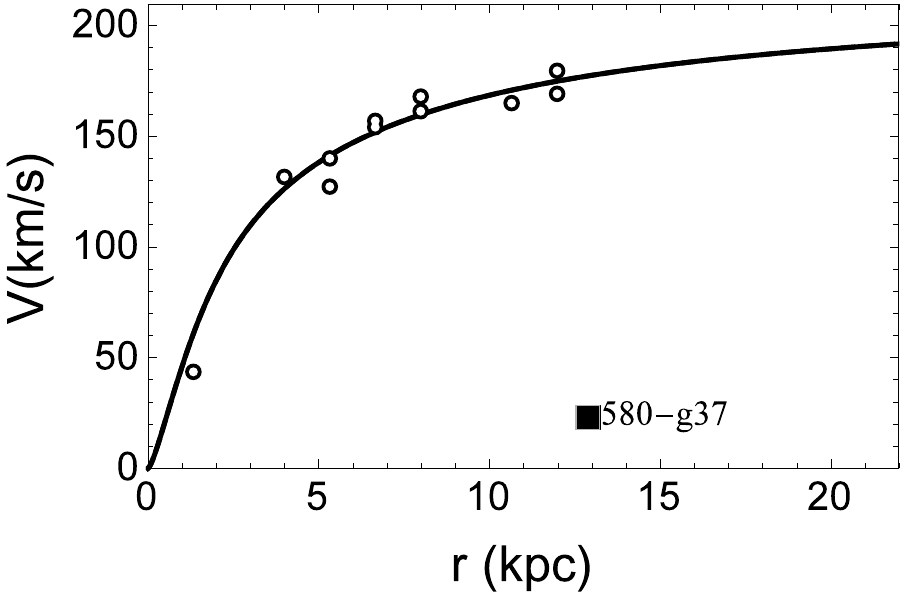}&
\includegraphics[width=0.50\textwidth]{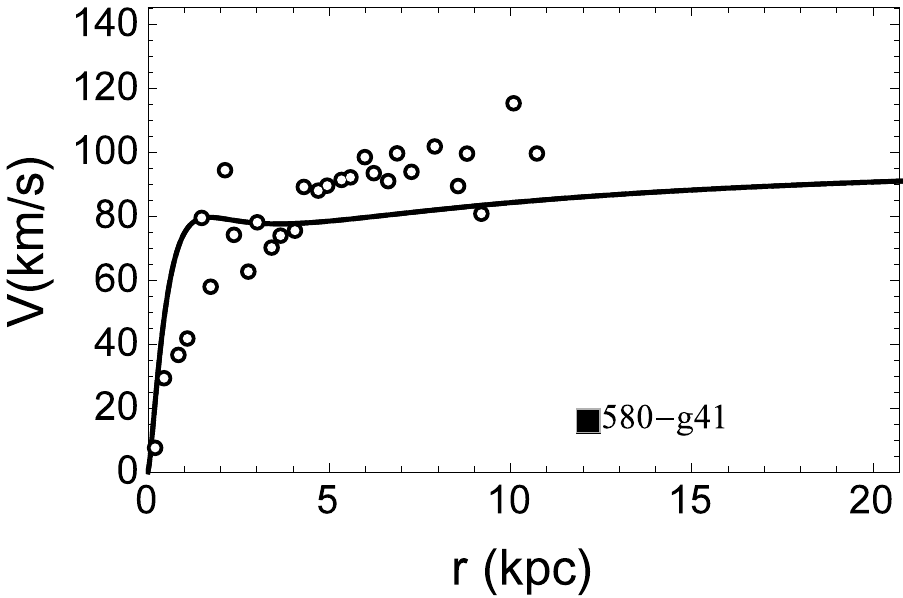}&
\includegraphics[width=0.50\textwidth]{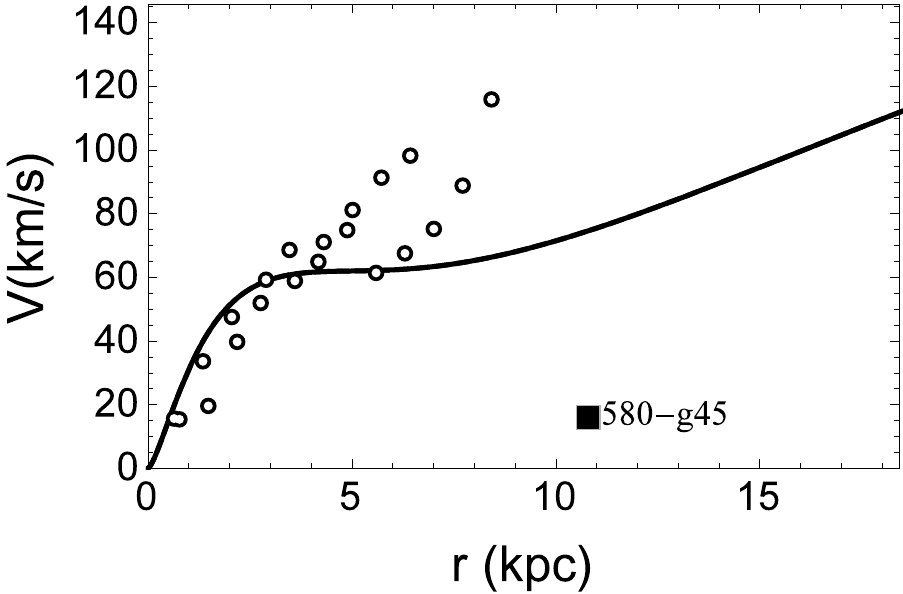}&
\includegraphics[width=0.50\textwidth]{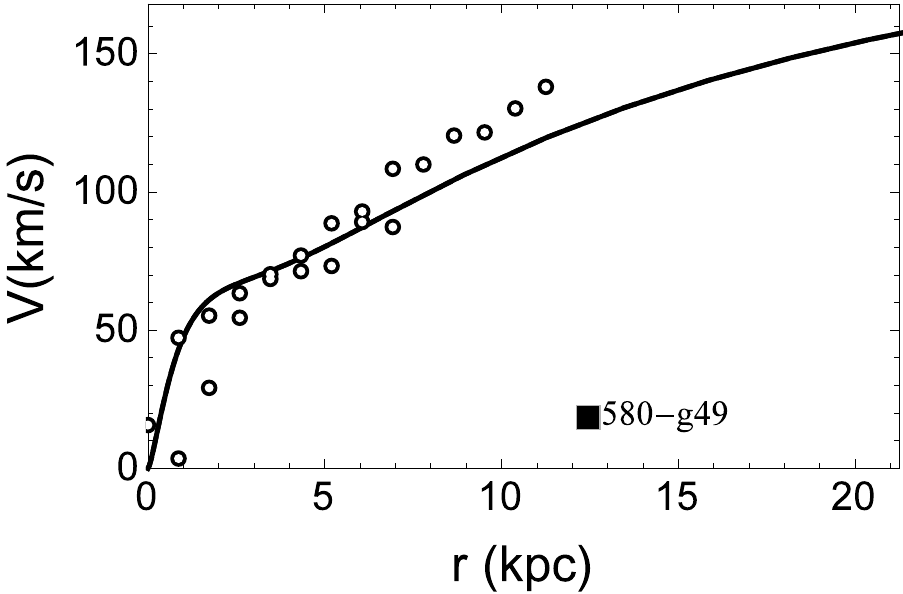}&
\includegraphics[width=0.50\textwidth]{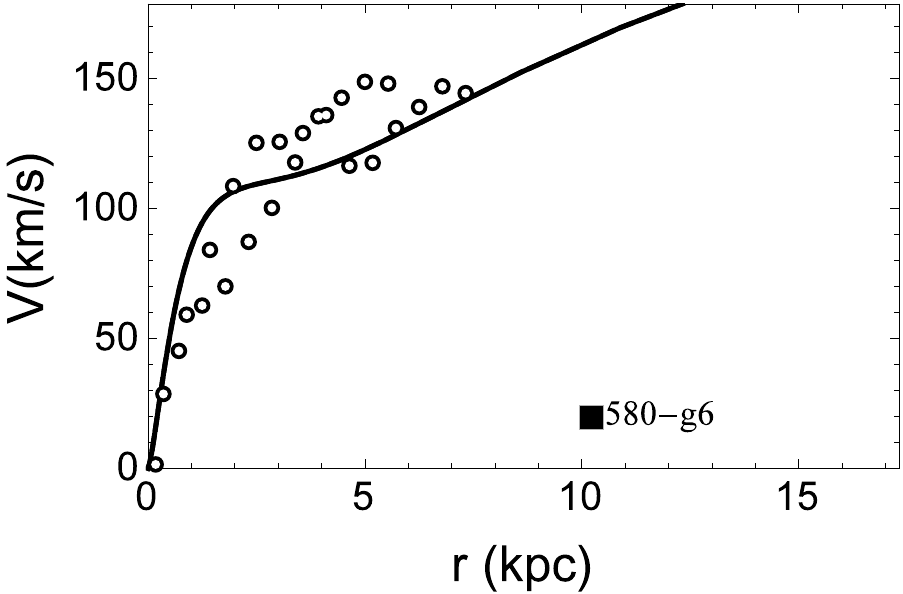}\\
\includegraphics[width=0.50\textwidth]{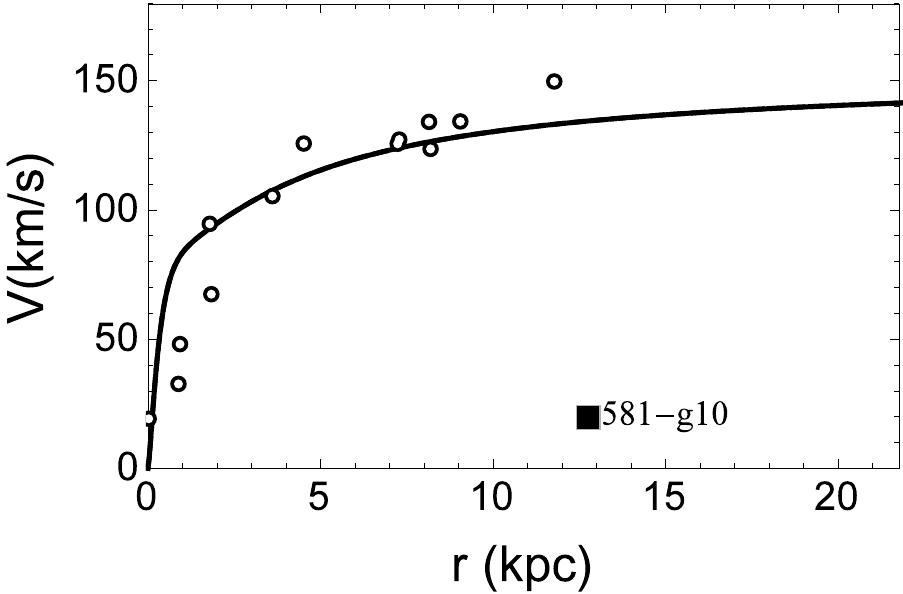}&
\includegraphics[width=0.50\textwidth]{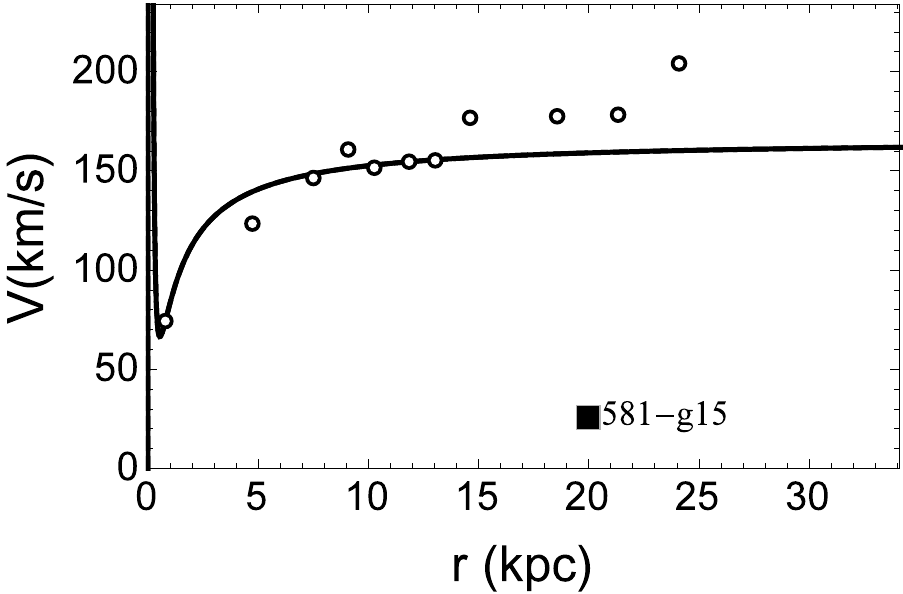}&
\includegraphics[width=0.50\textwidth]{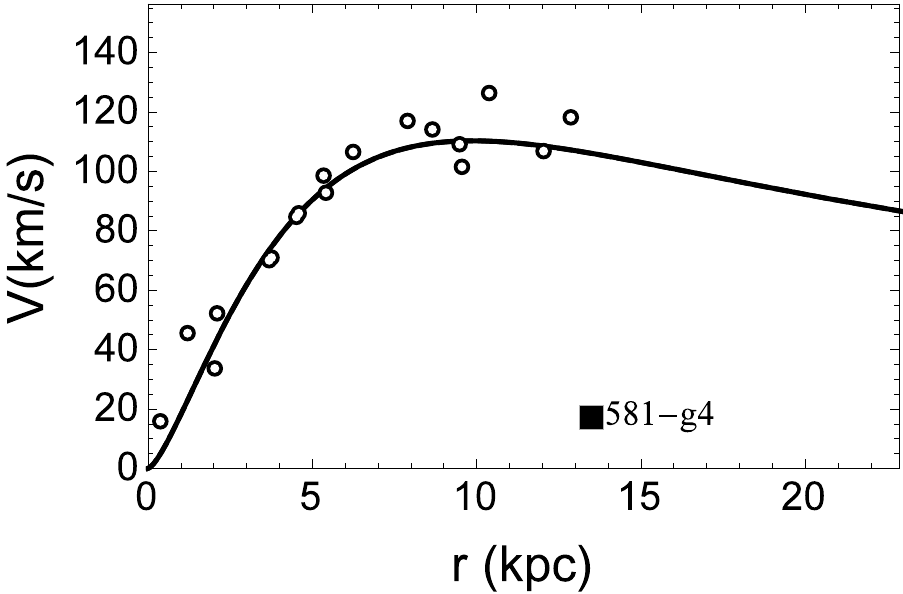}&
\includegraphics[width=0.50\textwidth]{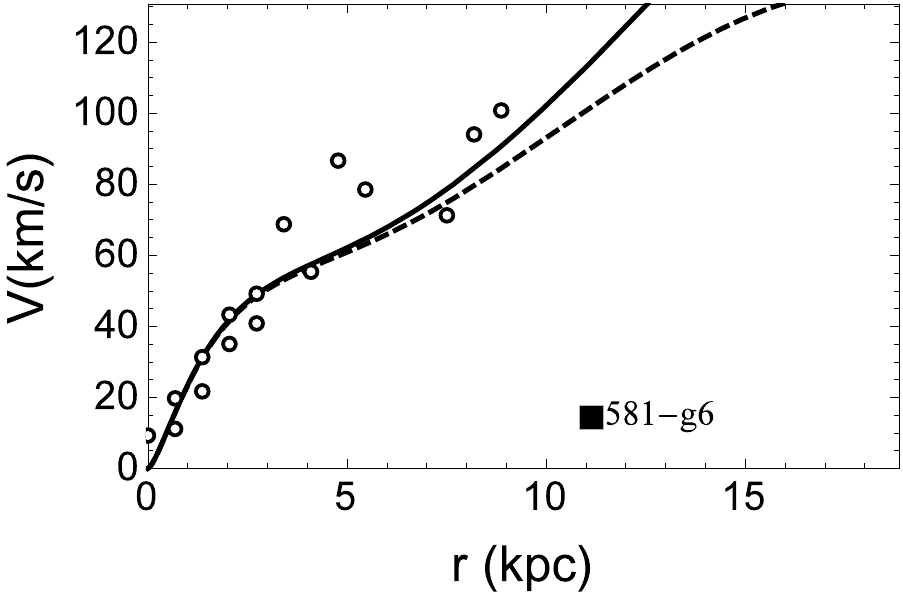}&
\includegraphics[width=0.50\textwidth]{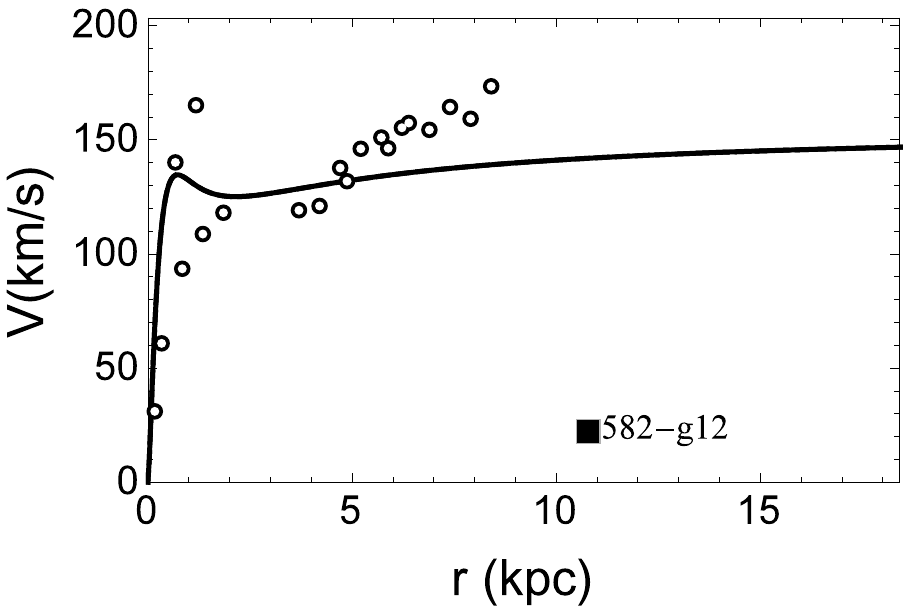}\\
\includegraphics[width=0.50\textwidth]{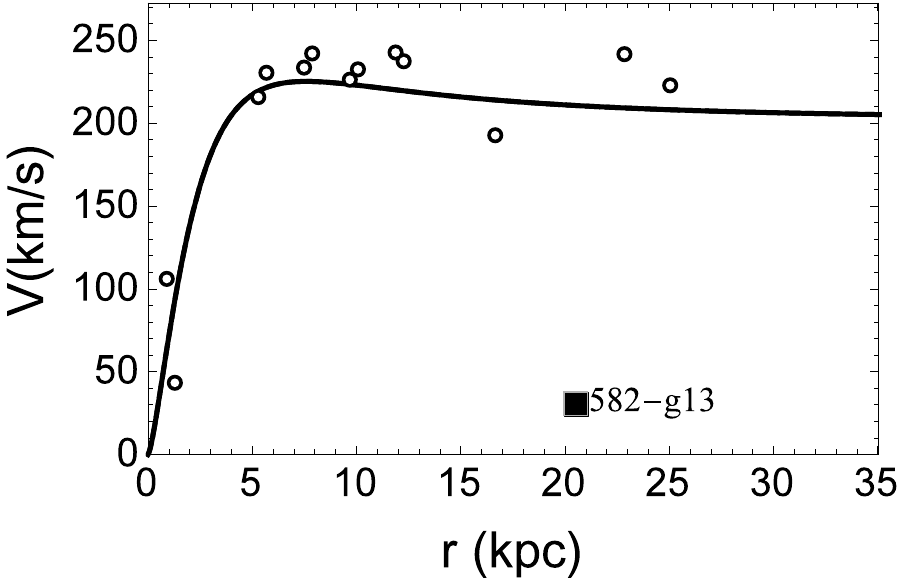}&
\includegraphics[width=0.50\textwidth]{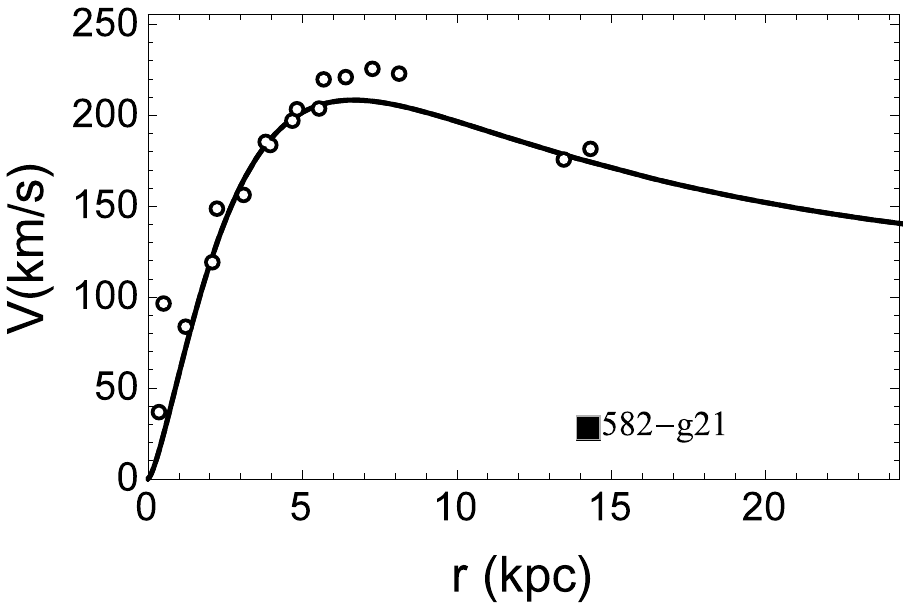}&
\includegraphics[width=0.50\textwidth]{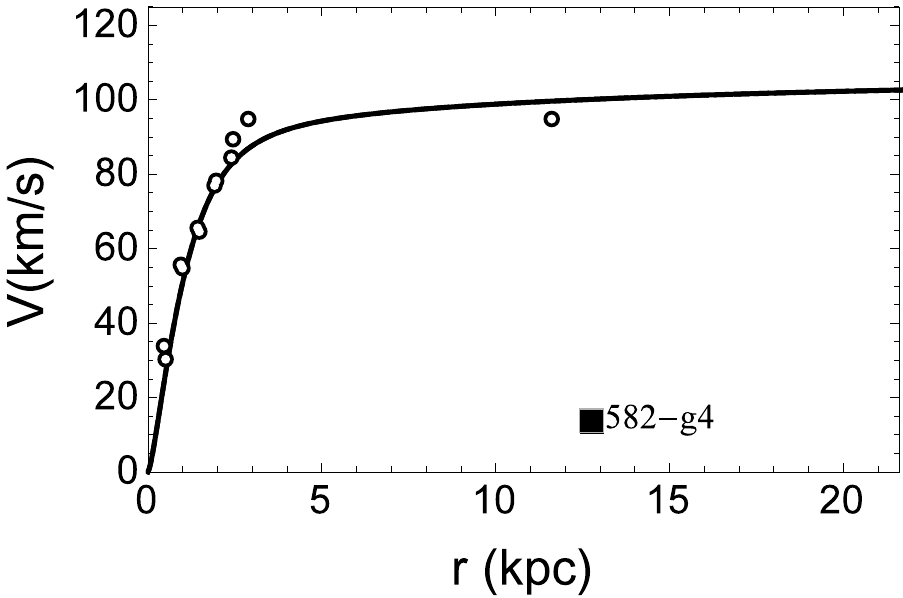}&
\includegraphics[width=0.50\textwidth]{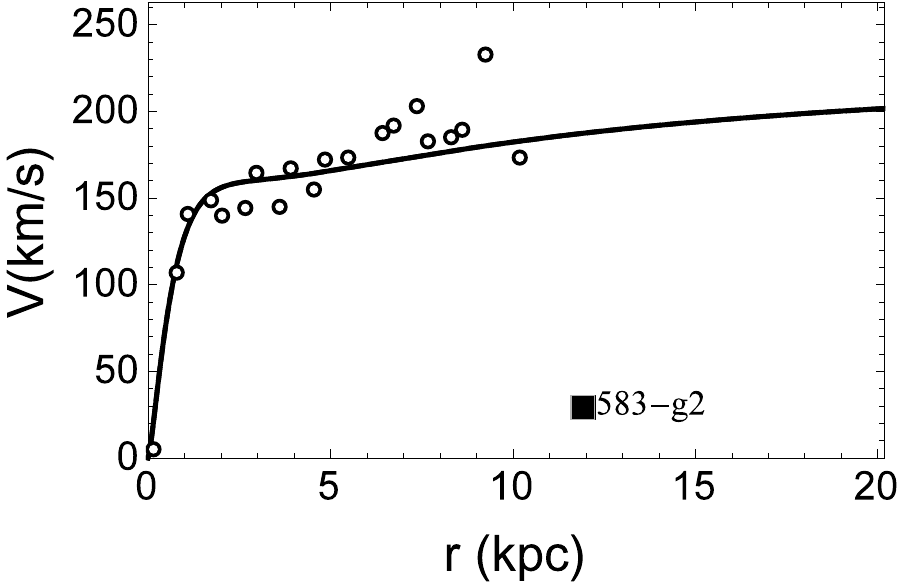}&
\includegraphics[width=0.50\textwidth]{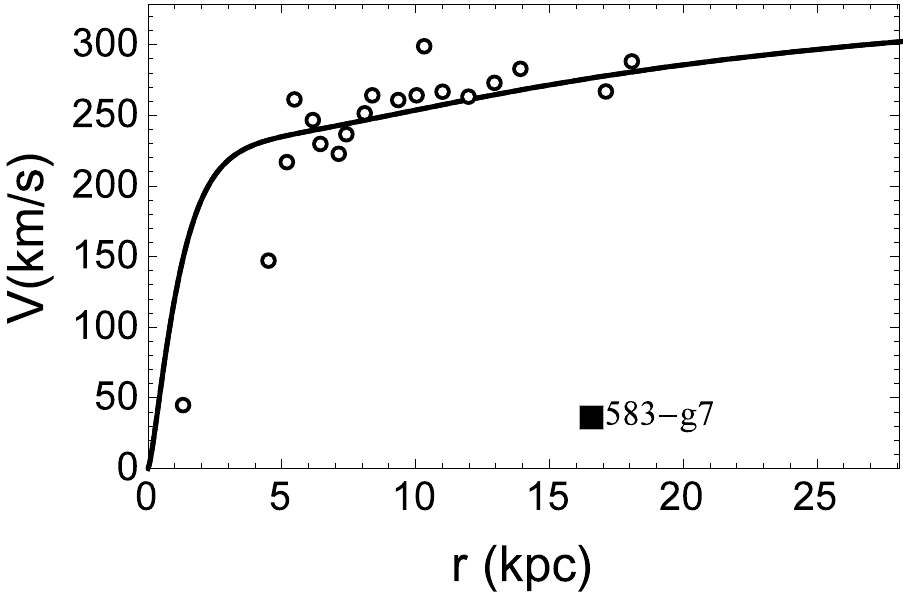}\\
\end{tabular}}}
\end{figure}

\clearpage

\begin{figure}
\flushleft
{\tiny\renewcommand{\arraystretch}{.95}
\resizebox{!}{.64\textwidth}{%
\begin{tabular}{lllll}
\includegraphics[width=0.50\textwidth]{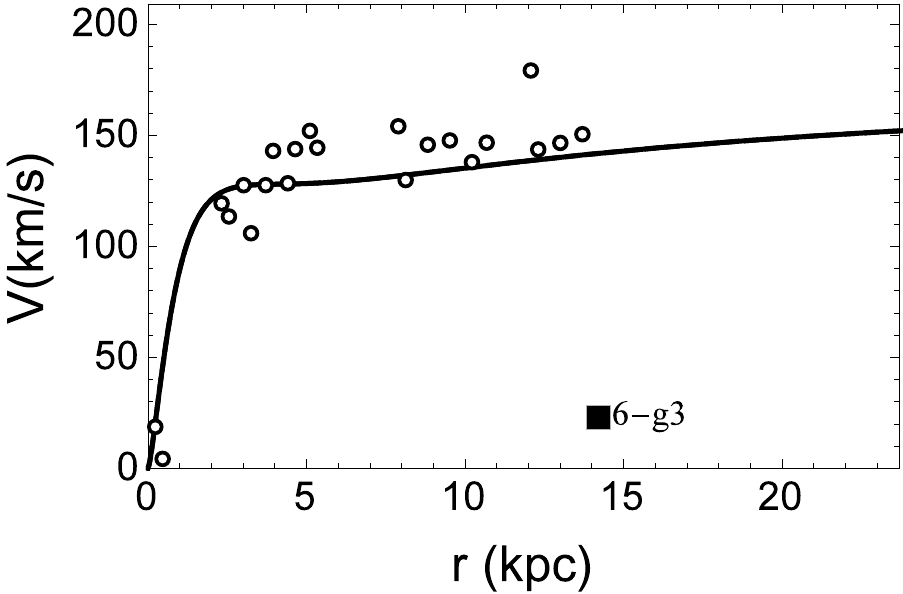}&
\includegraphics[width=0.50\textwidth]{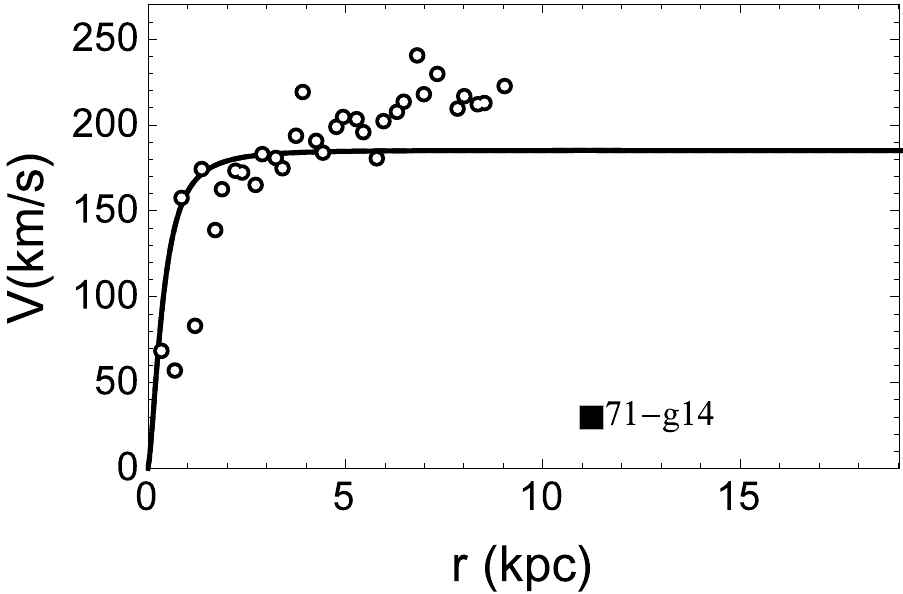}&
\includegraphics[width=0.50\textwidth]{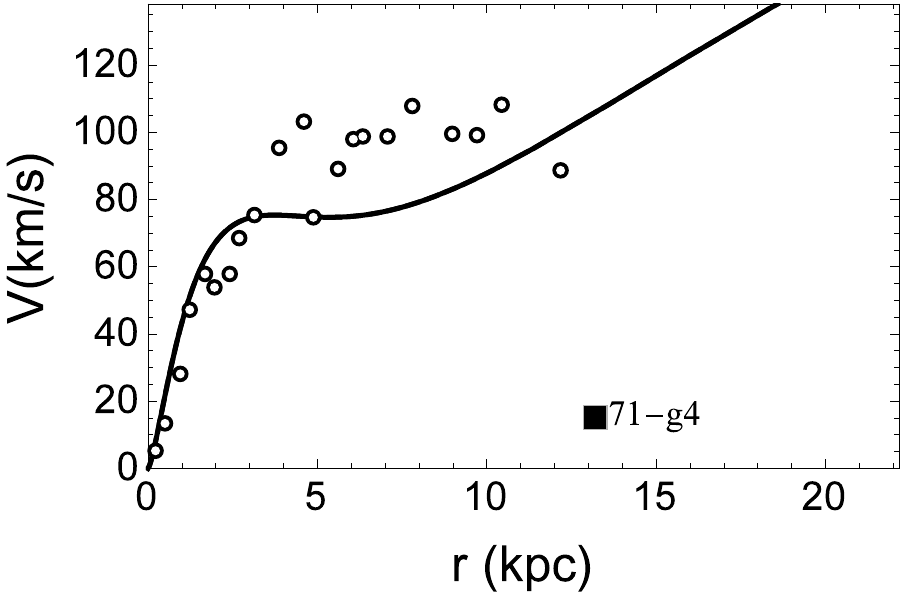}&
\includegraphics[width=0.50\textwidth]{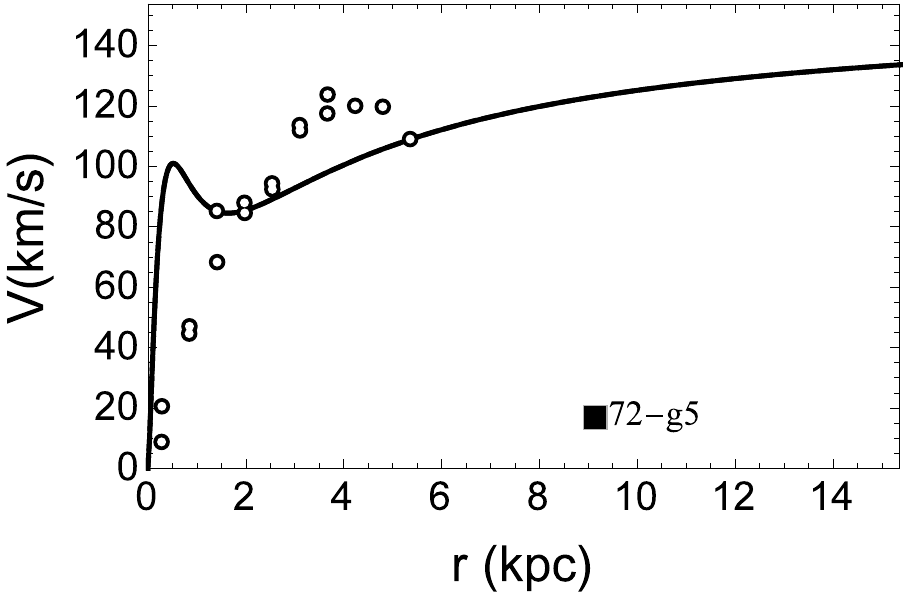}&
\includegraphics[width=0.50\textwidth]{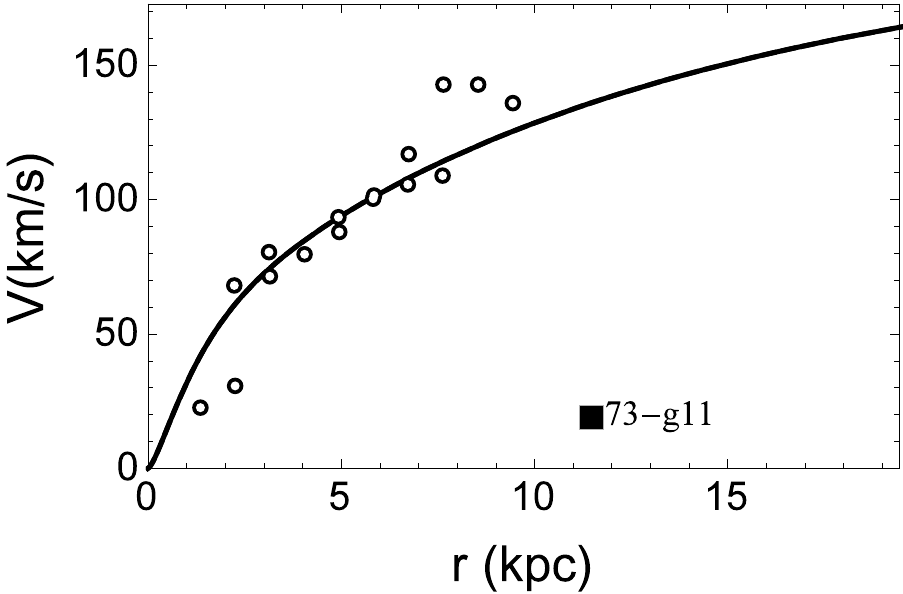}\\
\includegraphics[width=0.50\textwidth]{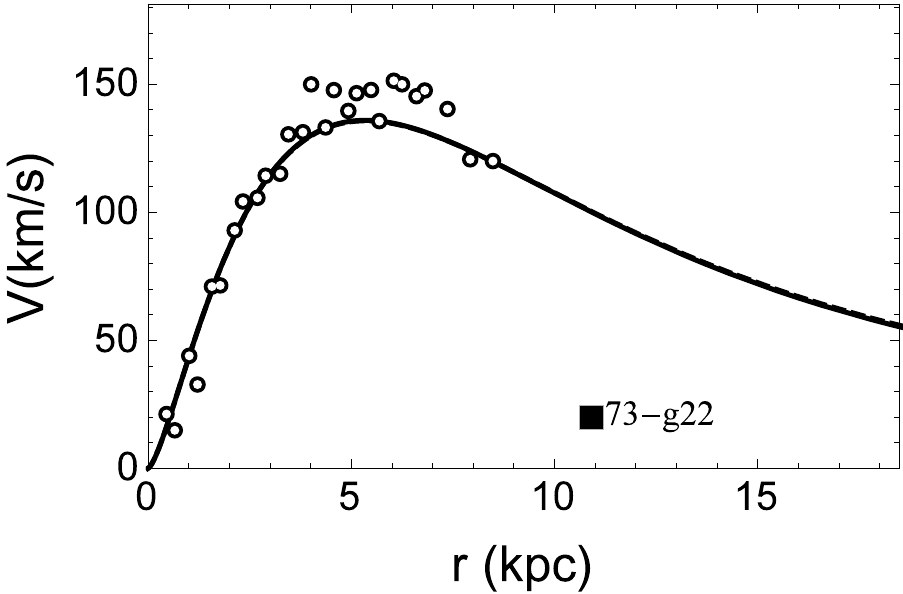}&
\includegraphics[width=0.50\textwidth]{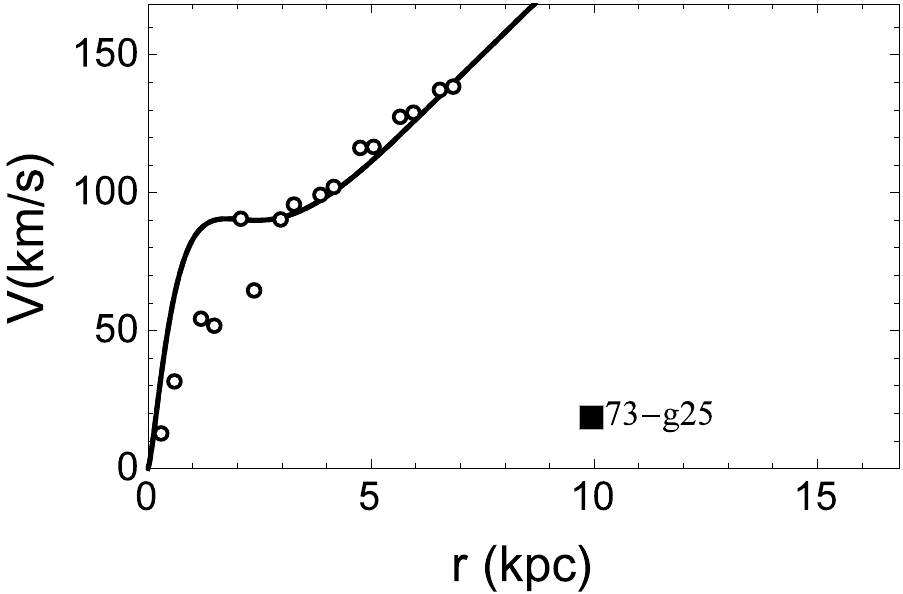}&
\includegraphics[width=0.50\textwidth]{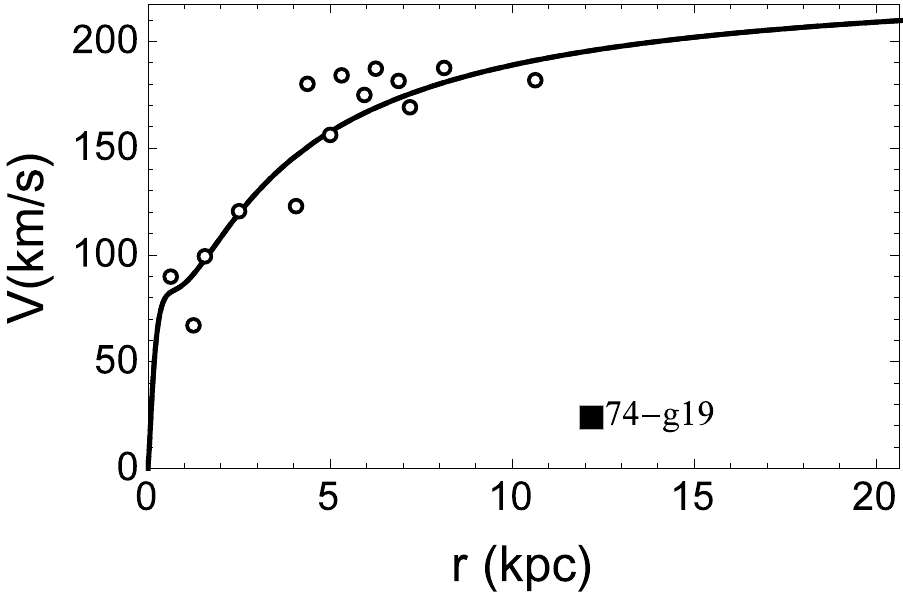}&
\includegraphics[width=0.50\textwidth]{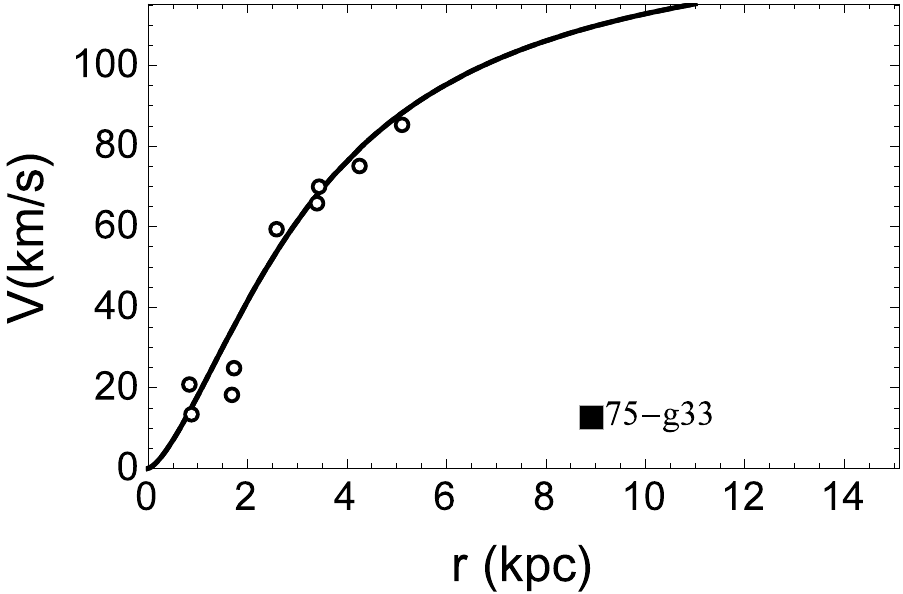}&
\includegraphics[width=0.50\textwidth]{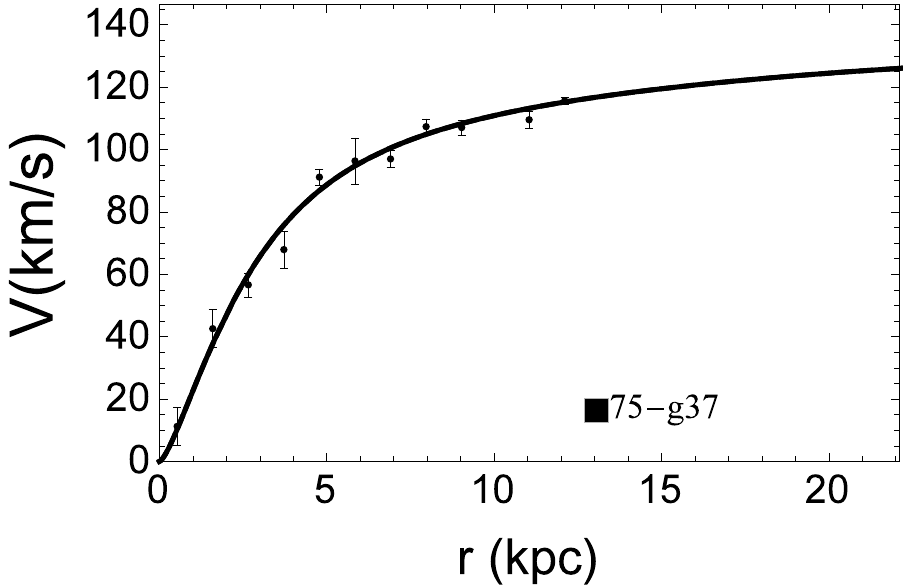}\\
\includegraphics[width=0.50\textwidth]{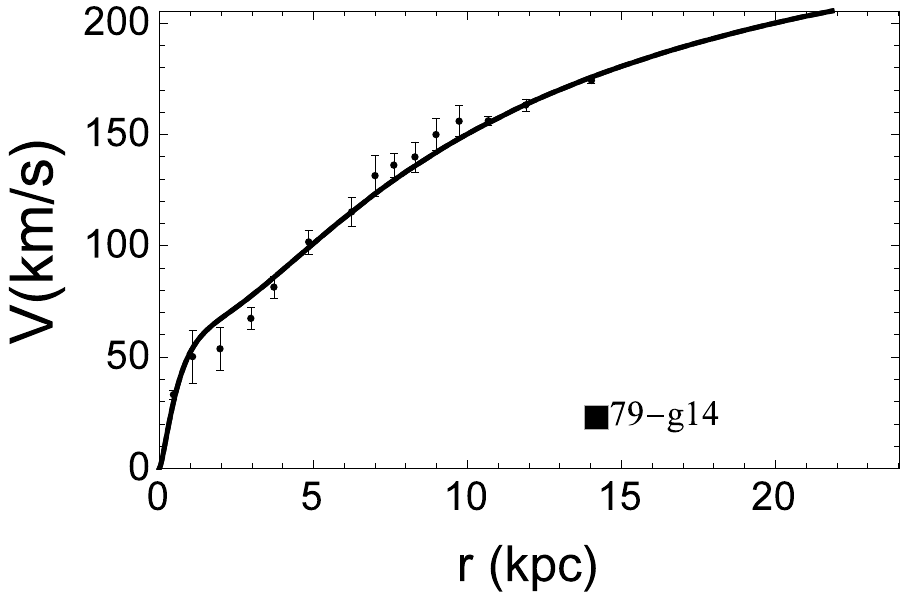}&
\includegraphics[width=0.50\textwidth]{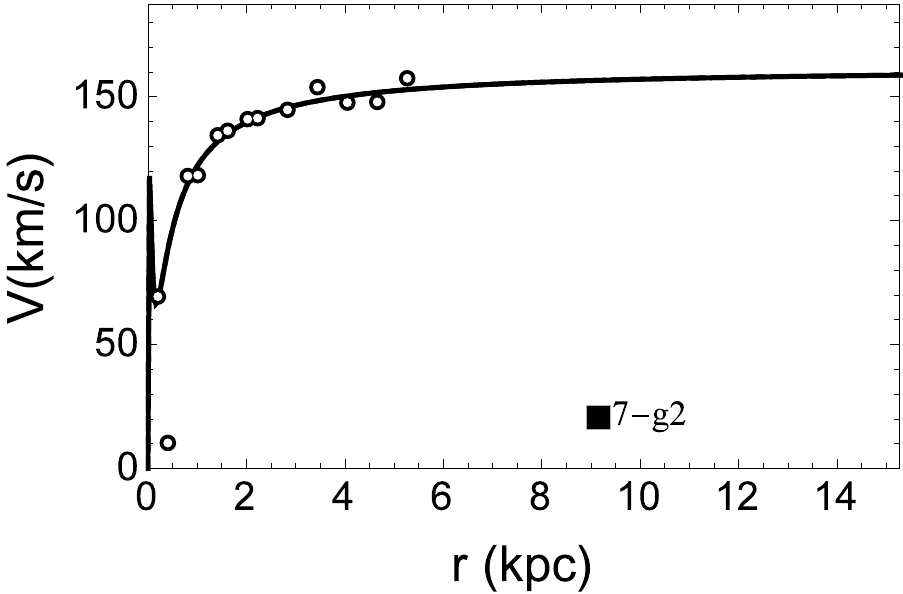}&
\includegraphics[width=0.50\textwidth]{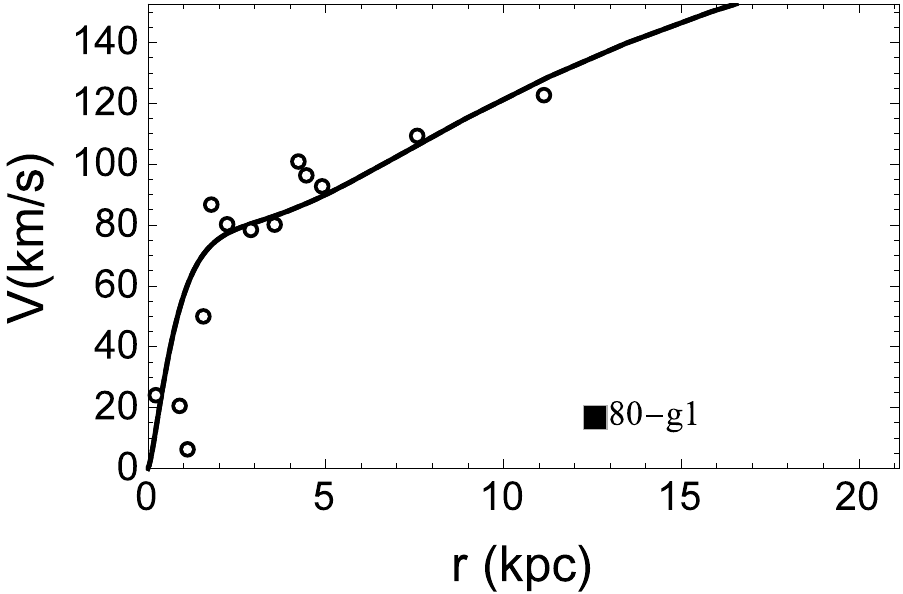}&
\includegraphics[width=0.50\textwidth]{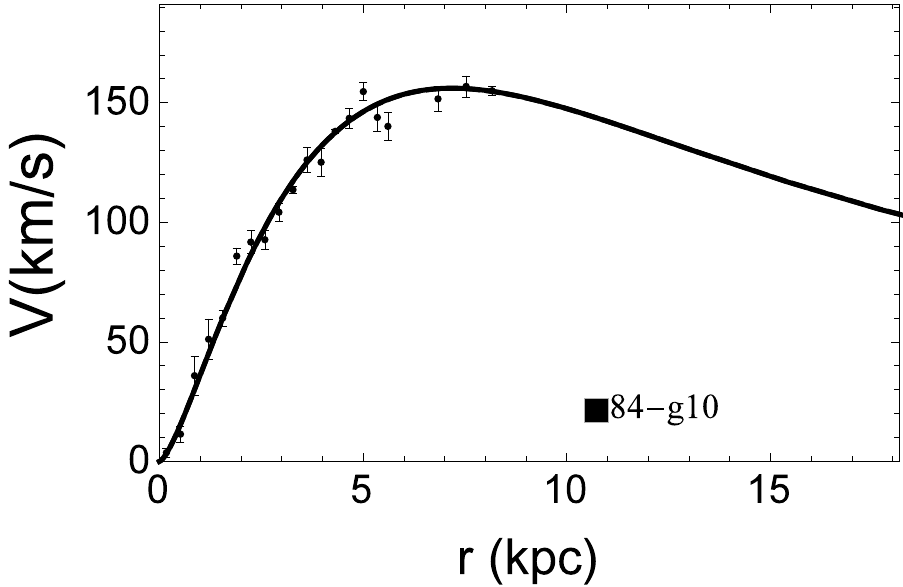}&
\includegraphics[width=0.50\textwidth]{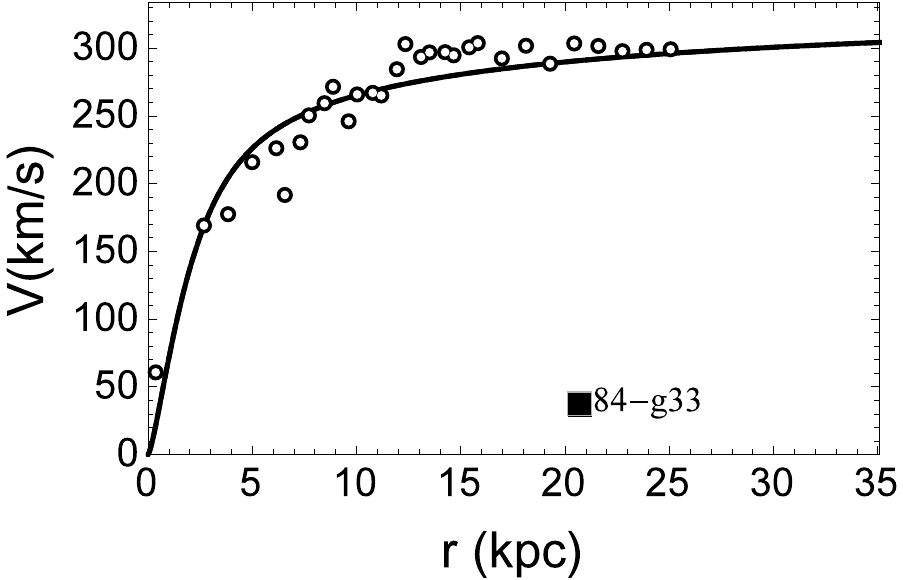}\\
\includegraphics[width=0.50\textwidth]{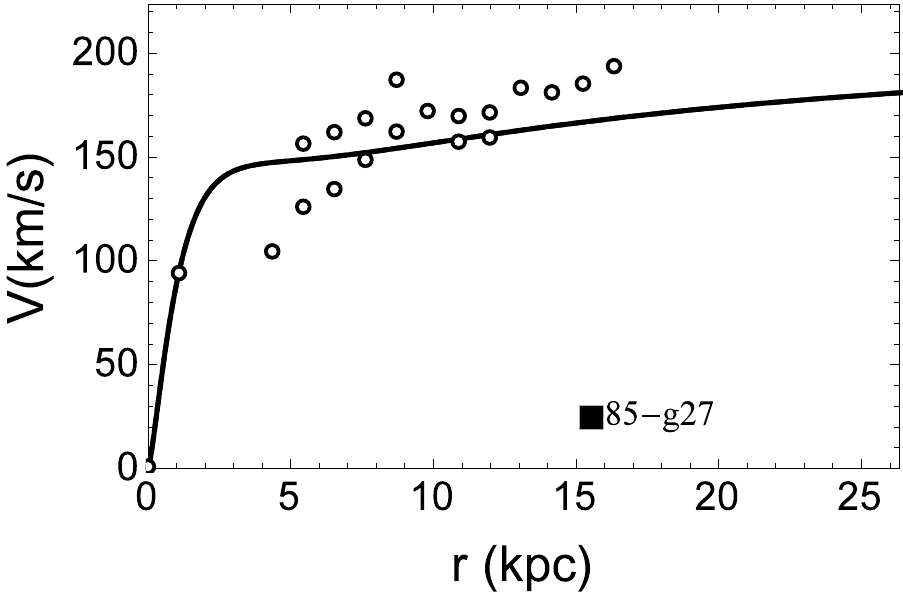}&
\includegraphics[width=0.50\textwidth]{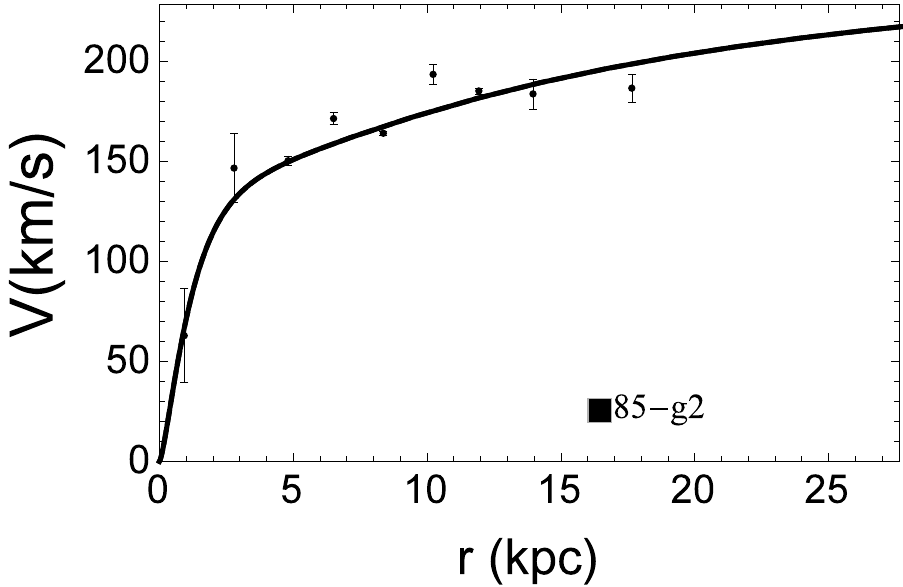}&
\includegraphics[width=0.50\textwidth]{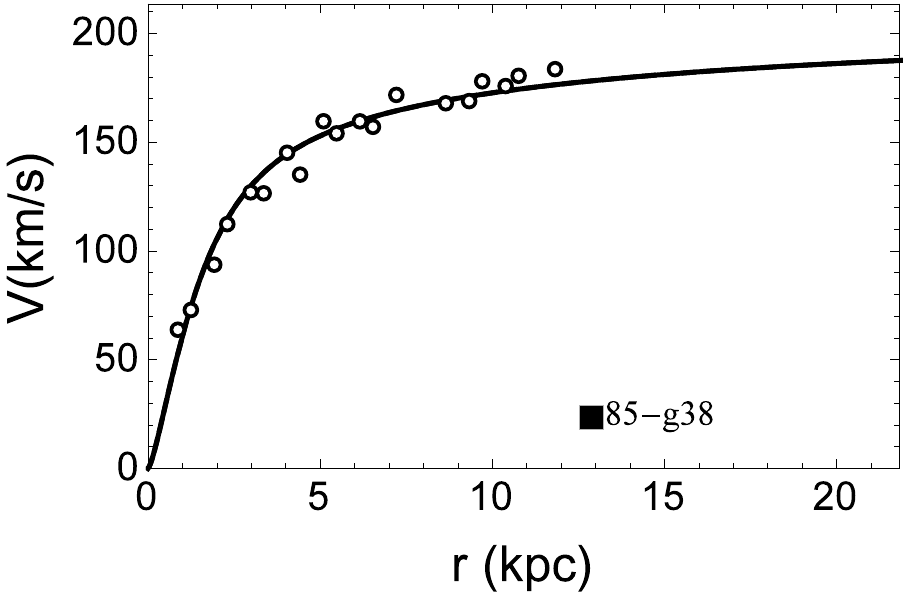}&
\includegraphics[width=0.50\textwidth]{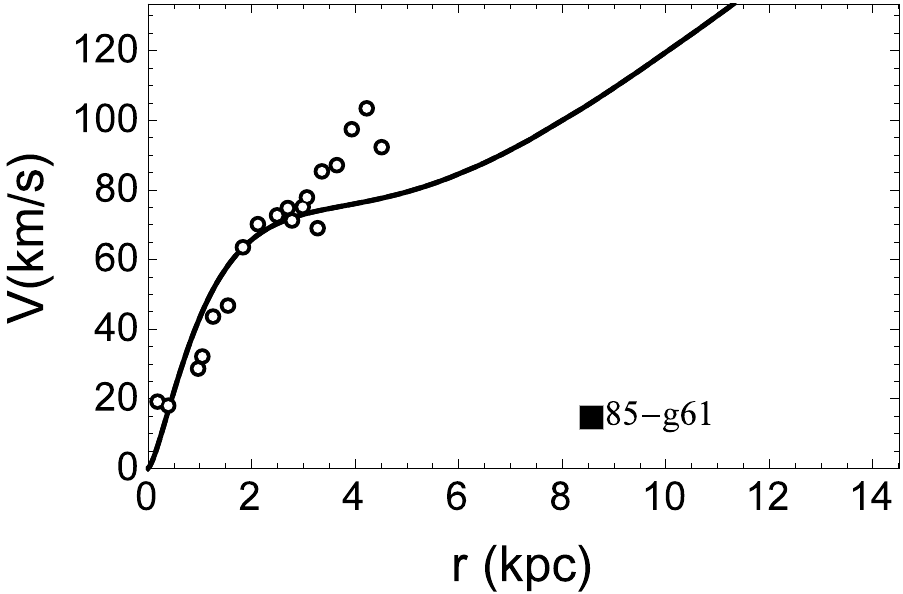}&
\includegraphics[width=0.50\textwidth]{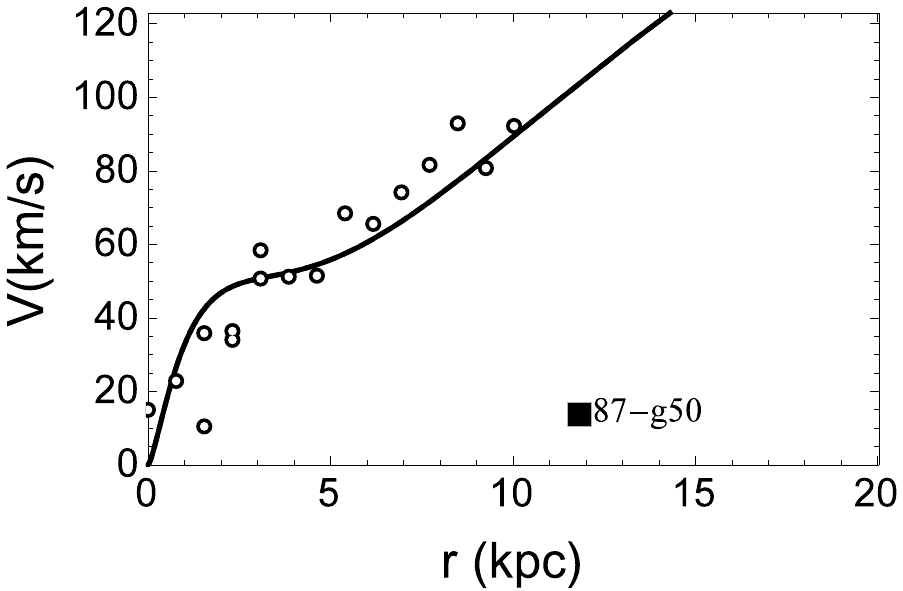}\\
\end{tabular}}}
\end{figure}

\clearpage

\begin{figure}
\flushleft
{\tiny\renewcommand{\arraystretch}{.95}
\resizebox{!}{.64\textwidth}{%
\begin{tabular}{lllll}
\includegraphics[width=0.50\textwidth]{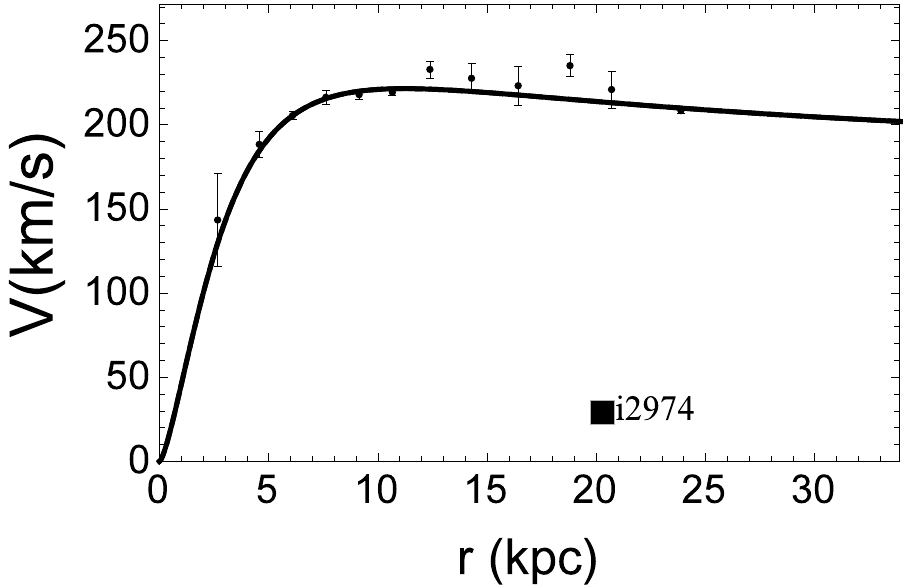}&
\includegraphics[width=0.50\textwidth]{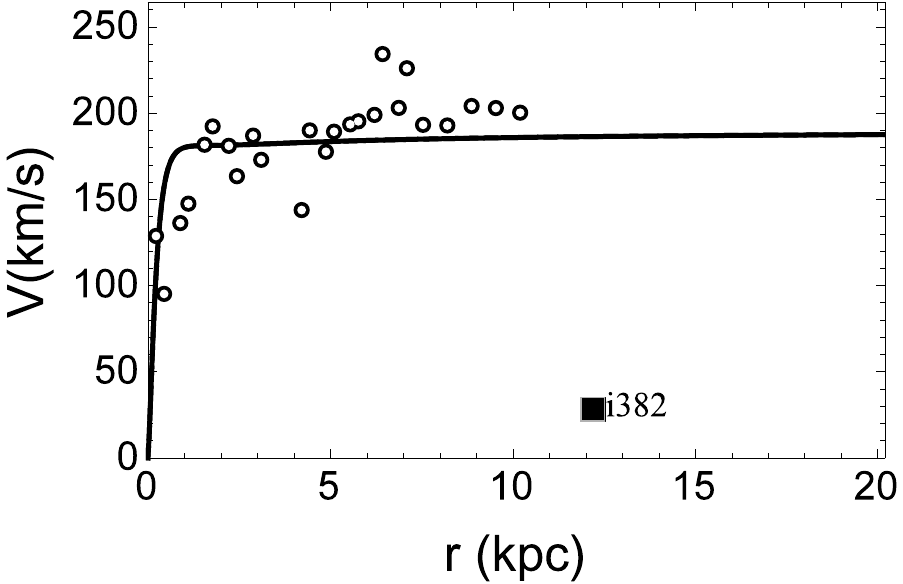}&
\includegraphics[width=0.50\textwidth]{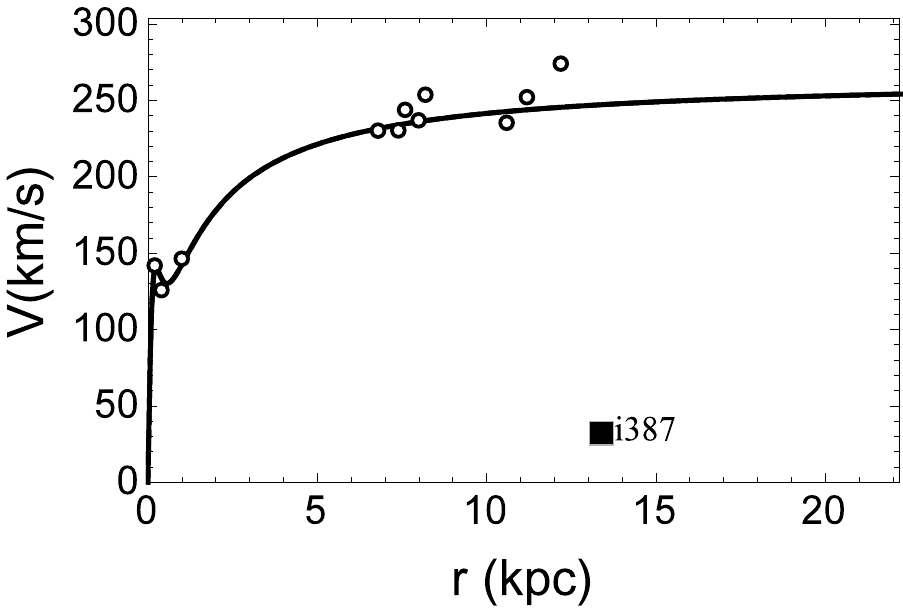}&
\includegraphics[width=0.50\textwidth]{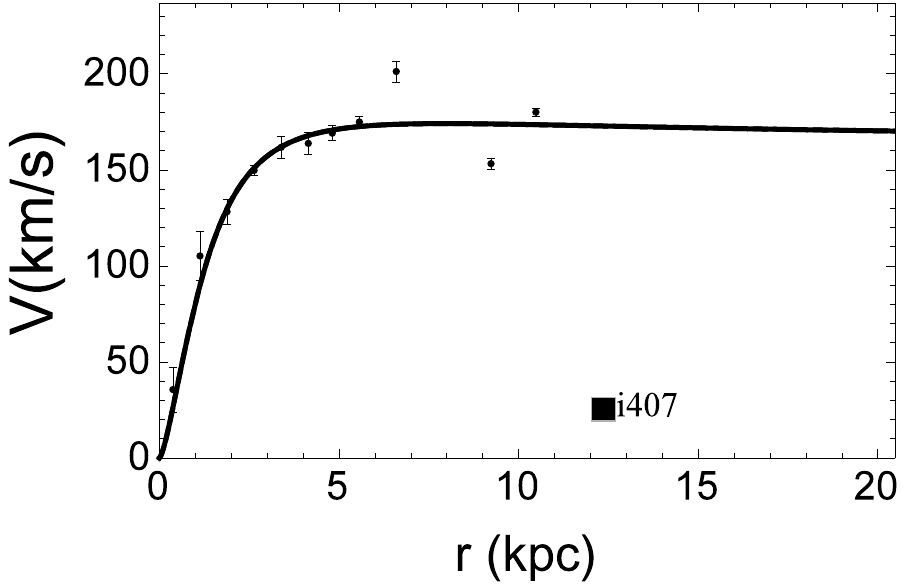}&
\includegraphics[width=0.50\textwidth]{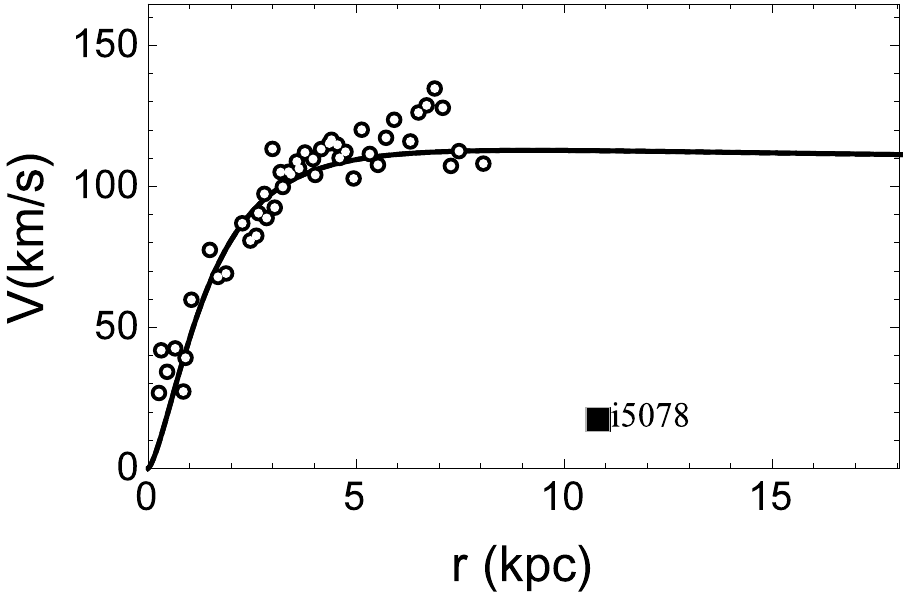}\\
\includegraphics[width=0.50\textwidth]{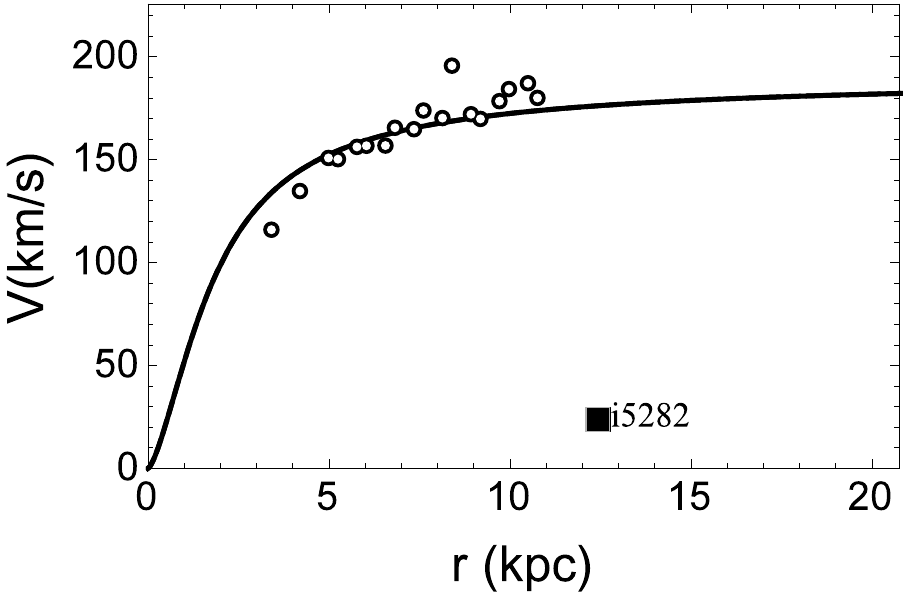}&
\includegraphics[width=0.50\textwidth]{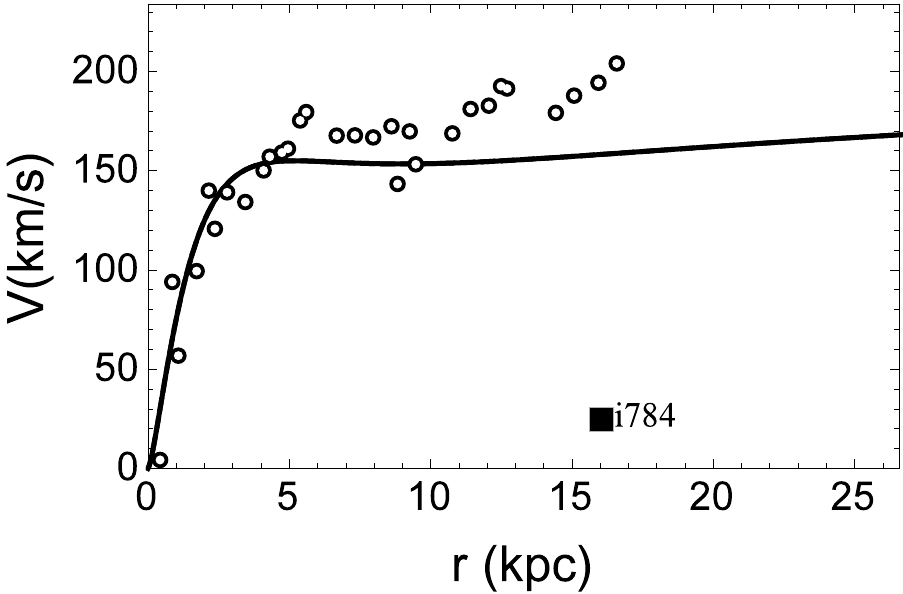}&
\includegraphics[width=0.50\textwidth]{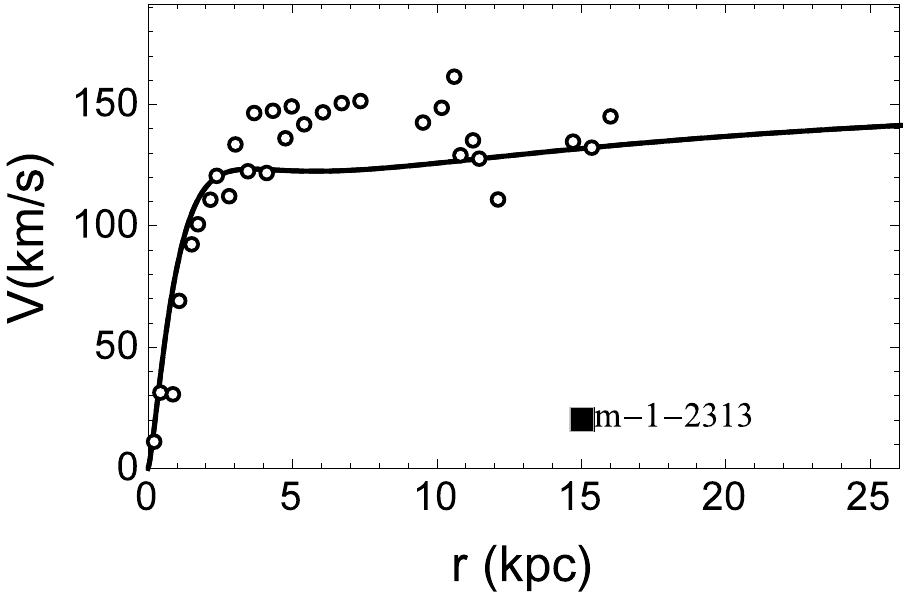}&
\includegraphics[width=0.50\textwidth]{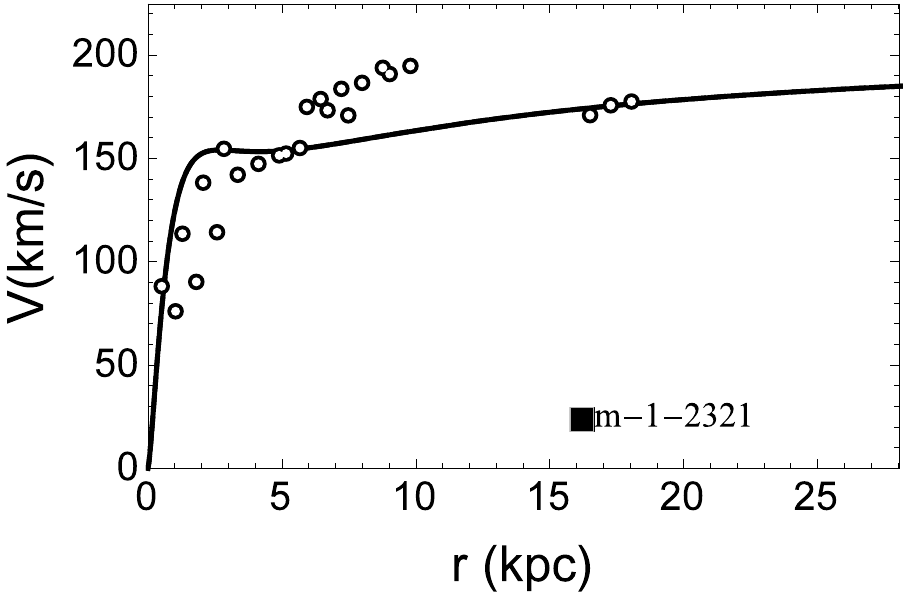}&
\includegraphics[width=0.50\textwidth]{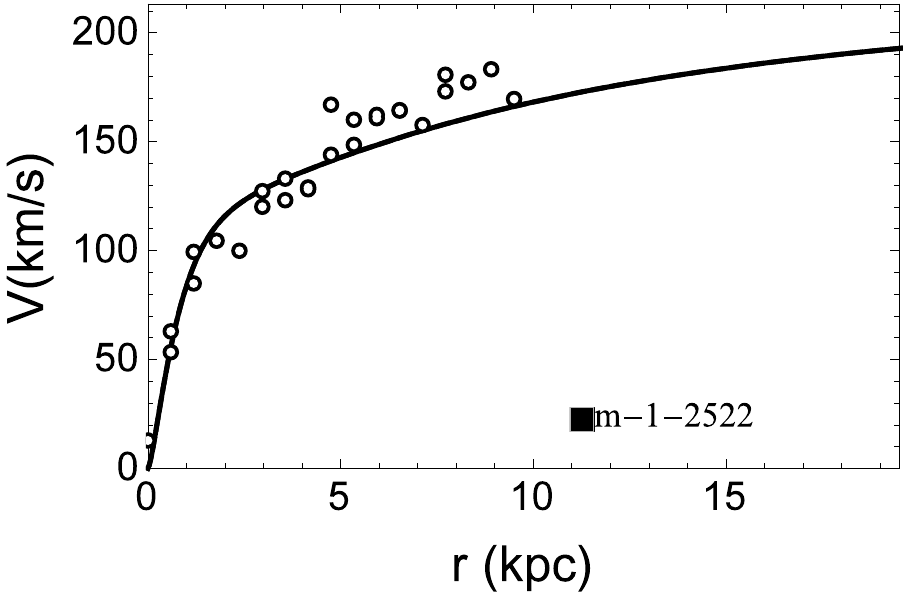}\\
\includegraphics[width=0.50\textwidth]{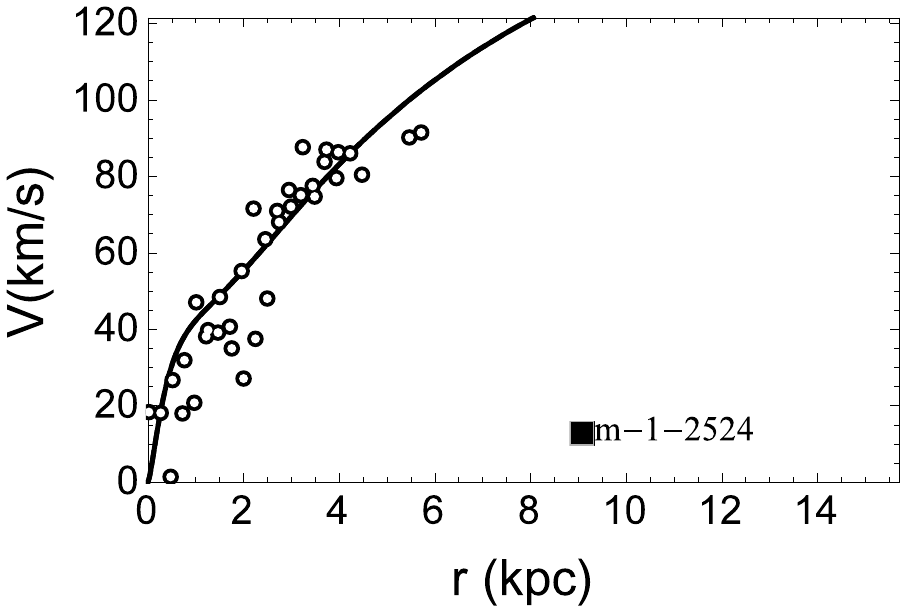}&
\includegraphics[width=0.50\textwidth]{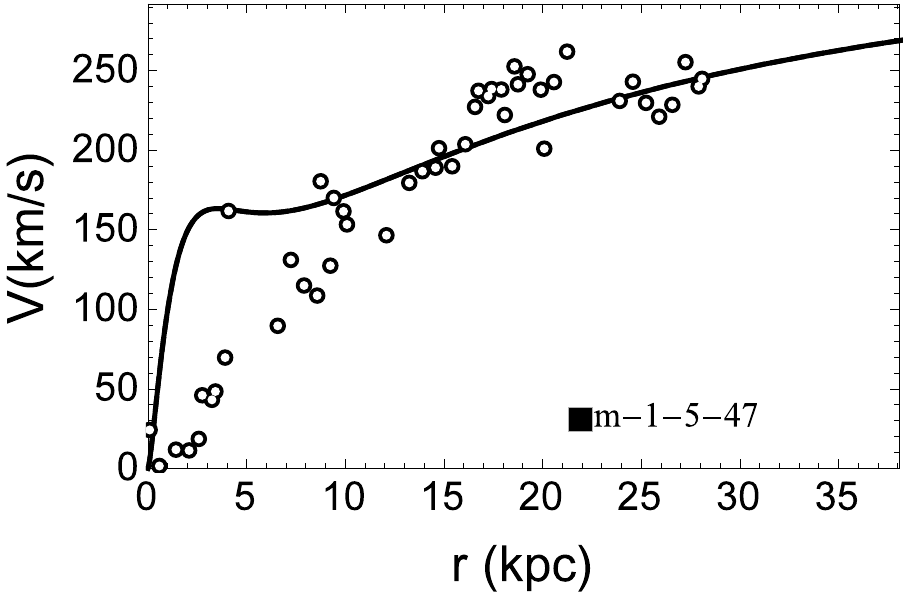}&
\includegraphics[width=0.50\textwidth]{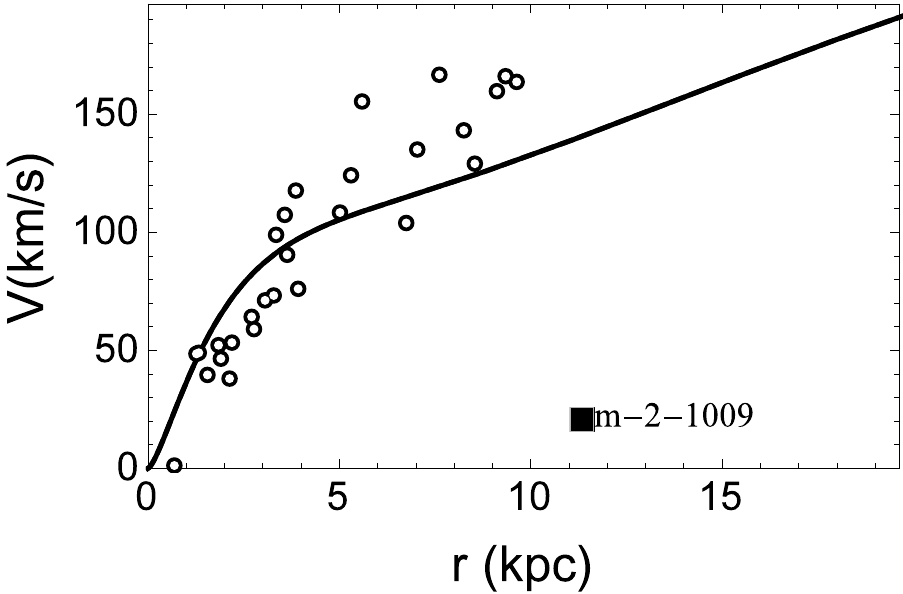}&
\includegraphics[width=0.50\textwidth]{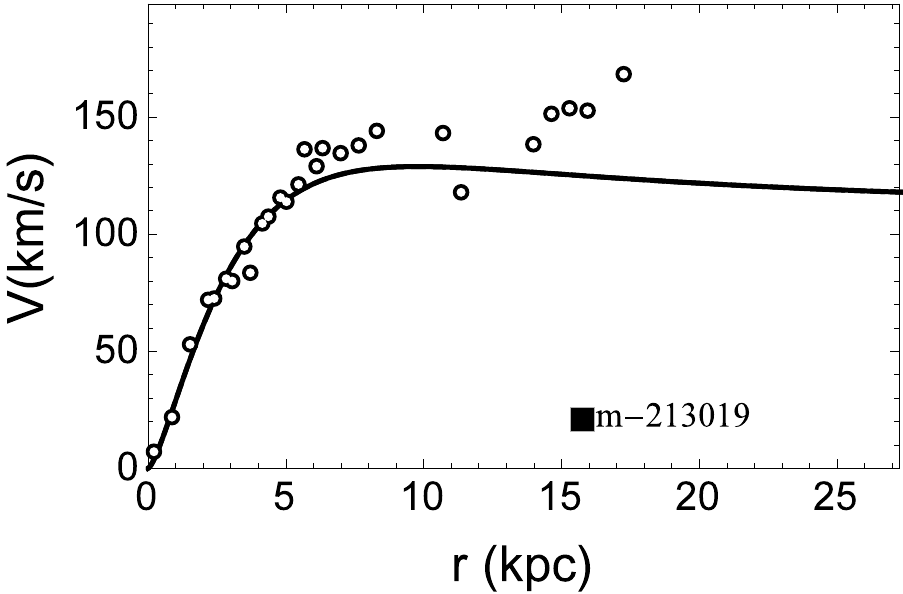}&
\includegraphics[width=0.50\textwidth]{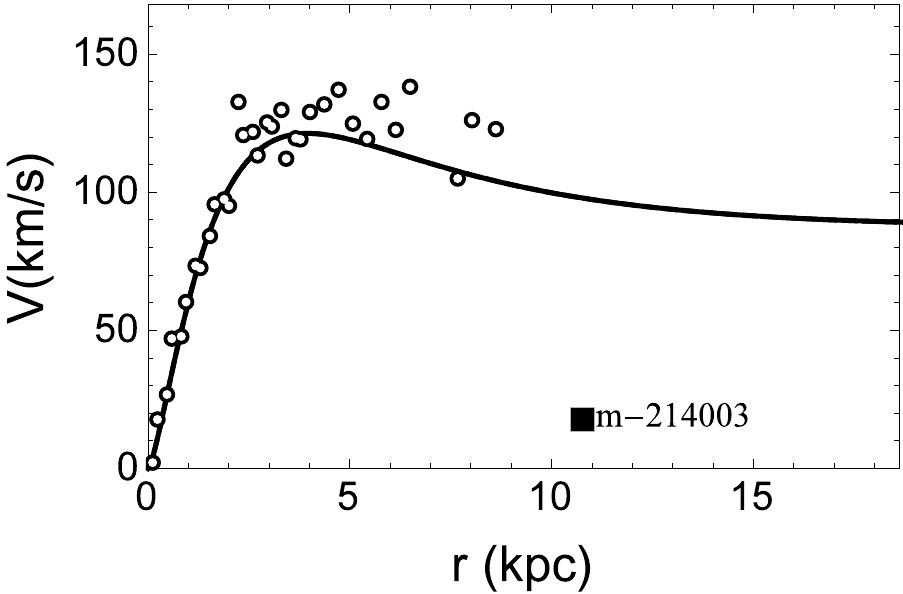}\\
\includegraphics[width=0.50\textwidth]{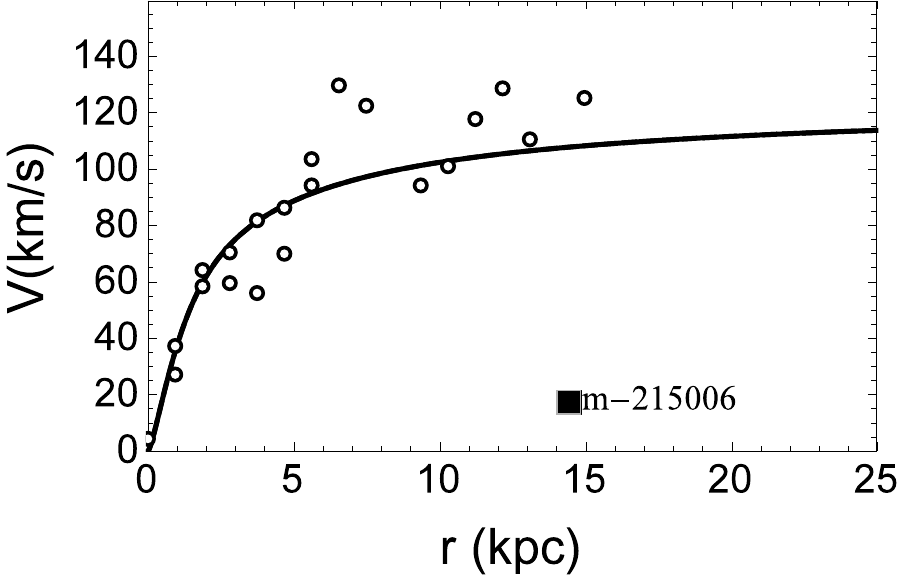}&
\includegraphics[width=0.50\textwidth]{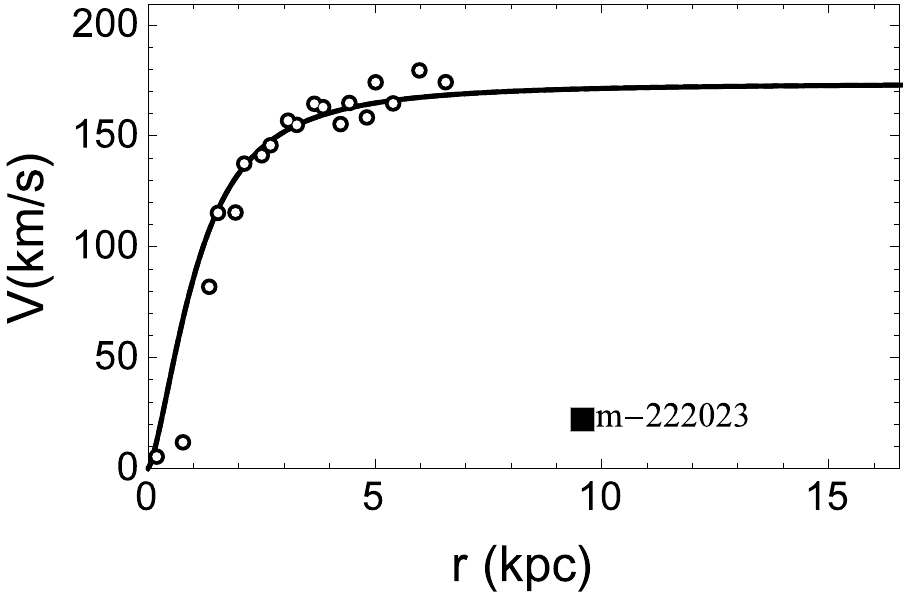}&
\includegraphics[width=0.50\textwidth]{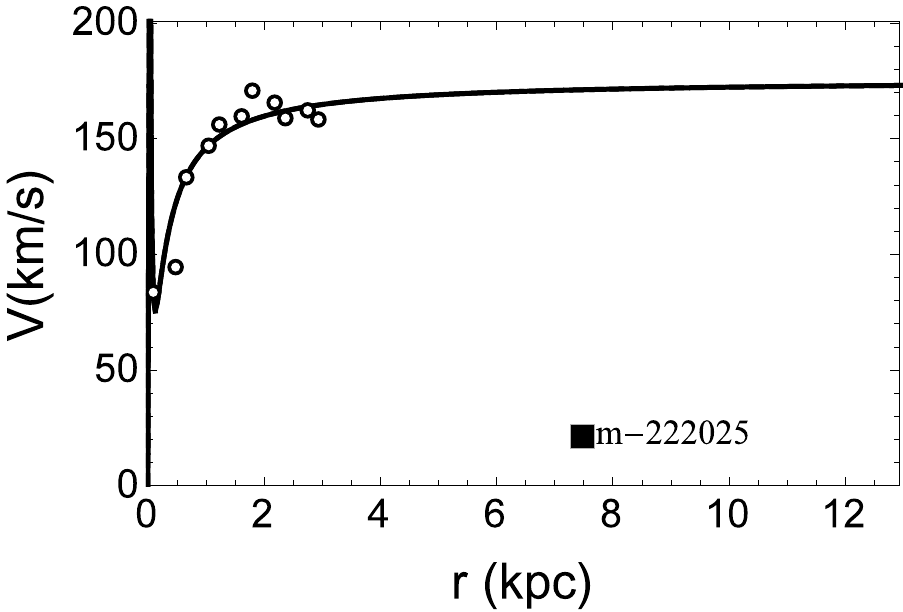}&
\includegraphics[width=0.50\textwidth]{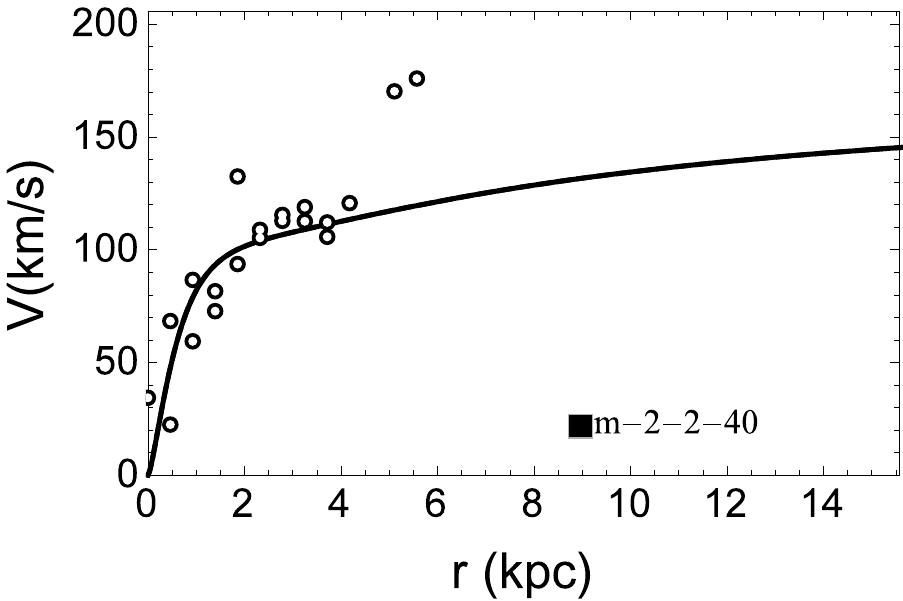}&
\includegraphics[width=0.50\textwidth]{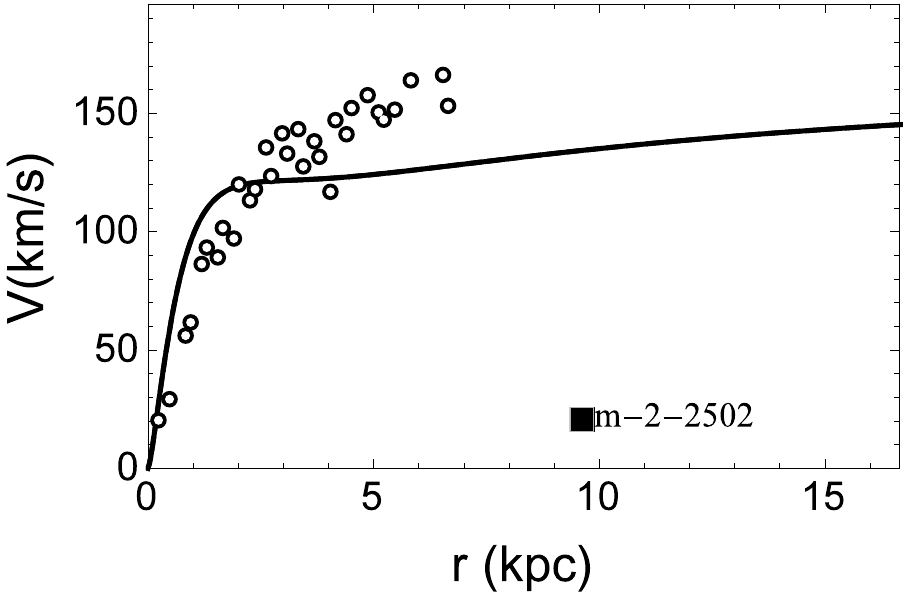}\\
\end{tabular}}}
\end{figure}

\clearpage

\begin{figure}
\flushleft
{\tiny\renewcommand{\arraystretch}{.95}
\resizebox{!}{.64\textwidth}{%
\begin{tabular}{lllll}
\includegraphics[width=0.50\textwidth]{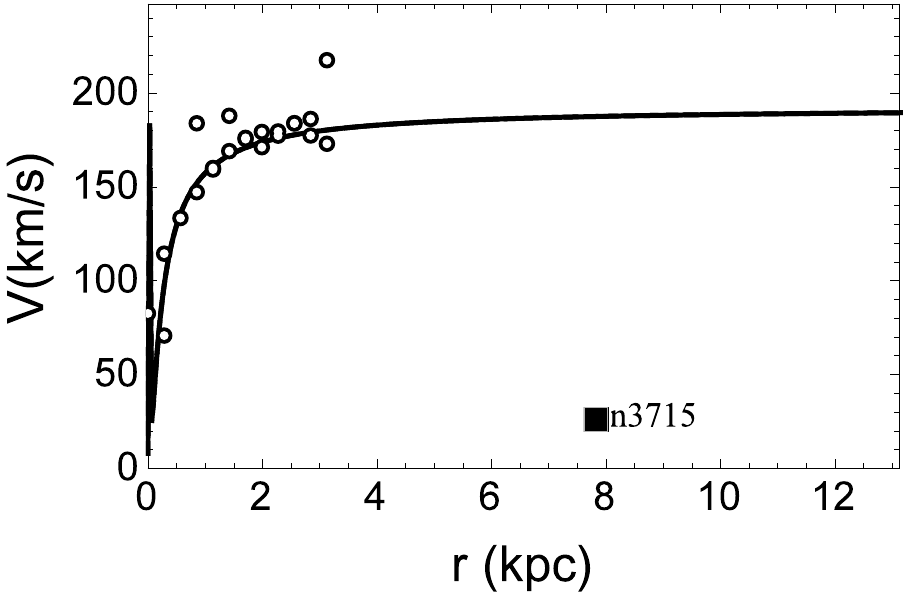}&
\includegraphics[width=0.50\textwidth]{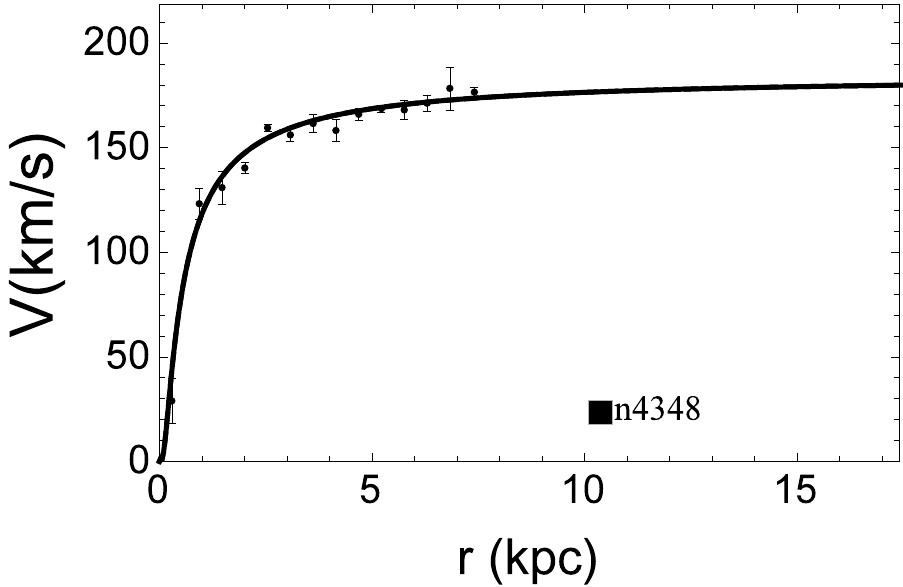}&
\includegraphics[width=0.50\textwidth]{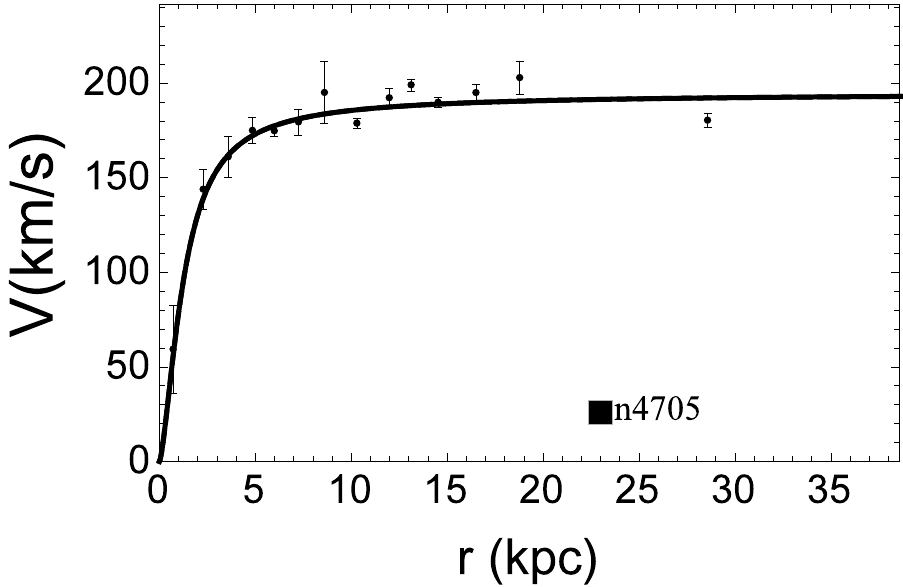}&
\includegraphics[width=0.50\textwidth]{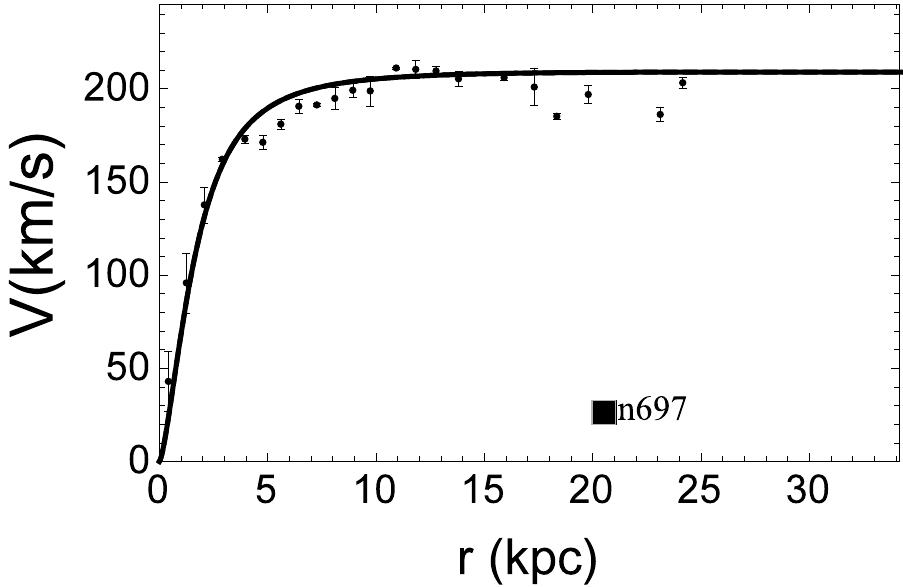}&
\includegraphics[width=0.50\textwidth]{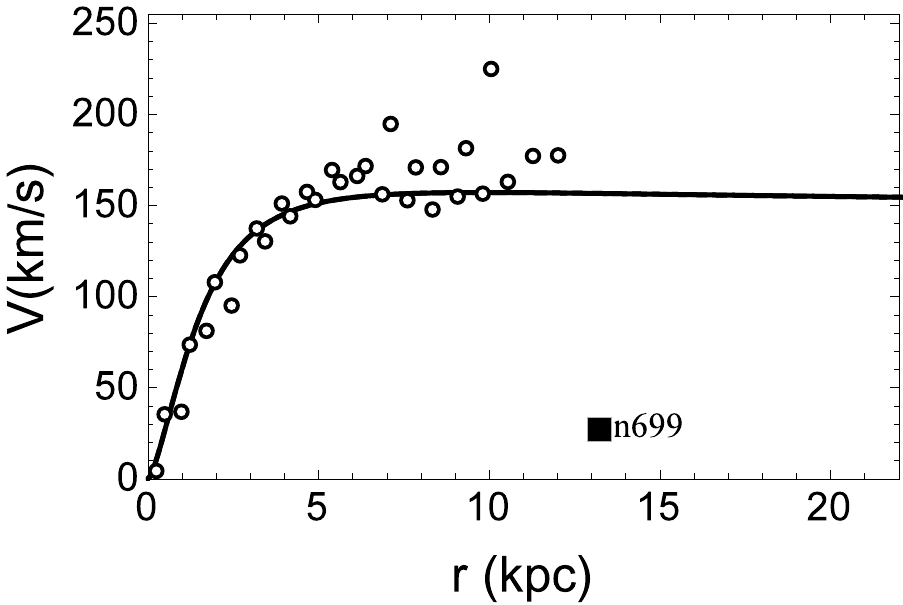}\\
\includegraphics[width=0.50\textwidth]{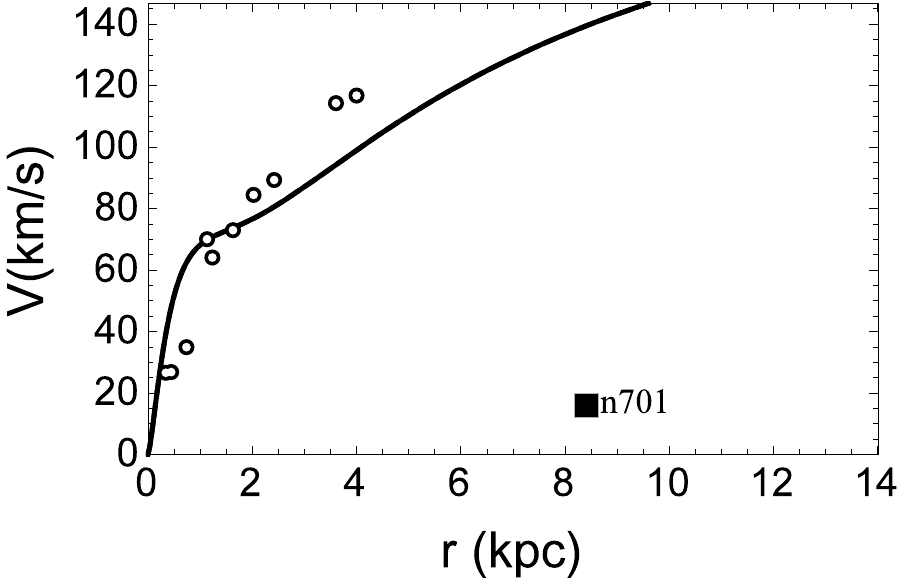}&
\includegraphics[width=0.50\textwidth]{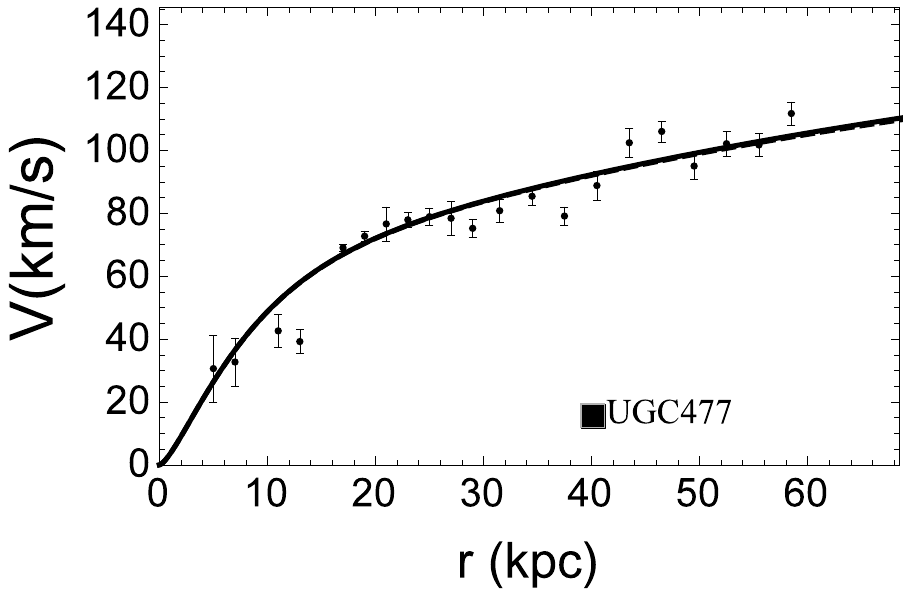}&
\includegraphics[width=0.50\textwidth]{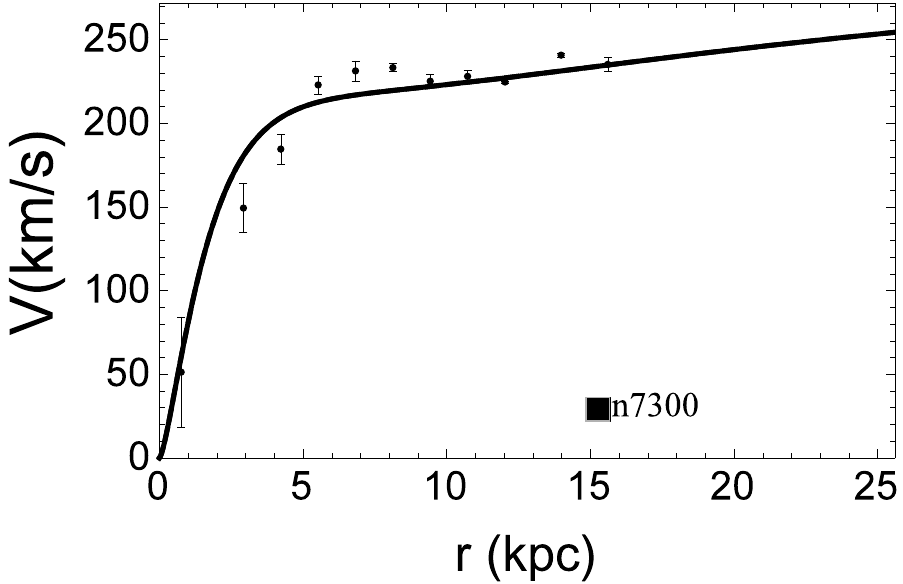}&
\includegraphics[width=0.50\textwidth]{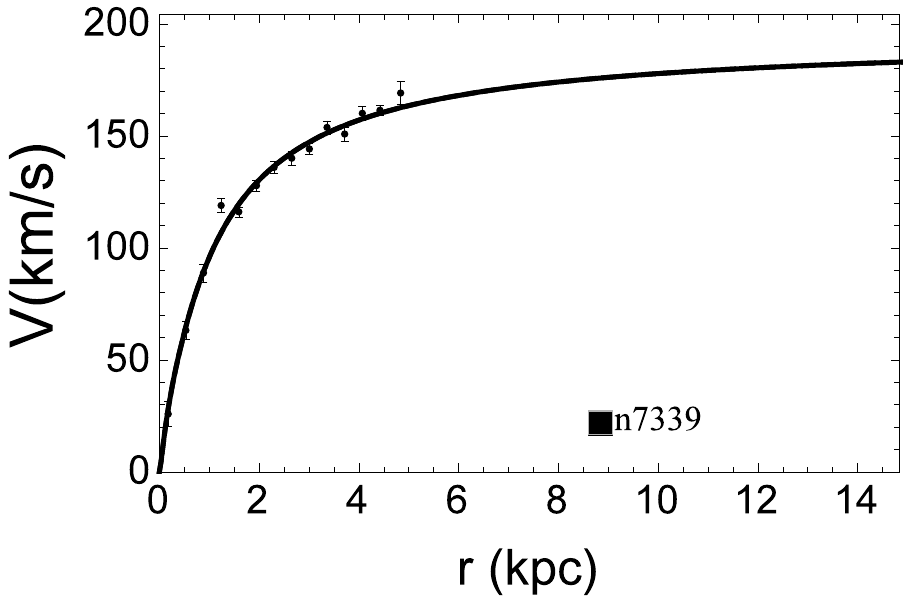}&
\includegraphics[width=0.50\textwidth]{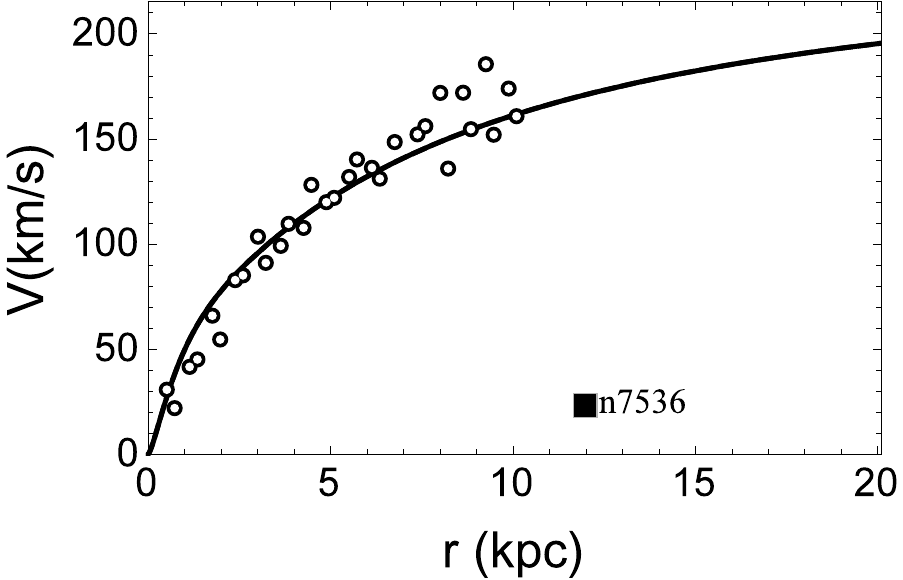}\\
\includegraphics[width=0.50\textwidth]{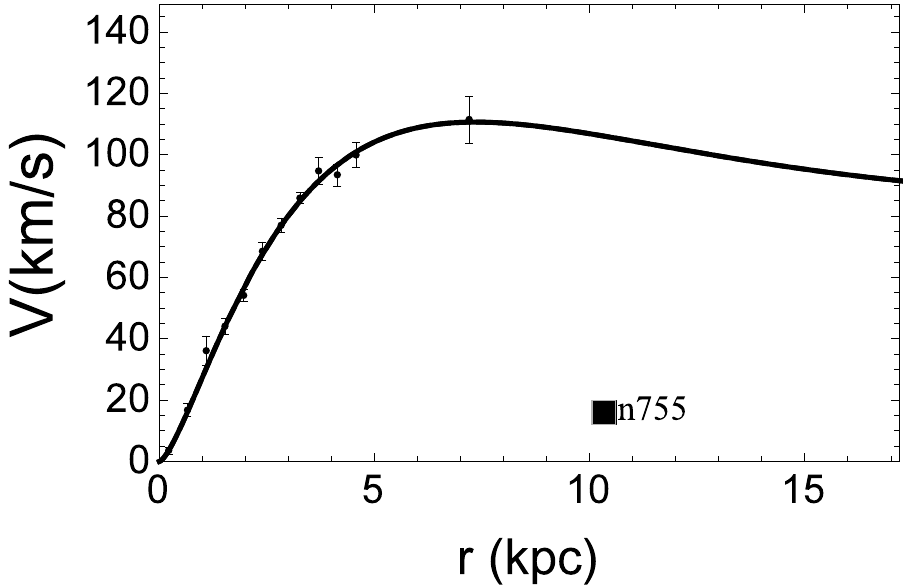}&
\includegraphics[width=0.50\textwidth]{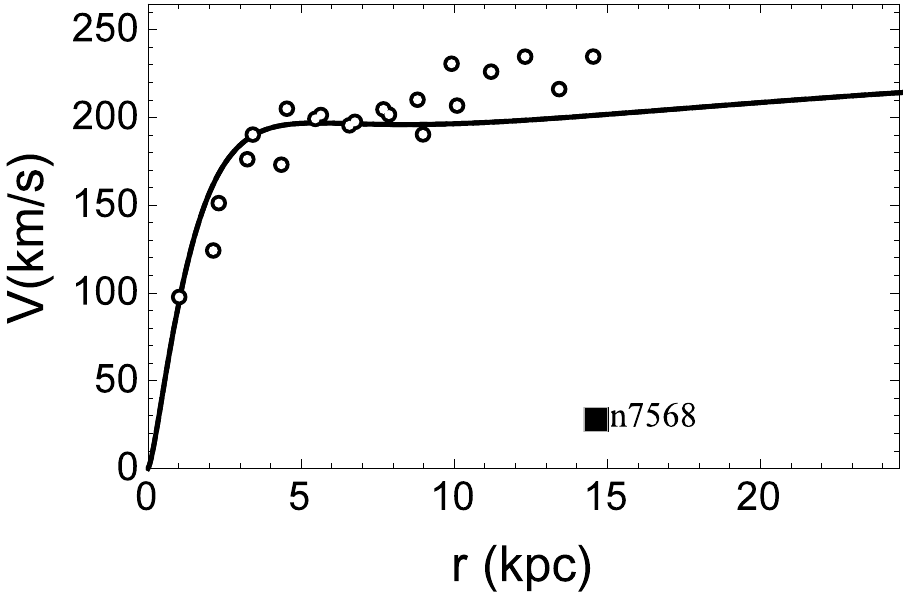}&
\includegraphics[width=0.50\textwidth]{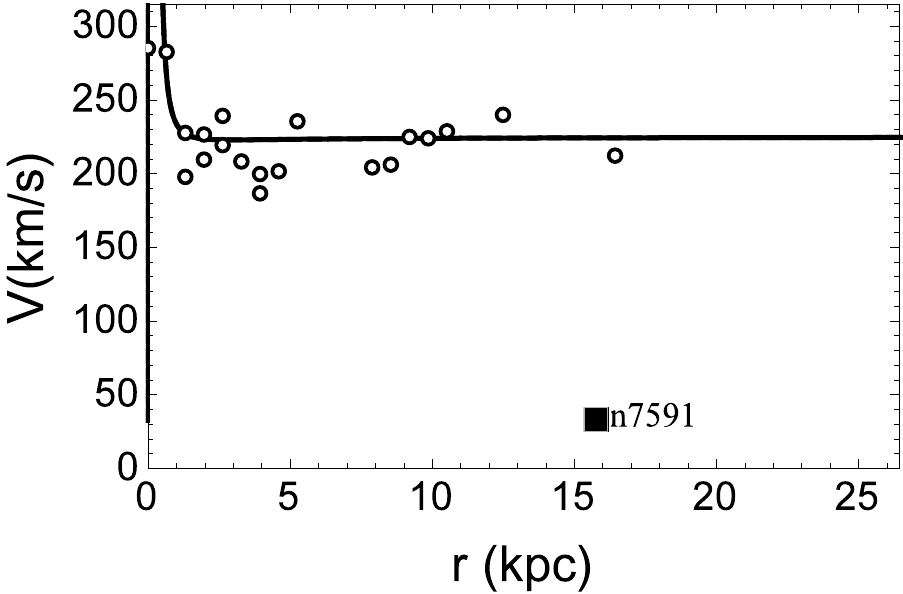}&
\includegraphics[width=0.50\textwidth]{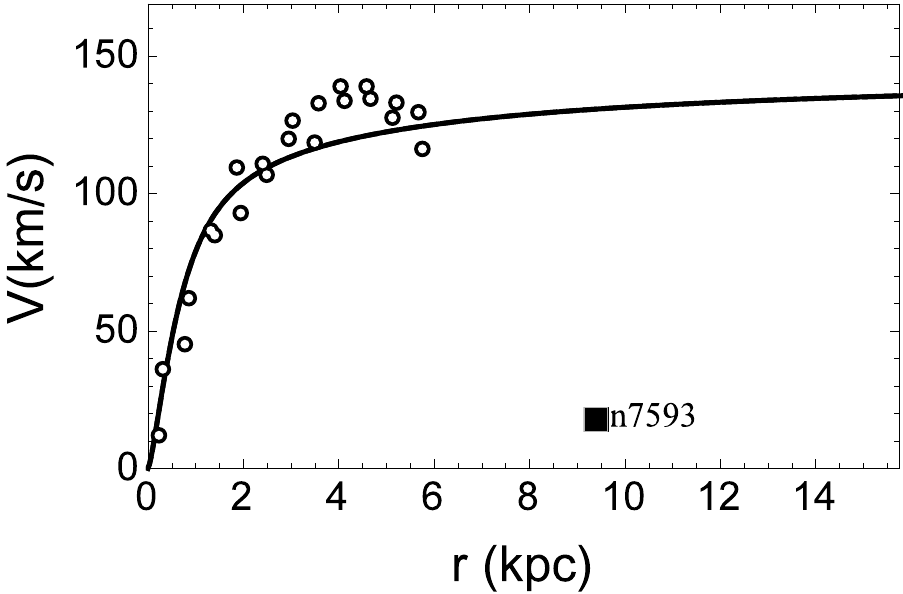}&
\includegraphics[width=0.50\textwidth]{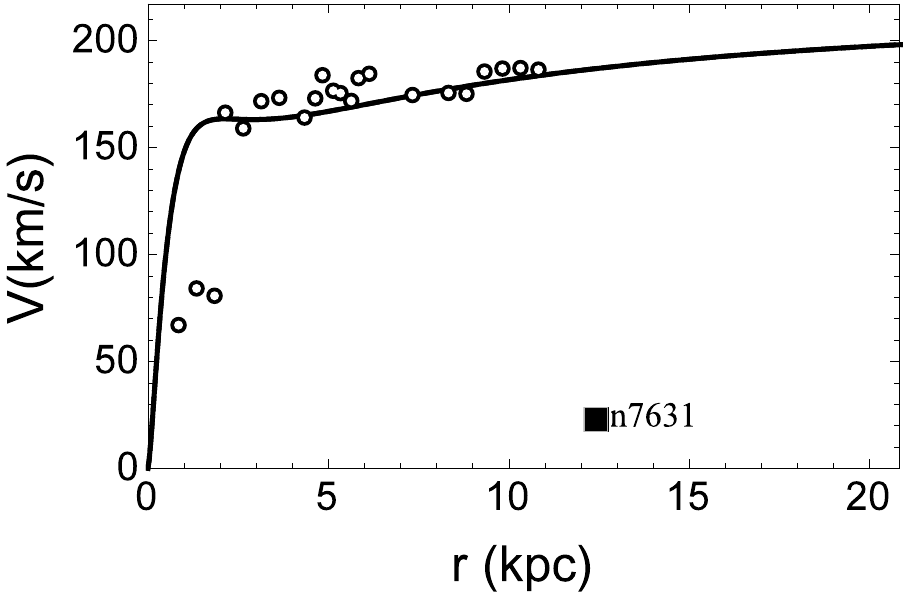}\\
\includegraphics[width=0.50\textwidth]{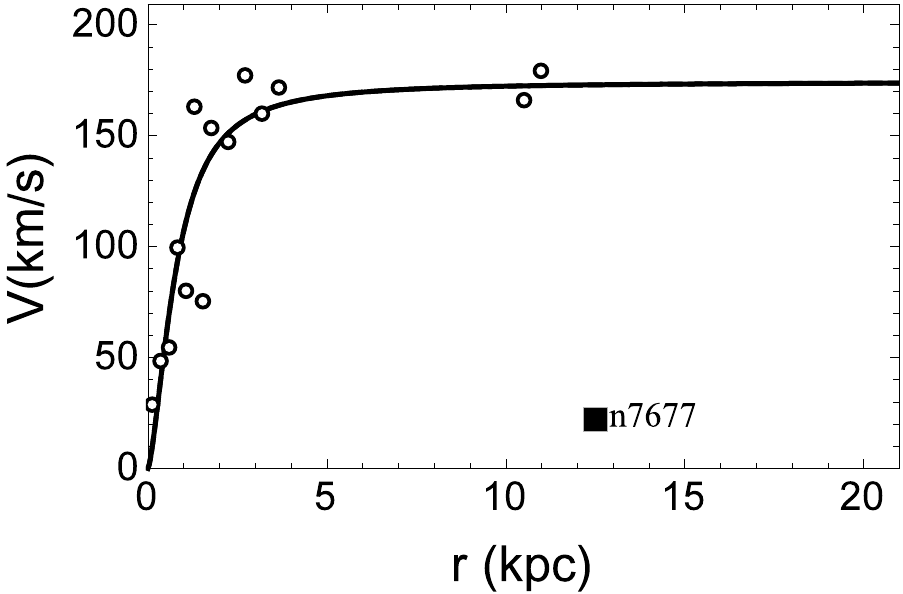}&
\includegraphics[width=0.50\textwidth]{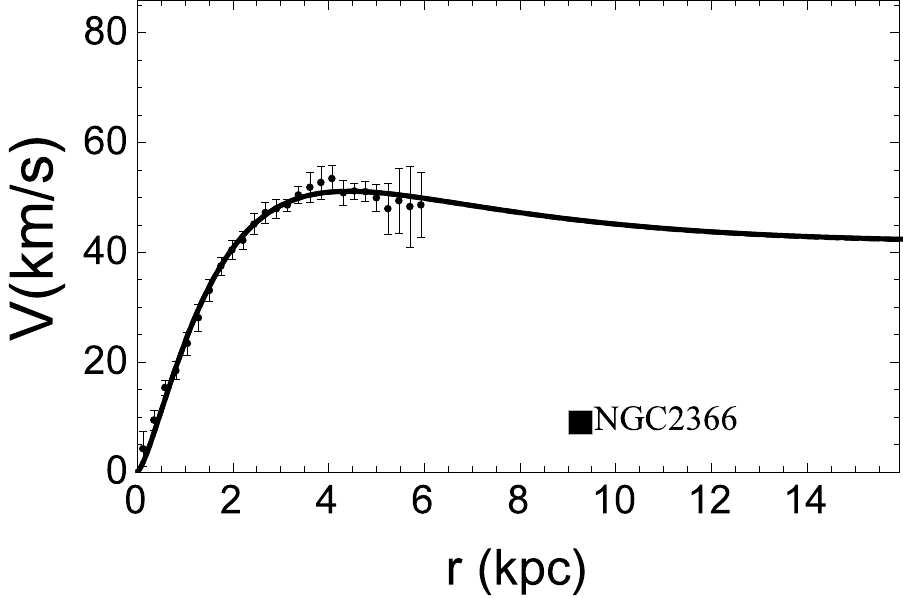}&
\includegraphics[width=0.50\textwidth]{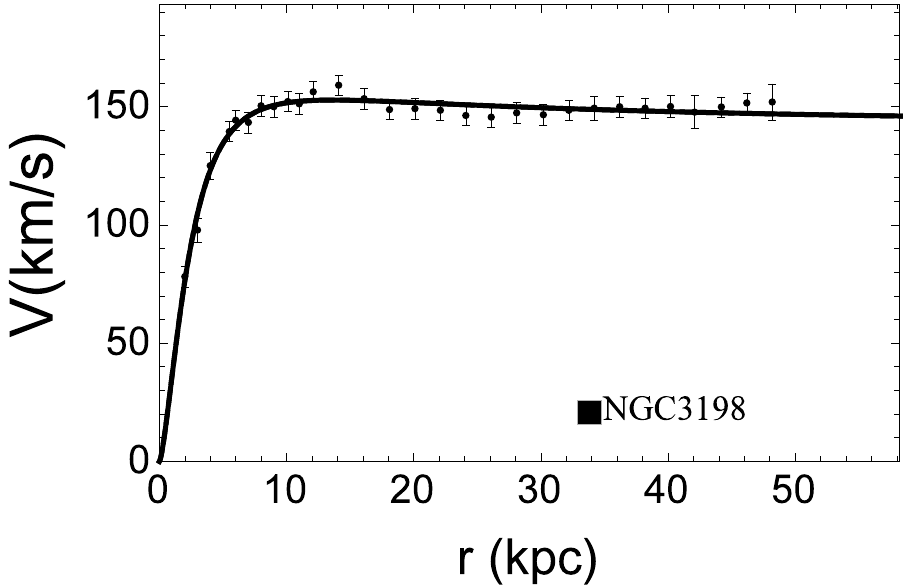}&
\includegraphics[width=0.50\textwidth]{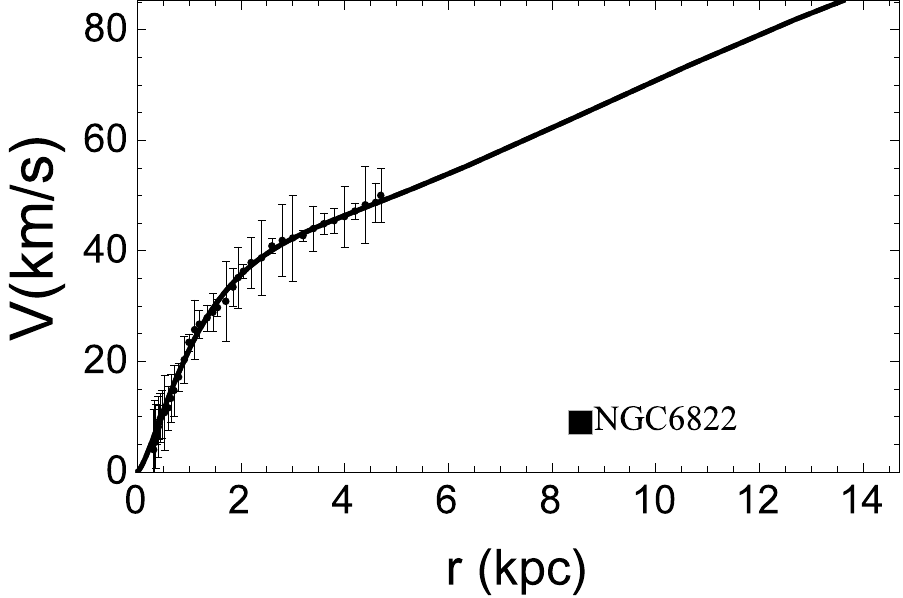}&
\includegraphics[width=0.50\textwidth]{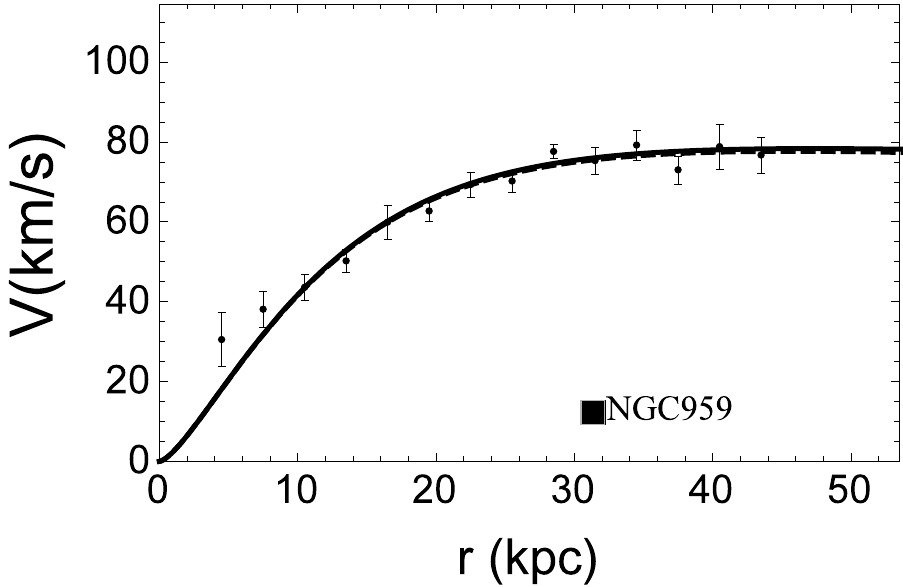}\\
\end{tabular}}}
\caption{Galaxy RCs best fittings using Eq.\eqref{Eq.2.2}.}\label{Fig.01}
\end{figure}

\clearpage

\begin{table}
\centering
{\tiny\renewcommand{\arraystretch}{.9}
\resizebox{\textwidth}{!}{%
\begin{tabular}{ccccccccc}
  \toprule
     Galaxy      &$\chi_{\text{red}}^{2}$ & $n$ & $\langle V_{0}\rangle$ (km s$^{-1}$) & $\langle r_{c}\rangle$ (kpc) & $d$ & $r_{\text{edge}}$ (kpc) & $M_{T}$ ($10^{11}\,\text{M}_{\odot}$) & Reference\\
     \hline
     \hline
101-g20	 & 	0.78	 & 	4	 & 	173.229	$\pm$	0.001	 & 	0.099	$\pm$	0.001	 & 	2.685	$\pm$	0.001	 & 	176.214	$\pm$	0.005	 & 	6.122	$\pm$	0.001	 & 	\cite{Persic:1995tc}	\\
101-g5	 & 	0.45	 & 	3	 & 	247.172	$\pm$	0.001	 & 	5.307	$\pm$	0.001	 & 	1.526	$\pm$	0.001	 & 	323.124	$\pm$	0.001	 & 	20.263	$\pm$	0.001	 & 	\cite{Persic:1995tc}	\\
102-g10	 & 	0.42	 & 	2	 & 	180.380	$\pm$	0.001	 & 	1.627	$\pm$	0.001	 & 	1.658	$\pm$	0.001	 & 	200.496	$\pm$	0.001	 & 	7.387	$\pm$	0.001	 & 	\cite{Persic:1995tc}	\\
102-g15	 & 	0.59	 & 	3	 & 	203.644	$\pm$	0.001	 & 	1.646	$\pm$	0.001	 & 	1.905	$\pm$	0.001	 & 	209.610	$\pm$	0.007	 & 	9.626	$\pm$	0.001	 & 	\cite{Persic:1995tc}	\\
103-g13	 & 	0.20	 & 	3	 & 	234.151	$\pm$	0.001	 & 	0.434	$\pm$	0.002	 & 	0.699	$\pm$	0.001	 & 	236.231	$\pm$	0.007	 & 	14.750	$\pm$	0.001	 & 	\cite{Persic:1995tc}	\\
103-g15	 & 	0.23	 & 	3	 & 	180.460	$\pm$	0.001	 & 	1.613	$\pm$	0.001	 & 	1.184	$\pm$	0.001	 & 	185.977	$\pm$	0.008	 & 	6.504	$\pm$	0.001	 & 	\cite{Persic:1995tc}	\\
103-g39	 & 	0.20	 & 	4	 & 	213.197	$\pm$	0.002	 & 	1.801	$\pm$	0.001	 & 	1.326	$\pm$	0.001	 & 	205.727	$\pm$	0.009	 & 	9.742	$\pm$	0.001	 & 	\cite{Persic:1995tc}	\\
\bottomrule
    \end{tabular} }}
\end{table}

\clearpage

\begin{table}
\centering
{\tiny\renewcommand{\arraystretch}{.9}
\resizebox{\textwidth}{!}{%
\begin{tabular}{ccccccccc}
  \toprule
     Galaxy      &$\chi_{\text{red}}^{2}$ & $n$ & $\langle V_{0}\rangle$ (km s$^{-1}$) & $\langle r_{c}\rangle$ (kpc) & $d$ & $r_{\text{edge}}$ (kpc) & $M_{T}$ ($10^{11}\,\text{M}_{\odot}$) & Reference\\
     \hline
     \hline
1547-02	 & 	1.85	 & 	1	 & 	206.744	$\pm$	0.003	 & 	3.075	$\pm$	0.004	 & 	1.028	$\pm$	0.003	 & 	290.927	$\pm$	0.001	 & 	14.166	$\pm$	0.002	 & 	\cite{Persic:1995tc}	\\
155-g6	 & 	0.31	 & 	5	 & 	422.955	$\pm$	0.001	 & 	6.826	$\pm$	0.001	 & 	1.350	$\pm$	0.003	 & 	428.039	$\pm$	0.004	 & 	68.748	$\pm$	0.002	 & 	\cite{Persic:1995tc}	\\
157-g20	 & 	0.92	 & 	1	 & 	111.562	$\pm$	0.001	 & 	0.223	$\pm$	0.001	 & 	1.938	$\pm$	0.001	 & 	131.242	$\pm$	0.008	 & 	1.904	$\pm$	0.001	 & 	\cite{Persic:1995tc}	\\
157-g38	 & 	0.10	 & 	3	 & 	141.833	$\pm$	0.002	 & 	0.327	$\pm$	0.001	 & 	0.986	$\pm$	0.001	 & 	142.920	$\pm$	0.009	 & 	3.266	$\pm$	0.001	 & 	\cite{Persic:1995tc}	\\
160-g2	 & 	0.19	 & 	4	 & 	402.901	$\pm$	0.001	 & 	1.331	$\pm$	0.001	 & 	1.202	$\pm$	0.001	 & 	401.843	$\pm$	0.005	 & 	72.603	$\pm$	0.001	 & 	\cite{Persic:1995tc}	\\
162-g15	 & 	0.32	 & 	4	 & 	160.632	$\pm$	0.001	 & 	1.697	$\pm$	0.001	 & 	1.478	$\pm$	0.001	 & 	169.781	$\pm$	0.005	 & 	4.523	$\pm$	0.001	 & 	\cite{Persic:1995tc}	\\
162-g17	 & 	0.52	 & 	4	 & 	126.004	$\pm$	0.008	 & 	2.044	$\pm$	0.001	 & 	1.466	$\pm$	0.001	 & 	135.809	$\pm$	0.001	 & 	2.099	$\pm$	0.001	 & 	\cite{Persic:1995tc}	\\
\bottomrule
    \end{tabular} }}
\end{table}

\clearpage

\vspace{-0.5cm}
\begin{table}
\centering
{\tiny\renewcommand{\arraystretch}{.9}
\resizebox{\textwidth}{!}{%
\begin{tabular}{ccccccccc}
  \toprule
     Galaxy      &$\chi_{\text{red}}^{2}$ & $n$ & $\langle V_{0}\rangle$ (km s$^{-1}$) & $\langle r_{c}\rangle$ (kpc) & $d$ & $r_{\text{edge}}$ (kpc) & $M_{T}$ ($10^{11}\,\text{M}_{\odot}$) & Reference\\
     \hline
     \hline
231-g29	 & 	0.28	 & 	3	 & 	134.655	$\pm$	0.001	 & 	2.947	$\pm$	0.001	 & 	1.494	$\pm$	0.001	 & 	170.667	$\pm$	0.001	 & 	3.150	$\pm$	0.001	 & 	\cite{Persic:1995tc}	\\
231-g6	 & 	0.53	 & 	3	 & 	135.294	$\pm$	0.001	 & 	4.130	$\pm$	0.001	 & 	1.452	$\pm$	0.001	 & 	173.828	$\pm$	0.009	 & 	3.076	$\pm$	0.001	 & 	\cite{Persic:1995tc}	\\
233-g25	 & 	0.30	 & 	3	 & 	355.865	$\pm$	0.004	 & 	15.208	$\pm$	0.009	 & 	1.144	$\pm$	0.001	 & 	418.525	$\pm$	0.009	 & 	44.611	$\pm$	0.001	 & 	\cite{Persic:1995tc}	\\
233-g36	 & 	0.30	 & 	3	 & 	184.129	$\pm$	0.002	 & 	2.104	$\pm$	0.001	 & 	1.363	$\pm$	0.001	 & 	188.759	$\pm$	0.005	 & 	6.778	$\pm$	0.001	 & 	\cite{Persic:1995tc}	\\
233-g41	 & 	1.06	 & 	1	 & 	233.473	$\pm$	0.001	 & 	1.297	$\pm$	0.001	 & 	2.289	$\pm$	0.001	 & 	239.951	$\pm$	0.007	 & 	15.458	$\pm$	0.001	 & 	\cite{Persic:1995tc}	\\
233-g42	 & 	0.41	 & 	3	 & 	111.599	$\pm$	0.001	 & 	1.424	$\pm$	0.001	 & 	1.695	$\pm$	0.001	 & 	128.209	$\pm$	0.001	 & 	1.718	$\pm$	0.001	 & 	\cite{Persic:1995tc}	\\
233-g43	 & 	0.59	 & 	2	 & 	77.486	$\pm$	0.002	 & 	2.610	$\pm$	0.001	 & 	0.993	$\pm$	0.001	 & 	71.1448	$\pm$	0.001	 & 	0.403	$\pm$	0.001	 & 	\cite{Persic:1995tc}	\\
\bottomrule
    \end{tabular} }}
\end{table}

\clearpage

\begin{table}
\centering
{\tiny\renewcommand{\arraystretch}{.9}
\resizebox{\textwidth}{!}{%
\begin{tabular}{ccccccccc}
  \toprule
     Galaxy      &$\chi_{\text{red}}^{2}$ & $n$ & $\langle V_{0}\rangle$ (km s$^{-1}$) & $\langle r_{c}\rangle$ (kpc) & $d$ & $r_{\text{edge}}$ (kpc) & $M_{T}$ ($10^{11}\,\text{M}_{\odot}$) & Reference\\
     \hline
     \hline
271-g4	 & 	0.43	 & 	6	 & 	97.394	$\pm$	0.002	 & 	0.292	$\pm$	0.001	 & 	1.569	$\pm$	0.001	 & 	96.367	$\pm$	0.008	 & 	1.001	$\pm$	0.001	 & 	\cite{Persic:1995tc}	\\
273-g6	 & 	0.14	 & 	2	 & 	168.488	$\pm$	0.002	 & 	2.721	$\pm$	0.001	 & 	1.557	$\pm$	0.001	 & 	166.865	$\pm$	0.002	 & 	5.199	$\pm$	0.001	 & 	\cite{Persic:1995tc}	\\
27-g17	 & 	0.57	 & 	4	 & 	183.918	$\pm$	0.001	 & 	0.901	$\pm$	0.001	 & 	1.341	$\pm$	0.001	 & 	195.926	$\pm$	0.009	 & 	7.272	$\pm$	0.001	 & 	\cite{Persic:1995tc}	\\
27-g24	 & 	0.61	 & 	2	 & 	176.780	$\pm$	0.001	 & 	1.755	$\pm$	0.001	 & 	1.697	$\pm$	0.001	 & 	187.971	$\pm$	0.001	 & 	6.637	$\pm$	0.001	 & 	\cite{Persic:1995tc}	\\
27-g8	 & 	8.17	 & 	1	 & 	117.720	$\pm$	0.002	 & 	3.650	$\pm$	0.001	 & 	3.597	$\pm$	0.003	 & 	182.792	$\pm$	0.009	 & 	3.419	$\pm$	0.001	 & 	\cite{Persic:1995tc}	\\
280-g13	 & 	0.14	 & 	4	 & 	396.151	$\pm$	0.001	 & 	3.727	$\pm$	0.004	 & 	1.178	$\pm$	0.001	 & 	378.585	$\pm$	0.004	 & 	60.712	$\pm$	0.001	 & 	\cite{Persic:1995tc}	\\
281-g38	 & 	0.16	 & 	1	 & 	173.711	$\pm$	0.001	 & 	1.990	$\pm$	0.004	 & 	2.355	$\pm$	0.001	 & 	185.437	$\pm$	0.001	 & 	6.733	$\pm$	0.001	 & 	\cite{Persic:1995tc}	\\
\bottomrule
    \end{tabular} }}
\end{table}

\clearpage

\begin{table}
\centering
{\tiny\renewcommand{\arraystretch}{.9}
\resizebox{\textwidth}{!}{%
\begin{tabular}{ccccccccc}
  \toprule
     Galaxy      &$\chi_{\text{red}}^{2}$ & $n$ & $\langle V_{0}\rangle$ (km s$^{-1}$) & $\langle r_{c}\rangle$ (kpc) & $d$ & $r_{\text{edge}}$ (kpc) & $M_{T}$ ($10^{11}\,\text{M}_{\odot}$) & Reference\\
     \hline
     \hline
320-g26	 & 	0.71	 & 	3	 & 	227.000	$\pm$	0.001	 & 	0.068	$\pm$	0.001	 & 	1.812	$\pm$	0.003	 & 	231.151	$\pm$	0.002	 & 	13.819	$\pm$	0.001	 & 	\cite{Persic:1995tc}	\\
321-g10	 & 	0.26	 & 	4	 & 	234.197	$\pm$	0.001	 & 	1.202	$\pm$	0.001	 & 	1.303	$\pm$	0.001	 & 	230.965	$\pm$	0.009	 & 	13.785	$\pm$	0.001	 & 	\cite{Persic:1995tc}	\\
321-g17	 & 	0.37	 & 	4	 & 	158.620	$\pm$	0.001	 & 	1.012	$\pm$	0.001	 & 	1.432	$\pm$	0.001	 & 	155.417	$\pm$	0.001	 & 	4.200	$\pm$	0.001	 & 	\cite{Persic:1995tc}	\\
321-g1	 & 	0.25	 & 	2	 & 	83.029	$\pm$	0.001	 & 	7.306	$\pm$	0.001	 & 	3.304	$\pm$	0.001	 & 	136.258	$\pm$	0.001	 & 	1.421	$\pm$	0.001	 & 	\cite{Persic:1995tc}	\\
322-g33	 & 	0.99	 & 	6	 & 	335.597	$\pm$	0.001	 & 	2.426	$\pm$	0.001	 & 	1.097	$\pm$	0.002	 & 	456.970	$\pm$	0.001	 & 	53.385	$\pm$	0.002	 & 	\cite{Persic:1995tc}	\\
322-g36	 & 	0.66	 & 	4	 & 	278.486	$\pm$	0.001	 & 	1.419	$\pm$	0.001	 & 	1.470	$\pm$	0.003	 & 	295.191	$\pm$	0.008	 & 	25.12	$\pm$	0.001	 & 	\cite{Persic:1995tc}	\\
322-g45	 & 	0.37	 & 	3	 & 	147.497	$\pm$	0.001	 & 	0.440	$\pm$	0.001	 & 	1.245	$\pm$	0.001	 & 	164.128	$\pm$	0.003	 & 	4.052	$\pm$	0.001	 & 	\cite{Persic:1995tc}	\\
\bottomrule
    \end{tabular} }}
\end{table}

\clearpage

\begin{table}
\centering
{\tiny\renewcommand{\arraystretch}{.9}
\resizebox{\textwidth}{!}{%
\begin{tabular}{ccccccccc}
  \toprule
     Galaxy      &$\chi_{\text{red}}^{2}$ & $n$ & $\langle V_{0}\rangle$ (km s$^{-1}$) & $\langle r_{c}\rangle$ (kpc) & $d$ & $r_{\text{edge}}$ (kpc) & $M_{T}$ ($10^{11}\,\text{M}_{\odot}$) & Reference\\
     \hline
     \hline
352-g50	 & 	0.71	 & 	1	 & 	111.812	$\pm$	0.001	 & 	1.607	$\pm$	0.001	 & 	2.500	$\pm$	0.001	 & 	136.900	$\pm$	0.001	 & 	2.079	$\pm$	0.001	 & 	\cite{Persic:1995tc}	\\
352-g53	 & 	1.76	 & 	2	 & 	264.541	$\pm$	0.003	 & 	0.267	$\pm$	0.001	 & 	0.718	$\pm$	0.001	 & 	278.414	$\pm$	0.002	 & 	22.495	$\pm$	0.001	 & 	\cite{Persic:1995tc}	\\
353-g14	 & 	0.42	 & 	2	 & 	132.212	$\pm$	0.001	 & 	1.665	$\pm$	0.001	 & 	1.493	$\pm$	0.007	 & 	150.920	$\pm$	0.002	 & 	2.939	$\pm$	0.001	 & 	\cite{Persic:1995tc}	\\
353-g2	 & 	0.25	 & 	1	 & 	78.664	$\pm$	0.002	 & 	1.734	$\pm$	0.001	 & 	2.732	$\pm$	0.002	 & 	99.8382	$\pm$	0.001	 & 	0.776	$\pm$	0.001	 & 	\cite{Persic:1995tc}	\\
354-g17	 & 	0.44	 & 	3	 & 	226.787	$\pm$	0.001	 & 	2.993	$\pm$	0.001	 & 	1.704	$\pm$	0.009	 & 	220.604	$\pm$	0.004	 & 	12.012	$\pm$	0.001	 & 	\cite{Persic:1995tc}	\\
354-g46	 & 	0.66	 & 	3	 & 	264.978	$\pm$	0.001	 & 	1.719	$\pm$	0.001	 & 	1.719	$\pm$	0.001	 & 	282.608	$\pm$	0.006	 & 	22.106	$\pm$	0.001	 & 	\cite{Persic:1995tc}	\\
354-g47	 & 	0.73	 & 	6	 & 	211.467	$\pm$	0.001	 & 	0.183	$\pm$	0.001	 & 	4.019	$\pm$	0.005	 & 	214.641	$\pm$	0.001	 & 	11.064	$\pm$	0.001	 & 	\cite{Persic:1995tc}	\\
\bottomrule
    \end{tabular} }}
\end{table}

\clearpage

\begin{table}
\centering
{\tiny\renewcommand{\arraystretch}{.9}
\resizebox{\textwidth}{!}{%
\begin{tabular}{ccccccccc}
  \toprule
     Galaxy      &$\chi_{\text{red}}^{2}$ & $n$ & $\langle V_{0}\rangle$ (km s$^{-1}$) & $\langle r_{c}\rangle$ (kpc) & $d$ & $r_{\text{edge}}$ (kpc) & $M_{T}$ ($10^{11}\,\text{M}_{\odot}$) & Reference\\
     \hline
     \hline
381-g51	 & 	0.43	 & 	4	 & 	367.358	$\pm$	0.009	 & 	2.505	$\pm$	0.001	 & 	1.565	$\pm$	0.001	 & 	374.114	$\pm$	0.005	 & 	54.311	$\pm$	0.004	 & 	\cite{Persic:1995tc}	\\
382-g32	 & 	1.36	 & 	3	 & 	267.394	$\pm$	0.001	 & 	9.782	$\pm$	0.001	 & 	1.793	$\pm$	0.001	 & 	341.539	$\pm$	0.001	 & 	23.790	$\pm$	0.001	 & 	\cite{Persic:1995tc}	\\
382-g4	 & 	0.35	 & 	4	 & 	156.290	$\pm$	0.001	 & 	1.667	$\pm$	0.001	 & 	1.394	$\pm$	0.001	 & 	164.632	$\pm$	0.004	 & 	4.131	$\pm$	0.001	 & 	\cite{Persic:1995tc}	\\
382-g58	 & 	0.11	 & 	1	 & 	174.920	$\pm$	0.001	 & 	7.152	$\pm$	0.001	 & 	4.680	$\pm$	0.001	 & 	287.152	$\pm$	0.001	 & 	13.389	$\pm$	0.001	 & 	\cite{Persic:1995tc}	\\
383-g2	 & 	1.24	 & 	1	 & 	162.716	$\pm$	0.001	 & 	2.277	$\pm$	0.001	 & 	2.322	$\pm$	0.001	 & 	175.255	$\pm$	0.001	 & 	5.611	$\pm$	0.001	 & 	\cite{Persic:1995tc}	\\
383-g55	 & 	0.21	 & 	2	 & 	280.254	$\pm$	0.001	 & 	1.486	$\pm$	0.001	 & 	1.177	$\pm$	0.001	 & 	309.815	$\pm$	0.004	 & 	27.585	$\pm$	0.001	 & 	\cite{Persic:1995tc}	\\
383-g67	 & 	0.31	 & 	4	 & 	284.548	$\pm$	0.001	 & 	2.821	$\pm$	0.002	 & 	1.236	$\pm$	0.001	 & 	280.854	$\pm$	0.003	 & 	23.232	$\pm$	0.001	 & 	\cite{Persic:1995tc}	\\
\bottomrule
    \end{tabular} }}
\end{table}

\clearpage

\vspace{-0.5cm}
\begin{table}
\centering
{\tiny\renewcommand{\arraystretch}{.9}
\resizebox{\textwidth}{!}{%
\begin{tabular}{ccccccccc}
  \toprule
     Galaxy      &$\chi_{\text{red}}^{2}$ & $n$ & $\langle V_{0}\rangle$ (km s$^{-1}$) & $\langle r_{c}\rangle$ (kpc) & $d$ & $r_{\text{edge}}$ (kpc) & $M_{T}$ ($10^{11}\,\text{M}_{\odot}$) & Reference\\
     \hline
     \hline

418-g15	 & 	0.15	 & 	2	 & 	169.970	$\pm$	0.001	 & 	2.052	$\pm$	0.001	 & 	1.872	$\pm$	0.001	 & 	170.888	$\pm$	0.008	 & 	5.584	$\pm$	0.001	 & 	\cite{Persic:1995tc}	\\
418-g1	 & 	0.55	 & 	4	 & 	233.473	$\pm$	0.001	 & 	1.661	$\pm$	0.001	 & 	1.439	$\pm$	0.001	 & 	240.474	$\pm$	0.006	 & 	14.017	$\pm$	0.001	 & 	\cite{Persic:1995tc}	\\
418-g8	 & 	0.23	 & 	5	 & 	394.119	$\pm$	0.008	 & 	1.405	$\pm$	0.001	 & 	1.126	$\pm$	0.004	 & 	410.229	$\pm$	0.004	 & 	69.784	$\pm$	0.006	 & 	\cite{Persic:1995tc}	\\
419-g3	 & 	0.61	 & 	4	 & 	267.989	$\pm$	0.001	 & 	4.435	$\pm$	0.001	 & 	1.505	$\pm$	0.001	 & 	250.659	$\pm$	0.001	 & 	17.051	$\pm$	0.001	 & 	\cite{Persic:1995tc}	\\
419-g4	 & 	0.81	 & 	1	 & 	165.781	$\pm$	0.001	 & 	1.195	$\pm$	0.001	 & 	1.851	$\pm$	0.002	 & 	182.106	$\pm$	0.004	 & 	5.869	$\pm$	0.001	 & 	\cite{Persic:1995tc}	\\
41-g9	 & 	1.76	 & 	2	 & 	248.775	$\pm$	0.006	 & 	7.596	$\pm$	0.003	 & 	1.371	$\pm$	0.004	 & 	237.383	$\pm$	0.001	 & 	14.967	$\pm$	0.007	 & 	\cite{Persic:1995tc}	\\
420-g3	 & 	0.28	 & 	1	 & 	159.143	$\pm$	0.001	 & 	0.992	$\pm$	0.001	 & 	1.191	$\pm$	0.001	 & 	161.587	$\pm$	0.002	 & 	4.721	$\pm$	0.001	 & 	\cite{Persic:1995tc}	\\
\bottomrule
    \end{tabular} }}
\end{table}

\clearpage

\vspace{-0.5cm}
\begin{table}
\centering
{\tiny\renewcommand{\arraystretch}{.9}
\resizebox{\textwidth}{!}{%
\begin{tabular}{ccccccccc}
  \toprule
     Galaxy      &$\chi_{\text{red}}^{2}$ & $n$ & $\langle V_{0}\rangle$ (km s$^{-1}$) & $\langle r_{c}\rangle$ (kpc) & $d$ & $r_{\text{edge}}$ (kpc) & $M_{T}$ ($10^{11}\,\text{M}_{\odot}$) & Reference\\
     \hline
     \hline
443-g38	 & 	0.79	 & 	3	 & 	310.618	$\pm$	0.001	 & 	3.989	$\pm$	0.002	 & 	1.464	$\pm$	0.003	 & 	310.958	$\pm$	0.003	 & 	31.512	$\pm$	0.001	 & 	\cite{Persic:1995tc}	\\
443-g41	 & 	0.58	 & 	1	 & 	217.644	$\pm$	0.001	 & 	1.462	$\pm$	0.002	 & 	2.717	$\pm$	0.002	 & 	225.249	$\pm$	0.002	 & 	12.787	$\pm$	0.001	 & 	\cite{Persic:1995tc}	\\
443-g42	 & 	0.31	 & 	4	 & 	493.089	$\pm$	0.001	 & 	3.783	$\pm$	0.001	 & 	0.952	$\pm$	0.001	 & 	485.562	$\pm$	0.007	 & 	122.999	$\pm$	0.002	 & 	\cite{Persic:1995tc}	\\
443-g59	 & 	0.11	 & 	4	 & 	199.496	$\pm$	0.001	 & 	1.961	$\pm$	0.001	 & 	1.263	$\pm$	0.001	 & 	190.437	$\pm$	0.002	 & 	7.727	$\pm$	0.001	 & 	\cite{Persic:1995tc}	\\
443-g80	 & 	0.85	 & 	3	 & 	93.805	$\pm$	0.001	 & 	1.203	$\pm$	0.003	 & 	1.698	$\pm$	0.003	 & 	109.108	$\pm$	0.003	 & 	1.033	$\pm$	0.001	 & 	\cite{Persic:1995tc}	\\
444-g14	 & 	0.23	 & 	5	 & 	306.505	$\pm$	0.001	 & 	2.002	$\pm$	0.001	 & 	1.315	$\pm$	0.003	 & 	295.349	$\pm$	0.001	 & 	28.826	$\pm$	0.001	 & 	\cite{Persic:1995tc}	\\
444-g21	 & 	0.21	 & 	4	 & 	201.751	$\pm$	0.001	 & 	2.550	$\pm$	0.001	 & 	1.314	$\pm$	0.001	 & 	189.06	$\pm$	0.002	 & 	7.561	$\pm$	0.001	 & 	\cite{Persic:1995tc}	\\
\bottomrule
    \end{tabular} }}
\end{table}

\clearpage

\vspace{-0.5cm}
\begin{table}
\centering
{\tiny\renewcommand{\arraystretch}{.9}
\resizebox{\textwidth}{!}{%
\begin{tabular}{ccccccccc}
  \toprule
     Galaxy      &$\chi_{\text{red}}^{2}$ & $n$ & $\langle V_{0}\rangle$ (km s$^{-1}$) & $\langle r_{c}\rangle$ (kpc) & $d$ & $r_{\text{edge}}$ (kpc) & $M_{T}$ ($10^{11}\,\text{M}_{\odot}$) & Reference\\
     \hline
     \hline
46-g8	 & 	0.45	 & 	4	 & 	219.458	$\pm$	0.001	 & 	2.745	$\pm$	0.002	 & 	1.293	$\pm$	0.001	 & 	216.521	$\pm$	0.003	 & 	10.337	$\pm$	0.001	 & 	\cite{Persic:1995tc}	\\
471-g2	 & 	0.37	 & 	1	 & 	180.659	$\pm$	0.001	 & 	6.295	$\pm$	0.002	 & 	2.765	$\pm$	0.001	 & 	197.046	$\pm$	0.008	 & 	8.560	$\pm$	0.001	 & 	\cite{Persic:1995tc}	\\
474-g39	 & 	0.06	 & 	2	 & 	208.348	$\pm$	0.001	 & 	2.777	$\pm$	0.001	 & 	1.146	$\pm$	0.001	 & 	204.916	$\pm$	0.007	 & 	9.627	$\pm$	0.001	 & 	\cite{Persic:1995tc}	\\
474-g5	 & 	0.32	 & 	3	 & 	152.009	$\pm$	0.001	 & 	4.971	$\pm$	0.001	 & 	1.682	$\pm$	0.001	 & 	159.410	$\pm$	0.001	 & 	3.485	$\pm$	0.001	 & 	\cite{Persic:1995tc}	\\
476-g15	 & 	8.72	 & 	2	 & 	170.336	$\pm$	0.004	 & 	1.541	$\pm$	0.001	 & 	1.023	$\pm$	0.002	 & 	238.804	$\pm$	0.005	 & 	7.756	$\pm$	0.001	 & 	\cite{Persic:1995tc}	\\
476-g16	 & 	0.96	 & 	3	 & 	234.095	$\pm$	0.001	 & 	5.473	$\pm$	0.001	 & 	1.828	$\pm$	0.001	 & 	305.483	$\pm$	0.002	 & 	17.388	$\pm$	0.001	 & 	\cite{Persic:1995tc}	\\
476-g25	 & 	0.08	 & 	2	 & 	197.258	$\pm$	0.001	 & 	3.880	$\pm$	0.001	 & 	2.016	$\pm$	0.001	 & 	198.060	$\pm$	0.009	 & 	8.693	$\pm$	0.001	 & 	\cite{Persic:1995tc}	\\
\bottomrule
    \end{tabular} }}
\end{table}

\clearpage

\vspace{-0.5cm}
\begin{table}
\centering
{\tiny\renewcommand{\arraystretch}{.9}
\resizebox{\textwidth}{!}{%
\begin{tabular}{ccccccccc}
  \toprule
     Galaxy      &$\chi_{\text{red}}^{2}$ & $n$ & $\langle V_{0}\rangle$ (km s$^{-1}$) & $\langle r_{c}\rangle$ (kpc) & $d$ & $r_{\text{edge}}$ (kpc) & $M_{T}$ ($10^{11}\,\text{M}_{\odot}$) & Reference\\
     \hline
     \hline
499-g26	 & 	0.69	 & 	5	 & 	401.061	$\pm$	0.001	 & 	6.484	$\pm$	0.001	 & 	1.210	$\pm$	0.001	 & 	369.067	$\pm$	0.005	 & 	51.420	$\pm$	0.001	 & 	\cite{Persic:1995tc}	\\
499-g39	 & 	0.59	 & 	1	 & 	176.223	$\pm$	0.001	 & 	0.689	$\pm$	0.001	 & 	1.543	$\pm$	0.001	 & 	190.814	$\pm$	0.007	 & 	6.892	$\pm$	0.001	 & 	\cite{Persic:1995tc}	\\
499-g4	 & 	0.50	 & 	9	 & 	143.231	$\pm$	0.001	 & 	0.144	$\pm$	0.001	 & 	3.566	$\pm$	0.002	 & 	159.971	$\pm$	0.002	 & 	3.735	$\pm$	0.001	 & 	\cite{Persic:1995tc}	\\
499-g5	 & 	7.49	 & 	1	 & 	148.020	$\pm$	0.002	 & 	1.097	$\pm$	0.001	 & 	1.927	$\pm$	0.005	 & 	214.268	$\pm$	0.005	 & 	5.503	$\pm$	0.001	 & 	\cite{Persic:1995tc}	\\
4-g19	 & 	0.86	 & 	3	 & 	133.628	$\pm$	0.001	 & 	1.078	$\pm$	0.001	 & 	1.843	$\pm$	0.001	 & 	150.652	$\pm$	0.009	 & 	2.986	$\pm$	0.001	 & 	\cite{Persic:1995tc}	\\
501-g11	 & 	0.14	 & 	1	 & 	82.9424	$\pm$	0.001	 & 	4.085	$\pm$	0.001	 & 	4.275	$\pm$	0.002	 & 	115.438	$\pm$	0.002	 & 	1.269	$\pm$	0.001	 & 	\cite{Persic:1995tc}	\\
501-g1	 & 	0.26	 & 	1	 & 	91.1524	$\pm$	0.001	 & 	2.099	$\pm$	0.001	 & 	2.122	$\pm$	0.001	 & 	112.883	$\pm$	0.002	 & 	1.136	$\pm$	0.001	 & 	\cite{Persic:1995tc}	\\
\bottomrule
    \end{tabular} }}
\end{table}

\clearpage

\vspace{-0.5cm}
\begin{table}
\centering
{\tiny\renewcommand{\arraystretch}{.9}
\resizebox{\textwidth}{!}{%
\begin{tabular}{ccccccccc}
  \toprule
     Galaxy      &$\chi_{\text{red}}^{2}$ & $n$ & $\langle V_{0}\rangle$ (km s$^{-1}$) & $\langle r_{c}\rangle$ (kpc) & $d$ & $r_{\text{edge}}$ (kpc) & $M_{T}$ ($10^{11}\,\text{M}_{\odot}$) & Reference\\
     \hline
     \hline
534-g24	 & 	0.58	 & 	4	 & 	306.565	$\pm$	0.001	 & 	2.550	$\pm$	0.001	 & 	1.460	$\pm$	0.001	 & 	310.971	$\pm$	0.002	 & 	30.796	$\pm$	0.001	 & 	\cite{Persic:1995tc}	\\
534-g31	 & 	0.45	 & 	4	 & 	269.543	$\pm$	0.002	 & 	4.286	$\pm$	0.006	 & 	1.494	$\pm$	0.001	 & 	252.241	$\pm$	0.001	 & 	17.482	$\pm$	0.003	 & 	\cite{Persic:1995tc}	\\
534-g3	 & 	0.06	 & 	4	 & 	226.489	$\pm$	0.007	 & 	1.049	$\pm$	0.001	 & 	1.144	$\pm$	0.001	 & 	223.752	$\pm$	0.008	 & 	12.534	$\pm$	0.001	 & 	\cite{Persic:1995tc}	\\
534-g9	 & 	0.80	 & 	3	 & 	206.611	$\pm$	0.001	 & 	2.732	$\pm$	0.001	 & 	1.939	$\pm$	0.001	 & 	288.750	$\pm$	0.001	 & 	13.537	$\pm$	0.001	 & 	\cite{Persic:1995tc}	\\
535-g15	 & 	0.27	 & 	4	 & 	258.423	$\pm$	0.001	 & 	0.727	$\pm$	0.001	 & 	1.354	$\pm$	0.001	 & 	258.812	$\pm$	0.009	 & 	19.397	$\pm$	0.001	 & 	\cite{Persic:1995tc}	\\
539-g14	 & 	0.47	 & 	4	 & 	286.078	$\pm$	0.001	 & 	4.724	$\pm$	0.001	 & 	1.415	$\pm$	0.001	 & 	287.439	$\pm$	0.001	 & 	22.459	$\pm$	0.001	 & 	\cite{Persic:1995tc}	\\
539-g5	 & 	0.14	 & 	2	 & 	172.677	$\pm$	0.001	 & 	1.987	$\pm$	0.002	 & 	1.633	$\pm$	0.006	 & 	172.715	$\pm$	0.009	 & 	5.765	$\pm$	0.002	 & 	\cite{Persic:1995tc}	\\
\bottomrule
    \end{tabular} }}
\end{table}

\clearpage

\begin{table}
\centering
{\tiny\renewcommand{\arraystretch}{.9}
\resizebox{\textwidth}{!}{%
\begin{tabular}{ccccccccc}
  \toprule
     Galaxy      &$\chi_{\text{red}}^{2}$ & $n$ & $\langle V_{0}\rangle$ (km s$^{-1}$) & $\langle r_{c}\rangle$ (kpc) & $d$ & $r_{\text{edge}}$ (kpc) & $M_{T}$ ($10^{11}\,\text{M}_{\odot}$) & Reference\\
     \hline
     \hline
554-g28	 & 	0.65	 & 	5	 & 	279.243	$\pm$	0.001	 & 	1.832	$\pm$	0.001	 & 	1.215	$\pm$	0.001	 & 	282.442	$\pm$	0.009	 & 	22.932	$\pm$	0.001	 & 	\cite{Persic:1995tc}	\\
554-g29	 & 	0.57	 & 	5	 & 	486.027	$\pm$	0.007	 & 	5.399	$\pm$	0.001	 & 	1.231	$\pm$	0.001	 & 	461.337	$\pm$	0.007	 & 	104.076	$\pm$	0.006	 & 	\cite{Persic:1995tc}	\\
554-g34	 & 	0.12	 & 	2	 & 	201.949	$\pm$	0.001	 & 	3.512	$\pm$	0.001	 & 	1.737	$\pm$	0.001	 & 	200.941	$\pm$	0.002	 & 	9.078	$\pm$	0.001	 & 	\cite{Persic:1995tc}	\\
555-g16	 & 	8.40	 & 	2	 & 	252.860	$\pm$	0.003	 & 	2.678	$\pm$	0.001	 & 	1.660	$\pm$	0.001	 & 	258.457	$\pm$	0.007	 & 	18.585	$\pm$	0.002	 & 	\cite{Persic:1995tc}	\\
555-g29	 & 	0.41	 & 	4	 & 	306.351	$\pm$	0.001	 & 	2.618	$\pm$	0.001	 & 	1.368	$\pm$	0.001	 & 	305.903	$\pm$	0.002	 & 	30.027	$\pm$	0.001	 & 	\cite{Persic:1995tc}	\\
555-g2	 & 	0.75	 & 	5	 & 	143.238	$\pm$	0.001	 & 	0.079	$\pm$	0.001	 & 	1.893	$\pm$	0.001	 & 	159.430	$\pm$	0.002	 & 	3.775	$\pm$	0.001	 & 	\cite{Persic:1995tc}	\\
555-g8	 & 	0.44	 & 	4	 & 	332.883	$\pm$	0.001	 & 	2.525	$\pm$	0.001	 & 	1.059	$\pm$	0.001	 & 	328.335	$\pm$	0.009	 & 	38.087	$\pm$	0.001	 & 	\cite{Persic:1995tc}	\\
\bottomrule
    \end{tabular} }}
\end{table}

\clearpage

\begin{table}
\centering
{\tiny\renewcommand{\arraystretch}{.9}
\resizebox{\textwidth}{!}{%
\begin{tabular}{ccccccccc}
  \toprule
     Galaxy      &$\chi_{\text{red}}^{2}$ & $n$ & $\langle V_{0}\rangle$ (km s$^{-1}$) & $\langle r_{c}\rangle$ (kpc) & $d$ & $r_{\text{edge}}$ (kpc) & $M_{T}$ ($10^{11}\,\text{M}_{\odot}$) & Reference\\
     \hline
     \hline
576-g51	 & 	0.79	 & 	3	 & 	188.435	$\pm$	0.003	 & 	2.566	$\pm$	0.001	 & 	1.786	$\pm$	0.001	 & 	194.271	$\pm$	0.003	 & 	7.353	$\pm$	0.001	 & 	\cite{Persic:1995tc}	\\
577-g1	 & 	1.14	 & 	3	 & 	174.978	$\pm$	0.004	 & 	1.807	$\pm$	0.001	 & 	1.766	$\pm$	0.001	 & 	239.521	$\pm$	0.006	 & 	8.101	$\pm$	0.001	 & 	\cite{Persic:1995tc}	\\
579-g9	 & 	0.41	 & 	1	 & 	85.219	$\pm$	0.001	 & 	2.049	$\pm$	0.001	 & 	2.817	$\pm$	0.002	 & 	91.510	$\pm$	0.001	 & 	0.857	$\pm$	0.001	 & 	\cite{Persic:1995tc}	\\
57-g80	 & 	0.76	 & 	3	 & 	173.212	$\pm$	0.001	 & 	3.650	$\pm$	0.001	 & 	1.849	$\pm$	0.001	 & 	232.279	$\pm$	0.008	 & 	7.346	$\pm$	0.001	 & 	\cite{Persic:1995tc}	\\
580-g29	 & 	0.51	 & 	3	 & 	67.286	$\pm$	0.002	 & 	6.554	$\pm$	0.001	 & 	2.677	$\pm$	0.001	 & 	94.161	$\pm$	0.001	 & 	0.477	$\pm$	0.001	 & 	\cite{Persic:1995tc}	\\
580-g37	 & 	0.22	 & 	2	 & 	217.735	$\pm$	0.002	 & 	2.224	$\pm$	0.001	 & 	1.115	$\pm$	0.001	 & 	215.813	$\pm$	0.004	 & 	11.246	$\pm$	0.001	 & 	\cite{Persic:1995tc}	\\
580-g41	 & 	0.84	 & 	3	 & 	100.734	$\pm$	0.009	 & 	1.479	$\pm$	0.003	 & 	1.849	$\pm$	0.001	 & 	97.952	$\pm$	0.001	 & 	1.052	$\pm$	0.001	 & 	\cite{Persic:1995tc}	\\
\bottomrule
    \end{tabular} }}
\end{table}

\clearpage

\begin{table}
\centering
{\tiny\renewcommand{\arraystretch}{.9}
\resizebox{\textwidth}{!}{%
\begin{tabular}{ccccccccc}
  \toprule
     Galaxy      &$\chi_{\text{red}}^{2}$ & $n$ & $\langle V_{0}\rangle$ (km s$^{-1}$) & $\langle r_{c}\rangle$ (kpc) & $d$ & $r_{\text{edge}}$ (kpc) & $M_{T}$ ($10^{11}\,\text{M}_{\odot}$) & Reference\\
     \hline
     \hline
6-g3	 & 	0.57	 & 	3	 & 	177.500	$\pm$	0.001	 & 	2.562	$\pm$	0.001	 & 	1.762	$\pm$	0.002	 & 	187.982	$\pm$	0.001	 & 	6.289	$\pm$	0.001	 & 	\cite{Persic:1995tc}	\\
71-g14	 & 	0.83	 & 	1	 & 	184.724	$\pm$	0.001	 & 	0.505	$\pm$	0.001	 & 	1.587	$\pm$	0.001	 & 	197.043	$\pm$	0.003	 & 	7.823	$\pm$	0.001	 & 	\cite{Persic:1995tc}	\\
71-g4	 & 	0.61	 & 	5	 & 	409.476	$\pm$	0.001	 & 	6.148	$\pm$	0.001	 & 	1.267	$\pm$	0.001	 & 	362.853	$\pm$	0.003	 & 	53.453	$\pm$	0.005	 & 	\cite{Persic:1995tc}	\\
72-g5	 & 	0.81	 & 	4	 & 	152.986	$\pm$	0.001	 & 	0.803	$\pm$	0.001	 & 	1.732	$\pm$	0.001	 & 	164.807	$\pm$	0.002	 & 	4.251	$\pm$	0.001	 & 	\cite{Persic:1995tc}	\\
73-g11	 & 	0.24	 & 	3	 & 	237.975	$\pm$	0.001	 & 	3.751	$\pm$	0.004	 & 	1.215	$\pm$	0.001	 & 	227.474	$\pm$	0.004	 & 	12.935	$\pm$	0.001	 & 	\cite{Persic:1995tc}	\\
73-g22	 & 	0.22	 & 	4	 & 	12.248	$\pm$	0.001	 & 	11.121	$\pm$	0.001	 & 	3.703	$\pm$	0.001	 & 	51.226	$\pm$	0.002	 & 	0.075	$\pm$	0.001	 & 	\cite{Persic:1995tc}	\\
73-g25	 & 	0.36	 & 	5	 & 	499.957	$\pm$	0.001	 & 	2.863	$\pm$	0.001	 & 	1.262	$\pm$	0.001	 & 	494.386	$\pm$	0.006	 & 	130.324	$\pm$	0.002	 & 	\cite{Persic:1995tc}	\\
\bottomrule
    \end{tabular} }}
\end{table}

\clearpage

\begin{table}
\centering
{\tiny\renewcommand{\arraystretch}{.9}
\resizebox{\textwidth}{!}{%
\begin{tabular}{ccccccccc}
  \toprule
     Galaxy      &$\chi_{\text{red}}^{2}$ & $n$ & $\langle V_{0}\rangle$ (km s$^{-1}$) & $\langle r_{c}\rangle$ (kpc) & $d$ & $r_{\text{edge}}$ (kpc) & $M_{T}$ ($10^{11}\,\text{M}_{\odot}$) & Reference\\
     \hline
     \hline
i2974	 & 	2.74	 & 	1	 & 	171.519	$\pm$	0.006	 & 	4.475	$\pm$	0.003	 & 	3.257	$\pm$	0.007	 & 	188.111	$\pm$	0.001	 & 	7.448	$\pm$	0.005	 & 	\cite{Persic:1995tc}	\\
i382	 & 	0.88	 & 	2	 & 	189.659	$\pm$	0.003	 & 	0.457	$\pm$	0.001	 & 	1.941	$\pm$	0.001	 & 	199.154	$\pm$	0.001	 & 	8.286	$\pm$	0.001	 & 	\cite{Persic:1995tc}	\\
i387	 & 	0.13	 & 	4	 & 	265.343	$\pm$	0.004	 & 	0.341	$\pm$	0.001	 & 	1.622	$\pm$	0.002	 & 	286.002	$\pm$	0.002	 & 	23.090	$\pm$	0.001	 & 	\cite{Persic:1995tc}	\\
i407	 & 	13.77	 & 	1	 & 	162.742	$\pm$	0.002	 & 	1.853	$\pm$	0.002	 & 	2.160	$\pm$	0.005	 & 	174.497	$\pm$	0.004	 & 	5.515	$\pm$	0.001	 & 	\cite{Persic:1995tc}	\\
i5078	 & 	0.28	 & 	1	 & 	105.856	$\pm$	0.002	 & 	2.087	$\pm$	0.001	 & 	2.134	$\pm$	0.003	 & 	110.215	$\pm$	0.002	 & 	1.498	$\pm$	0.001	 & 	\cite{Persic:1995tc}	\\
i5282	 & 	0.53	 & 	1	 & 	190.451	$\pm$	0.002	 & 	1.981	$\pm$	0.003	 & 	1.060	$\pm$	0.001	 & 	195.453	$\pm$	0.009	 & 	8.093	$\pm$	0.001	 & 	\cite{Persic:1995tc}	\\
i784	 & 	0.75	 & 	3	 & 	204.083	$\pm$	0.001	 & 	4.187	$\pm$	0.001	 & 	1.813	$\pm$	0.002	 & 	279.136	$\pm$	0.003	 & 	12.278	$\pm$	0.001	 & 	\cite{Persic:1995tc}	\\
\bottomrule
    \end{tabular} }}
\end{table}

\clearpage

\begin{table}
\centering
{\tiny\renewcommand{\arraystretch}{.9}
\resizebox{\textwidth}{!}{%
\begin{tabular}{ccccccccc}
  \toprule
     Galaxy      &$\chi_{\text{red}}^{2}$ & $n$ & $\langle V_{0}\rangle$ (km s$^{-1}$) & $\langle r_{c}\rangle$ (kpc) & $d$ & $r_{\text{edge}}$ (kpc) & $M_{T}$ ($10^{11}\,\text{M}_{\odot}$) & Reference\\
     \hline
     \hline
n3715	 & 	0.32	 & 	12	 & 	192.428	$\pm$	0.002	 & 	0.020	$\pm$	0.001	 & 	2.417	$\pm$	0.001	 & 	195.728	$\pm$	0.005	 & 	8.390	$\pm$	0.001	 & 	\cite{Persic:1995tc}	\\
n4348	 & 	2.77	 & 	13	 & 	184.594	$\pm$	0.002	 & 	0.037	$\pm$	0.002	 & 	1.263	$\pm$	0.002	 & 	192.431	$\pm$	0.006	 & 	7.551	$\pm$	0.003	 & 	\cite{Persic:1995tc}	\\
n4705	 & 	3.79	 & 	1	 & 	194.989	$\pm$	0.004	 & 	1.497	$\pm$	0.001	 & 	1.241	$\pm$	0.001	 & 	197.949	$\pm$	0.004	 & 	8.678	$\pm$	0.001	 & 	\cite{Persic:1995tc}	\\
n697	 & 	31.37	 & 	1	 & 	207.744	$\pm$	0.001	 & 	2.240	$\pm$	0.002	 & 	1.657	$\pm$	0.001	 & 	212.349	$\pm$	0.002	 & 	10.714	$\pm$	0.001	 & 	\cite{Persic:1995tc}	\\
n699	 & 	0.27	 & 	1	 & 	147.800	$\pm$	0.001	 & 	2.206	$\pm$	0.001	 & 	2.121	$\pm$	0.001	 & 	167.378	$\pm$	0.002	 & 	4.380	$\pm$	0.001	 & 	\cite{Persic:1995tc}	\\
n701	 & 	0.62	 & 	4	 & 	227.265	$\pm$	0.001	 & 	1.550	$\pm$	0.001	 & 	1.368	$\pm$	0.001	 & 	235.746	$\pm$	0.004	 & 	13.047	$\pm$	0.001	 & 	\cite{Persic:1995tc}	\\
UGC477	 & 	4.69	 & 	3	 & 	183.986	$\pm$	0.003	 & 	23.777	$\pm$	0.003	 & 	1.359	$\pm$	0.005	 & 	170.217	$\pm$	0.003	 & 	2.759	$\pm$	0.001	 & 	\cite{KuziodeNaray:2006wh, KuziodeNaray:2007qi}	\\
\bottomrule
    \end{tabular} }}
\caption{Best fitting dynamical parameters $\langle V_{0}\rangle,\,\langle r_{c}\rangle, d,\,n,\,r_{\text{edge}}$ and the total mass $M_{T}$ given by Eq.\eqref{Eq.20}.}\label{Tab.2}
\end{table}









\bsp	
\label{lastpage}
\end{document}